\documentclass[twoside]{articlek}

\usepackage[normal]{caption}

\renewcommand{\refname}{REFERENCES}

\def\BibTeX{{\rm B\kern-.05em{\sc i\kern-.025em b}\kern-.08em
		T\kern-.1667em\lower.7ex\hbox{E}\kern-.125emX}}

\usepackage{graphicx}

\renewcommand{\refname}{REFERENCES}

\def\BibTeX{{\rm B\kern-.05em{\sc i\kern-.025em b}\kern-.08em
		T\kern-.1667em\lower.7ex\hbox{E}\kern-.125emX}}

\usepackage{graphicx}

\begin{document}

\begin{flushleft}
	{\large\bf Low lying isomers in the region of superheavy nuclei
		}
	\vspace*{25pt}

	{\bf Fritz Peter He\ss berger$^{1,2,}\footnote{E-mail: \texttt{f.p.hessberger@gsi.de}}$} \\
	\vspace{5pt}
	{$^1$GSI - Helmholtzzentrum f\"ur Schwerionenforschung GmbH, Planckstra\ss e 1, 64291 Darmstadt, Germany\\
		$^2$Helmholtz Institut Mainz, Staudingerweg 18, 55128 Mainz, Germany}\\

\end{flushleft}
          
   Version: October, 07, 2025 \\

\abstract{In the present study we want to give an overview on low lying
isomeric states in the heaviest nuclei. After a short report on
the early history on the discovery of nuclear isomerism and
attempts to understand their physical nature, decay probabilities
and structure of all low lying isomeric states in heaviest nuclei
with half-lives typically larger than one microsecond are presentet.
Special emphasisis laid on cases where the above mentioned
properties are still unclear or under discussion. We do not
claim to have solved the problems in that cases, we rather
want to give som hints for further discussions.
	}

\section{1. Introduction}

The expression 'isomer' is of Greek origin, it is composed from the words 'isos' ('equal') and 'meros' ('part'). It was introduced
in the 1820s for chemical substances having the same molecular formula, but different steric configurations, resulting in different
chemical and/or physical properties.
In the early 1930ies the name 'isomer' was also adopted for atomic nuclei having the same nuclear charge (Z) and mass number (A), but
different nuclear structure, resulting in different radioactive decay properties. As the first isomer was found within the 
radioactive decay chain of $^{238}$U (named 'UI' at that time) when little was known about nuclear structure of the chain members,
and all members of the decay chain were unstable, differences in the nuclear structure were seen as a criterion for isomerism and
'stability'. The latter rather came into focus after 'isomers' were discovered also in nuclei which were stable in their ground state
and 'isomers' were seen as 'meta-stable' states.
As at the time of discovery 
of isomeric states in atomic nuclei it was assumed that nuclear levels had lifetimes $<$1$\times$10$^{-13}$s \cite{DrW16}, 
it thus seemed justified to denote nuclear levels with lifetimes $\tau$$>$10$^{-13}$ s as 'long-lived' or 'isomeric'. Recently G.D. Dracoulis et al. \cite{DrW16} have expressed it in the following way:
'on the technical side any state which has a directly measurable lifetime in the sense of an electronic measurement, even down to the subnanosecond  region could be
termed an isomer, or more properly a metastable state'. We will follow this wording in the present paper.\\
Here we will distinguish (different from  \cite{WaD99} )four kinds of isomers:\\
\\
a) spin isomers\\
b) seniority isomers \\
c) K isomers\\
d) shape isomers\\

In the present paper, after a short 'history' on discovery and early theoretical considerations, for a more detailed presentation see 
'On the history of study of isomerism' by A.P. Grinberg \cite{Grin80}, we will give an overview 
on essentially 'low lying' spin isomers in transuranium nuclei. We will restrict to 'long-lived' isomers with
half-lives typically larger than one microsecond.\\
\\
\\
\\
\section{2. History}
\vspace{-5mm}
Possible existence of nuclei having the same nuclear charge and mass number but different decay properties
was first discussed by F. Soddy in 1917 \cite{Soddy1917}, who noted 'the mass ({\it{atomic nucleus}}) depends on the gross number of '+' and '-'
charges in the nucleus, chemical properties depend
on the difference between the gross number of '+' and '-' charges. But the radioactive properties depend not only on the gross number
of charges, but on the constitution of the nucleus. We can have isotopes with identity of atomic weight, as well as chemical
character, which are different in the stability and mode of breaking up.'
As a possible candidate for such a structure he considered the decay product of ThC and argued in the following way 'As has been long known
thorium-C, an isotope of bismuth, disintegrates dually. For 35 per cent of the atoms disintegrating, an $\alpha$ ray is expelled, followed by a
$\beta$-ray. For the remaining 65 per cent the $\beta$-ray is first expelled, and is followed by the $\alpha$-ray. The two products are isotopes
of lead, and both have the same atomic weight, but they are not the same. More energy is expelled in the changes of the 65 per cent than in those of the 
35 per cent. Unless they are both completely stable a difference of period of change is to be anticipated.' \cite{Soddy1917}. Although the conclusion
was sound it was not correct as the decay path of $^{212}$Bi was only insufficiently known at that time. For illustration the decay of 
$^{212}$Bi into $^{208}$Pb is sketched in fig. 1a. Evidently the $\beta^{-}$ - decay of $^{208}$Tl, the $\alpha$ - decay product of $^{212}$Bi
populates the I$^{\pi}$ = 3$^{-}$ - level at E$^{*}$ = 2614.551 keV in $^{208}$Pb. The high energy transition 3$^{-}$ $\rightarrow$ 0$^{+}$ (ground state)
eventually was not measured at that time, resulting in a lower decay energy for the path $^{212}$Bi $^{\alpha}_{\rightarrow}$ $^{208}$Tl  
$^{\beta}_{\rightarrow}$ $^{208}$Pb than for the path $^{212}$Bi $^{\beta}_{\rightarrow}$ $^{212}$Po  
$^{\alpha}_{\rightarrow}$ $^{208}$Pb.\\ 
The idea of a possible existence of nuclei, having the same atomic mass and charge, as well as the same chemical properties, but
different decay properties was taken up short
time later by St. Meyer \cite{Meyer1918}, who called them 'Isotope h\"oherer Ordnung' ('isotopes of higher order'). Specifically branchings
in the 'natural' decay chains were suspected to be candidates. Careful analyses of the decay of ThC, which F. Soddy regarded as a candidate, 
however, showed that both decay branches led to the same end product ThD ('S\"amtliche Untersuchungen und Betrachtungen lassen es als
wahrscheinlich erscheinen, da\ss \hspace{1mm} sowohl der Teil der Atome, der sich aus ThC \"uber ThC' in ThD verwandelt, wie derjenige, der \"uber ThC'' entsteht zu stabilem 
Thorblei f\"uhrt. Ein Nachweis zweier ThD - Arten, Isotopen mit gleichem Atomgewicht, aber verschiedener Zerfalls-wahrscheinlichkeit konnte bisher nicht
erbracht werden.'\cite{Meyer1918}). He thus concluded, that there is so far no cogent reason for the assumption there could exist 'isotopes of higher' order,
having the same atomic weight but different decay probabilities ('Es liegt sonach bisher in keinem Falle ein zwingender Grund für die Annahme vor, 
es k\"onnte Isotope höherer Ordnung mit gleichem Atomgewicht geben, die trotzdem verschiedene Zerfallswahrscheinlichkeiten bes\"a\ss en' \cite{Meyer1918}). \\
The breakthrough in the discovery of nuclear isomerism was the observation of weak 6.7 h - activity within the decay chain of UI ($^{238}$U) by O. Hahn
 \cite{Hahn1921}. This new activity was found to have the chemical properties of protactinium and was denoted as 'UZ'. Careful analyses of the observations
showed that UZ was produced by decay of UX1 and decayed into UII. Thus a 'dual' decay mode of UX1 was assumed: a) UX1 $\rightarrow$ UX2 $\rightarrow$ UII and
b) UX1 $\rightarrow$ UZ $\rightarrow$ UII. Remarkable, however, was the fact that eventually UX2 and UZ had different lifetimes. So it was a feature that never had
observed before within the 'natural' decay chains. (We should here remark that due to the short half-life of $^{234m}$Pa (UX2, T$_{1/2}$ = 1.17 m) UX1 and UX2 could
not be separated by chemical means, both were distinguished on the basis of the different $\beta$ - ray energies. UX1 was attributed to 'soft' $\beta$ rays, UX2 was
attributed to 'hard' $\beta$ rays.) For illustration the decay scheme of UX1 ($^{234}$Th) is shown in fig. 1b. On this basis it is clear that O. Hahn indeed discovered
the ground state decay of $^{234}$Pa, which is populated weakly by internal transitions of the isomer $^{234m}$Pa. Insofar O. Hahn was not correct in assuming
a dual decay branch of UX1.\\
O. Hahn concluded that the origin of the new activity was either the result of a dual decay branch of UX1 having properties so far never observed in the radioactive
elements or member of a new uranium decay - chain, whose members range as isotopes in the known uranium - radium - chain. 
('Als Mutterubstanz kommt nur das UX1 oder ein neues UX1 - Isotop ähnlicher Lebensdauer in Frage. Im ersteren Fall erlitte das UX1 einen dualen Zerfall in der Art, wie
er bisher bei den Radioelementen noch nicht beobachtet wurde. Im anderen Fall w\"are die wahrscheinlichste Annahme die Existenz einer neuen Uran-Zerfallsreihe geringer
Strahlungsintensit\"at, deren einzelne Glieder sich als Isotope in die bekannte Uran - Radiumreihe einreihen'.)
As a possible origin of a new uranium decay chain, an isotope with the mass number A = 240 (UIII) was suspected.\\
\begin{figure*}
	\resizebox{0.75\textwidth}{!}{
		\includegraphics{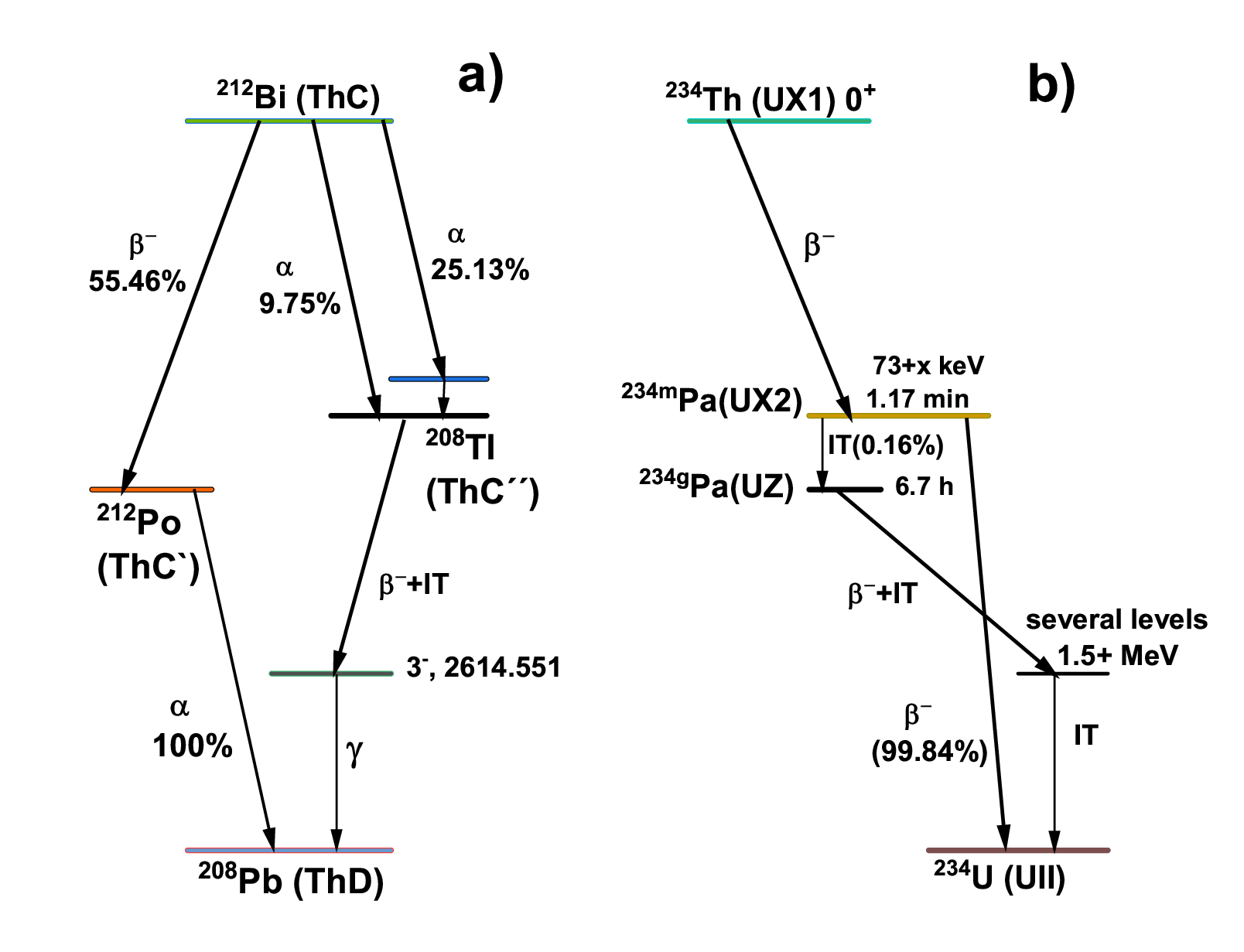}	
	}
	\caption{Simplified decay schemes for $^{212}$Bi and $^{234}$Pa; the decay branches shown for $^{212}$Bi do not 
	add up to 100$\%$ as some weaker $\beta^{-}$ and $\alpha$ decays are not shown in this simplified decay scheme.};
	
	\label{fig:1}       
\end{figure*}
\begin{figure*}
	\resizebox{0.75\textwidth}{!}{
		\includegraphics{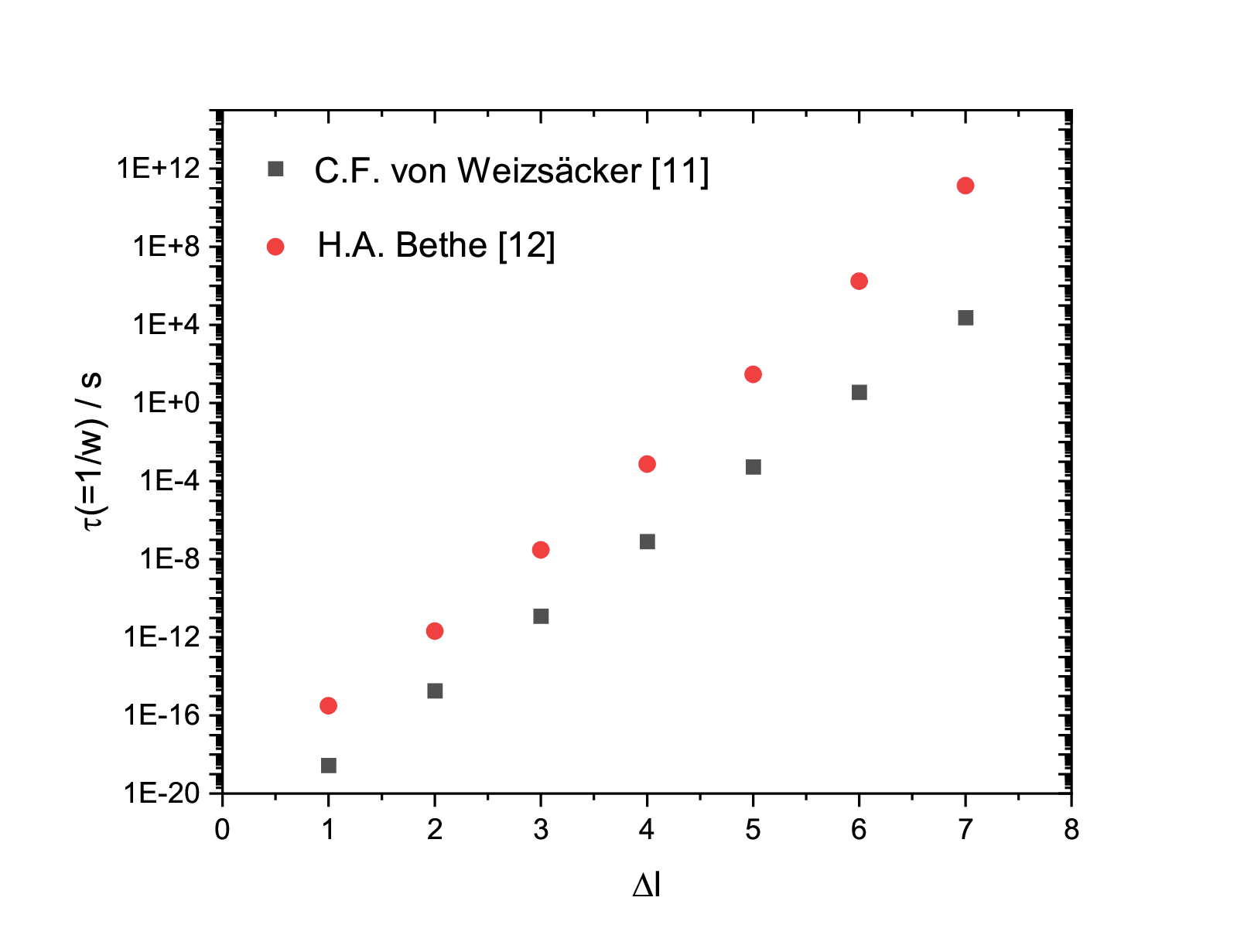}	
	}
	\caption{Lifetimes for 2$^{l}$ multipole radiation according to C.F. von Weizs\"acker
		\cite{Weizs1936} (black squares) and H.A. Bethe \cite{Bethe1937} (red dots) for a nucleus 
		of Z\,=\,90 and $\Delta$E\,=\,\,500\, keV.};
	
	\label{fig:2}       
\end{figure*}
The expression 'isomer' in context with atomic nuclei was to our knowledge used for the first time by G. Gamow \cite{Gamow1934a,Gamow1934b,Gamow1935}, who also attempted
to give a physical explanation for their existence. To understand his explanation one has to respect the postulation of at that time hypothetical negative protons (today 
named 'antiprotons') in atomic nuclei. An interesting feature of this idea was the explanation of nuclear isomerism and thus shall be discussed in the following in more
detail.\\
It should be noted first that at the time when G. Gamow presented his theory (1934/35) already 
the existence of the positron (positive 
electron) and the neutron were experimentally confirmed.
The nucleus was now seen as a composition of protons and neutrons.
An annihilation of positive protons and negative protons was not considered when existing together in
an atomic nucleus. Concerning the structure of the atomic nucleus following a suggestion of W. Heisenberg \cite{Heis32} a strong interaction between protons and neutrons and a weaker
interaction between nucleons of the same type, i.e. between protons and protons as well as between neutrons and neutrons was assumed. The interaction between
protons and neutrons was assumed to be strong at 'short' distances (in modern language: nuclear dimensions) and decreasing steeply at increasing distances ('short 
range force'). At 'shorter' distances it should become repulsive. The interaction between negative protons and negative protons or between negative protons and
neutrons was assumed to be identical to the interaction between positive protons and positive protons and positive protons and neutrons, respectively.
The total binding energy of an atomic nucleus was then given by the 'positive' contribution of the attractive force between protons and neutrons and the 'negative'
binding energy due to the repulsive Coulomb force between the protons. Qualitatively it was assumed that the 'positive' contribution was maximum when
the number of protons and neutrons was approximately equal, i.e. N$_{p+}$ $\approx$ A/2, N$_{n}$ $\approx$ A/2, with A being the atomic mass number, while the
'negative' contribution should be lowest for N$_{p+}$ = 0. As a consequence in atomic nuclei it was assumed that N$_{n}$ $>$ N$_{p+}$, 
while the ratio N$_{n}$ / N$_{p+}$ should 
increase with the atomic number (nuclear charge number). Introducing now a possible existence of negative protons in a nucleus and an attractive force between
positive and negative protons one would obtain N$_{n}$ = A/2, N$_{p+}$ $\approx$ A/4, and N$_{p-}$ $\approx$ A/4. This, however, was in contradiction with
'reality' as no nuclei with an atomic number Z = 0 exist. It thus was assumed that the interaction between positive and negative protons was strongly repulsive,
which meant that no nuclei with 'large' numbers of positive and negative protons could exist, however existence of nuclei with a small number of negative
protons was not excluded. The atomic mass number of such nuclei then would be A = N$_{n}$ + N$_{p+}$ + N$_{p-}$ and the atomic number would be 
Z = N$_{p+}$ - N$_{p-}$. On the basis of these considerations  the existence of isomers could be explained by replacing a pair of neutrons by a positive and 
a negative proton. The mass number and the atomic number would be the same, however the binding energy and the spin would be different. The isomer of higher 
energy will be unstable and transform into the other isomer by transformation 
of two particles  p$^{+}$ $\rightarrow$ n,  p$^{-}$ $\rightarrow$ n or n $\rightarrow$ p$^{+}$, n $\rightarrow$ p$^{-}$ \cite{Gamow1935}.
The probability for such double transformations, however, were assumed to be extremely small resulting in long lifetimes, 
so the isomer of the higher energy would be metastable.
In that context G. Gamow stated \cite{Gamow1935} 'Thus the isomeric nucleus will differ widely from an ordinary excited state of a nucleus
for which the emission of surplus energy in form of a $\gamma$-quantum usually takes place in a very short fraction of a second ($\approx$10$^{-16}$ sec)'.\\
The view of G. Gamow that excited nuclear levels could not be the origin of isomeric states because of their 
short lifetimes was disproved by C.F. von Weizs\"acker \cite{Weizs1936}, who theoretically investigated the (upper limit) of the 
emission probability of 2$^{l}$ multipole radiation using the relation \\
\\
w(l) $\le$ 6x10$^{18}$ $\times$ ($\beta$/137)$^{2l}$ $\times$ $\alpha^{2l+1}$ $\times$ Z$^{2+2l/3}$\\
\\
with 'w(l)' being the emissions per second, 'l' denoting the angular momentum difference between the initial (emitting) and final
nuclear state and 'Z' the atomic number. The result for a  nucleus Z\,=\,90
and an energy difference of $\Delta$E\,=\,500 keV ($\alpha$=0.5) between
the initial and final state is shown in fig.2 using the relation $\tau$ = 1/w(l); evidently the lower limit of the lifetime
increases from $\approx$3x10$^{-19}$ s from an angular momentum
difference $\Delta$l\,=\,1 to  $\approx$2.4x10$^{4}$ s
($\approx$7 h) for $\Delta$l\,=\,7.
Although this estimate was certainly quite crude, it became clear, that
large angular momentum differences between an excited state and the 
ground state could be the origin of long-lived metastable nuclear states,
provided, that there were no states of lower angular momentum difference inbetween. No exotic nucleon configurations were needed to explain the 
occurrence of isomeric states in nuclei.\\
C.F. von Weizs\"ackers idea on low transition rates between levels of large
l-difference, i.e. of $\gamma$ rays of high multipole order was picked up
by H.A. Bethe, whose calculations resulted in the relation between 
l-difference and lifetime $\tau$ given below \cite{Bethe1937}:\\
\\
 $\tau$ / s = 5$\times$10$^{-21}$ $\times$ (l!)$^{2}$ $\times$ (20/$\hbar\omega$)$^{2l+1}$ \\
\\
which was dependent on the energy $\hbar\omega$ of the transition, but not 
explicitly on the atomic number Z and the mass number of the nucleus. \\
The
results are compared in fig.2 with the lower limits for the lifetimes estimated by C.F. von Weizs\"acker. 
Evidently the calculations of H.A. Bethe result in longer lifetimes.\\
The discovery of the neutron, the installation of first particle accelerators and also the discovery of nuclear fission
enhanced the possibilities to produce new radioactive nuclei and to investigate their decay properties. Consequently these
investigations lead to the discovery of a considerable number of isomeric states. An overview of the situation at the beginning of
the 1940ies was given by  J. Mattauch \cite{Mattauch1941}. At this time 25 cases of isomeric states in nuclei were known,
16 isomers in odd-mass nuclei, 8 cases in odd-odd nuclei, while in one case, a xenon isotope with mass number A\,$>$\,131 
(later identified as $^{135}$Xe \cite{Fire96}) the assignment was uncertain. No isomeric state had been identified in an
even - even nucleus. The existence of isomeric states in even - even nuclei, however, was not excluded. J. Mattauch stated,
that probably half-lives of isomers in even - even nuclei might be too short or too long or emitted $\gamma$ rays
might be too 'soft' (too low energetic) to be detected with the experimental techniques available at that time \cite{Mattauch1941}.
Spins of the ground states were known only in six cases, I\,=\,9/2 for $^{83}$Kr, $^{87}$Sr, $^{93}$Nb, $^{113}$In, $^{115}$In and 
I\,=\,1/2 for $^{107,109}$Ag (here the mass assignment was unclear; indeed both isotopes have a ground state spin of I\,=\,1/2
and isomeric states). Spin differences of $\Delta$I\,=\,3-4 were assumed for $^{83}$Kr, $^{107,109}$Ag and $\Delta$I\,=\,5 for
the other cases from experimental half-lives and excitation energies and theoretical half-lives
calculated using the formula suggested by H.A. Bethe \cite{Bethe1937}. J. Mattauch commented in this context, that due to
large conversion coefficients (he used the expression 'innere Umwandlung'), which were not respected in H.A. Bethe's formula,
theoretical lifetimes should be even lower and spin differences should be even higher.
Spin differences $\Delta$I\,=\,5 or $\Delta$I\,$>$\,5 even at a ground state spin  I\,=\,9/2 were seen as problematic, 
as they would require spins of I\,$\ge$\,19/2 for the
isomeric state, which were seen as quite unlikely \cite{Mattauch1941}. S. Fl\"ugge \cite{Fluegge1941} commented to that case,
that it seems in that case necessary to assume one more selection rule, i.e. to explain only part of the hindrance (or 'forbiddenness')
by a spin difference $\Delta$I\,=\,4 and to explain the rest by a further symmetry property of the nucleus.
('Man wird daher zwangsl\"aufig vor die Notwendigkeit gestellt, in diesen F\"allen noch eine andere Auswahlregel au\ss er der Drehimpulsregel
hinzunehmen, also etwa nur ein Teil des Verbots mit der Spindifferenz zu erkl\"aren (was eine zu kurze Lebensdauer ergeben w\"urde) und den 
Rest durch irgendeine weitere Symmetrieeigenschaft des Kerns zu verstehen. Ob es sich dabei um die Symmetrie bez\"uglich der
Spiegelung am Nullpunkt handelt, wie dies Cork und Lawson (l.c.) im Falle des Indiums angenommen haben oder um eine andere Eigenschaft,
bleibe dahingestellt \cite{Fluegge1941}.)\\
Indeed in all nuclei $^{83}$Kr, $^{87}$Sr, $^{93}$Nb, $^{113}$In, $^{115}$In with ground state spin and parity (not mentioned in \cite{Mattauch1941})
I$^{\pi}$\,=\,9/2$^{+}$ the isomeric state has been identified as I$^{\pi}$\,=\,1/2$^{-}$ \cite{Fire96}.\\
Concerning the non-obeservation of isomeric states in even-even nuclei S. Fl\"ugge remarked, that in even-even nuclei
the first excited states might be too high in energy to be metastable \cite{Fluegge1941}.\\
More rigorous calculations for lifetimes of transitions in dependence of 
l-difference, respecting also internal conversion were performed 
by N. Koyenuma \cite{Koyenuma1941}.\\
\\
\\
\vspace{55mm}
\begin{figure*}
	\resizebox{0.75\textwidth}{!}{
		\includegraphics{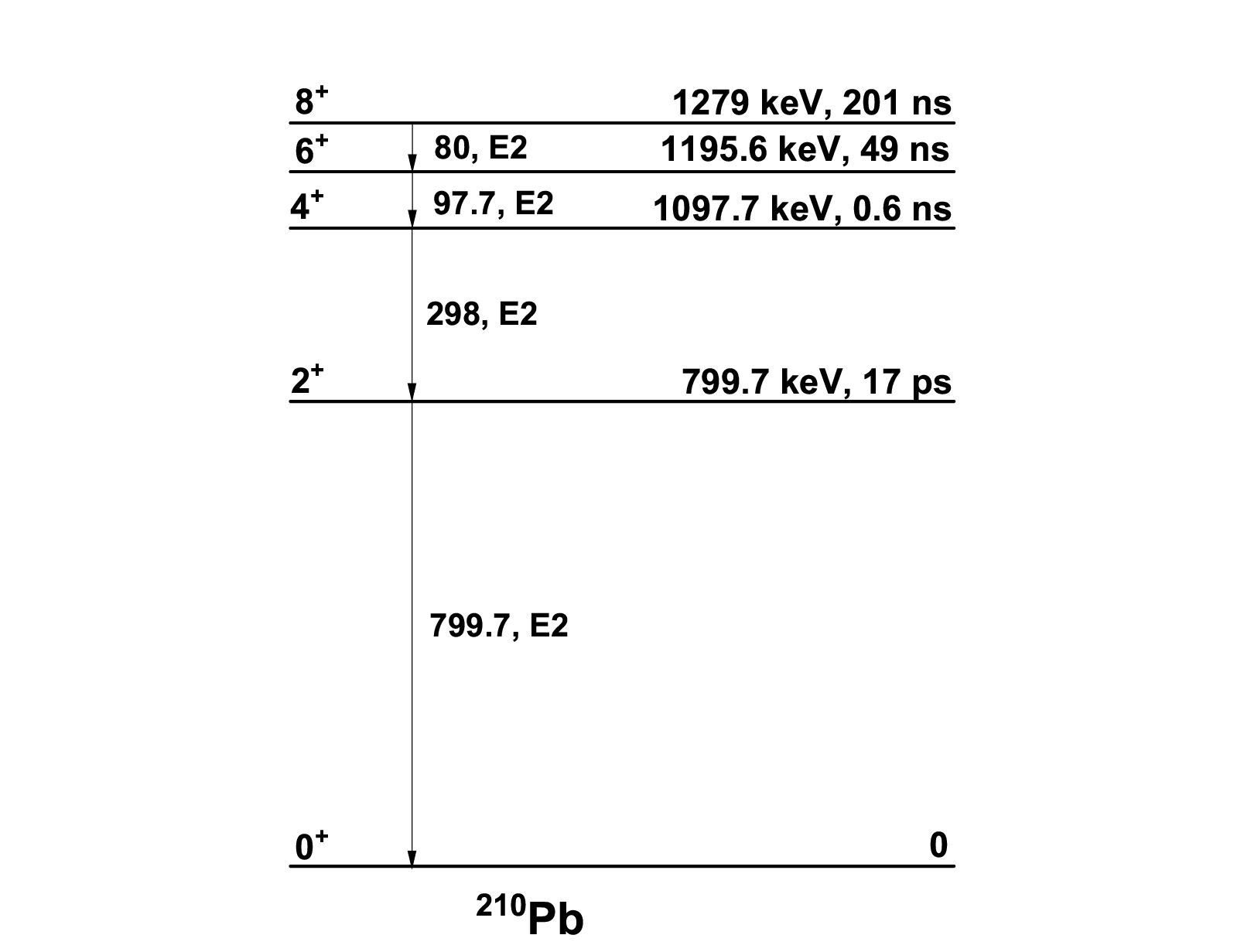}	
	}
	\caption{Partial level scheme of $^{210}$Pb; data taken from \cite{Fire96}. Transition energies are given in keV.};
	
	\label{fig:2}       
\end{figure*}
\section{3. Types of Isomers}
\vspace*{-60mm}
\textbf{Spin isomers}\\
As has been discussed in the previous section, isomers had been interpreted as excited nuclear states exhibiting
large angular momenta differences to lower lying levels, thus transitions between the initial (isomeric) state
and the final state are characterized by high multipolarities and thus low transition probabilities resulting in
long lifetimes for the initial state. But not only the spin differences characterize the lifetimes, but also
changes in nucleonic configuration, nuclear structure and nuclear deformations play a role.
B. Maheshwari and K. Nomura \cite{Mahesh2022} express that item in the following way: 'The hindrance to the isomer decay
is mostly due to three reasons: different structural matrix elements of initial and final states, small decay energies
and large change in angular momentum. In simple words, the excited states adopting different configurations from lower lying
states result in meta-stable states, that is, isomeric states'.\\
So we find aside 'simple' spin-isomers, whose lifetimes can be, at least qualitatively, described by the Weisskopf - estimate
(see e.g. \cite{Fire96}), in addition in spherical nuclei 'seniority isomers', in strongly deformed nuclei 'K isomers' and 'shape isomers'.\\
\vspace*{10mm}
\\
\textbf{Seniority isomers}\\
The concept of 'seniority' was first introduced by G. Racah in 1943 \cite{Racah43} in context of the description of 
atomic electron shells to explain the atomic spectra. A couple of years later \cite{Racah50} it was adopted to atomic nuclei
where it turned out be more useful than for the atomic electron shell, specifically for the case of jj-coupling.
The 'seniority' quantum number $\nu$ denotes the number of nucleons which are not coupled to pairs with resulting spin of I = 0.
Examples for such configurations are low lying excited states in even-even nuclei, formed by breaking a nucleon pair. Such states 
will have the seniority $\nu$ = 2. If, for example, the nucleons are in the g$_{9/2}$ orbital, the states of $\nu$ = 2 will have
spins I = 2, 4, 6, 8. As E2 - transitions between states of the same seniority are 'forbidden', transitions will be strongly hindered,
thus the states will be isomeric. As an example in fig. 3 the partial level scheme of $^{210}$Pb is given.
As seniority isomers are a phenomenon in 'semi magical' nuclei, they do not play a role in the region of strongly deformed nuclei
at Z$\ge$100, but may occur in nuclei close to the spherical 'superheavy shells', expected at Z = 114, 120 or 126 and N = 184 or 172.
For further details see e.g. \cite{Walker2020}.\\

\begin{figure*}
	\resizebox{0.99\textwidth}{!}{
		\includegraphics{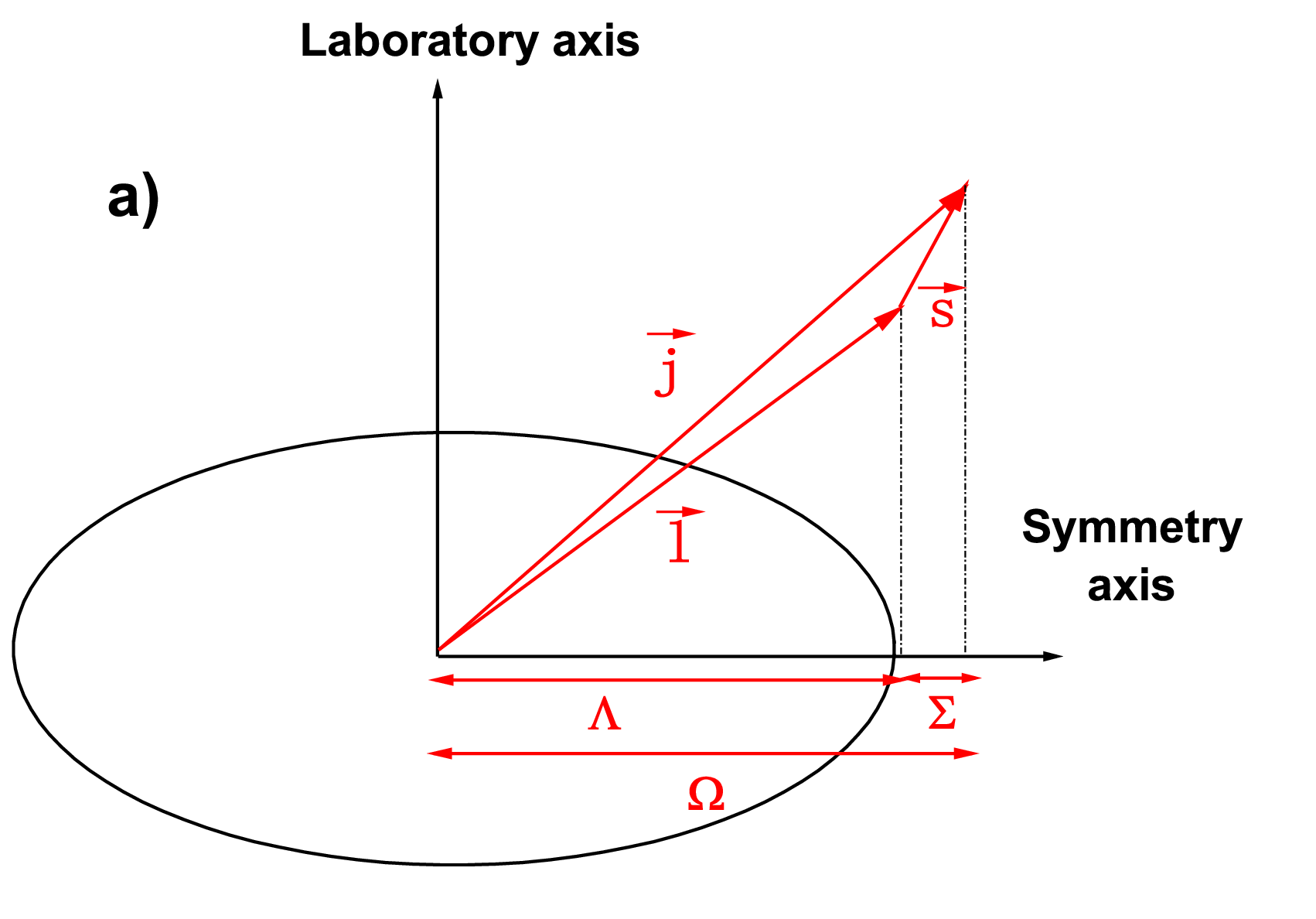}
		\includegraphics{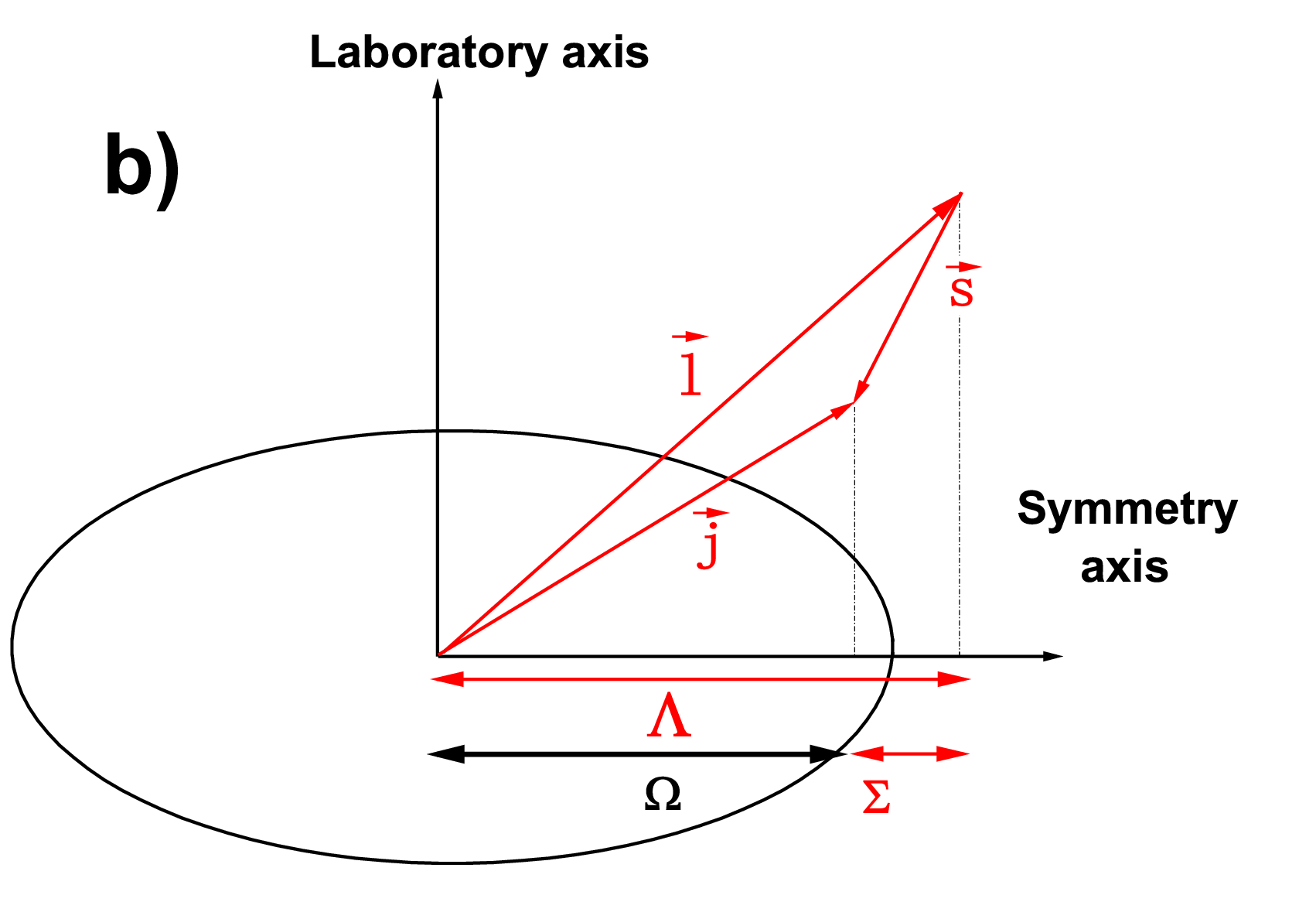}
		\includegraphics{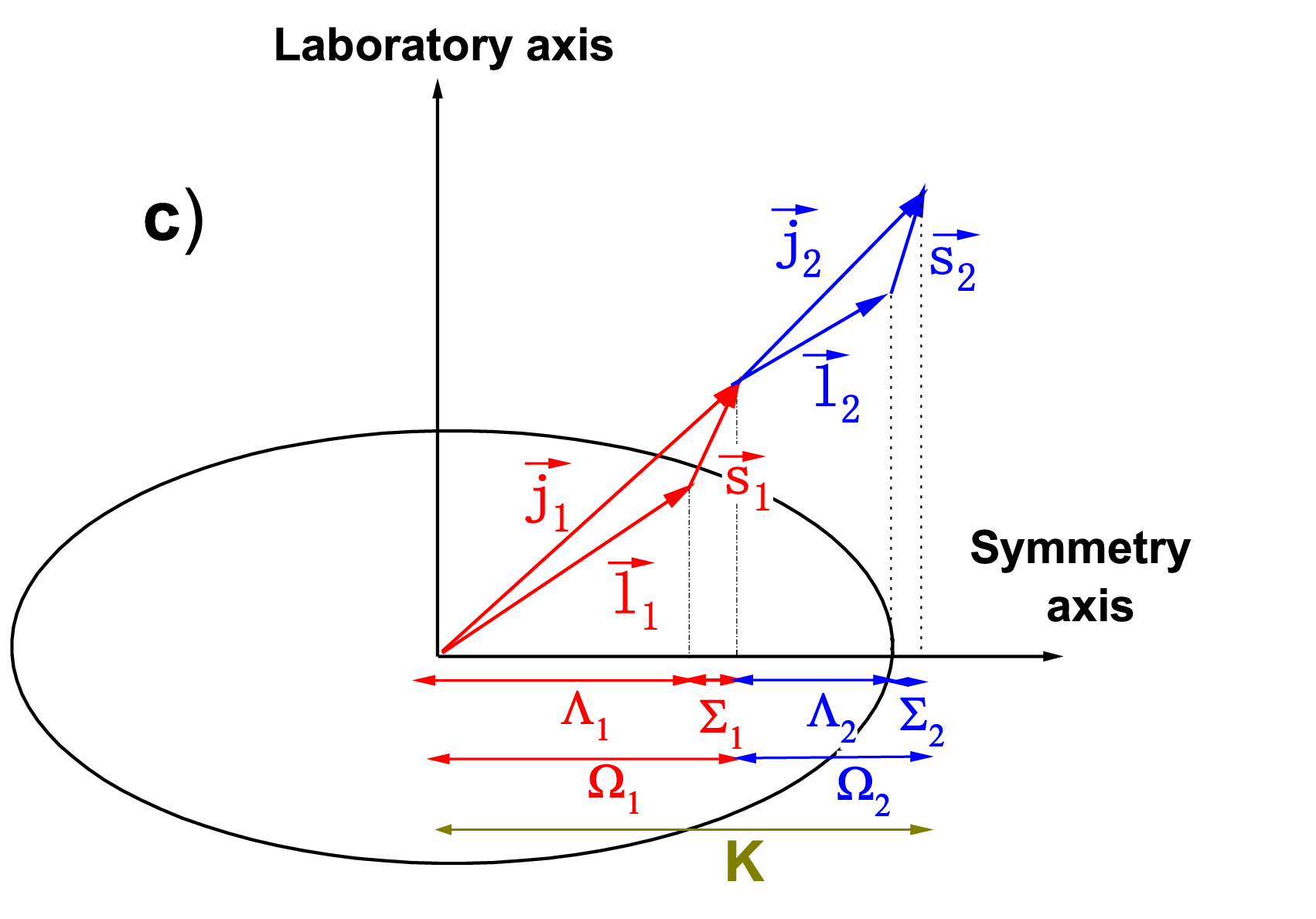}
	}
	\caption{Angular momentum coupling schemes; a) single unpaired nucleon with nucleon orbital angular momentum and spin vectors in 'parallel';
		b) single unpaired nucleon with nucleon orbital angular momentum and spin vectors 'anti-parallel'; c) coupling scheme for two unpaired 
		nucleons (for nonrotating nuclei). }
	\label{fig:4}       
\end{figure*}

\textbf{K isomers} \\
Another case of specific spin isomers are K isomers which occur in strongly prolate deformed nuclei. 
But also some evidence for K isomers in oblate deformed nuclei has been found \cite{Walker24}.
For illustration angular momentum 
coupling schemes are shown in fig. 4. The quantum number K  denotes here the projection of the total angular momentum onto the 
symmetry axis, which is shown for the coupling of two nucleons in fig. 4c. However, one has to note, that formally (as seen in figs. 4a, 4b)
in the case of one unpaired nucleon the equation $\Omega$\,=\,K is valid.
In case of internal transition from an initial state {\it{i}} characterized by the quantum numbers (K$_{i}$ J$_{i}$ $\pi_{i}$) to a final 
state {\it{f}} characterized
by (K$_{f}$ J$_{f}$ $\pi_{f}$) some selection rules have to be fulfilled \cite{Alaga55}:\\
\\
a) {\it{$\mid$$J_{i}-J_{f}$$\mid$\,$\le$$\lambda$\,$\le$J$_{i}$+J$_{f}$}}\\

b) $\pi$\,=\,$\pi_{i}$$\pi_{f}$\\

c) {\it{$\mid$$K_{i}-K_{f}$$\mid$\,=\,$\Delta$K\,$\le$\,$\lambda$}}\\	
\\
where $\lambda$ denotes the multipolarity of the transition.\\

Rules a) and b) are strict, while transitions violating rule c) are rather 'hindered' than forbidden, leading to isomeric states,
denoted as 'K isomers'. The degree of forbiddenness (or hindrance) $\nu$ can be expressed by $\nu$\,=\,$\Delta$K-$\lambda$. For a more
detailed discussion of that feature see \cite{Hessberg23}. Here it just should be noted that due to the equation $\Omega$\,=\,K, 
'single particle' K isomers will also occur at low excitation energies for high spin states, as in such cases $\Delta$K\,=\,$\mid$$K_{i}-K_{f}$$\mid$ 
is always larger than  
the 'spin difference' $\Delta$I between the initial state and the members of the rotational band built up on the state with the bandhead
K$_{f}$, which means that transitions into members of the rotational band will have lower multipolarities than transitions into the
band-head, but are K-hindered.\\

\textbf{Shape isomers} \\
In view of the liquid drop model the atomic nuclei have a spherical shape in its ground state.
At deformation the total energy increases steadily reaching a maximum at a certain elongation,
denoted as 'fission barrier', and decreases afterwards. As a consequence the nucleus will 
disrupt, 'undergo fission' after a certain deformation, denoted as 'scission' point', is reached. 
The shell structure of the atomic nucleus modifies the 
evolution of the change of the total energy of the nucleus. One consequence is, that
the nucleus may not be spherical in its ground state, but the energy minimum will occur
at a certain deformation. The other consequence is, that the increase of the total
energy as a function of deformation will not be 'steady' any more, but gets a 
structure, often exhibiting one (or even more) local minima between the ground
state deformation and the scission point (see e.g. \cite{Nilsson69}), as sketched 
in fig. 5a. In this sense the fission barrier becomes 'double humped'. As another 
consequence a nucleus may be 'trapped' during the deexcitation process in 
the 'second' minimum. Transitions into a state above the ground state deformation
is hindered due to a large change in the nuclear shape, leading to isomeric states,
which in actinide nuclei often, but not necessarily,
decay to a certain degree by spontaneous fission (SF) and therefore 
these shape isomers are 
commonly called 'fission isomers'\footnote{'fission isomers' should not be mixed up 
	with beta-delayed fission, occuring when $\beta$ decay populates excited states 
	at ground state deformation above the fission barrier.}. 
First fission isomers were observed by S.M. Polikanov et al. \cite{Poli62,Poli62a} in 
1962.\\
SF of shape isomers is not necessarily the dominating decay mode. There are examples,
e.g. $^{238}$U as shown in fig. 5b \cite{Fire96}, where 'back decay' by emission
of $\gamma$ radiation into states in the 'first minimum' (at ground state deformation)
dominate. \\
The occurrence of 'fission isomers' is a wide spread phenomenon in actinides
up to berkelium (Z\,=\,97). In transeinsteinium nuclei (Z\,$>$\,99) they have 
not been observed so far.\\

\begin{figure*}
	\resizebox{0.75\textwidth}{!}{
		\includegraphics{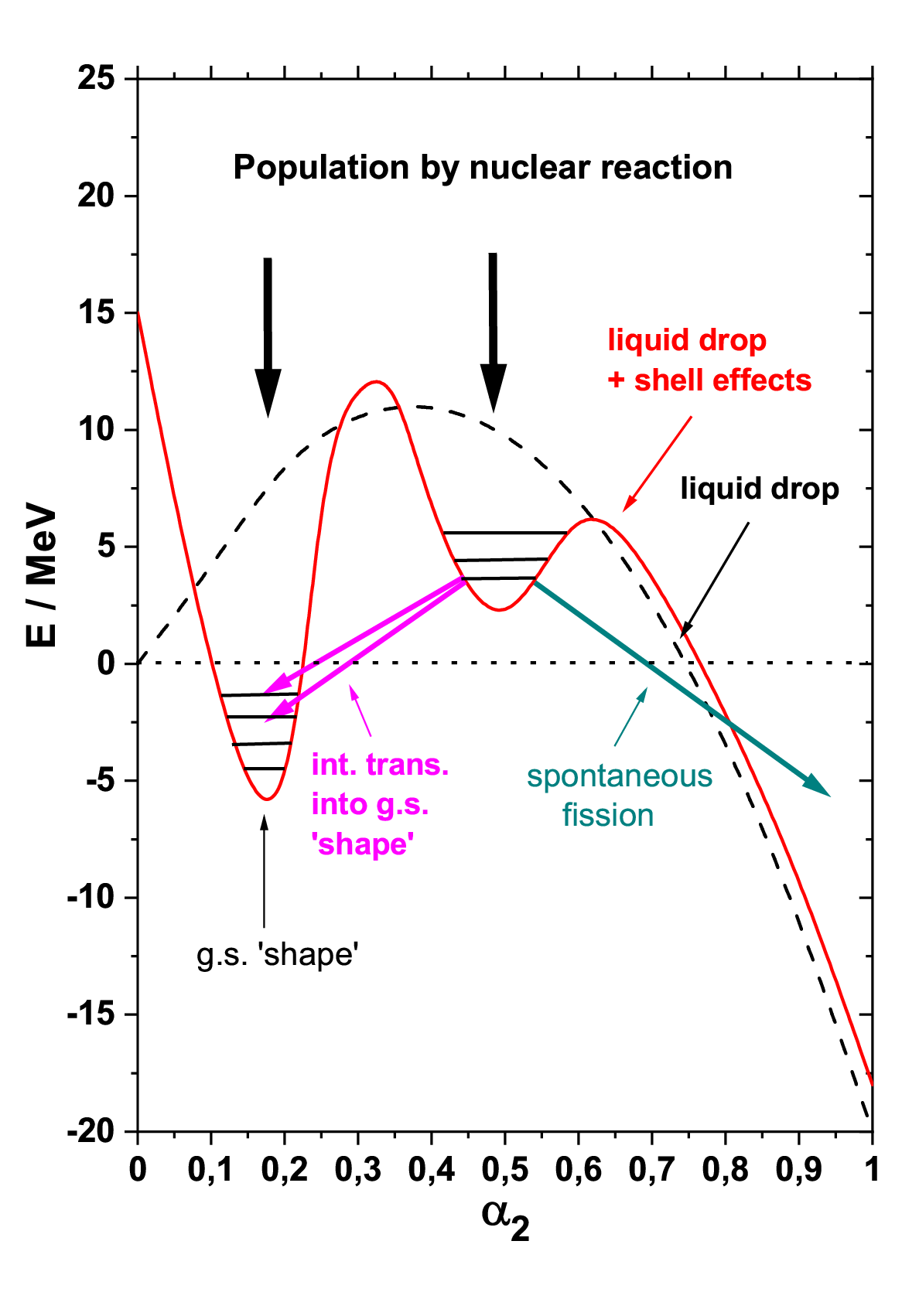}	
	    \includegraphics{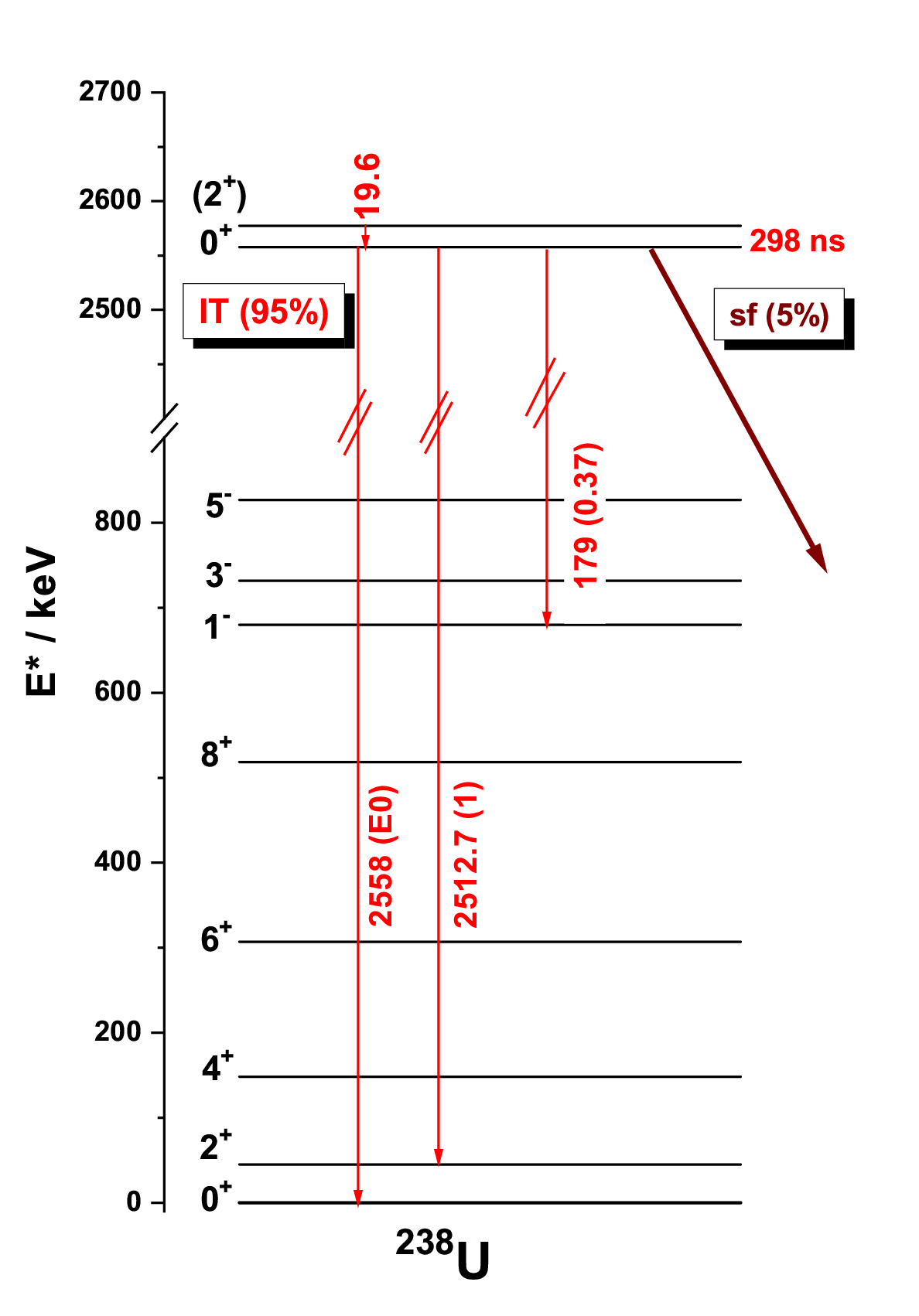}	
	}
	\caption{a) Sketch of a shape isomer, b) decay scheme of the shape isomer in $^{238}$U.};
	
	\label{fig:5}       
\end{figure*}

\begin{figure*}
	\resizebox{0.7\textwidth}{!}{
		\includegraphics{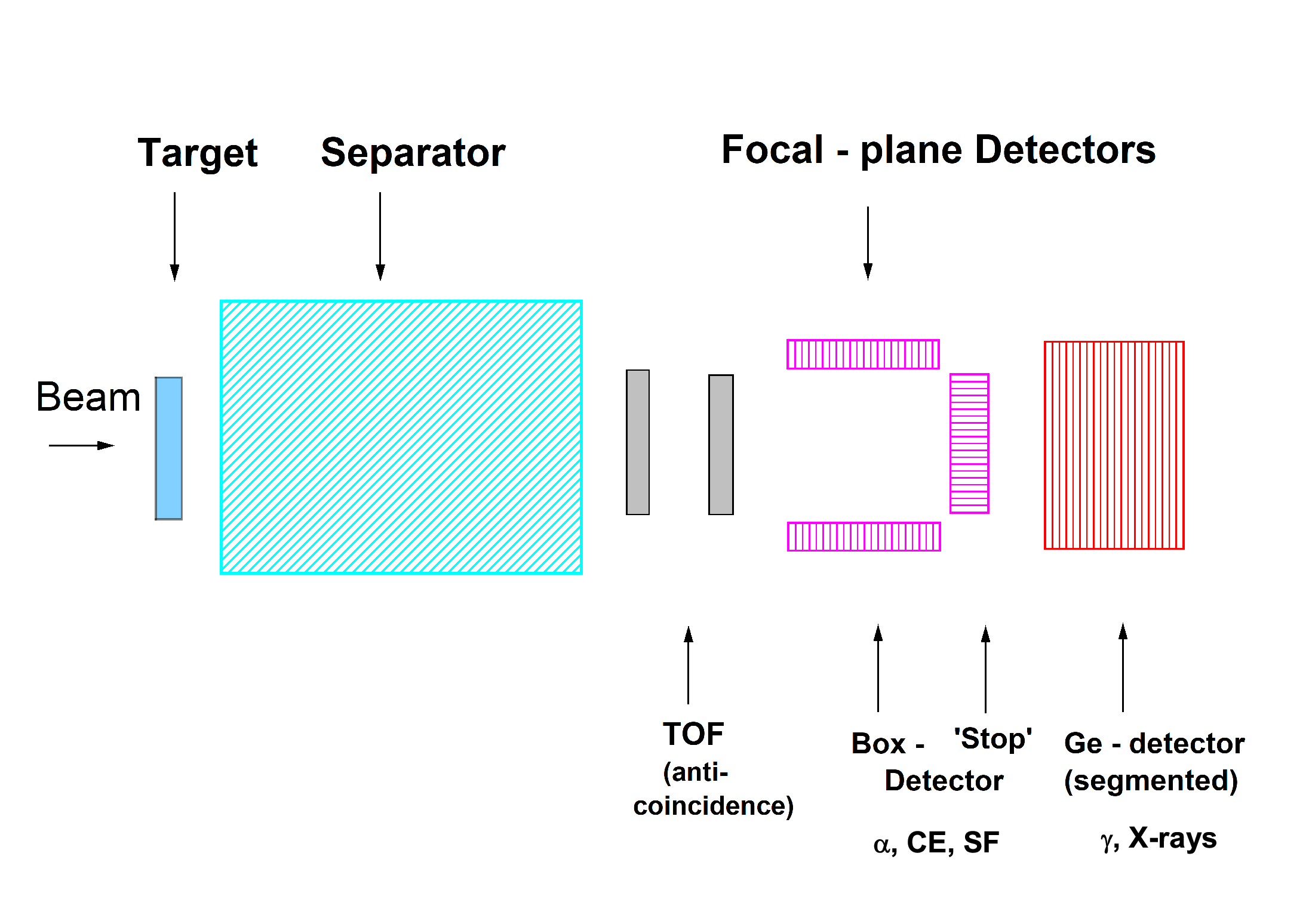}	
	}
	\caption{Schematic experimental set-up for spectroscopy of 
		heaviest nuckei. };
	
	\label{fig:5}       
\end{figure*}
\begin{figure*}
	\resizebox{0.7\textwidth}{!}{
		\includegraphics{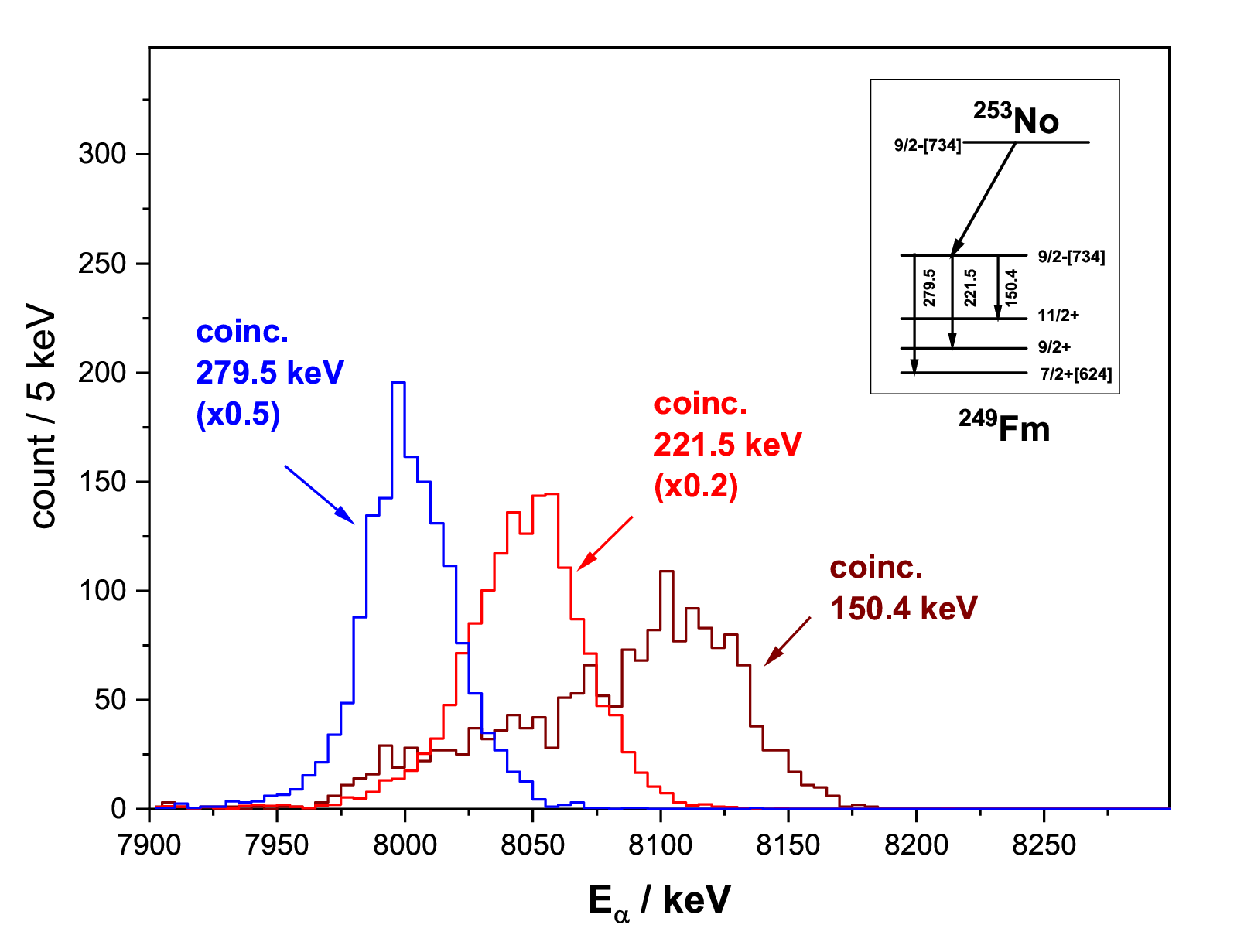}
	}
	\caption{Summing of $\alpha$ particles and CE for $^{253}$No; insert: simplified decay scheme of $^{253}$No.};
	
	\label{fig:7}       
\end{figure*}

\begin{figure*}
	\resizebox{0.8\textwidth}{!}{
		\includegraphics{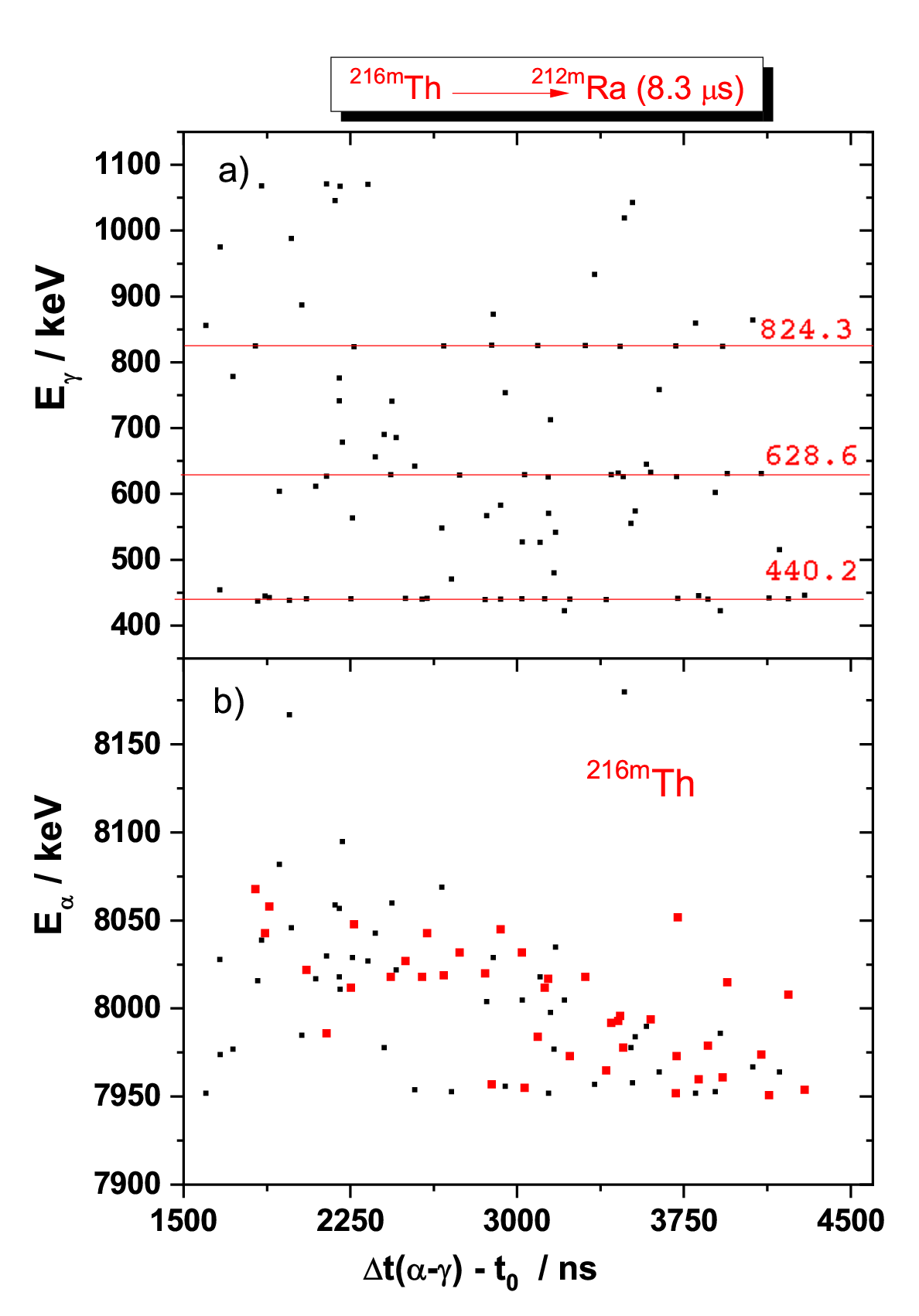}
		\includegraphics{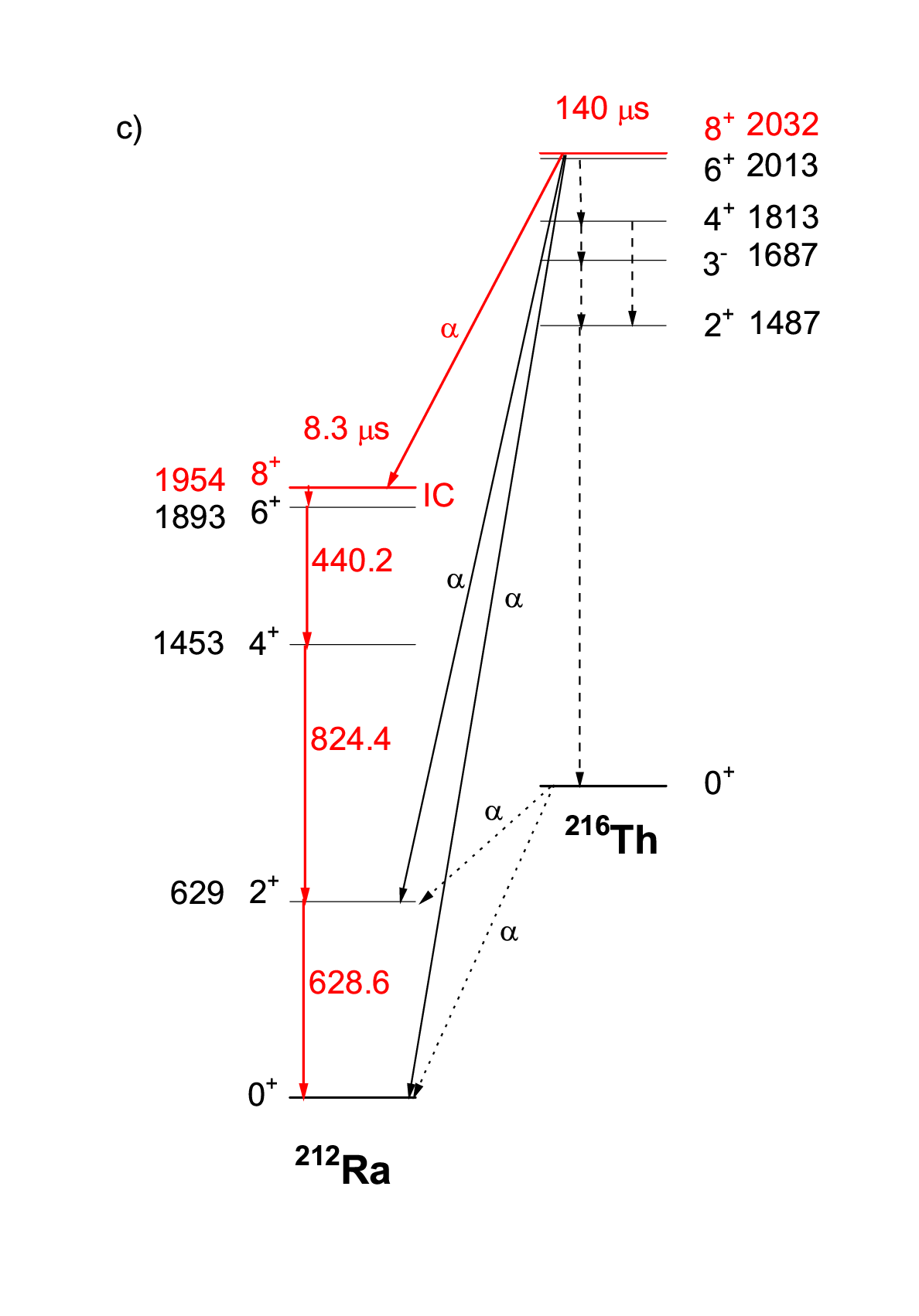}	
	}
	\caption{Decay scheme of $^{216}$Th \cite{Kuus06}; a) time distribution of the $\gamma$ - lines representing the decay $^{212m}$Ra $\rightarrow$ $^{212g}$Ra,
		the horizontal lines are given for better presentation,
		b) energy distribution of $\alpha$ decays of $^{216m}$Th into the 8$^{+}$ level of $^{212}$Ra in the time interval 1500-4500 ns, black points: $\alpha$s in prompt coincidence with $\gamma$ rays
		(any $\gamma$ energy), red points; $\alpha$s in prompt coincidence with $\gamma$ rays E = 440.2, 628.6 or 820.3 keV; c) decay schemes of $^{216m}$Th and
		$^{212m}$Ra.}
	
	\label{fig:8}       
\end{figure*}

\section{4. Experimental features}
The nuclei of interest in the region Z\,$\ge$\,100 are produced by complete fusion reactions. Due to the short lifetimes
mostly lower than a few seconds, the desired reaction products ('evaporation residues' (ER)) have to be separated rapidly (in-flight) 
from the primary beam and are focussed 
on to a detector system suited for $\alpha$-, $\gamma$- and conversion electron (CE) - spectroscopy and detection of
spontaneous fission products \footnote{$\beta^{-}$, $\beta^{+}$ - spectroscopy presently plays a minor role in nuclear structure investigations of
nuclei in the transfermium region}. \\
In the following for better understanding of the discussion the main features of the experimental method will be presented. For more details see, e.g.
\cite{Acker2017}.\\
A typical set-up for nuclear structure investigations is sketched in fig. 6. The ER are produced by complete fusion reactions and are separated 
in-flight by an electro-magnetic device or a gas-filled separator and are guided to the detection system.
The latter usually consists of a time-of-flight (TOF) device, a silicon detector arrangement for measuring $\alpha$- particles, CE and registering 
SF products, and a germanium detector arrangement for measuring $\gamma$- and X-rays. The 'heart' of the detector system is the focal plane detector 
('stop detector') in which
the ERs are implanted, and where $\alpha$ - particles, CE and SF products are measured. As the range of $\alpha$ - particles in silicon is larger 
($>$50 $\mu$m for E$_{\alpha}$ $>$8 MeV) than the implantation depth of the ER (typically $<$10 $\mu$m) only (50-60) $\%$ of the $\alpha$ particles
 are completely stopped in the
focal plane detector, while the rest leaves the detector losing only a fraction $\Delta$E$_{stop}$ of their energy. To measure the energy of these $\alpha$
particles the 'stop detector' is surrounded in 'backward' direction by an arrangement of silicon detectors ('box detector'), where those escaping
$\alpha$ particles are stopped with a probability of typically $>$50$\%$, depending on the geometry of the 'box detector', and their residual energy
E$_{box}$ is measured. The full energy of the $\alpha$ particles thus can be reconstructed as the sum E$_{\alpha}$ = $\Delta$E$_{stop}$ + E$_{box}$.
The energy resolution of these 'reconstructed' events is typically $\Delta$E(FWHM) $>$ 60 keV, considerably worse than for events fully stopped in the
focal plane detector, for which values $\Delta$E(FWHM) $<$ 25 keV can be reached.\\
The role of the TOF is twofold: a) it serves as an anti-coincidence device to discriminate particles passing the separator and being implanted into
the 'stop detector' and radioactive decays in the 'stop detector', and b) by a kinetic energy measurement (implantation energy into the 'stop detector') 
and a TOF measurement a rough mass estimate of the particle can be obtained and thus a discrimination between ERs and 'unwanted' particles (scattered projectiles, 
scattered target nuclei, products from reactions others than complete fusion) passing the separator can be achieved.\\
Implantation of nuclei into a silicon detector has the consequence that the 'source' and the detector are not separated spatially.
Therefore two features have to be considered:\\
a) the measured '$\alpha$ particle energy' is indeed the sum of the '$\alpha$ particle energy' and part (typically $\approx$1/3 \cite{Hof82}) of the recoil
energy transferred by the $\alpha$ particle to the residual nucleus, i.e. E$_{\alpha}$(apparent) =  E$_{\alpha}$ + $\Delta$E$_{recoil}$. This has to be considered 
by a proper energy calibration and also leads to a slight decrease ($\le$ 5 keV) of the $\alpha$ energy resolution of the detector.\\
b) in case of populating an excited level by the $\alpha$ decay which decays by prompt CE emission the $\alpha$ particle energy adds up with the full or 
a part of the energy of the CE, depending if the CE is stopped in the silicon detector or not \cite{Hes89}. This leads not only to a shift of the
$\alpha$ energy but also to a broad energy distribution, which might be quite severe, if decay into the ground state occurs in several steps.\\
As an example in fig. 7 the energy distribution of $\alpha$ decays of $^{253}$No in coincidence with the $\gamma$ transitions is shown. Data are taken from
\cite{Hessb12}; a simplified decay scheme of $^{253}$No is given in the insert. Alpha decay of $^{253}$No (ground state configuration 9/2$^{-}$[734]) populates
(predominantly) the corresponding state in $^{249}$Fm, which decays by three E1 $\gamma$ transitions into the ground state of $^{249}$Fm (configuration 
7/2$^{+}$[624]) or the 9/2$^{+}$ or the 11/2$^{+}$ members of the ground state rotational band.
The $\alpha$s in coincidence  with the E$_{\gamma}$ = 279.5 keV are not
influenced by energy summing with CE, the $\alpha$ line has a width  $\Delta$E(FWHM)\,=\,39 keV, 
the $\alpha$ particles in coincidence with the E$_{\gamma}$ = 221.5 keV line 
are summed with the CEs from the 9/2$^{+}$ $\rightarrow$ 7/2$^{+}$[624] transition.
The line is shifted by $\Delta$E = 47 keV compared the $\alpha$ line in coincidence with the
E$_{\gamma}$ = 279.5 keV - line and is broader ($\Delta$E(FWHM) = 50 keV)
\footnote{It should be noted that the energy shift of $\Delta$E = 47 keV is larger than the energy
of the CE (E = 30.6 keV). The reason is that not only the CE contribute to the energy summing
but also Auger electrons and low energy X-rays from the deexcitation of the atomic shell.}.
The situation is more complex for the $\alpha$ particles in coincidence with E$_{\gamma}$ = 150.4 keV.
The 11/2$^{+}$ state may decay directly into the 7/2$^{+}$ ground state, but the two step process 11/2$^{+}$ $\rightarrow$ 9/2$^{+}$ $\rightarrow$ 7/2$^{+}$ has a much higher intensity. So one has the following
possibilities for energy summing:\\
a) E$_{\alpha}$ + E(CE1) + E(CE2), i.e. both CEs are stopped in the detector\\
b) E$_{\alpha}$ + E(CE1) + $\Delta$E(CE2), i.e. CE1 is stopped in the detector,  CE2 escapes \\
c) E$_{\alpha}$ +  $\Delta$E(CE1) + E(CE2), i.e. CE2 is stopped in the detector, CE1 escapes \\
d) E$_{\alpha}$ + $\Delta$E(CE1) + $\Delta$E(CE2), i.e. both CEs escape.\\
This leads to the broad distribution seen in fig. 7. However a maximum is visible for case a) which is due to a high probability
for both CEs to be stopped due to their low energy (30.6 keV, 43.6 keV).\\
Evidently the effect of energy summing of $\alpha$ particles and CE is dependent on the time difference between the $\alpha$ decay and the emission of the CE, as shown in fig. 8.
This behaviour can be simply qualitatively explained: energy summing means pile-up of the CE - pulse
with the $\alpha$ - particle pulse. The effect will be strongest if the CE is emitted promptly to the $\alpha$ particle
when the CE pulse is sitting on the top of the $\alpha$ pulse. At delayed emission the CE pulse will sit on the
falling flank of the $\alpha$ pulse and its influence on the total pulse height will decrease with increasing time
difference.\\
The 8$^{+}$ seniority isomer in $^{216}$Th has a small $\alpha$ branch b$_{\alpha}$ = 0.028 $\pm$ 0.009 \cite{Kuus06}.
It also decays partly into the 8$^{+}$ isomer of $^{212}$Ra having a half-life of 8.3 $\mu$s. The latter decays
via four E2 transitions into the ground state as shown in fig. 8c. Specifically the 8$^{+}$ $\rightarrow$ 6$^{+}$ transition
is strongly converted and the CE sum up with the $\alpha$ particles. Fig. 8a
shows the time distribution of the three intense $\gamma$ lines (440.2, 628.6, 824.4 keV) in the time interval 
$\Delta$t = (1.5\,-\,4.5) $\mu$s, fig. 8b shows the apparent energy of $\alpha$ decays from $^{216m}$Th
in coincidence with the $\gamma$- events. A clear decrease of the 'apparent' $\alpha$ energies at increasing time differences  is evident.\\

\section{5. Isomers in even-Z nuclei} 
\vspace{-5mm}
\subsection{{\bf 5.1 Isomers in N\,=\,143 isotones}} 
{\bf{$^{233}$Th, $^{235}$U, $^{237}$Pu}}\\
The ground state of  $^{233}$Th was assigned as 1/2$^{+}$[631] \cite{Fire96}, while 
the 5/2$^{+}$[631] (E$^{*}$=6.04 keV) and 7/2$^{-}$[743] (E$^{*}$=6.06 keV)
were identified as low lying levels. \\
In $^{235}$U the 7/2$^{-}$[743] - state was assigned as ground state,
while the 1/2$^{+}$[631] level was assigned as an isomeric state with a half-life of $\approx$25 min at
E$^{*}$ = 7.68 keV \cite{Fire96}. \\
In $^{237}$Pu the ground state is also assigned as 7/2$^{-}$[743],
while the 1/2$^{+}$[631] level is an isomer with a half-life of T$_{1/2}$\,=\,0.18$\pm$0.02 s at 
E$^{*}$\,=\,145 keV \cite{Fire96}.\\

{\bf{$^{239}$Cm, $^{241}$Cf}}\\
	Little information is available for $^{239}$Cm and $^{241}$Cf.
	As based on systematics the ground states of
	$^{239}$Cm and $^{241}$Cf are expected as 7/2$^{-}$[743],
	in both cases also the 1/2$^{+}$[631] level can be assumed as isomeric being populated by unhindered
	$\alpha$ emission of $^{243}$Cf and $^{245}$Fm, respectively, for which 1/2$^{+}$[631]
	can be assumed as ground state \cite{Hess06,Khuy20}. 
	Indeed both precursers $^{243}$Cf \cite{Hess06} and $^{245}$Fm \cite{Khuy20} decay by unhindered
	$\alpha$ transitions, neither in prompt or delayed coincidence with $\gamma$ rays nor significantly influenced
	by energy summing with conversion electrons (CE). Thus it can be assumed that the 1/2$^{+}$[631] level 
	in both nuclei, $^{239}$Cm and $^{241}$Cf is indeed isomeric with a half-life larger than some
	microseconds. A direct identification of the isomeric states could not be obtained so far, however.\\
	
	{\bf{$^{243}$Fm}}\\
	The isotope $^{243}$Fm was first identified by G. M\"unzenberg et al. \cite{MunH81}. Its decay properties
	were later studied by J. Khuyagbaatar et al. \cite{Khuy20} in detail. No indication for a isomeric state was found so far.
	
\subsection{{\bf 5.2 Isomers in N\,=\,145 isotones}}
No low lying isomeric states with half-lives T$>>$ 1 $\mu$s have been identified in even-Z N\,=\,145 isotones so far.\\

\subsection{{\bf 5.3 Isomers in N\,=\,147 isotones}}

\begin{figure*}
	\resizebox{0.90\textwidth}{!}{
		\includegraphics{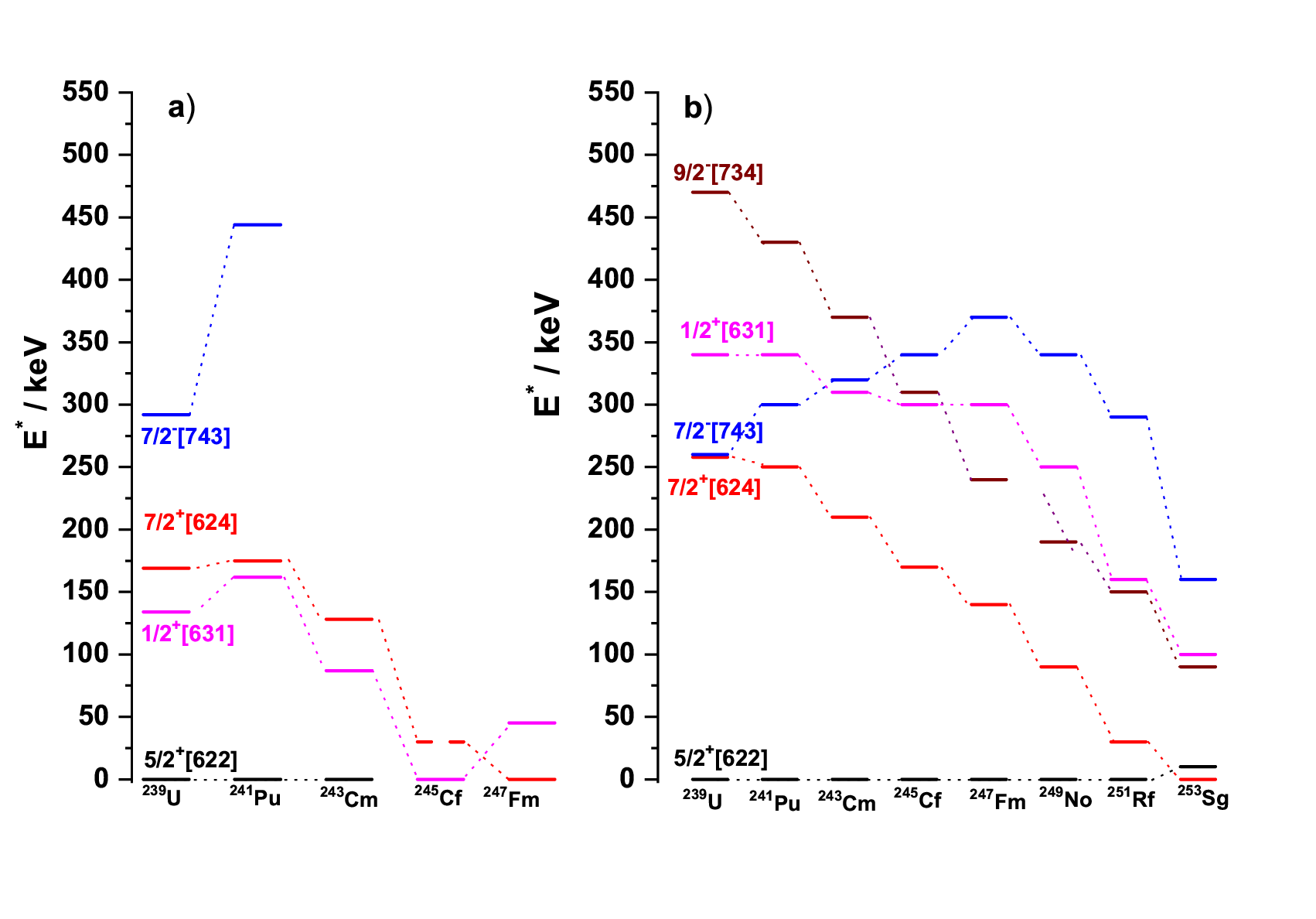}
	}
	\caption{Level schemes N = 147; a) experiment, b) calculation \cite{ParS05}}
	
	\label{fig:9}       
\end{figure*}

Experimentally assigned levels in even Z\,$\ge$\,92, N\,=\,147 isotones are shown in fig. 9a, 
while results of theoretical calculations 
performed by A. Parkhomenko and A. Sobiczewski \cite{ParS05} are presented in  fig. 9b.
The 5/2$^{+}$[622] Nilsson level is predicted as the ground state up to $^{251}$Rf.
Other low lying states 7/2$^{+}$[624], 1/2$^{+}$[631], and 9/2$^{-}$[734] are
predicted at E$^{*}$ $>$ 250 keV for $^{239}$U and to decrease gradually
at increasing atomic numbers with the 7/2$^{+}$[624] level finally becoming 
ground state in $^{253}$Sg. A somewhat different behaviour exhibits the 7/2$^{-}$[743]
level, which increases in energy up to Z\,=\,100 and then starts to decrease.
According to the calculated level schemes shown in fig. 9b, no 
long-lived low lying isomeric states
are expected for the N\,=\,147 isotones in the region of prediction,
as the level with the lowest spin (1/2$^{+}$[631]) could decay into the ground state
(5/2$^{+}$[622]) by an E2 - transition, while the state of the highest spin
(9/2$^{-}$[734]) could decay into the 7/2$^{+}$[624] ($\rightarrow$ 5/2$^{+}$[622]) or 
7/2$^{-}$[743] ($\rightarrow$  5/2$^{+}$[622]) levels by an E1 or M1 transition, respectively.
Indeed in $^{239}$U, $^{241}$Pu, and $^{243}$Cm the 1/2$^{+}$[631] is a 
short-lived isomer, the half-lives, however, are in-line the lifetimes for E2 single particle
transitions according to the Weisskopf estimate (see \cite{Fire96}). \\
The experimental data, however, show a quite different situation as shown 
in fig. 9a. The 5/2$^{+}$[622]  level is ground state only for $^{239}$U, $^{241}$Pu, and $^{243}$Cm,
while ground states for $^{245}$Cf and $^{247}$Fm are 1/2$^{+}$[631] and 7/2$^{+}$[624], respectively.
As these levels are the lowest ones in both nuclei, low lying isomeric states are expected due to the
spin difference $\Delta$I = 3.\\       
\\

{\bf{$^{239}$U}}\\
The 1/2$^{+}$[631] state is located at E$^{*}$\,=\,133.799 keV and a half-life of T$_{1/2}$ = 0.78$\pm$0.04 $\mu$s
is reported \cite{Fire96}. The Weisskopf estimate delivers a half-life T$_{1/2, Weisskopf}$ = 0.149 $\mu$s.\\ 

{\bf{$^{241}$Pu}}\\
The 1/2$^{+}$[631] state is located at E$^{*}$\,=\,161.6 keV and a half-life of T$_{1/2}$ = 0.88$\pm$0.05 $\mu$s
is reported \cite{Fire96}. The Weisskopf estimate delivers a half-life T$_{1/2, Weisskopf}$ = 0.057 $\mu$s.\\ 

{\bf{$^{243}$Cm}}\\
The 1/2$^{+}$[631] state is located at E$^{*}$\,=\,87 keV and has a half-life of T$_{1/2}$ = 1.08$\pm$0.03 $\mu$s
\cite{Fire96}. The Weisskopf estimate delivers a half-life T$_{1/2, Weisskopf}$ = 1.19 $\mu$s.\\ 

\vspace{5mm}

{\bf{$^{245}$Cf}}\\
Intense decay studies of $^{245}$Cf show that $\alpha$ decay essentially populates the ground state of $^{241}$Cm by an
unhindered transition \cite{MaS96,HeH04}. It thus is straightforward to assign the ground state of $^{245}$Cf to the same
level as the ground state of $^{241}$Cm. The latter has been established as 1/2$^{+}$[631].
The existence of an isomeric state in $^{245}$Cf is still unknown. Alpha - decay of $^{249}$Fm was found to populate
by an unhindered transition an excited state at E$^{*}$ $\approx$ 55 keV in $^{245}$Cf corresponding to the ground state
of $^{249}$Fm (7/2$^{+}$[624]) \cite{Hessb12}. As the $\alpha$ line is strongly influenced with CE the lifetime of the 
7/2$^{+}$[624] state in $^{245}$Cf must be lower than some microseconds. 
In conclusion: a low lying isomeric state in $^{245}$Cf is yet not identified. \\

{\bf{$^{247}$Fm}}\\
The isotope $^{247}$Fm was first identified by G.N. Flerov et al. \cite{FlP67}.
They observed three $\alpha$ lines of 
a) E$_{\alpha 1}$ = 7.87$\pm$0.05 MeV, T$_{1/2}$ = 32$\pm$7 s,
b) E$_{\alpha 2}$ = 7.93$\pm$0.05 MeV, T$_{1/2}$ = 35$\pm$10 s,
c) E$_{\alpha 3}$ = 8.18$\pm$0.03 MeV, T$_{1/2}$ = 9.2$\pm$3.2 s.
Due to the same half-lives the $\alpha$ lines  E$_{\alpha 1}$ and E$_{\alpha 2}$
were assigned to the same activity, the ground state decay of $^{247}$Fm, 
while E$_{\alpha 3}$ was assigned to an isomeric state. No excitation energy of the isomer
and no spin assignments were presented.\\
The spin-assignment of $^{247m}$Fm could be clarified in a detailed decay study of $^{251}$No \cite{Hess06}.
It was shown that $^{247m}$Fm was populated by an unhindered $\alpha$ transition of $^{251m}$No.
$^{247m}$Fm on the other hand decayed by an unhindered $\alpha$ transition into $^{243}$Cf. This simply meant
that both $\alpha$ decays connect identical levels. Further it was shown that the decay Q-value of $^{247m}$Fm 
was larger than that of $^{247g}$Fm, and since no $\gamma$ rays were observed in coincidence 
with $\alpha$ decays of $^{247m}$Fm it was concluded that the $\alpha$ decay populates the ground state
of $^{243}$Cf. As the ground state of $^{243}$Cf was assigned as 1/2$^{+}$[631], the same configuration follows for $^{247m}$Fm.
This interpretation is in-line with the 1/2$^{+}$[631] assignment of $^{251m}$No on the basis of systematics. 
As this isomer decays by an unhindered $\alpha$ transition into $^{247m}$Fm, the same configuration as for
$^{247m}$Fm has to be assumed for $^{251m}$No.
The excitation energy of $^{247m}$Fm was estimated in \cite{Hess06} as E$^{*}$\,=\,45$\pm$7 keV, 
while the half-life is given as T$_{1/2}$\,=\,5.1$\pm$0.2 s.\\

{\bf{$^{249}$No}}\\
The isotope $^{249}$No was first safely identified by J. Khuyagbaatar et al.\cite{Khuy21} via $\alpha$ decay of $^{253}$Rf 
and by A.I. Svirikhin et al. \cite{Sviri21} via 'direct' production in the reaction $^{204}$Pb($^{48}$Ca,3n)$^{249}$No.
(A former claim by A.V. Belozerov et al. \cite{Beloz03} turned out to be erroneous.) While J. Khuyagbaatar et al. observed
only one decay, A.I. Svirikhin et al. registered a larger amount of decays, a single $\alpha$ line (E$_{\alpha}$ = 9129 keV
within the detector resolution) that 
could be assigned to a single half-life  of T$_{1/2}$ = 38.1$\pm$2.5 ms. Also only a single $\alpha$ - line was
observed in a decay study of $^{253}$Rf by A. Lopez-Martens et al. \cite{LopH22a}. 
So there is presently no indication for an isomeric state in $^{249}$No.\\

\subsection{{\bf 5.4 Isomers in N\,=\,149 isotones}}

\begin{figure*}
	\resizebox{0.9\textwidth}{!}{
		\includegraphics{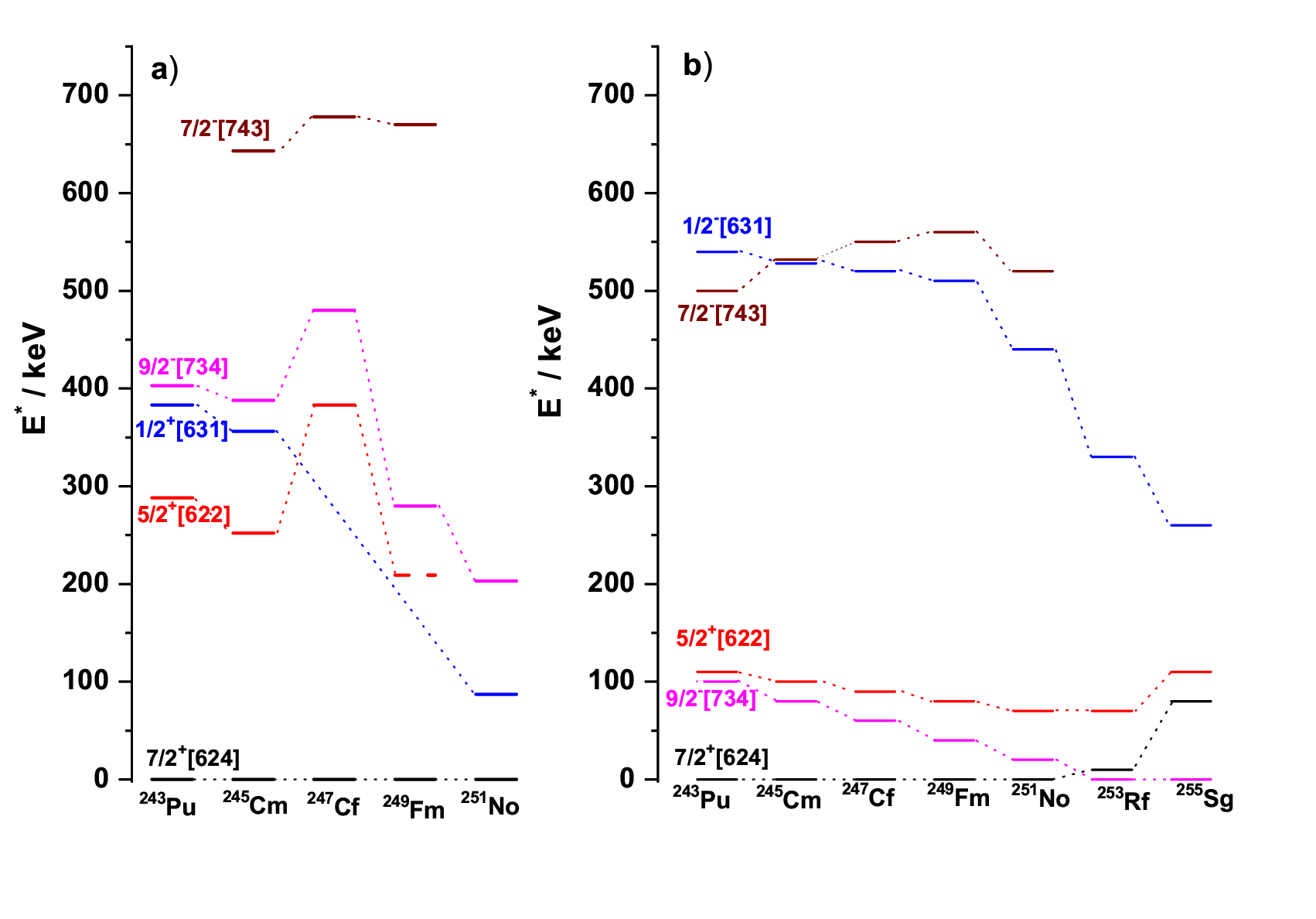}
	}
	\caption{Level scheme N = 149; a) experiment, b) calculation \cite{ParS05}}
	
	\label{fig:10}       
\end{figure*}

Experimentally assigned levels in the N\,=\,149 isotones and predicted ones  \cite{ParS05} are shown in figs. 10a (experimental)
and 10b (calculated). The ground states are predicted as 7/2$^{+}$[624] up to $^{251}$No and changes to 9/2$^{-}$[734] for $^{253}$Rf \cite{ParS05}. As low lying levels besides the 7/2$^{+}$[624] one the Nilsson levels
9/2$^{-}$[734] and 5/2$^{+}$[622] are predicted at E$^{*}$ $<$ 150 keV, while as higher lying levels the 1/2$^{+}$[631] and 7/2$^{-}$[743] Nilsson
states are predicted. The 5/2$^{+}$[622] state varies only slightly in energy while the 9/2$^{-}$[734] is predicted to decrease slightly
becoming ground state for $^{253}$Rf. The 1/2$^{+}$[631] and 7/2$^{-}$[743] levels are predicted at around E$^{*}$ $\approx$ 500 keV for the light
isotones. While the 7/2$^{-}$[743] state varies only slightly with the atomic number, the 1/2$^{+}$[631] state starts to decrease steeply for Z$>$100.
The predicted level scheme does not suggest low lying long lived isomeric states in the range of Z\,=\,(94-106).
Experimentally short lived 1/2$^{+}$[631] isomeric states have been identified in $^{243}$Pu and $^{245}$Cm decaying by E2 transitions into the  5/2$^{+}$[622] level. As shown in fig. 10a the 1/2$^{+}$[631] level 
has been identified at E$^{*}$ = (350-400) keV in $^{243}$Pu and $^{245}$Cm located roughly 100 keV above the 5/2$^{+}$[622] one. According to Weisskopf estimates \cite{Fire96}, lifetimes up to
1 $\mu$s can be expected. In $^{247}$Cf and $^{249}$Fm so far no such an isomeric state was observed, while in $^{251}$No
the 1/2$^{+}$[631] was assigned to an isomeric state at E$^{*}$ $\approx$ 105 keV. In $^{253}$Rf it already might have
become the ground state as assumed from decay studies of this isotope (see below),
which indicates that the 1/2$^{+}$[631] starts earlier to decrease and decreases more steeply than 
predicted.\\

{\bf{$^{243}$Pu}}\\
In $^{243}$Pu an isomeric state with the configuration 1/2$^{+}$[631] has
been identified at an excitation energy E$^{*}$ = 383.64 keV \cite{iaea24}.
It has a half-life of T$_{1/2}$ = 0.33$\pm$0.03 $\mu$s and decays via an E2 transition into
the 5/2$^{+}$[622]  Nilsson level.\\

\vspace{5mm}

{\bf{$^{245}$Cm}}\\
In $^{245}$Cm an  isomeric state with the configuration 1/2$^{+}$[631] has
been identified at an excitation energy E$^{*}$ = 355.95 keV \cite{iaea24}.
It has a half-life of T$_{1/2}$ = 0.29$\pm$0.02 $\mu$s and decays via an E2 transition into
the 5/2$^{+}$[622]  Nilsson level.\\

{\bf{$^{251}$No}}\\
The isotope $^{251}$No was first identified by A. Ghiorso et al. \cite{GhS67} who produced this isotope  in the reaction $^{244}$Cm($^{12}$C,5n)$^{251}$No.
They assigned two $\alpha$ lines of E$_{\alpha 1}$ = 8.60 MeV (i$_{rel}$ = 0.8), E$_{\alpha 2}$ = 8.68 MeV (i$_{rel}$ = 0.2) to that isotope and gave a half-life 
T$_{1/2}$ = (0.8$\pm$0.2) s. Later the isotope was also produced at SHIP, GSI, either indirectly by $\alpha$ decay of $^{255}$Rf \cite{Hess85a,Hess97,Hess06} 
or by the reaction $^{206}$Pb($^{48}$Ca,3n)$^{251}$No \cite{Hess06}. While in the 'indirect' production only the E$_{\alpha}$ = 8.60 MeV line was observed,
both lines were observed in the 'direct' production. This indicated that the E$_{\alpha}$ = 8.68 MeV line was emitted from a state not populated by the
	$\alpha$ decay of $^{255}$Rf.  As $\alpha$ decay of $^{255}$Rf populates the 9/2$^{-}$[734] level in $^{251}$No, which decays via internal transitions
	into the 7/2$^{+}$[624] ground state \cite{Hess06}, the latter $\alpha$ line had to stem from the decay of an isomeric state. On the basis of the
	predicted level scheme (see fig. 10b), the most probable candidate for the isomeric state was the 1/2$^{+}$[631] Nilsson state. This assignement was
	corroborated by the observation, that decay of $^{251m}$No populated the T$_{1/2}$ = 5.1 s 1/2$^{+}$[631]  isomer in $^{247}$Fm \cite{Hess06}
	(see sect. 5.3).\\
	On the basis of the decay energies the excitation energy of the isomer was settled at E$^{*}$ $\approx$ 105 keV. This value was later
	confirmed by a direct mass measurement at SHIPTRAP, which resulted in E$^{*}$ = 105 $\pm$ 3 keV \cite{Kaleja20}.\\
	The half-life was reported as T$_{1/2}$\,=\,1.02$\pm$0.3 s in \cite{Hess06}.\\
	
{\bf{$^{253}$Rf}}\\
	The isotope $^{253}$Rf was first identified by F.P. He\ss berger et al. \cite{Hess97}. Two SF activities of T$_{1/2}$ = 48$^{+17}_{-10}$ $\mu$s
	and T$_{1/2}$ = 11$^{+6}_{-3}$ ms were observed in a bombardment of $^{206}$Pb targets with $^{50}$Ti projectiles. As it could not completely 
	excluded that the ms-SF - activity stemmed from $^{256}$Rf (T$_{1/2}$ = (6.2 $\pm$ 0.2) ms) produced in reactions with target impurities of $^{207}$Pb
	(or even $^{208}$Pb) it was not assigned to $^{253}$Rf.  The existence of the T$_{1/2}$ = 11 ms - activity was recently confirmed 
	by J. Khuyagbaatar et al. \cite{Khuy21}
	and A. Lopez-Martens et al. \cite{LopH22a}, but now definitely assigned this activityto $^{253}$Rf, which meant the existence of two low lying states in 
	$^{253}$Rf decaying by spontaeous fission. Differences between the two publications, however, occured in the interpretation, i.e. in the level assignment
	of the two activities. While J.Khuyagbaatar et al. assigned the short-lived activity to a low spin state (1/2$^{+}$[631]) following the idea that low spin states in a nucleus
	will have lower fission barriers and thus a lower spontaneous fission half-life as already predicted by J. Randrup in 1973 \cite{Rand73} and is observed for a couple of cases 
	\cite{Hess17}, A. Lopez-Martens et al. came to a different conclusion. They found the short-lived activity in delayed coincidence with CE, which they 
	ascribed to the decay of a K isomer with a half-life of T$_{1/2}$ = 0.66 ms located at E$^{*}$ $\ge$ 1.02 MeV. As it seemed meaningful that the decay of
a high-spin K isomer rather populates a high-spin low lying state than a low-spin one, they assigned the short-lived SF activity (they gave a half-life of 52.8 $\mu$s)
to the 7/2$^{+}$[624] Nilsson level, while the longer lived SF activity (they gave a half-life of 9.9 ms) is attributed to the 1/2$^{+}$[631] state, which is also 
assumed as the ground state. If this interpretation will be confirmed one here would have not only a case where the high-spin state has a lower fission barrier 
and thus a lower half-life but also a change in the ground state configuration from  7/2$^{+}$[624]  ($^{251}$No) to  1/2$^{+}$[631] ($^{253}$Rf). The
latter scenario cannot be excluded as one observes a decrease of the 1/2$^{+}$[631] - level from E$^{*}$ $\approx$ 350 keV for $^{245}$Cm to 
E$^{*}$ $\approx$ 105 keV for $^{251}$No, i.e. by $\approx$ 80 keV per two units of the atomic number Z. At a continuation of this trend, 
it might be already the ground state of $^{253}$Rf.\\
The conclusions drawn by A. Lopez-Mertens concerning spin and parity assingments of both
activities were confirmed by a study of decay of $^{257}$Sg performed by P. Mosat et al. \cite{Mosat25}
at the TASCA separator at GSI, Darmstadt. They observed one $\alpha$ decay of $^{257}$Sg 
followed by a fission event after 11 $\mu$s. On the basis of systematics the ground state of 
$^{257}$Sg can be assigned as 9/2$^{-}$[734]. It is known for N=151 isotones (see e.g. \cite{Hess06} for the case of $^{255}$Rf)
that they decay by $\alpha$ emission predominantly into the corresponding level in the daughter nuclei,
which then decay by E1-transitions into 7/2$^{+}$[624] ground state or the 9/2$^{+}$ and 11/2$^{+}$ 
members of the rotational band built up on the ground state, while  $\alpha$ decay into the 1/2$^{+}$[631]  
was not observed \cite{Hess06}.
So this decay chain strongly suggests
that the T$_{1/2}$ = 52.8 $\mu$s  fission activity stems from the 7/2$^{+}$[624] level
which, however, must not necessarily be the ground state of $^{253}$Rf. 
Due to the observed and predicted steep decrease of the 1/2$^{+}$[631] level at
increasing proton number, the latter one which is isomeric in the neighbouring
N\,=\,149 isotone might be the ground state in $^{253}$Rf.\\
In addition to sf an $\alpha$ activity of T$_{1/2}$\,=\,5.7$^{+3.1}_{-1.5}$ ms 
(E$_{\alpha}$=9.31$\pm$0.02 MeV, i$_{rel}$$\approx$0.57 and E$_{\alpha}$=9.21$\pm$0.02 MeV, i$_{rel}$$\approx$0.43)
was obsered by A. Lopez-Martens et al.\cite{LopH22a}.

\vspace{5mm}
\subsection{{\bf 5.5 Isomers in N\,=\,151 isotones}}
The N\,=\,151 isotones are characterized by low lying isomeric states with half-lives in the range of  20-50 $\mu$s. 
These isomers are due to
low lying 5/2$^{+}$[622] Nilsson states, located above the 9/2$^{-}$[734] ground states, requiring an M2 transition (with possible E3 admixture) 
into the ground state. From theory these isomers are not expected, as for all isotones the 7/2$^{+}$[624] level is predicted below
the 5/2$^{+}$[622]  one\cite{ParS05}, allowing transitions 5/2$^{+}$[622] $^{M1}_{\rightarrow}$ 7/2$^{+}$[624] $^{E1}_{\rightarrow}$ 9/2$^{-}$[734].
From experimental side only for $^{245}$Pu the 7/2$^{+}$[624] level is placed below the 5/2$^{+}$[622] one \cite{Asai11}. In this isotope, however,
as also in $^{247}$Cm and $^{251}$Fm, a submicrosecond isomer has been identified, which is attributed to the 1/2$^{+}$[622] Nilsson level \cite{Asai11}.
As the 7/2$^{+}$[624] state increases in energy from $^{245}$Pu to $^{249}$Cf already for $^{247}$Cm the 5/2$^{+}$[622] state becomes isomeric. Its 
energy changes only slightly with increasing atomic number without a pronounced trend to increase or to decrease in the range E$^{*}$ $\approx$ (135\,-\,250) keV
from $^{247}$Cm (Z\,=\,96) to $^{255}$Rf (Z\,=\,104).\\
As mentioned above in several N\,=\,151 isotones low lying 1/2$^{+}$ isomeric states have been identified, decaying by E2 transitions 
into the 5/2$^{+}$[622] level as shown fig. 11. In $^{245}$Pu, $^{247}$Cm, $^{249}$Cf (identified within this study) they are assigned
to the 1/2$^{+}$[620] level, in $^{251}$Fm it is assigned to the 1/2$^{+}$[631] level.\\

\begin{figure*}
	\resizebox{0.85\textwidth}{!}{
		\includegraphics{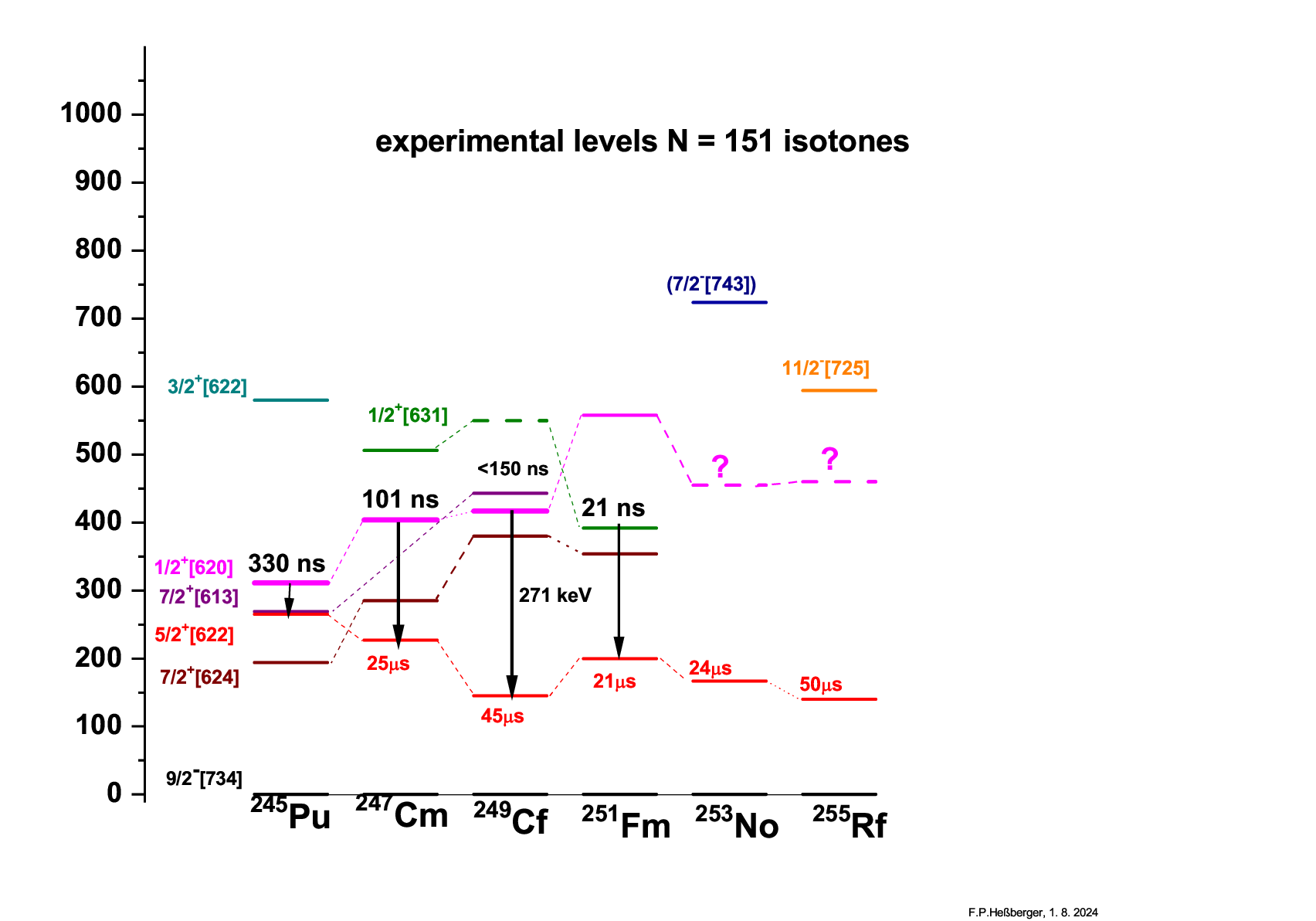}
	}
	\caption{1/2$^{+}$ isomeric levels in N\,=\,151 isotones.}
	
	\label{fig:9}       
\end{figure*}

\begin{figure*}
	\resizebox{0.75\textwidth}{!}{
		\includegraphics{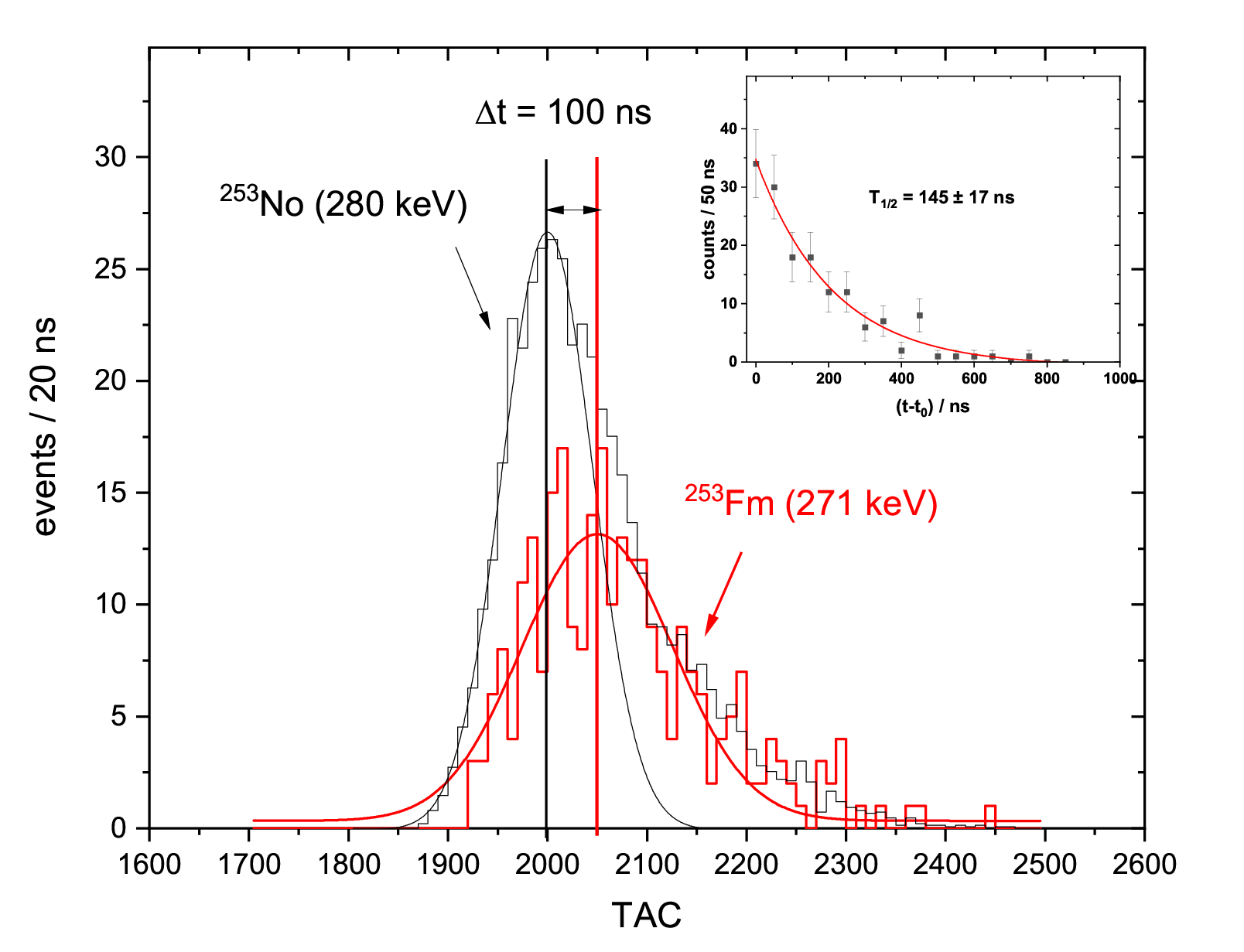}
	}
	\caption{TAC - spectra for the E\,=\,271 keV - line (1/2$^{+}$ $\rightarrow$ 5/2$^{+}$ transition in $^{249}$Cf),
		and the E\,=\,280 keV - line (9/2$^{-}$ $\rightarrow$ 7/2$^{+}$ transition in $^{249}$Fm).}
	
	\label{fig:9}       
\end{figure*}


{\bf{$^{245}$Pu}}\\
Contrary to the heavier N\,=\,151 isotones the 5/2$^{+}$[622] level is placed above the 7/2$^{+}$[624] one \cite{Asai11}, so it is not isomeric. 
However, a low lying isomeric state at E$^{*}$ = 311 keV with a half-life T$_{1/2}$ = 0.33$\pm$0.02 $\mu$s was identified \cite{Asai11}. It was assigned
to the 1/2$^{+}$[620] level, decaying by an E2 transition into the 5/2$^{+}$[620] Nilsson state.\\

{\bf{$^{247}$Cm}}\\
The 5/2$^{+}$[622] level populated by $\alpha$ decay of $^{251}$Cf \cite{Ahmad03} (in earlier investigations 
also population by $\beta^{-}$ decay  of $^{247}$Am  is
reported, see \cite{Fire96}) has been identified at E$^{*}$ = 227.38 keV. A half-life of T$_{1/2}$ = 26.3$\pm$0.3 $\mu$s was measured.
In that study \cite{Ahmad03} a second isomeric state at E$^{*}$ = 404.9 keV with a half-life of T$_{1/2}$ = 100.6$\pm$0.6 ns was identified and assigned as in
$^{245}$Pu to the 1/2$^{+}$[620] level, decaying by an E2 transition into the 5/2$^{+}$[622] Nilsson state.\\

{\bf{$^{249}$Cf}}\\
The 5/2$^{+}$[622] level populated by $\alpha$ decay of $^{253}$Fm or by EC decay of $^{249}$Es has been identified at E$^{*}$ = 144.96 keV. 
A half-life of T$_{1/2}$ = 45$\pm$5 $\mu$s was measured \cite{Fire96}.\\
While in $^{245}$Pu, $^{247}$Cm (1/2$^{+}$[620]) and $^{251}$Fm (1/2$^{+}$[631]) 1/2$^{+}$ isomers
are known, decaying via an E2 transtion into the 5/2$^{+}$[622] level, the 1/2$^{+}$[620] $\rightarrow$
5/2$^{+}$[622] transition in $^{249}$Cf (E$_{\gamma}$ = 271.8 keV) has been known
for a quite a long time \cite{Ahmad67}, a half-life, however, was not measured so far.
In experiments on spectroscopy of $^{253}$No and its decay products \cite{AntH11,Hessb12}
the 1/2$^{+}$[620] in $^{249}$Cf was populated by $\alpha$ decay of $^{253}$Fm, 
produced by EC decay of $^{253}$No ($^{253}$No $^{EC}_{\rightarrow}$ $^{253}$Md $^{EC}_{\rightarrow}$
$^{253}$Fm). To get information about a finite (measurable) lifetime of the 1/2$^{+}$[620] - state
the time (TAC) spectra between $\alpha$ decays of $^{253}$Fm (E$_{\alpha}$ = 6675 keV) and 
the 1/2$^{+}$[620] $\rightarrow$ 5/2$^{+}$[622] transition in $^{249}$Cf (E$_{\gamma}$ = 272.0 keV) 
were investigated \cite{Hess24}.
The result is shown in fig. 12 in comparison with the TAC spectrum registered for the prompt
coincidences between $\alpha$ decays of $^{253}$No and the E$_{\gamma}$ = 279.5 keV	(9/2$^{+}$[734] $\rightarrow$ 7/2$^{+}$[624]) transition in $^{249}$Fm. Evidently the 
TAC - spectrum for the E$_{\gamma}$ = 272.0 keV transition is shifted by $\approx$100 ns compared
to the  one for E$_{\gamma}$ = 279.5 keV transition, proving a finite lifetime of the
former one. Fitting an exponential decay curve to the falling flank (TAC values $>$2050 channels) of
the distribution for the  E$_{\gamma}$ = 272.0 keV events  one obtains a half-life of
T$_{1/2}$ = 145$\pm$17 ns. This value, however, has to be taken with care.
First, the used set-up is not optiumum for measuring short lifetimes; the width of the
TAC values for the  E$_{\gamma}$ = 279.5 keV transition is $\approx$360 ns (FWHM);
second, a tail is present at higher TAC values for the E$_{\gamma}$ = 279.5 keV transition;
this has to be considered for the E$_{\gamma}$ = 272.0 keV transition as it certainly effects the
half-life measurement. So the 'real' half-life might be lower. 
Thus we presently state, a) the 1/2$^{+}$[620] has a measurable lifetime and thus 
has to be considered as isomeric, b) its half-life is T$_{1/2}$ $\le$ 150 ns.\\

{\bf{$^{251}$Fm}}\\
The  5/2$^{+}$[622] isomeric state was first identified by C.E. Bemis, Jr., P.F. Dittner et al. in 
a decay study of $^{255}$No performed at ORNL aimed to prove the atomic
number of that isotope by measuring $\alpha$ - K X-rays coincidences \cite{Bemis71,Dittner71}. Part of the K X-rays were found 
in delayed coincidence with the $\alpha$ decays and were interpreted to stem from an isomeric state. From the time distribution
of the X-ray events a half-life of T$_{1/2}$ = 15.2$\pm$2.3 $\mu$s was obtained. It was assigned to the 5/2$^{+}$[622] Nilsson
level, the excitation energy was settled at E$^{*}$ = 191 keV on the basis of the energy difference between the $\alpha$ decays 
populating the ground state (E$_{\alpha}$ = 8312 keV) and those populating the isomeric state (E$_{\alpha}$ = 8121 keV).
A more precise value for the excitation energy was obtained by F.P. He\ss berger et al. \cite{Hess06a} in an experiment performed at SHIP, GSI  by measuring 
the $\gamma$ - transition  5/2$^{+}$[622] $\rightarrow$ 9/2$^{-}$[734] by delayed coincidences between $\alpha$ decays and $\gamma$ events.
A value E$^{*}$ = 199.9$\pm$0.3 keV and a half-life T$_{1/2}$ = 21$\pm$3 $\mu$s were measured. These data later were confirmed by M. Asai et al. \cite{Asai11} who
obtained E$^{*}$ = 200.09 keV and  T$_{1/2}$ = 21.1$\pm$1.9 $\mu$s.\\
In addition M. Asai et al. could identify a second isomeric at E$^{*}$ = 392 keV with a half-life of T$_{1/2}$ = 22 ns, which 
contrary to the cases $^{245}$Pu and $^{247}$Cm was assigned to the 1/2$^{+}$[631] Nilsson level.\\
From the ratio of gammas and K X-rays a K-conversion coefficient of $\alpha_{K}$ = 8.3$\pm$2.9 was obtained in \cite{Hess06a},
for the 199.9-keV transition which was not 
in-line with the conversion coefficient expected for a pure M2 - transition.
Using the relation 
$\alpha_{K}$ = a$\times \alpha_{K}$(M2)+(1$-$a)$\times$$\alpha_{K}$(E3)
and conversion coefficients from \cite{HaS68} an E3 admixture 
a $\approx$ 0.5 was obtained. A similar result is obtained using the more precise
code BrIcc \cite{Kib08} for calculating the M2 (14.73) and E3 (0.225) 
K conversion coefficients. One obtains a similar value a $\approx$ 0.56. \\

{\bf{$^{253}$No}}\\
The  5/2$^{+}$[622] isomeric state was first identified by C.E. Bemis, Jr. et al. \cite{Bemis73} in an experiment performed at ORNL aimed at 
'X-Ray identification of Element 104' by measuring  coincidences between $\alpha$-decays attributed to $^{257}$Rf and K X-rays.
Ten of thirteen K X-rays were found 
in delayed coincidence with the $\alpha$ decays and were interpreted to stem from an isomeric state.
The half-life was measured as T$_{1/2}$ = 31.1$\pm$4.1 $\mu$s. The excitation energy was estimated as E$^{*}$ $\approx$ 300 keV.
Later studies using $\gamma$-spectroscopic methods, either producing the isomer directly using the reaction $^{207}$Pb($^{48}$Ca,2n)$^{253}$No and 
measuring the $\gamma$ decay in delayed coincidence with evaporation residues implanted in the 'stop detector' system behind the in-flight separator
\cite{Hess07,Hess07a,LoH07} or registering the gammas in delayed coincidence with $\alpha$ decays \cite{Hess07,Hess07a,StH10} resulted in a lower excitation 
energy of E$^{*}$ = 167 keV. A half-life of  T$_{1/2}$ = 24$\pm$2 $\mu$s was obtained in \cite{Hess07}.\\

{\bf{$^{255}$Rf}}\\
The 5/2$^{+}$[622] isomeric state in $^{255}$Rf was identified
in a decay study of $^{259}$Sg \cite{AntH15} by triple correlations
ER - $\alpha$($^{259}$Sg) - CE. From the time distribution of the
conversion electrons (CE) a half-life T$_{1/2}$ = 50$\pm$17 $\mu$s
was obtained. No $\gamma$ rays or K X-rays that could be attributed to the 
isomeric state were observed. The latter means that its excitation energy
is below the K binding energy in element 104 (156.2880 keV \cite{Fire96}).
This is in-line with the excitation energy E$^{*}$ $\approx$ 135 keV 
extracted from the energy distribution of the conversion electrons \cite{AntH15}.
The excitation energy of the 5/2$^{+}$[622] isomeric state was essentially 
confirmed by a study of R. Chakma et al. \cite{Chakma23}, who populated
the isomer directly by the reaction $^{207}$Pb($^{50}$Ti,2n)$^{255}$Rf and 
reported a value E$^{*}$ = 150 keV.\\

\subsection{{\bf 5.6 Isomers in N\,=\,153 isotones}}

In the N\,=\,153 isotones ($^{249}$Cm, $^{251}$Cf) a 7/2$^{+}$[613] isomeric state
has been identified, decaying into rotational band members built up on the 1/2$^{+}$[620] ground
state. For the heavier isotones this isomeric state has been so far not identified.
Throughout the isotone range from $^{249}$Cm to $^{257}$Rf the 11/2$^{-}$[725] state was
found as an isomeric one. The energy of that state is quite stable (E$^{*}$ $\approx$ 350-400 keV)
for $^{249}$Cm, $^{251}$Cf and $^{253}$Fm and then starts to drop rapidly, being at
E$^{*}$ = 74 keV in $^{257}$Rf and finally becoming the ground state in $^{259}$Sg.\\

{\bf{$^{249}$Cm}}\\
In $^{249}$Cm two low lying isomeric states are known, a 7/2$^{+}$[613] one at E$^{*}$ = 48.74 keV
with a half-life of T$_{1/2}$ $\approx$ 23 $\mu$s \cite{Fire96}, decaying into the 3/2$^{+}$  member 
of the rotational band built up on the 1/2$^{+}$[620] ground state.\\
A second isomeric state with the configuration 11/2$^{-}$[725] was identified 
by T. Ishii et al. \cite{Ishii08} at E$^{*}$ = 375 keV.
It has a half-life of T$_{1/2}$ = 19$\pm$1 ns. It decays into the 9/2$^{+}$ and 7/2$^{+}$ members
of the rotational band built up on the 7/2$^{+}$[613]  Nilsson level. \\

{\bf{$^{251}$Cf}}\\
In $^{251}$Cf two low lying isomeric states have been identified, one at E$^{*}$ = 106 keV 
with a half-life T$_{1/2}$ = 37$\pm$2 ns attributed to the 7/2$^{+}$[613] Nilsson level by F. Asaro et al. 
\cite{Asaro64}. It decays into the 3/2$^{+}$, 5/2$^{+}$, 7/2$^{+}$ members of the 
rotational band built up on the 1/2$^{+}$[620] ground state with similar intensities.
Existence and decay data of this isomer were later confirmed by I. Ahmad et al. \cite{Ahmad71}
who reported T$_{1/2}$ = 38$\pm$2 ns and E$^{*}$ = 106.3 keV. They also identified a second isomer,
located at E$^{*}$ = 370.39 keV with a half-life T$_{1/2}$ = 1.3$\pm$0.1 $\mu$s.
It was attributed to the 11/2$^{-}$[725] Nilsson level and decays by two transitions of nearly equal 
intensity into the  9/2$^{+}$ and 11/2$^{+}$ members
of the rotational band built up on the 7/2$^{+}$[613] Nilsson level and by a considerably weaker 
transition into the band head.\\

{\bf{$^{253}$Fm}}\\
The 11/2$^{-}$ isomer in $^{253}$Fm was identified by S. Antalic et al. \cite{AntH11} within a decay study of $^{253}$No and
its daughter products where it was populated by EC decay of $^{253}$Md. A half-life of T$_{1/2}$ = 0.56$\pm$0.06 $\mu$s
was measured. It decays by two E1 transitions  into the  9/2$^{+}$ and 7/2$^{+}$ members
of the rotational band built up on the 7/2$^{+}$[613]  Nilsson level. As the decay of that level into lower lying levels or into
the ground state could not be established unambiguously the excitation energy could not be given precisely. Only limits
E$^{*}$ = (341\,-\,361) keV were reported. \\

{\bf{$^{255}$No}}\\
The 11/2$^{-}$ isomer in $^{255}$No  was identified by A. Bronis et al. \cite{Bronis22} within a decay study of $^{255}$No.
A half-life of T$_{1/2}$ = 109$\pm$9 $\mu$s was measured. The decay path of the isomer could not be satisfactorily established,
so only an excitation energy range E$^{*}$ = (240\,-\,300) keV could be given. \\
Similar, but slightly different results are presented by K. Kessaci \cite{Kessaci22,Kessaci24} who gave values of  T$_{1/2}$ = 86$\pm$8 $\mu$s
and E$^{*}$ $\approx$ 200 keV. \\

{\bf{$^{257}$Rf}}\\
The isotope $^{257}$Rf was first synthesized by A. Ghiorso et al. \cite{GhN69}
by the reaction $^{249}$Cf($^{12}$C,4n)$^{257}$Rf, who obtained a quite complex 
structure of the $\alpha$ spectrum. Their result was later confirmed by C.E. Bemis et al. \cite{Bemis73},
who measured in addition K X-rays in coincidence with the $\alpha$ particles and thus identified the
atomic number as Z\,=\,104.\\
At SHIP, GSI, this isotope was produced 'directly' by the reaction $^{208}$Pb($^{50}$Ti,n)$^{257}$Rf \cite{Hess85a}
and 'indirectly' via $\alpha$ decay of $^{265}$Hs ($^{265}$Hs $^{\alpha}_{\rightarrow}$ $^{261}$Sg  $^{\alpha}_{\rightarrow}$  $^{257}$Rf)
produced in the reaction  $^{208}$Pb($^{58}$Fe,n)$^{265}$Hs \cite{Hess97,Hofm95}. 
It was found that the two $\alpha$ lines of highest energy (E$_{\alpha}$ = 9021 keV, E$_{\alpha}$ = 8968 keV) were missing in the
'indirect' production. This observation was interpreted in the way, that both decays stem from a level not being populated by 
$\alpha$ decay of $^{261}$Sg. Neglecting these two lines, the $\alpha$ decay pattern resembled those of the N\,=\,153
isotones $^{253}$Fm and $^{255}$No, thus these two $\alpha$ lines were attributed to the decay of an isomeric state in $^{257}$Rf.
The ground state was assigned as 1/2$^{+}$[620] as in the lighter isotones, while on the basis of a calculated
level scheme for $^{257}$Rf \cite{Cwiok94} the configuration of the isomer was assigned as 11/2$^{-}$[725].
	The excitation energy was settled by the energy difference between the assumed ground state to ground state decay
	of $^{257}$Rf (8903 keV) ($^{257g}$Rf  $^{\alpha}_{\rightarrow}$ $^{253g}$No)  and the  E$_{\alpha}$ = 9021 keV - line, which 
	was interpreted as decay into the ground state of $^{253}$No ($^{257m}$Rf  $^{\alpha}_{\rightarrow}$ $^{253g}$No) \cite{Hess97}
	at E$^{*}$ = 118 keV  \cite{Hess97}. This value later had to be corrected to E$^{*}$ = 70 keV \cite{StH10} after in a study
	with 'higher statistics' a weak line of E$_{\alpha}$ = 8950 keV was registered, now interpreted as the ground state to ground state
	transition, that had not been observed before. The value was essentially confirmed in a recent study of K. Hauschild et al. \cite{HaL22},
	who gave a value of E$^{*}$ = 74 keV.\\
	It should be emphasized that in the early experiments using the 'direct' production \cite{Bemis73,Hess85a,GhN69} ground state decay
	and isomeric decay could not be disentangled because of the similar half-lives. After identification of the isomer some effort has been
	undertaken to estimate these values. The results of these analyses are summarized in \cite{HaL22}, who gave
	values of T$_{1/2}$ = 6.2$^{+1.2}_{-1.0}$ s for $^{257g}$Rf and T$_{1/2}$ = 4.37 $\pm$0.05 s for $^{257m}$Rf in agreement
	with the values reported earlier \cite{StH10,Hess16a}. \\

{\bf{$^{259}$Sg}}\\
First report on the synthesis of $^{259}$Sg came from Yu.Ts. Oganessian et al. \cite{Ogan74}, who
observed an SF activity of T$_{1/2}$ = (4-10) ms in irradiations of $^{206,207,208}$Pb with
$^{54}$Cr. A more precise value of  T$_{1/2}$ = 7.5 ms was later given by G.N. Flerov \cite{Flerov76}.\\
These 'early' Dubna results were disproved by later Dubna experiments performed by A.G. Demin et al. \cite{Demin84},
who obtained a lower limit for the half-life of $^{259}$Sg of T$_{1/2}$\,$\ge$\,0.1 s and in experiments performed at
SHIP, GSI by G. M\"unzenberg et al. \cite{Muenz85} who identified the isotope as an $\alpha$ emitter (E$_{\alpha}$ = 9.62$\pm$0.03 MeV)
with a half-life  T$_{1/2}$ = 0.48$^{+0.28}_{-0.13}$ s. These data were later confirmed by C.M. Folden III et al. \cite{Folden09}, who in
addition registered a broad distribution of $\alpha$ events  in the range E$_{\alpha}$ = (9.0-9.47) MeV and by F.P. He\ss berger et al. \cite{Hessb09},
who observed, besides also a broad energy distribution in the range (9.0-9.5) MeV two $\alpha$ lines (E$_{\alpha}$ = 9607$\pm$10 keV, 9550$\pm$10 keV).
The two $\alpha$ lines suggested similar to case in the neighboring N\,=\,153 isotone $^{257}$Rf decay of two different states, but
the quality of the data was not sufficient to draw definite conclusions. In a follow-up study at SHIP a factor of about twenty higher
number of $\alpha$ decays was obtained \cite{AntH15}. It could be shown, that half-lives of both $\alpha$ transitions 
are different.\\
E$_{\alpha}$ = 9614 keV (formerly 9607 keV): T$_{1/2}$ = (402$\pm$56) ms.\\ 
E$_{\alpha}$ = 9550 keV: T$_{1/2}$ = (226$\pm$27) ms.\\ 
The E$_{\alpha}$ = 9550 keV line was attributed to the decay of the 1/2$^{+}$[620] level as it a) feeds the 
5/2$^{+}$[622] isomeric state in $^{255}$Rf, similar as it is the case in the decay of the lighter N\,=\,153 isotones
$^{257}$Rf, $^{255}$No, $^{253}$Fm, $^{251}$Cf, b) an SF activity of T$_{1/2}$ = (229$\pm$147) ms
was observed, which seemed rather identical to the T$_{1/2}$ = (226$\pm$27) ms $\alpha$ activity than
to the T$_{1/2}$ = (402$\pm$56) ms one. As fission barriers of low spin states are estimated to be lower
than those of high spin states and thus usually low spin states have higher fission probabilities, the SF activity
was assigned to the low spin state.\\
Consequently the  E$_{\alpha}$ = 9614 keV line was attributed to the decay of the 11/2$^{-}$[725] level.
This assignment had consequences for the level ordering. Similar to the case of $^{257}$Rf it can be 
assumed that $\alpha$ decay of the 11/2$^{-}$[725] level populates predominantly the 9/2$^{-}$[734] 
ground state in $^{255}$Rf, so the total decay energy is E\,=\,9614 keV, while the total decay energy
of the 1/2$^{+}$[620] is the sum of the $\alpha$ energy feeding the 5/2$^{+}$[622] plus its excitation energy,
i.e. E = 9550 keV +E$^{*}$(5/2$^{+}$) = 9550 keV + ($\approx$135 keV) $\approx$ 9685 keV thus
E$^{*}$(1/2$^{+}$)\, $>$\,E(11/2$^{-}$). In other words, the 11/2$^{-}$[725] level is the ground state of
$^{259}$Sg, the 1/2$^{+}$[620] level is assigned to the isomeric state.\\
We note here a not only a change of the ground state configuration from $^{257}$Rf to $^{259}$Sg, but also a change
of the character of the isomer. While the 11/2$^{-}$[725] isomer in $^{257}$Rf has the character of
single particle K isomer, the 1/2$^{+}$[620] isomer in $^{259}$Sg has the character of a 'simple'
spin isomer.\\

\subsection{{\bf 5.7 Isomers in N\,=\,155 isotones}}
{\bf{$^{259}$Rf}}\\
The identification of $^{259}$Rf was first reported by A. Ghiorso et al. \cite{GhN69}, who 
produced the isotope in the reaction $^{249}$Cf($^{13}$C,3n)$^{259}$Rf. They reported
two $\alpha$ - lines (E$_{\alpha 1}$ = 8.86 MeV (i$_{rel}$\,=\,0.4), E$_{\alpha 2}$ = 8.77 MeV (i$_{rel}$=0.6))
and T$_{1/2}$ $\approx$ 3 s. The data were confirmed by C.E. Bemis, Jr. et al. \cite{BeD81}, who reported
	(E$_{\alpha 1}$ = 8.87$\pm$0.02 MeV (i$_{rel}$$\approx$0.4), E$_{\alpha 2}$ = 8.77$\pm$0.02 MeV (i$_{rel}$$\approx$0.6)).
	Both $\alpha$ lines were also reported in experiments on the synthesis of $^{263}$Sg, where $^{259}$Rf 
	was produced by $\alpha$ decay \cite{GhN74,GrL94}.\\
	In a later study a half-life T$_{1/2}$ = 2.5$^{+0.4}_{-0.3}$ s was reported by J. Gates et al. \cite{Gates2008}, whose study was
	focussed on EC - decay of $^{259}$Rf. No results on $\alpha$ decay properties of this isotope were given, 
	it was just noted, that the properties were in-line with the known data from previous studies.
	The authors, however, remarked, that no decays ($\alpha$ decays or spontaneous fission) of $^{259}$Lr were observed
	when $^{259}$Rf was produced by $\alpha$ decay of $^{263}$Sg, which possibly indicated the presence of
	an isomeric state that is formed only in direct production of $^{259}$Rf, which does not undergo EC decay.\\ 
	Indeed a completely different result was obtained when $^{259}$Rf was produced within the $\alpha$ - decay chain
	starting from $^{271}$Ds ($^{271}$Ds $^{\alpha}_{\rightarrow}$ $^{267}$Hs $^{\alpha}_{\rightarrow}$ 
	$^{263}$Sg $^{\alpha}_{\rightarrow}$ $^{259}$Rf $^{\alpha}_{\rightarrow}$ ) \cite{Hof170,Ack25,Mori04,Fold04,Ginter03}.
	It had already been realized that practicly only the 'high' energy $\alpha$ line of $^{263}$Sg was observed
	within the $\alpha$ decay chain of $^{271}$Ds, indicating the existence of a long-lived isomeric state
	(see section 5.8). An analysis of the correlated $\alpha$ decays $^{263}$Sg $^{\alpha}_{\rightarrow}$ 
	$^{259}$Rf $^{\alpha}_{\rightarrow}$ showed that practically only the 'high' energy line of $^{259}$Rf
	is observed in correlation with $\alpha$ decays of $^{263}$Sg, as shown in fig. 13.
	Evidently there are also in $^{259}$Rf two longlived isomeric states.
	From the correlated events one obtains an $\alpha$ energy of  E$_{\alpha 1}$ = 8.89$\pm$0.02 MeV
	and a half-life T$_{1/2}$ = 2.25$^{+0.75}_{-0.45}$ s. This activity will be denoted as
	$^{259}$Rf (2) in the following. Evidently the two longlived states 
	have very similar half-lives and thus can only be distinguished on the basis of different 
	production mechanisms (direct production, production by $\alpha$ decay).\\
	A theoretical half-life T$_{\alpha}$ = 0.44 s and thus a hindrance factor $\approx$5 is obtained, 
		which may be still regarded as an unhindered transition.\\
		The other activity, denoted as $^{259}$Rf(1) in the following is ascribed to the E$_{\alpha 2}$ = 8.77$\pm$0.02 MeV 
		$\alpha$ decay. The half-life is still uncertain as the values given above represent a mixture between $^{259}$Rf(1)
		and $^{259}$Rf(2), but as in direct production it was observed as the stronger component, it certainly is close to the
		values given there. We will here do to give a value T$_{1/2}$ $\approx$ 3 s.\\
		\begin{figure*}
			\resizebox{0.75\textwidth}{!}{
				\includegraphics{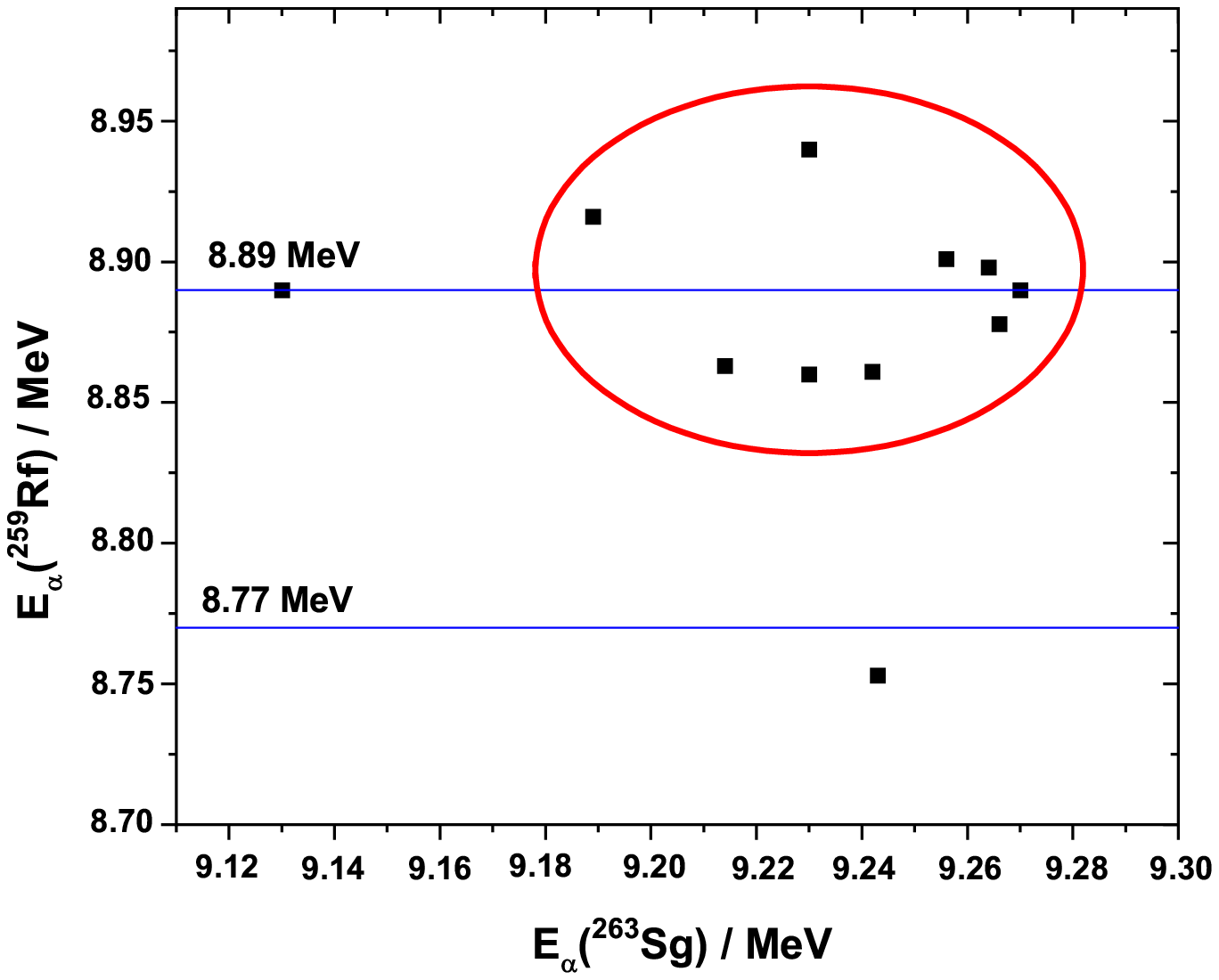}	 
			}
			\caption{$\alpha$ - $\alpha$ correlation plot $^{263}$Sg - $^{259}$Rf from decay of $^{271}$Ds.}	
			\label{fig:13}       
		\end{figure*}

{\bf{$^{261}$Sg}}\\
So far $^{261}$Sg is the only N\,=\,155 isotone for which the existence of an isomer is published .
It was reported by J.S. Berryman et al. \cite{Berry10}.
They gave a half-life of T$_{1/2}$ = 9.0$^{+2.0}_{-1.5}$ $\mu$s. It should have an 
excitation energy E$^{*}$\,$\approx$\,200 keV and  decay via 
internal transitions. Tentatively it was attributed to
the  11/2$^{-}$[725] Nilsson level. Theoretical level schemes were predicted by 
A. Parkhomenko and A. Sobiczewski \cite{ParS05} as well as by S. Cwiok et al. \cite{Cwiok94}.
The ground state of $^{261}$Sg was predicted as 1/2$^{+}$[620] in \cite{Cwiok94}, the
3/2$^{+}$[622] - state at E$^{*}$ = 20 keV, the
11/2$^{-}$[725] - state at E$^{*}$ = 30 keV. In \cite{ParS05} the
1/2$^{+}$[622] and 3/2$^{+}$[622] - states are degenerated in energy at E$^{*}$ = 0 keV,
while the 11/2$^{-}$[725] - state is predicted at E$^{*}$ = 10 keV. 
Experimentally the ground state of $^{261}$Sg was established as 3/2$^{+}$[622] \cite{StH10}.
(for more details see sect. 5.8).
Although the 
11/2$^{-}$[725] state could be expected as an isomeric state, the calculated level schemes are
not in-line with the experimental results as the K - difference between the isomer
and ground state would be $\Delta$K = 4, so decays into the ground state or states of
the rotational band build upon the ground state would be strongly K - hindered, resulting in a lifetime
$\tau$ $>>$ 1 $\mu$s. So the situation could be expected similar to $^{257}$Rf,
where the  11/2$^{-}$[725] isomeric state decays by $\alpha$ emission. In such a case
decay of an 11/2$^{-}$[725] isomer can be expected to decay by $\alpha$ emission
as in the case of $^{257}$Rf and populating
the low lying 11/2$^{-}$[725] - state in $^{257}$Rf. The decay of latter, however, was not
observed within the $\alpha$-decay study of $^{261}$Sg \cite{StH10}. So it is
likely, as already discussed in \cite{Berry10}, that the 7/2$^{+}$[613] level is 
located below the  11/2$^{-}$[725] one. In that case, a decay scenario
11/2$^{-}$[725] $^{M2}_{\rightarrow}$ 7/2$^{+}$[613] $^{E2}_{\rightarrow}$ 3/2$^{+}$[622] 
is likely, while an M2 - transition would be in-line with a half-life of 9 $\mu$s.\\
\\

\subsection{{\bf 5.8 Isomers in N\,=\,157 isotones}}
{\bf{$^{261}$Rf}}\\
The isotope $^{261}$Rf was first synthesized by A. Ghiorso et al. \cite{GhN70}
who reported one activity with an $\alpha$ - decay energy of E$_{\alpha}$ = 8280$\pm$20 keV
and a half-life of T$_{1/2}$ = 65$\pm$10 s.
In later experiments  this isotope was produced either directly \cite{LaL00,Asai05,Haba11}
or by $\alpha$ decay of $^{265}$Sg (or precursers $^{269}$Hs, $^{273}$Ds, $^{277}$Cn),
\cite{TuD98,HoN96,HoH02,MoM07,DuB02,TuD03,DvK05,DvB06,DvB08,Haba12}.
These studies gave indication of two activities that had to be attributed to this isotope.
The data, however, often suffered from limited $\alpha$ - energy resolution of
the detectors, low number of registered events, similar decay energies of 
$^{261}$Rf and its daughter isotope $^{257}$No. 
Alpha energy and time distributions of the decays of both states 
using the data from \cite{Haba11,HoH02,MoM07,DvB08,Haba12}
are shown in fig 14.\\
For $^{261}$Rf(1) an $\alpha$ - energy of E$_{\alpha}$ = 8.27$\pm$0.01 MeV and 
a half-life T$_{1/2}$ = 55$\pm$9 s is obtained.\\
For $^{261}$Rf(2) an $\alpha$ - energy of E$_{\alpha}$ = 8.51$\pm$0.01 MeV is measured.
Half-lives for SF and $\alpha$ decay vary considerably as seen in fig. 15. For the  SF events  
a half-life T$_{1/2}$ = 2.57$^{+0.66}_{-0.44}$ s is obtained,
for the $\alpha$ decays a value T$_{1/2}$ = 5.95$^{+2.07}_{-1.22}$ s,
the combined half-life of fission and $\alpha$ decays is
T$_{1/2}$ = 3.80$^{+0.72}_{-0.52}$ s.\\
It should be emphasized, that the half-lives obtained for the SF events is lower than the
value obtained from the $\alpha$ events.This excludes the possibility that the different
half-lives are due to an SF contribution of $^{261}$Rf(1).
The time distributions for the fission (SF) events and $\alpha$ events are shown
in fig. 15a (SF events), 15b ($\alpha$ events) and 15c ($\alpha$\,+\,SF events).
Although the difference by more than a factor of two may suggest different sources 
for the SF and $\alpha$ acticities, the low number of events and the broad straggling
of the decay times indicate that such an assumption is not necessarily straightforward.
So we presently assign both, $\alpha$ decays and SF events to the same activity
and give a half-life T$_{1/2}$ = 3.80$^{+0.72}_{-0.52}$ s.\\
A detailed decay study of $^{261}$Rf using $\alpha$ - $\gamma$ - spectroscopy 
was performed by M. Asai \cite{Asai05}, who used the production reaction
$^{248}$Cm($^{18}$O,5n)$^{261}$Rf. In this experiment only the long-lived
activity $^{261}$Rf(1) was observed, no experimental evidence for $^{261}$Rf(2),
in-line with the results of previous studies \cite{GhN70,LaL00},
but in contradiction with the later study of H. Haba et al. \cite{Haba11}, who reported
for the same reaction nearly equal production rates of $^{261}$Rf(1) and $^{261}$Rf(2).
On the basis of $\alpha$ - $\gamma$ coincidence measurements M. Asai assigned
$^{261}$Rf(1) to the 9/2$^{+}$[615] Nilsson level ('most probable') or 
the 11/2$^{-}$[725] ('possible') one and interpreted it as the isomeric state.\\ 
The existence of a low lying isomeric state is expected from theoretical 
predictions. A. Parkhomenko and A. Sobiczewski \cite{ParS05} predict the
11/2$^{-}$[725] Nilsson level as the ground state of $^{261}$Rf and the 
3/2$^{+}$[622] (E$^{*}$\,=\,50 keV), 7/2$^{+}$[613] (E$^{*}$\,=\,60 keV), 
1/2$^{+}$[620] (E$^{*}$\,=\,80 keV) levels as low lying states at E$^{*}$ $<$ 100 keV.\\
Given the assignment of M. Asai for $^{261}$Rf(1), possible states for
$^{261}$Rf(2) would then be 3/2$^{+}$[622] or 1/2$^{+}$[620]. 
Such an assignment could also explain the high fission branch of 
$^{261}$Rf(2) (b$_{SF}$ = 0.73$\pm$0.06 \cite{Haba11})
as commonly low spin states have lower fission barriers than high spin states
in the same isotope.
(see e.g. \cite{Rand73}).  \\
\begin{figure*}
	\resizebox{0.75\textwidth}{!}{
		\includegraphics{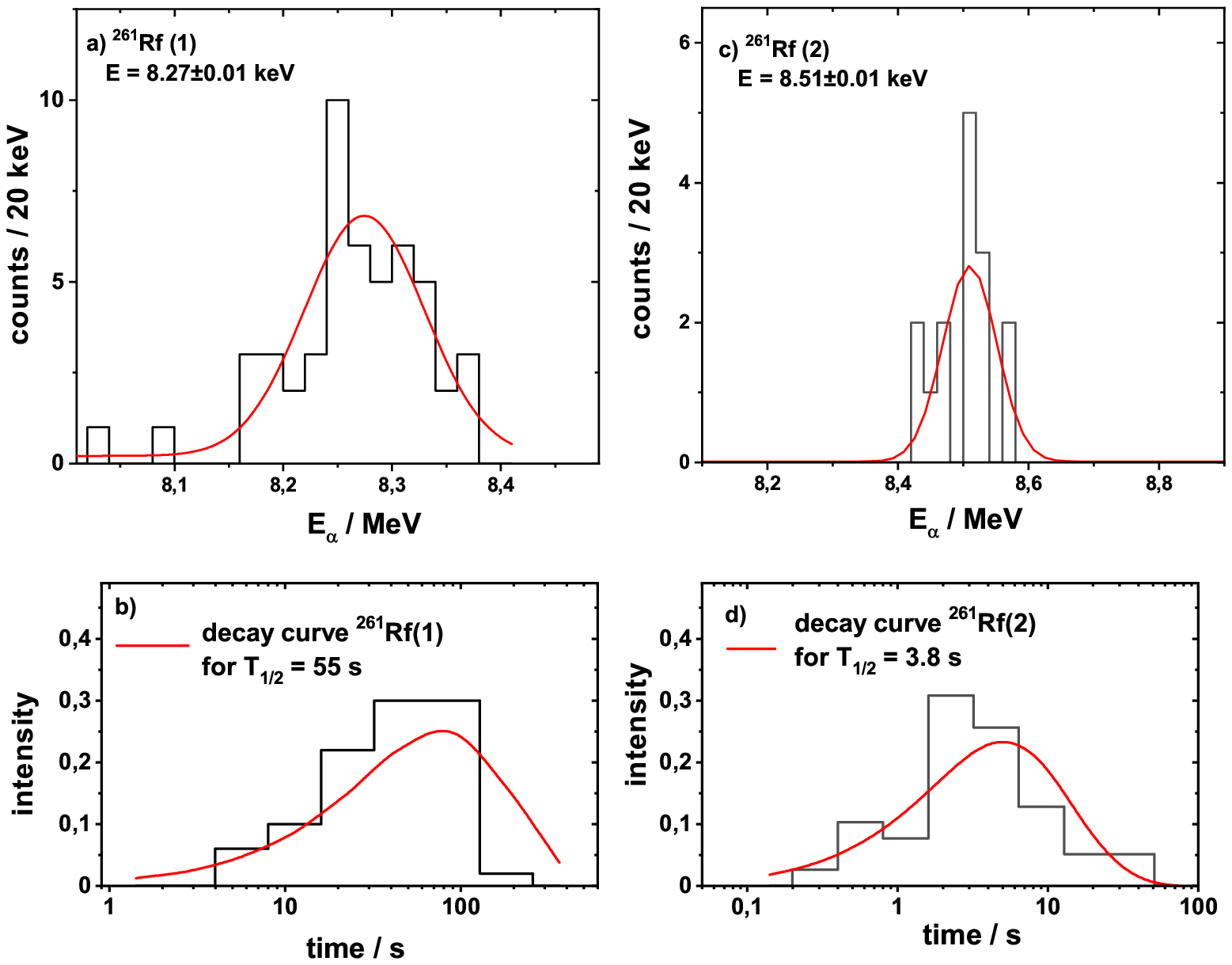}	
	}
	\caption{$\alpha$ spectra and time distribution of events assigned to $^{261}$Rf(1) and $^{261}$Rf(2).}	
	\label{fig:14}       
\end{figure*}
\begin{figure*}
	\resizebox{0.75\textwidth}{!}{
		\includegraphics{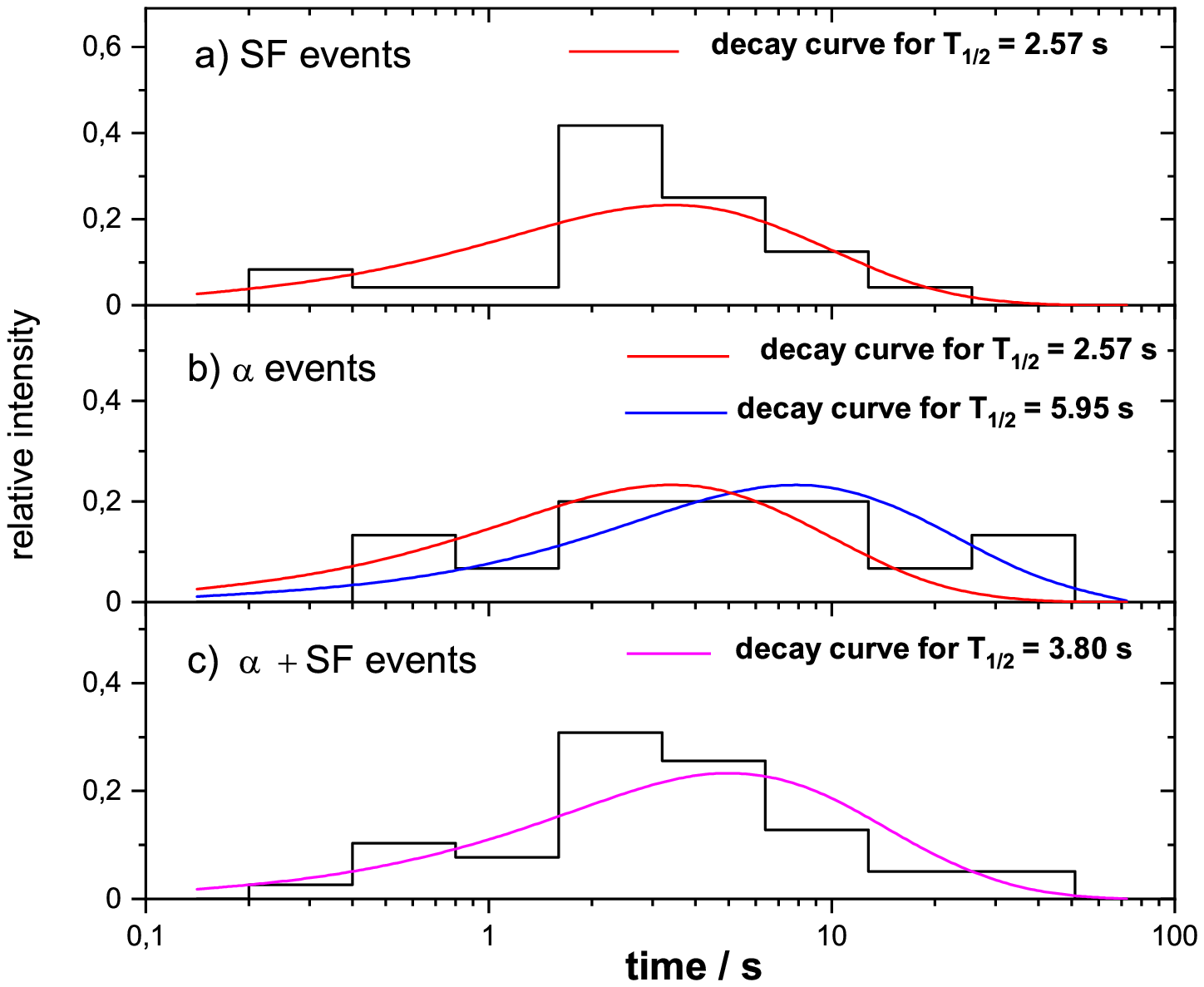}	
	}
	\caption{Time distributions of $\alpha$ decays and SF events assigned to $^{261}$Rf(2);
		a) time distribution of SF events, b) time distribution of $\alpha$ events, 
		c) time distribution of $\alpha$ plus SF events.}	
	\label{fig:15}       
\end{figure*}

{\bf{$^{263}$Sg}}\\
The isotope $^{263}$Sg was first synthesized by A. Ghiorso et al.\cite{GhN74} in the reaction
$^{249}$Cf($^{18}$O,4n)$^{263}$Sg. They observed a 'principal decay energy' of E$_{\alpha}$\,=\,9.06$\pm$0.4 MeV
and a weaker line at  E$_{\alpha}$\,=\,9.25 MeV. For both activities a common half-life of T$_{1/2}$\,=\,0.9$\pm$0.2 s is given.
The result was later confirmed by K.E. Gregorich et al. \cite{GrL94} and also in production by the reaction
$^{238}$U($^{30}$Si,5n)$^{263}$Sg both $\alpha$ lines were observed \cite{GrG06,NiH06}.\\
The low energy activity  was also observed when $^{263}$Sg was produced by $\alpha$ decay of $^{267}$Hs, synthesized 
in the reaction $^{238}$U($^{34}$S,5n)$^{267}$Hs \cite{NiH10}.\\
The situation appeared completely different when $^{263}$Sg was produced by $\alpha$ decay within the decay chain of 
$^{271}$Ds ($^{271}$Ds $^{\alpha}_{\rightarrow}$ $^{267}$Hs $^{\alpha}_{\rightarrow}$ $^{263}$Sg). In this case 
(except one case) only the E$_{\alpha}$\,=\,9.25$\pm$0.02 MeV line is observed \cite{Hof170,Ack25,Mori04,Fold04,Ginter03}.
It was so far straightforward to assume the existence of a long-lived isomeric state in $^{263}$Sg. 
The half-life obtained from the recorded events is
T$_{1/2}$ = 0.33$^{+0.10}_{-0.06}$ s. From the decays of E$_{\alpha}$ $<$ 9.15 MeV reported in 
\cite{GrL94,GrG06,NiH06,NiH10} one obtains  E$_{\alpha}$  = 9.05$\pm$0.04 MeV and T$_{1/2}$ = 0.79$^{+0.37}_{-0.19}$ s,
in agreement with the results reported in \cite{GhN74}. As it is presently unclear which $\alpha$ line represents
the decay of the ground state and which the decay of the isomer, the 9.05-MeV activity will be denoted as $^{263}$Sg(1) 
and the 9.25-MeV activity as $^{263}$Sg(2). Both $\alpha$ transitions are unhindered. Using the half-life formula suggested
by D.N. Poernaru et al. \cite{PoI80}, one obtains T$_{\alpha}$ = 0.73 s for the 9.05-MeV activity and 
T$_{\alpha}$ = 0.197 s for the 9.25-MeV activity. This indicates that they represent decays between analogous levels 
in the mother and daughter nuclei.
The existence of an isomeric state is not unexpected
from predicted low lying levels in $^{263}$Sg \cite{ParS05} as seen in fig. 16. The ground state is predicted as 
11/2$^{-}$[725], while the first excited Nilsson level is 3/2$^{+}$[622]. Spin difference of $\Delta$I\,=\,4
(or even $\Delta$I\,=\,5, if the 1/2$^{+}$[620] level is located below the  3/2$^{+}$[622] one) suggests long
halflives for internal transition ($>>$1 s) and thus the existence of an isomeric state decaying by $\alpha$ emission.
The low lying level scheme predicted for $^{259}$Rf (see fig. 16) suggests also the existence of a longlived
isomer in the daughter nucleus $^{259}$Rf and explains the observed $\alpha$ - $\alpha$ correlations $^{263}$Sg - $^{259}$Rf
(see sect. 5.7). 
As the $\alpha$ transitions are unhindered and internal transitions 11/2$^{-}$[725] $\rightarrow$ 3/2$^{+}$[622] (1/2$^{+}$[620])
are unprobable due to the long lifetimes, one expects, assigning the 9.25-MeV transition
to  $^{263}$Sg(2) only transitions $^{263}$Sg(2)  $\rightarrow$ $^{259}$Rf(2). \\
\begin{figure*}
	\resizebox{0.75\textwidth}{!}{
		\includegraphics{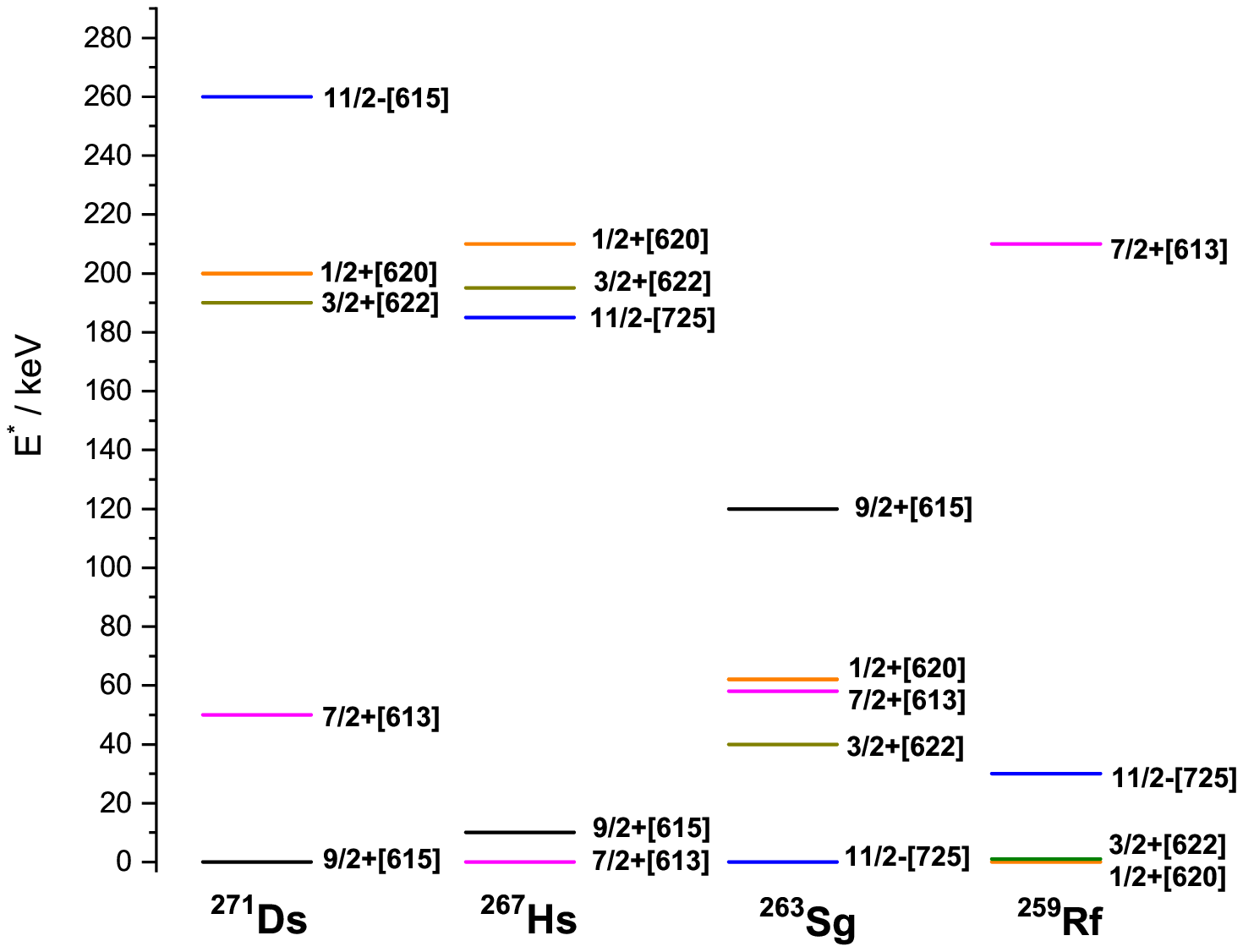}
	}
	\caption{Calculated level schemes for $^{271}$Ds, $^{267}$Hs, $^{263}$Sg, $^{259}$Rf. \cite{ParS05}.}	
	\label{fig:16}       
\end{figure*}
{\bf{$^{265}$Hs}}\\
The isotope $^{265}$Hs was first observed in march 1984 in an irradiation of $^{208}$Pb with $^{58}$Fe at
SHIP, GSI, which marked the discovery of element 108 ('hassium') \cite{MuA84}. An alpha-decay energy of E$_{\alpha}$ = 10.36$\pm$0.03 MeV
and a half-life of T$_{1/2}$ = 1.8$^{+2.2}_{-0.7}$ ms were obtained. More detailed decay data were obtained in experiments at SHIP in
1994, 1997 \cite{Hofm95,HoH97}, 1999 and 2002. The sum of the results of all these experiments were presented in \cite{Hessb09}.  
The analysis clearly shows the existence of two components in the complex $\alpha$ spectrum (see fig. 17) and table 1.\\

\begin{figure*}
	\resizebox{0.75\textwidth}{!}{
		\includegraphics{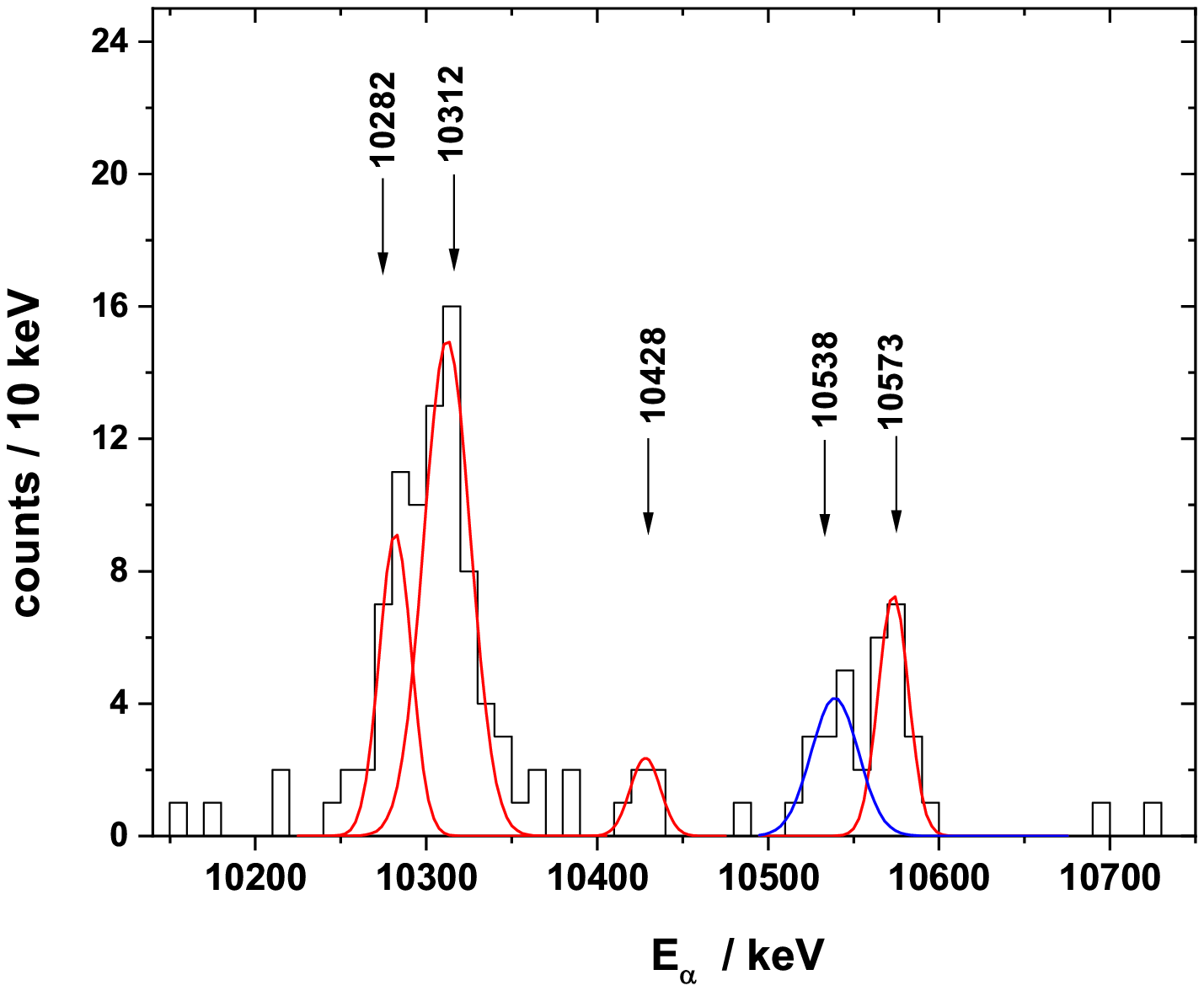}
	}
	\caption{$\alpha$ spectrum observed for  $^{265}$Hs \cite{Hessb09}.}	
	\label{fig:17}       
\end{figure*}
\begin{figure*}
	\resizebox{0.75\textwidth}{!}{
		\includegraphics{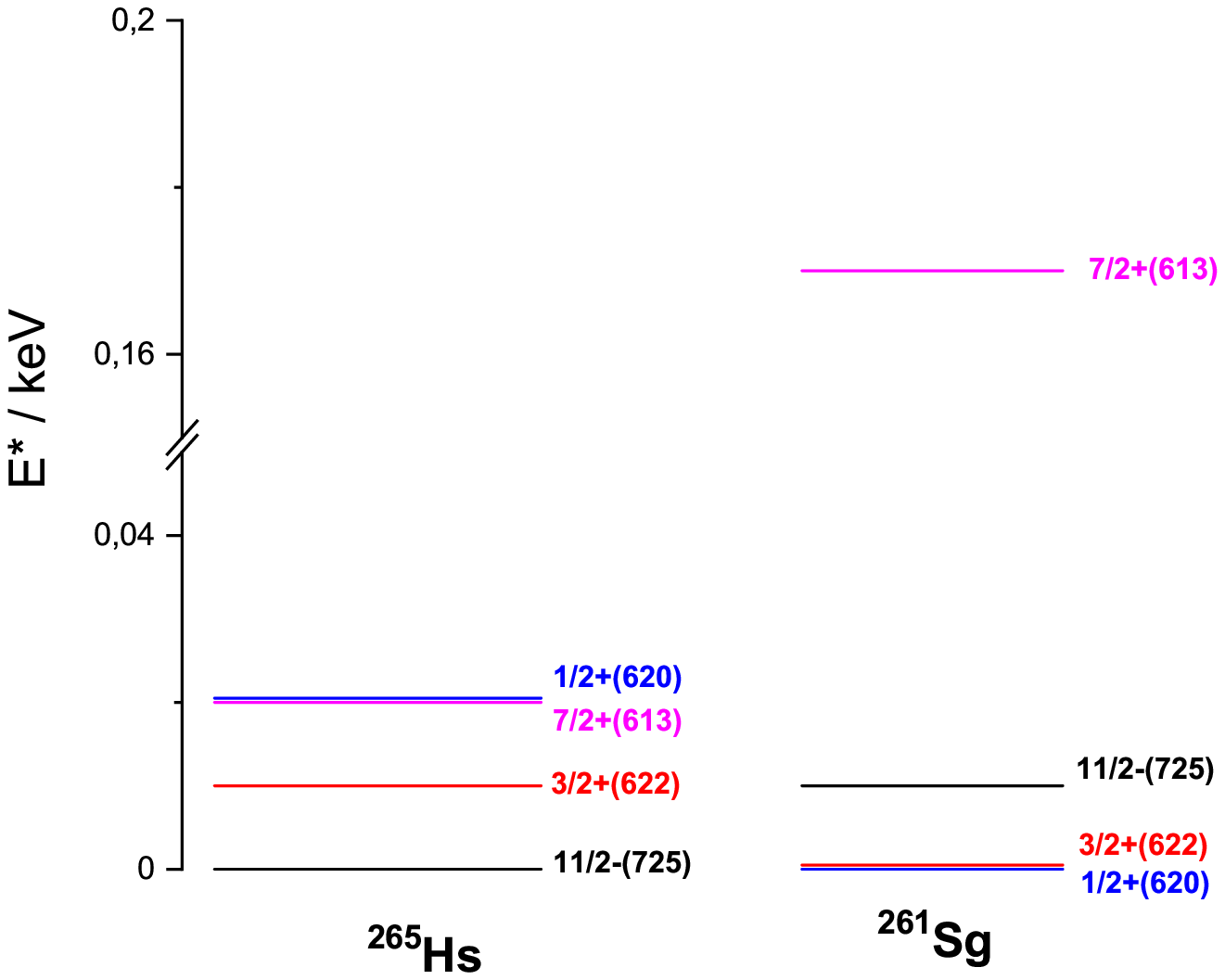}
	}
	\caption{Predicted low lying levels in $^{265}$Hs and $^{261}$Sg \cite{ParS05}.}	
	\label{fig:18}       
\end{figure*}

\begin{table}
	\caption{Decay properties of $^{265}$Hs.}
	\label{tab:1}       
	\begin{tabular}{llll}
      	\hline\noalign{\smallskip}
		Isotope &  E$_{\alpha}$ / keV & T$_{1/2}$ / ms &  T$_{1/2}$ / ms (average)\\
			\hline\noalign{\smallskip}
		$^{265}$Hs(1)   &  10282$\pm$15 & 1.4$^{+0.3}_{-0.2}$ &          \\
		&  10312$\pm$15 & 2.1$^{+0.4}_{-0.3}$ &          \\
		&  10428$\pm$15 & 1.7$^{+1.7}_{-0.6}$ &          \\
		&  10573$\pm$15 & 1.1$^{+0.4}_{-0.2}$ &          \\
		& 10282-10573 &   & 1.9$\pm$0.2   \\
		\hline\noalign{\smallskip}     
		$^{265}$Hs(2)   &  10.538$\pm$15 & 0.3$^{+0.2}_{-0.1}$  &       \\ 		
	\end{tabular}
\end{table}  
The existence of an isomeric state can be understood on the basis of the level scheme
predicted by A. Parkhomenko and A. Sobiczewski \cite{ParS05} (see fig. 18). The ground state of 
$^{265}$Hs is predicted as 11/2$^{-}$[725], the first excited state is 3/2$^{+}$[622]
which was assigned as the ground state in $^{261}$Sg \cite{StH10}, whereas 
according to the prediction the 3/2$^{+}$[622] and 1/2$^{+}$[620] states are
degenerated in energy. So, one decay branch could be interpreted as
$^{265}$Hs (3/2$^{+}$[622]) $\rightarrow$ $^{261}$Sg (3/2$^{+}$[622]),
the other one as $^{265}$Hs (11/2$^{-}$[725])  $\rightarrow$ 
$^{261}$Sg (11/2$^{-}$[725]). As the 11/2$^{-}$[725] level,
however, is predicted in $^{261}$Sg at E$^{*}$ = 10 keV,
due to the large angular momenta difference  $\Delta$L = 4,5 (and thus strong K hindrance) to
the lower lying states it could be expected as a
longlived isomeric state decaying by $\alpha$ emission.
Alpha transitions  $^{261}$Sg (11/2$^{-}$[725])  $\rightarrow$ 
$^{257}$Rf (11/2$^{-}$[725]), however, are not observed. 
In context to the 9$\mu$s isomer in $^{261}$Sg it was suspected
that contrary to the prediction the 11/2$^{-}$[725] might be located
above the 7/2$^{+}$[613] and a decay path 
11/2$^{-}$[725] $^{M2}_{\rightarrow}$ 7/2$^{+}$[613] $^{E2}_{\rightarrow}$ 3/2$^{+}$[622] 
was considered (see sect. 5.7). The non-observation of an $\alpha$ transition between
the 11/2$^{-}$[725]  states in $^{261}$Sg and $^{257}$Rf seems to corroborate 
this assumption.\\
The interpretation of the $\alpha$ spectrum is not straightforward.
A clear assignment to transitions between certain levels in the mother and daughter nuclei
cannot be given on the quality of the present data. This has to be left for further studies.
We will do here with giving some remarks to be taken into account for further studies,
which have to show if they are confirmed or 'falsified'.\\
The theoretical half-life according to \cite{PoI80} for the E$_{\alpha}$ = 10538 keV - line
is T$_{\alpha}$ = 0.31 ms, which results in a hindrance factor HF = 1, i.e. the decay 
represents an unhindered $\alpha$ transition between analogous states in $^{265}$Hs and 
$^{261}$Sg.\\
Relative intensities (i$_{rel}$) for the lines attributed to $^{265}$Hs(1) are 0.78 for E$_{\alpha}$ = 10282/10312 keV - lines
(below reasons why here these two lines are treated as one transition are given), 0.05 for the 
E$_{\alpha}$ = 10428 keV - line, 0.17 for the E$_{\alpha}$ = 10573 keV - line.
Theoretical half-lives \cite{PoI80} and hindrance factors are T$_{\alpha}$ = 1.33 ms, HF = 1.83 for
the E$_{\alpha}$ = 10282/10312 keV - lines, T$_{\alpha}$ = 0.57 ms, HF = 67 for 
the E$_{\alpha}$ = 10428 keV - line, and T$_{\alpha}$ = 0.25 ms, HF = 45 for
the E$_{\alpha}$ = 10573 keV - line using the relation HF = T$_{1/2}$ / (i$_{rel}$ $\times$ T$_{\alpha}$).
So, the E$_{\alpha}$ = 10282/10312 keV - lines represent an unhindered $\alpha$ transition between 
analogous states in $^{265}$Hs and $^{261}$Sg; this could be the transition 
$^{265}$Hs (11/2$^{-}$[725]) $\rightarrow$ $^{261}$Sg (11/2$^{-}$[725]). Assuming the  
11/2$^{-}$[725] as an isomeric one at E$^{*}$ $\approx$ 200 keV, the 
E$_{\alpha}$ = 10538 keV - line could then represent the unhindered transition
$^{265}$Hs (3/2$^{+}$[622]) $\rightarrow$ $^{261}$Sg (3/2$^{+}$[622]). As the
11/2$^{-}$[725] state in $^{261}$Sg was assumed to decay into the 7/2$^{+}$[613] one
predominanty by internal conversion, the E$_{\alpha}$ = 10282 keV - line could represent
the $\alpha$ decay into the 11/2$^{-}$[725]  level, while the  E$_{\alpha}$ = 10312 keV - line
could be the sum line E$_{\alpha}$ = 10312 keV + E$_{CE}$ (11/2$^{-}$[725] $\rightarrow$ 7/2$^{+}$[613] ).
Due to the lifetime of the 11/2$^{-}$[725]  level of 13 $\mu$s the summing effect would be rather small.
An assignment of the  E$_{\alpha}$ = 10428 keV and  E$_{\alpha}$ = 10573 keV - lines, however, is
quite uncertain. One could assume direct $\alpha$ decay 
$^{265}$Hs (11/2$^{-}$[725]) $^{\alpha}_{\rightarrow}$ $^{261}$Sg (7/2$^{+}$[613]). 
Such a transition, however, requires a parity change, for which hindrance factors HF $>$ 100 are expected \cite{SeaL90},
which is in contradiction with the observed hindrance factors for these transitions. 
So an assignment has to be left for further studies.\\
Decay of $^{265}$Hs was recently also studied by N. Sato et al. \cite{Sato11} at the GARIS separator
at RIKEN (Wako, Japan). The data, however, did not deliver deeper insight into the decay of $^{265}$Hs(1)
and $^{265}$Hs(2).\\

\subsection{{\bf 5.9 Isomers in N\,=\,159 isotones}}

{\bf{$^{265}$Sg}}\\
The isotope $^{265}$Sg was first observed by Yu.A. Lazarev et al. \cite{LaL94} who produced it in the reaction $^{248}$Cm($^{22}$Ne,5n)$^{265}$Sg.
They reported a broad $\alpha$-energy distribution  in the range E$_{\alpha}$ = (8.72\,-\,8.91) MeV. A half-life was not obtained.
A few years later using the same reaction A. T\"urler et al. \cite{TuD98} principally confirmed the results of \cite{LaL94}. They reported also a broad 
alpha-energy distribution E$_{\alpha}$ = (8.69\,-\,8.97) MeV, which could be subdivided into four groups (the numbers in brackets give the relative
intensities), E$_{\alpha 1}$ = 8.94 MeV (0.23), E$_{\alpha 2}$ = 8.84 MeV (0.46), E$_{\alpha 3}$ = 8.76 MeV (0.23), E$_{\alpha 4}$ = 8.69 MeV (0.08), 
and a half-life of T$_{1/2}$ = 7.4$^{+3.3}_{-2.7}$ s.
Different decay data were observed, when $^{265}$Sg was produced by $\alpha$ decay of $^{269}$Hs \cite{HoN96,HoH02,MoM07,DuB02,TuD03,DvK05,DvB06,DvB08}.
The energy distribution of the $\alpha$ decays, observed in 'direct' production was not reproduced in the 'indirect' production (via $\alpha$ decay of $^{269}$Hs); also 
in indirect production $\alpha$ decay of $^{265}$Sg was followed to a large extent by spontaneous fission of $^{261}$Rf, which was not the case in 'direct' production.
These observations indicated the existence of two longlived levels in $^{265}$Sg which are populated with different intensities in 'direct' or 'indirect'
production.\\
The data published up to 2008 were reexamined by Ch. D\"ullmann and A. T\"urler \cite{DuT08}, who concluded each two longlived states in $^{265}$Sg and $^{261}$Rf.
The result of their analysis is shown in fig. 19 (upper part).
$^{265}$Sg(1) (E$_{\alpha}$ = (8.8-8.9) MeV, T$_{1/2}$ = 8.9$^{+2.7}_{-1.9}$s) decays predominantly (i$_{rel}$ = 0.8)  into 
$^{261}$Rf(1) (E$_{\alpha}$ = 8.28 MeV, T$_{1/2}$ = 68 s), with a small decay branch $^{265}$Sg(1) $\rightarrow$ $^{261}$Rf(2) (i$_{rel}$ = 0.2),
while $^{265}$Sg(2) (E$_{\alpha}$ = 8.69MeV, T$_{1/2}$ = 16.2$^{+4.7}_{-3.5}$ s) decays predominantly (i$_{rel}$ = 0.88)  into 
$^{261}$Rf(2) (E$_{\alpha}$ = 8.51 MeV, T$_{1/2}$ = 3 s), with a small decay branch $^{265}$Sg(2) $\rightarrow$ $^{261}$Rf(1) (i$_{rel}$ = 0.12).\\
Another detailed decay study of $^{265}$Sg was later performed by H. Haba et al. \cite{Haba12}, which confirmed the conclusions drawn in \cite{DuT08} (see fig. 19, lower part).\\
\begin{figure*}
	\resizebox{0.75\textwidth}{!}{
		\includegraphics{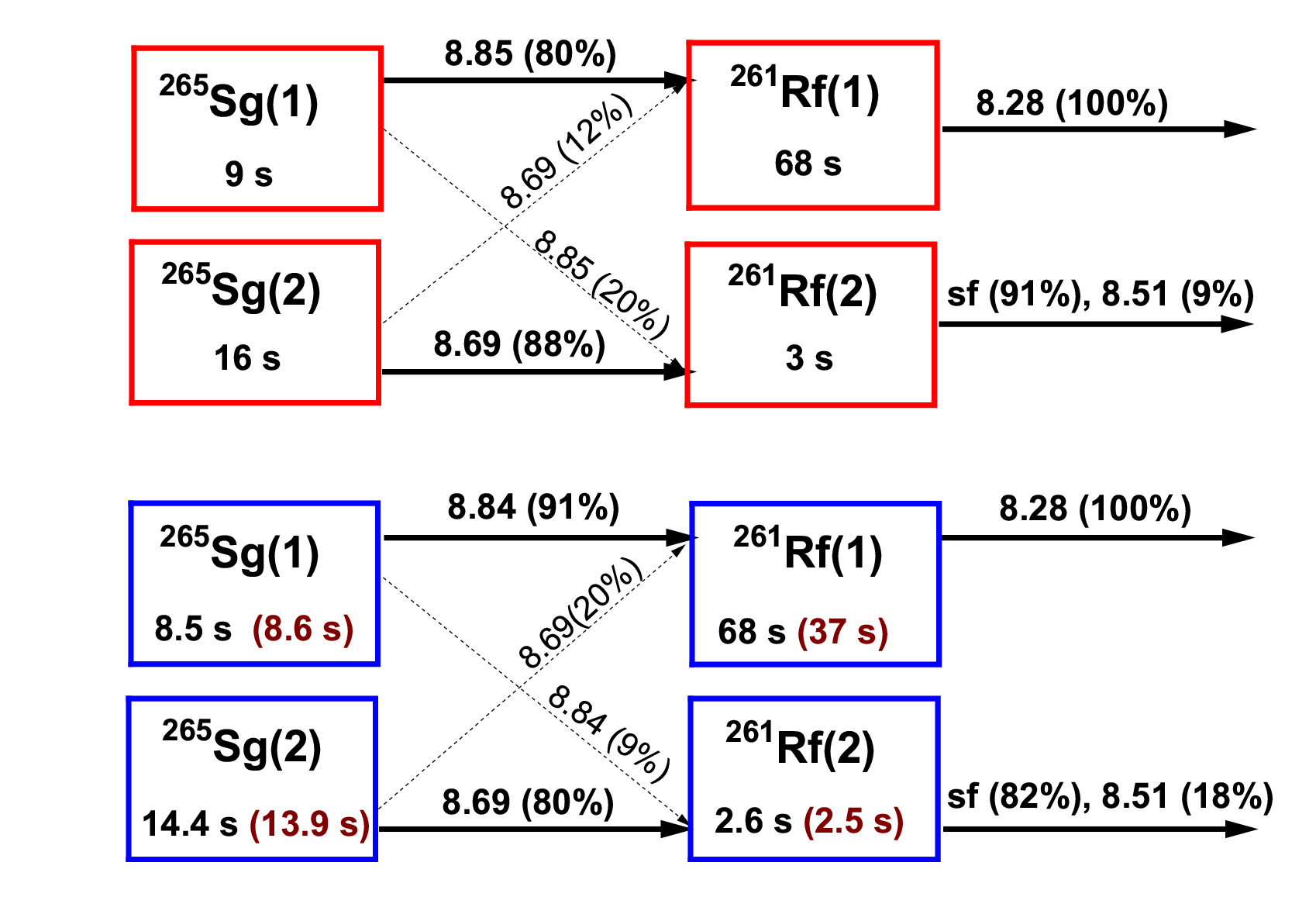}	
	}
	\caption{Decay schemes of $^{265}$Sg(1) and $^{265}$Sg(2); upper part: decay scheme as suggested by 
		Ch. D\"ullmann and A. T\"urler \cite{DuT08}; lower part: decay scheme as suggested by 
		H. Haba et al. \cite{Haba12}.}	
	\label{fig:19}       
\end{figure*}

\begin{figure*}
	\resizebox{0.75\textwidth}{!}{
		\includegraphics{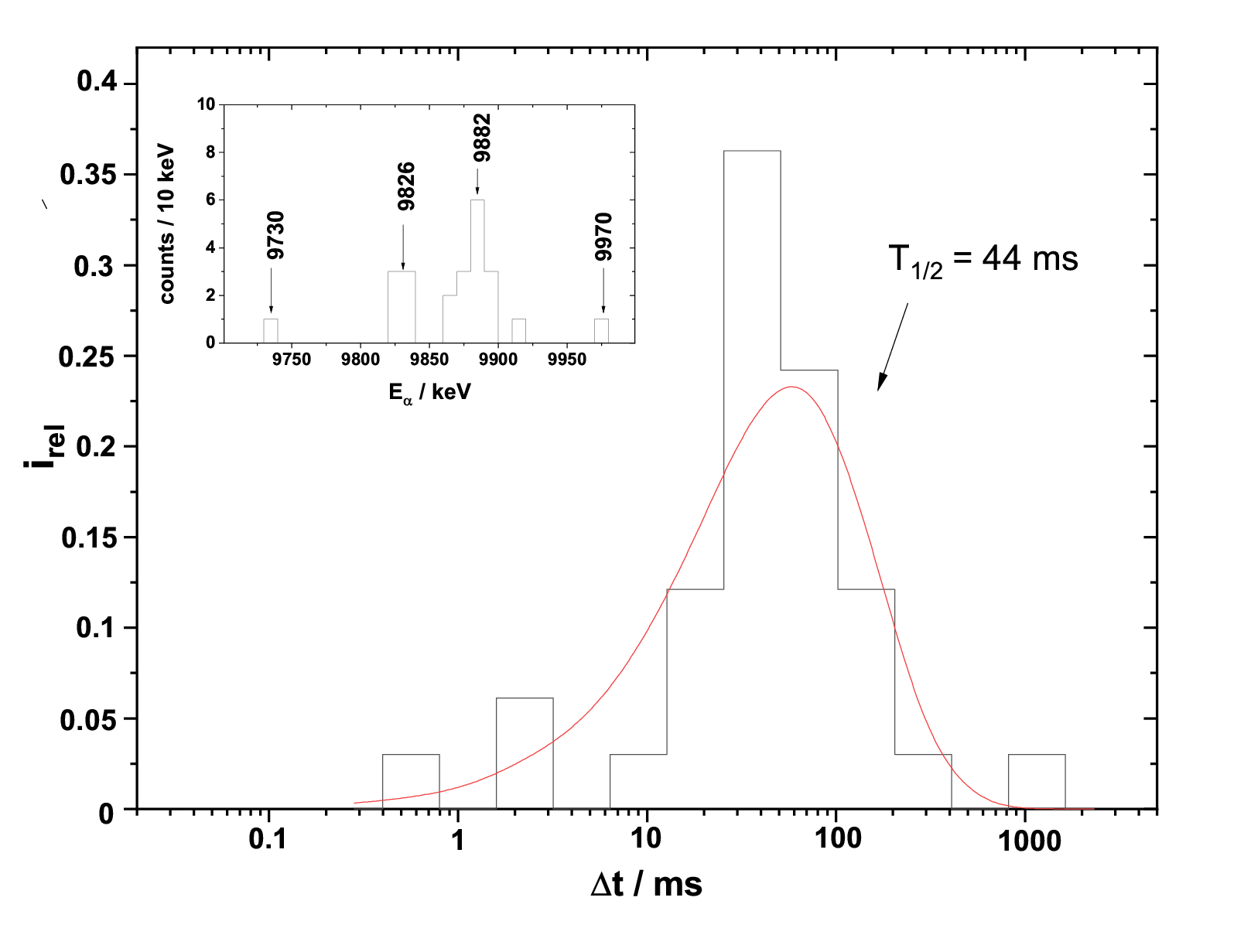}	
	}
	\caption{Time distribution of events assigned to $^{267}$Hs; insert $\alpha$ - spectrum}	
	\label{fig:20}       
\end{figure*}

{\bf{$^{267}$Hs}}\\
Evidence for the existence of an isomeric state in $^{267}$Hs is reported by K. Morita et al. \cite{Mori04} on the basis of a
single $\alpha$ decay chain observed in the reaction $^{64}$Ni + $^{208}$Pb leading to the evaporation residue $^{271}$Ds. The chain consisted
of three $\alpha$ particles interpreted as:\\
$^{271}$Ds (E$_{\alpha}$\,=\,10440 keV, $\Delta$t\,=\,46 ms) $\rightarrow$ $^{267}$Hs (E$_{\alpha}$\,=\,9730 keV, $\Delta$t\,=\,1158.6 ms)$\rightarrow$
$^{263}$Sg (E$_{\alpha}$\,=\,9310 keV, $\Delta$t\,=\,5.999 s). Assuming, however,  that the single event for $^{263}$Sg (9.310 MeV)
	 might be still in-line with the mean energy E$_{\alpha}$\,=\,9.25$\pm$0.02 MeV) the large discrepancy between the time
	 differece and the half-life by a factor of 16 (probability $<$10$^{-5}$) is still striking. \\ 
The $\alpha$ energy of the mother event $^{271}$Ds is extremely low (see sect.
5.10 for more exhaustive discussion), the $\alpha$ energy and lifetime 
 of the second chain member is significantly different (for the lifetime a factor of 26) to the mean values for $^{267}$Hs
(see fig. 20) and the 
$\alpha$ energy of $^{263}$Sg is somewhat higher, the lifetime is considerably longer
than the literature values reported for $^{263}$Sg(1) and $^{263}$Sg(2) (see sect. 5.8).
Remarkebly all three decays seem to be strongly hindered, as can be seen in table 2.
Although it is quite problematic to estimate the half-life on the basis of one event as 
T$_{1/2}$\,=\,ln2$\times$$\Delta$t as done here, but it should be noted that the 
'experimental half-life' is more than an order of magnitude longer than the
calculated one \cite{PoI80,Rur84} in \textbf{all} three cases. 
Calculated level schemes for $^{271}$Ds, $^{267}$Hs and $^{263}$Sg
\cite{ParS05} are shown in fig. 16. 
According to the calculations isomeric states in $^{271}$Ds and $^{267}$Hs can be expected
if the 1/2$^{+}$[620] level is placed below the 3/2$^{+}$[622] level which certainly not 
seems to be unrealistic, while in $^{263}$Sg the 3/2$^{+}$[622] level is expected to be isomeric.
Thus, provided, the 1/2$^{+}$[620] level in $^{271}$Ds being isomeric the following $\alpha$ decay
sequence is possible:\\
$^{271}$Ds (1/2$^{+}$) $^{\alpha 1}_{\rightarrow}$ $^{267}$Hs (1/2$^{+}$) $^{\alpha 2}_{\rightarrow}$ 
$^{263}$Sg (1/2$^{+}$) $^{IT (M1)}_{\rightarrow}$ $^{263}$Sg (3/2$^{+}$).\\
However, in that scenario the decays $\alpha 1$ and $\alpha 2$ would be unhindered 
in contrast to the experimental finding. So the assignment of the (10.44 MeV/46 ms) - transition
of $^{271}$Ds is not straightforeward. Further experimental data are needed to clarify the situation.\\
\begin{table}
	\caption{Hindrance factors for decays of $^{271}$Ds (10.44 MeV), $^{267}$Hs (9.73 MeV), $^{263}$Sg (9.30 MeV) }
	\label{tab:3}       
	\begin{tabular}{llllll}
		\noalign{\smallskip}\hline\noalign{\smallskip}
		Isotope & E$_{\alpha}$/MeV & $\Delta$t / ms &  T$_{1/2}$ / ms & T$_{\alpha}$ / ms & Hf\\  
		\hline\noalign{\smallskip}     
		$^{271}$Ds & 10.44 & 46 & 32 & 2.2 &  14.5 \\
		\hline\noalign{\smallskip}     
		$^{267}$Hs & 9.73 & 1158.6 & 803 & 38 &  21.2 \\
		\hline\noalign{\smallskip}     
		$^{263}$Sg & 9.31 & 5999 & 4100 & 1220 &  34.1 \\
		\hline\noalign{\smallskip} 
	\end{tabular}
\end{table}  

\vspace{5mm}
\subsection{{\bf 5.10 Isomers in N\,=\,161 isotones}}
{\bf $^{267}$Sg}\\
The isotope $^{267}$Sg has been so far only observed as $\alpha$ decay daughter of  $^{271}$Hs, produced in the reaction
$^{248}$Cm($^{26}$Mg,3n)$^{271}$Hs by J. Dvorak et al. \cite{DvB06,DvB08} and as $\alpha$ decay granddaughter of  $^{275}$Ds,
produced in the reaction
$^{232}$Th($^{48}$Ca,5n)$^{275}$Ds by Yu. Ts. Oganessian et al. \cite{OgU24}. The results of the $^{248}$Cm - irradiations were summarized in \cite{DvB08} as
follows: E$_{\alpha}$ = (8.02$\pm$0.05) MeV, T$_{1/2}$ = 80$^{+60}_{-20}$ s, b$_{\alpha}$\,=\,0.17, b$_{sf}$\,=,0.83.\\
The complete results are listed in table 3.
The results of \cite{DvB08} were principally reproduced in \cite{OgU24} but interpreted in a different way. \\
The results of Yu.Ts. Oganessian clearly show two groups of $\alpha$ decays of $^{271}$Hs, which is not so significant in \cite{DvB06,DvB08} due
to worse energy resolution (see sect. 5.11 for further discussion). The group of E$_{\alpha}$ $>$ 9.30 MeV is correlated to $\alpha$ decays of $^{267}$Sg,
that of E$_{\alpha}$ $<$ 9.30 MeV is correlated to SF. Half-life for the $\alpha$ decays (Ch1,Ch6,Ch7,Ch8) is T$_{1/2}$ = 465$^{+465}_{-155}$ s.
Theoretical half-life according to \cite{PoI80} is T$_{\alpha}$ = 300 s; thus the decay has to be regarded as unhindered.\\
For the fission events a half-life T$_{1/2}$ = 84$^{+51}_{-23}$ s is obtained\footnote{Disregarding the extreme low time differences in Ch10, Ch11 one obtains
	a somewhat larger value T$_{1/2}$ = 117$^{+94}_{-36}$s}, which suggests that $\alpha$ decays and SF represent different activities.\\
\begin{table}
	\caption{Summary of correlations $^{271}$Hs $\rightarrow$ $^{267}$Sg. }
	\label{tab:3}       
	\begin{tabular}{lllll}
		\hline\noalign{\smallskip}
		\noalign{\smallskip}\hline\noalign{\smallskip}
		&   $^{271}$Hs &   &  $^{267}$Sg &   \\
		No. / Reference &  E$_{\alpha}$/MeV  &  $\Delta$t / s & E$_{\alpha}$/MeV  &  $\Delta$t / s  \\
		Ch1 \cite{DvB06} & 9.30  &    & 8.20  & 149  \\
		Ch2 \cite{DvB08} & 9.14  &    & (SF)  & 47.9  \\
		Ch3 \cite{DvB08} & 9.16  &    & (SF)  & 142. \\
		Ch4 \cite{DvB08} & 9.02  &    & (SF)  & 30.4 \\
		Ch5 \cite{DvB08} & 9.23  &    & (SF)  & 264. \\
		\hline\noalign{\smallskip}
		Ch6 \cite{OgU24} & 9.396$\pm$0.040 & 2.0483 & 8.281$\pm$0.019 & 761.6 \\
		Ch7 \cite{OgU24} & 9.375$\pm$0.037 & 142.74 & 8.258$\pm$0.020 & 805.5 \\
		Ch8 \cite{OgU24} & 9.318$\pm$0.018 & 4.3221 & (5.387) & 967.49 \\
		Ch9 \cite{OgU24} & 9.058$\pm$0.042 & 17.5283 & (SF) & 358.126 \\
		Ch10 \cite{OgU24} & 9.053$\pm$0.020 & 8.7446 & (SF) & 1.635\\
		Ch11 \cite{OgU24} & 9.018$\pm$0.020 & 4.3221 & (SF) & 1.1811\\
	\end{tabular}
\end{table}  

{\bf $^{269}$Hs}\\
On the basis of two decay chains observed in irradiations of $^{238}$U with $^{40}$Ar and attributed the decay of the ground state of $^{273g}$Ds (11/2$^{-}$[725])
and an isomeric state $^{273m}$Ds (1/2$^{+}$[620]) Yu.Ts. Oganessian et al. \cite{OgU24} postulated 
a 1/2$^{+}$[620] isomer in $^{269}$Hs at E$^{*}$ $\approx$ 20 keV. As their postulation
of an isomeric 1/2$^{+}$[620] - state in $^{273}$Ds is seemingly dubious (see sect. 5.11 for detailed discussion) the claim of  an isomer $^{269}$Hs seems 
presently dubious too.\\

{\bf $^{271}$Ds}\\
The only case in N\,=\,161 even-Z isotones for which real evidence for an isomeric state was found so far is $^{271}$Ds. It was first produced at SHIP, GSI in the reaction
$^{64}$Ni + $^{208}$Pb. Besides the assumed ground state decay with a half-life of T$_{1/2}$ = 1.1$^{+0.6}_{-0.3}$ ms, strong evidence for a second activity of T$_{1/2}$ = 56$^{+270}_{-26}$ ms
was found \cite{Hofmann1995a,Hofmann2000}. Further decay data on that activities were obtained in SHIP experiments performed later \cite{Hess2024,Acker2024}, at the BGS, Berkeley 
\cite{Ginter03,Fold04} and at the GARIS - Separator at RIKEN \cite{Mori04}. Alpha - decay energies are plotted versus the decay times (time difference between
implantation of the evaporation residue (ER) and $\alpha$ decay of $^{271}$Ds, $\Delta$t(ER-$\alpha$)) are shown in fig. 21a.
\begin{figure*}
	\resizebox{0.75\textwidth}{!}{
		\includegraphics{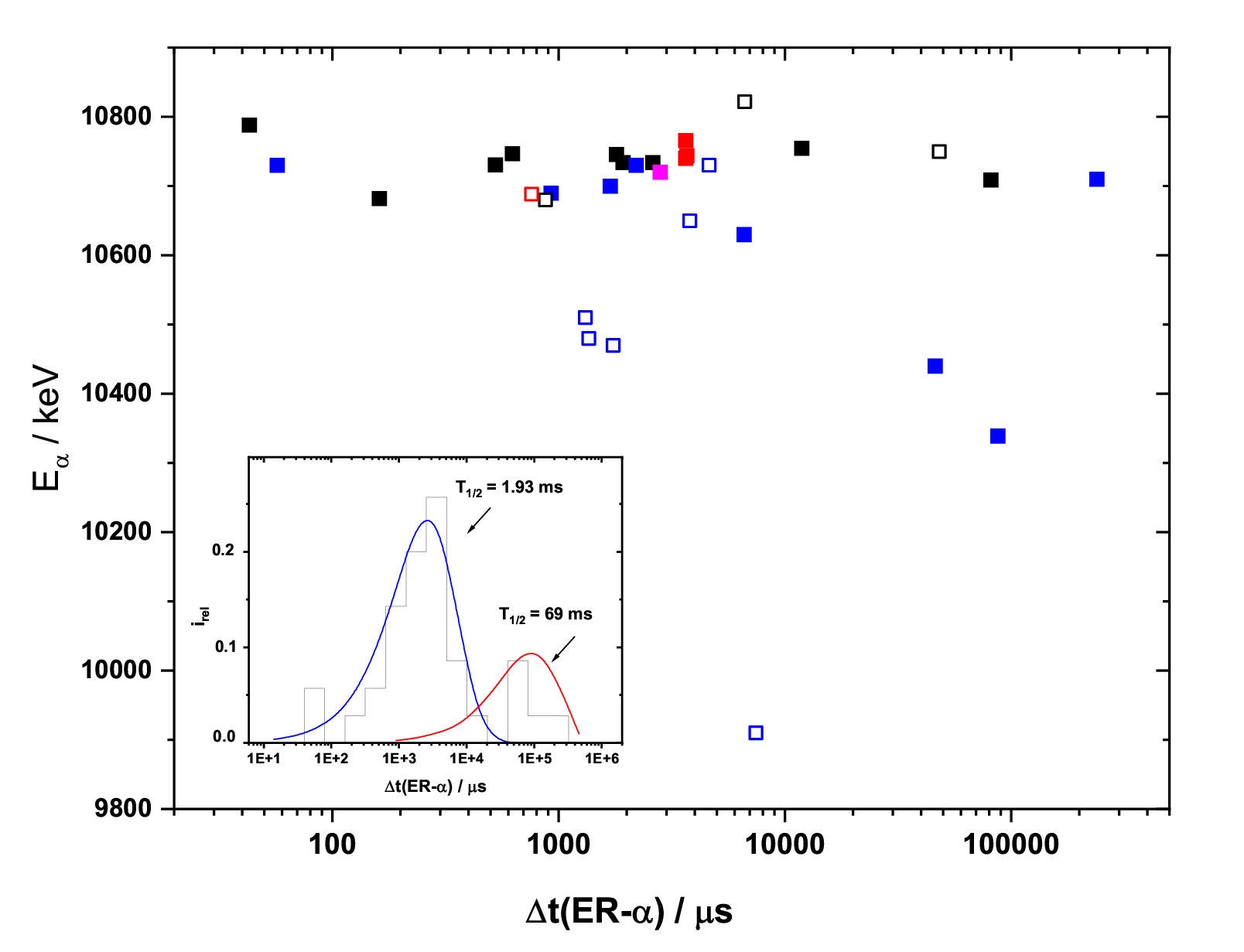}	
	}
	\caption{Two dimensional plot $\Delta$t(ER-$\alpha$) versus E$_{\alpha}$ of events attributed to the decay 
	of $^{271}$Ds; black: SHIP data \cite{Hofmann1995a,Hofmann2000,Hess2024,Acker2024}; 
	magenta: BGS data \cite{Ginter03}; red: BGS data \cite{Fold04}; blue: GARIS data \cite{Mori04}; insert:
	Time distribution ($\Delta$t(ER-$\alpha$)) of events attributed to decays of $^{271}$Ds; blue smooth line:
		time distribution for a half-life of T$_{1/2}$\,=\,1.93 ms, red smooth line:
		time distribution for a half-life of T$_{1/2}$\,=\,69 ms. 
	}
	
	\label{fig:21}       
\end{figure*}
\begin{figure*}
	\resizebox{0.80\textwidth}{!}{
		\includegraphics{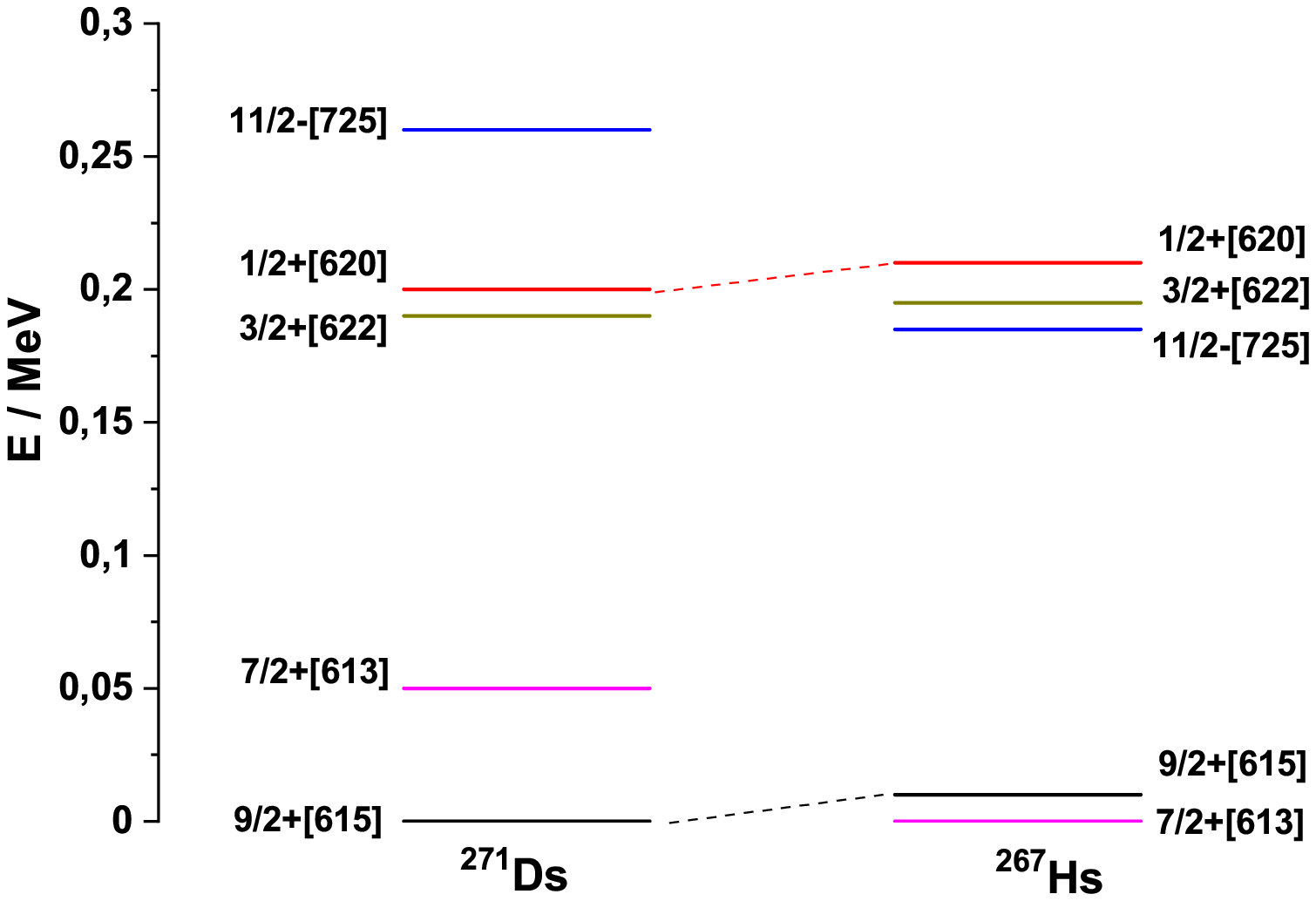}	
	}
	\caption{Low lying Nilsson levels in $^{271}$Ds and $^{267}$Hs predicted by A. Parkhomenko and A. Sobiczewski \cite{ParS05}.
	}
	
	\label{fig:22}       
\end{figure*}
Seemingly there is no real connection between the $\alpha$ energies and the time differences except for two cases reported by 
K. Morita et al. \cite{Mori04} (E$_{\alpha}$\,=\,10.44 MeV, $\Delta$t(ER-$\alpha$)\,=\,46 ms; E$_{\alpha}$\,=\,10.33 MeV,  $\Delta$t(ER-$\alpha$)\,=\,87.1 ms). 
The latter event, however, was observed as a single event, i.e. no subsequent decays of $^{267}$Hs, $^{263}$Sg, $^{259}$Rf, $^{255}$No were observed, while for the
first event energies and time differences for $^{267}$Hs (E$_{\alpha}$\,=\,9.73 MeV, $\Delta$t(ER-$\alpha$)\,=\,1158.6 ms), and $^{263}$Sg
(E$_{\alpha}$\,=\,9.310 MeV, $\Delta$t(ER-$\alpha$)\,=\,5998.9 ms) differ significantly from the decay values of these isotopes in the other decay chains
and $\alpha$ decay of $^{259}$Rf or $^{255}$No was not observed. Thus the assignment of both decay chains appears questionable.   
(Here only events with full energy release in the
stop detector are considered due to worse energy resolution for 'sum events').
The time distribution of the events is shown in fig. 21b; clearly a long lived component is visible and clearly separated from the 'short' lived events.
Taking all events $\Delta$t\,$<$\,12 ms as the 'short-lived' component, one obtains a half-life T$_{1/2}$\,=\,1.93$^{+0.43}_{-0.30}$ ms. For the 
long-lived component one obtains  T$_{1/2}$\,=\,85$^{+116}_{-31}$ ms taking all events, or  T$_{1/2}$\,=\,69$^{+52}_{-21}$ ms, disregarding the 10.44 MeV and 
10.33 MeV events. Tentatively we assume a single energy for 'all' events resulting in E$_{\alpha}$\,=\,10.74$\pm$0.01 MeV; the theoretical half-life for
$^{271}$Ds events at this energy is T$_{\alpha}$\,=\,0.44 ms according to the formula suggested by D.N. Poenaru et al. \cite{PoI80,Rur84}. 
The low energy level schemes calculated by A. Parkhomenko and A. Sobiczewski \cite{ParS05} are shown in fig. 22. Provided the 1/2$^{+}$[620] Nilsson level lying
below the 3/2$^{+}$[622] level, indeed due to an angular momentum difference $\Delta$l\,=\,3,4 to the lower lying 7/2$^{+}$[613] and 9/2$^{+}$[615] levels a low
lying isomeric 1/2$^{+}$ state could be expected. However, such an isomer could decay via an unhindered $\alpha$ transition into the corresponding level
in $^{267}$Hs, as indicated by the dashed line in fig. 22. This, however, is in contradiction with the long half-life of the isomeric state. Even for an
$\alpha$ transition energy of 10.40 MeV one expects an $\alpha$ half-life  of 2.85 ms \cite{PoI80,Rur84}. This, however, means, a 1/2$^{+}$ isomer may exist in $^{271}$Ds, but it is expected to have similar half-life (and probably also a similar $\alpha$ decay energy)
as the ground state, but certainly is in conflict with the long lifetime. Altough the existence of an isomer in $^{271}$Ds has to be regarded as certain,
its configuration is unclear. Thus it is tempting to speculate on the existence of a 3-quasi-particle K isomer, decaying by
internal transitions to the ground state.This would be the heaviest K isomer observed so far. It thus seems of high interest to investigate the decay of $^{271}$Ds in more detail.\\
\subsection{{\bf 5.11 Isomers in N\,=\,163 isotones}}
{\bf $^{271}$Hs}\\
The isotope $^{271}$Hs was first observed as the 3n - deexcitation channel in bombardments of  $^{248}$Cm with $^{26}$Mg 
by J. Dvorak et al. \cite{DvB06,DvB08}.
The observed events are listed in table 3. Two $\alpha$ lines (E$_{\alpha 1}$ = 9.13$\pm$0.05 MeV, E$_{\alpha 2}$ = 9.30$\pm$0.05 MeV) were  
attributed to this isotope. The half-life was not measured, a value of $\approx$4 was estimated. More decays of $^{271}$Hs, produced by 
$\alpha$ decay of $^{275}$Ds,  have recently been
reported by Yu.Ts. Oganessian et al. \cite{OgU24}. The data are listed in table 3. The two $\alpha$ lines reported earlier were confirmed, 
although energies are somewhat different ( E$_{\alpha 1}$ = 9.05$\pm$0.02 MeV, E$_{\alpha 2}$ = 9.34$\pm$0.02 MeV)
in addition the half-life could be established. Values of  T$_{1/2}$ = 7.1$^{+8.4}_{-2.5}$s for  E$_{\alpha 1}$ and   T$_{1/2}$ = 46$^{+56}_{-16}$ s
for E$_{\alpha 2}$ were obtained. It should, however, be noted that the latter value was mainly due to the extreme long correlation time of Ch7 in table 3. 
Neglecting this value a half-life of $\approx$ 2.2 s can be estimated. Nevertheless on the basis of the correlated daughter events (see sect. 5.10)
it seems legal to assign them to different activites.\\
The decay pattern suggested in \cite{OgU24} is shown in Fig.23. The ground state of $^{275}$Ds is assumed as 3/2$^{+}$[622].
The $\alpha$ decay populates the correspondig level in $^{271}$Hs, which is isomeric in that isotope. It decays partly by $\alpha$ emission into
the correponding level in $^{267}$Sg, which decays by internal transition into the isomeric 1/2$^{+}$[620] state, that undergoes spontaneous fission.
A second decay mode of the 3/2$^{+}$[622] isomeric state in $^{271}$Hs is decay by internal transitions (probably in several steps) into the
11/2$^{-}$[725] ground state,  which then decays by $\alpha$ emission into the corresponding state in $^{267}$Sg, followed by an internal
transition into the ground state, which then undergoes $\alpha$ decay.\\

\begin{figure*}
	\resizebox{0.90\textwidth}{!}{
		\includegraphics{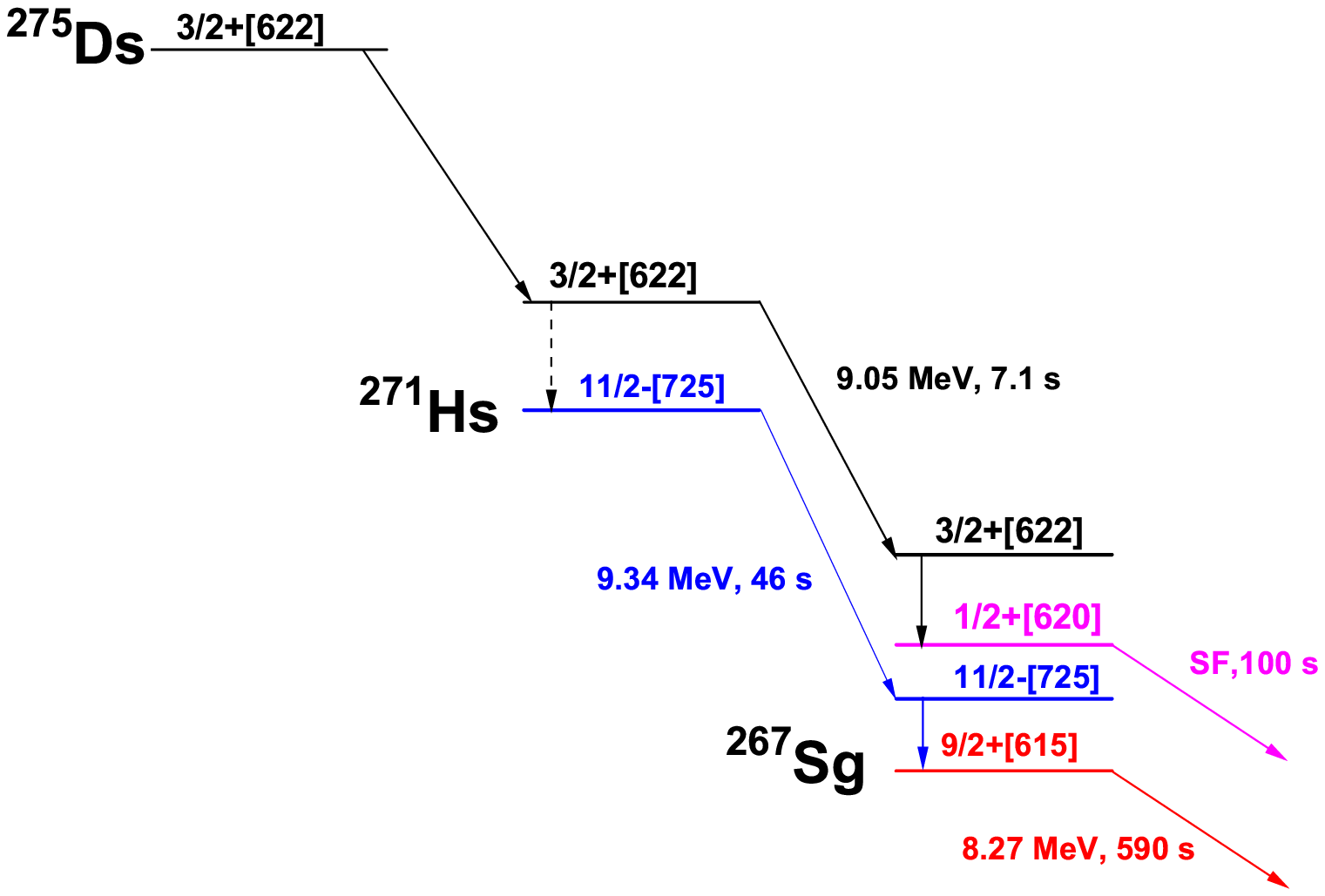}	
		 }
	\caption{Decay pattern of $^{275}$Ds and its daughter products as suggested in \cite{OgU24}.
	}
	
	\label{fig:23}       
\end{figure*}

{\bf $^{273}$Ds}\\
The isotope $^{273}$Ds was first observed unambiguously as $\alpha$-decay daughter of $^{277}$Cn in bombardments of
$^{208}$Pb with $^{70}$Zn ions at SHIP, GSI \cite{HoN96,HoH02}. Additional data using the same production method were 
obtained by K. Morita et al. \cite{MoM07} and by T. Sumita et al. \cite{Sumi13}.The data are presented in table 4.  Summarizing the results of these experiments
one obtains an $\alpha$ - decay energy E$_{\alpha}$ = 11.14 $\pm$0.05 MeV and a half-life T$_{1/2}$ = 0.19$^{+0.14}_{-0.06}$ ms.\\
In a recent experiment, performed at FLNR - JINR, Dubna Yu. Ts. Oganessian et al. \cite{OgU24} observed
in a bombardment of $^{238}$U with
$^{40}$Ar - projectiles two decay chains which they attributed to the decay of $^{273}$Ds (Chains Ch6 and Ch7 in table 4).\\
These  chains, however, are not free of ambiguities. The $\alpha$ energies attributed to $^{265}$Sg are significantly lower than the literature 
values of E$_{\alpha}$ = 8.840 MeV ($^{265}$Sg(1)) and E$_{\alpha}$ = 8.690 MeV ($^{265}$Sg(2)) (see sect. 5.9). Also the decay time of $^{273}$Ds (41.703 m) 
in chain 7 is a factor $\approx$290 longer than the theoretical $\alpha$ half-life T$_{\alpha}$ = 0.145 ms \cite{PoI80}.\\
Nevertheless Yu.Ts. Oganessian et al. assign chain 6 to an isomer with the configuration 1/2$^{+}$[620], settled at an excitation energy of 
E$^{*}$ $\approx$ 40 keV decaying into the corresponding level in $^{269}$Hs (this isomer would be populated by the decay of $^{277}$Cn), while 
despite the strong hindrance they assign chain 7  to the transition $^{273}$Ds (11/2$^{-}$[725]) $^{\alpha}_{\rightarrow}$ 
$^{269}$Hs (11/2$^{-}$[725]) $^{IT}_{\rightarrow}$ $^{269}$Hs (9/2$^{+}$[615]) with the latter interpreted as the ground state as predicted
by A. Parkhomenko and A. Sobiczewski \cite{ParS05}. Such an $\alpha$ transition, however, is unhindered thus in conflict with a hindrance factor HF $\approx$ 290.
On the other hand, the $\alpha$ decay in chain 7 cannot represent the decay into the ground state of $^{269}$Hs 
($^{273}$Ds (11/2$^{-}$[725]) $^{\alpha}_{\rightarrow}$ 
$^{269}$Hs (9/2$^{+}$[615]) or the into the 7/2$^{+}$[613] level also predicted below the 11/2$^{-}$[725] state \cite{ParS05}, as those 
transitions require a change in parity and a spin-flip expecting hindrance factors HF $>>$ 1000. \\
Under these considerations not only the chains themselves but also the assignment to decays of the ground state and an isomeric state are
quite dubious.\\

\begin{table}
	\caption{decay chains attributed to start either from $^{277}$Cn or from $^{273}$Ds.}
	\label{tab:4}       
	\begin{tabular}{lllllllllll}
		\hline\noalign{\smallskip}
		\noalign{\smallskip}\hline\noalign{\smallskip}
		&  & $^{277}$Cn & &  $^{273}$Ds &  &  $^{269}$Hs & &  $^{265}$Sg & &  $^{261}$Rf   \\
		No. / Reference &  E$_{\alpha}$ & T$_{1/2}$/ms & E$_{\alpha}$ & T$_{1/2}$/ms & E$_{\alpha}$ & T$_{1/2}$/s & E$_{\alpha}$ & T$_{1/2}$/s & E$_{\alpha}$ & T$_{1/2}$/s  \\
		Ch1 \cite{HoH02} & 11.45 & 0.280 & 11.08 & 0.110 & 9.23 & 19.7 & (4.60) &7.4 & 8.52 & 4.7 \\
		Ch2 \cite{HoH02} & 11.17 & 1.406 & 11.20 & 0.310 & 9.18 & 22.0 & (0.2) & 18.8 & (SF) & 14.5 \\
		Ch3 \cite{MoM07} & 11.09 & 1.10 & 11.14 & 0.520 & 9.17 & 14.2 & 8.71 & 23.0 & (SF) 2.97 \\
		Ch4 \cite{MoM07} & 11.32 & 1.22 & 11.15 & 0.0399 & 9.25 & 0.270 & 8.70 & 79.0 & (SF) 8.30 \\
		Ch5 \cite{Sumi13} & 11.07 & 0.370 & 11.03 & 0.373 & 9.15 & 36.0 & 8.66 & 13.8 & (SF) & 3.73 \\
		Ch6 \cite{OgU24} &   &   & 11.017 & 0.184 & - & - & 8.397 & 9.0766 & (SF) & 3.0344 \\
		Ch7 \cite{OgU24} &   &   & 10.929 & 41.703 & (0.342) & 4.0978 & 8.509 & 58.19 & (SF) & 0.3923 \\
	\end{tabular}
\end{table}  
\subsection{{\bf 5.13 Isomers in odd-mass isotopes of even-Z elements Z\,$>$110 }}

{\bf $^{289}$Fl}\\
Data on $^{289}$Fl are scarce. First claim on the synthesis of this isotope came from an irradiation of $^{244}$Pu with
$^{48}$Ca projectiles performed at the DGFRS, FLNR-JINR, Dubn by Yu.Ts. Oganessian et al. \cite{OgU99}, who 
observed one decay chain and assigned a decay energy E$_{\alpha}$ = 9.71 MeV and a decay time (time difference between
the implantation of the ER and the $\alpha$ decay) $\Delta$t\,=\, 30.4 s to this isotope. This decay chain, however, never could be
reproduced nor confirmed.\\
A couple of years later Yu.Ts. Oganessian et. al. \cite{OgU04} on the basis of eight observed decay chains reported
different decay data for this isotope: E$_{\alpha}$ = 9.82$\pm$0.06 MeV, T$_{1/2}$ = 2.7$^{+1.4}_{-0.7}$ s
(see also table 5).\\
Few years later, in an irradiation of $^{244}$Pu with $^{48}$Ca at the gas-filled separator TASCA at GSI, Darmstadt,
three decay chains attributed to the decay of $^{289}$Fl were obsered. The results are given in table 5.
In two further experiments \cite{HoH12,MoM14} $^{289}$Fl was produced by $\alpha$ decays of $^{293}$Lv, synthesized
in bombardments of $^{248}$Cm with $^{48}$Ca. In the SHIP experiment at GSI \cite{HoH12} 
one decay chain of $^{293}$Lv was observed\footnote{In \cite{HoH12} two decay chains of $^{293}$Lv were reported. One of them, however, later was assigned to
$^{291}$Lv \cite{HoH16}.} In the GARIS experiment at RIKEN two decay chains were reported. The results are given in
table 5.

\begin{table}
	\caption{Decay properties of $^{289}$Fl reported before 2023}
	\label{tab:5}       
	\begin{tabular}{llllllll}
		\hline\noalign{\smallskip}
		\noalign{\smallskip}\hline\noalign{\smallskip}
		DGFRS \cite{OgU04} &  & TASCA \cite{DuS09,Gates2011} &   & SHIP \cite{HoH12} &    & GARIS \cite{MoM14} & \\  
		\hline\noalign{\smallskip}     
		E$_{\alpha}$/MeV & T$_{1/2}$/s & E$_{\alpha}$/MeV & T$_{1/2}$/s &	E$_{\alpha}$/MeV & $\Delta$t/s & 
			E$_{\alpha}$/MeV & $\Delta$t/s \\
				\hline\noalign{\smallskip}     
		9.82$\pm$0.06 & 2.7$^{+1.4}_{-0.7}$ & 9.87$\pm$0.03 & 0.97$^{+0.97}_{-0.032}$ & 
	9.818 & 0.118 & 9.89 & 3.97 \\
	&  &  & & & & 9.72 & 0.66 \\
		\hline\noalign{\smallskip}     
	\end{tabular}
\end{table}  

Evidently there are signifikant differences in the energies and half-lives between the different publications,
which, however, cannot necessarily be regarded as striking differences, as at this low number of registered events
large fluctuations the half-lives may occur, while energy differences may be due to calibration purposes.\\
In a recent experiment at the gas-filled separator TASCA at GSI, $^{289}$Fl was again procuced by the reaction
$^{244}$Pu($^{48}$Ca,3n)$^{289}$Fl \cite{CoS23}. In this experiment 15 decay chains (in three chains, however, the 
$\alpha$ decay of $^{289}$Fl were missing) were attributed to start from $^{289}$Fl \cite{CoS23}. The energies were essentially 
distributed in the energy range E\,=\,(9.79\,-\,9.91) MeV, with two cases of significant deviations, E(1)\,=\,9.56$\pm$0.01 MeV
and E(2)\,=\,9.99$\pm$0.02 MeV, with the latter energy reconstructed from an energy loss signal in the implantation detector and 
the residual energy measured with the 'box detector'. The difference of the latter energy value from the 'mean' value 
E$_{\alpha}$\,=\,9.80$\pm$0.2 MeV. 
(In \cite{CoS23} the Q$_{\alpha}$(value), Q$_{\alpha}$\,=\,9.94$\pm$0.02, is given.)
was regarded as striking and ascribed to the decay of a different state, as also the $\alpha$ 
energy of the daughter nuclues $^{285}$Cn
was E\,=\,9.30$\pm$0.02 MeV thus  higher than the 'bulk' of the 
decay energies, which were in the range E\,=\,(9.09\,-\,9.18) MeV.
(Also a value reconstructed from an energy loss signal in the implantation detector and 
the residual energy measured in the 'box detector'.)
So it was not incorporated into the main decay branch of $^{289}$Fl.\\
On the other side in \cite{CoS23} four decays of 'higher' energies from the previous studies \cite{Gates2011}
due the similar half-life (see table 5)
disregarding probable discrepancies in calibration, were regarded as possible decays of the isomeric 
 state located at E$^{*}$ $\approx$ 70 keV. Due to the significant difference in the energy, however,
 we will present here the decay properties as given in \cite{CoS23} (see also fig. 3b there):\\
$^{289g}$Fl: E$_{\alpha}$ = 9.80$\pm$0.02 MeV, T$_{1/2}$ = 2.5$^{+0.8}_{-0.5}$ s,\\
$^{289m}$Fl: E$_{\alpha}$ = 9.99$\pm$0.02 MeV, T$_{1/2}$ = 1.1$^{+1.1}_{-0.4}$ s.\\
As possible spins and parities I$^{\pi}$ = 15/2$^{-}$ for the ground state and I$^{\pi}$ = 3/2$^{+}$ or I$^{\pi}$ = 1/2$^{+}$
were considered. However, with respect to possible problems with calibration,
the assumption of an isomeric state in $^{289}$Fl seems presently rather tentative.\\

\section{6. Isomers in odd-mass odd-Z nuclei}
\subsection{{\bf 6.1 Isomeric states in odd-mass mendelevium isotopes}}

{\bf $^{245}$Md}\\
The isotope $^{245}$Md was first synthesized by V. Ninov et al. \cite{NinH96} in the reaction $^{209}$Bi($^{40}$Ar,4n)$^{245}$Md 
at SHIP, GSI. Two activities were observed: a) a spontaneous fission activity of T$_{1/2}$ = 0.90$\pm$0.025 ms, and
b) an $\alpha$ activity of E$_{\alpha 1}$ = 8680$\pm$20 keV, E$_{\alpha 2}$ = 8620$\pm$20 keV, and a half-life of  
T$_{1/2}$ = 0.35$^{+0.23}_{-0.16}$ s. Based on calculations of J. Randrup et al. \cite{Rand73}, which predicted lower 
fission barriers for low-spin states in odd-mass nuclei and on calculations of low lying Nilsson levels in mendelevium isotopes
by S. Cwiok et al. \cite{Cwiok94}, the spontaneous fission activity was assigned to the ground state with the configuration
1/2$^{-}$[521], and the $\alpha$ activity to an isomeric state with the configuration 7/2$^{-}$[514].\\
The results of \cite{NinH96} were confirmed in an experiment at TASCA, GSI by J. Khuyagbaatar et al. \cite{Khuy20a}, who 
produced it in the reaction $^{197}$Au($^{50}$Ti,2n)$^{245}$Md. On the basis of the existing data configuration assigment of
\cite{NinH96} are still valid. Since, however, detailed decay studies of heavier odd-mass mendelevium isotopes resulted in the 
Nilsson level 7/2$^{-}$[514] as the ground state \cite{Hess05}, it seems meaningful also to assign the latter to the
ground state of $^{245}$Md and the 1/2$^{-}$[521] level to the isomer.\\

{\bf $^{247}$Md}\\
The isotope $^{247}$Md was first observed in an irradiation of $^{209}$Bi with $^{40}$Ar at SHIP, GSI.
On the basis of five decay events an $\alpha$ energy of E$_{\alpha}$ = 8424$\pm$25 keV and a half-life 
T$_{1/2}$ = 2.9$^{+1.7}_{-1.2}$ s were reported \cite{Muen81}. 
More detailed decay data were obtained in later studies, including $\gamma$ spectroscopic investigations,
also performed at SHIP \cite{Hess05,Hofm94,AntH10,Hess22}. Besides the activity reported in \cite{Muen81}, 
which was assigned to the 7/2$^{-}$[514] ground state (E$_{\alpha}$\,=\,8421$\pm$1 keV + some weaker lines, 
T$_{1/2}$ = 1.20$\pm$0.12 s\footnote{A reanalysis of the data published in \cite{Muen81} showed that the 
long half-life published there was due to one event with an accidentially long correlation time
$\Delta$t(ER-$\alpha$).},  b$_{sf}$ = 0.0086$\pm$0.0010),  on the basis of $\alpha$ decay and $\alpha$ - $\gamma$ coincidence measurements
a second activity with a half-life T$_{1/2}$ = 0.23$\pm$0.3 s and an SF branch
b$_{sf}$ = 0.20$\pm$0.02 was observed. It was attributed to an isomeric state with the configuration
1/2$^{-}$[521]. Besides the main $\alpha$ transitions, a rather broad line centered at E$_{\alpha}$\,=\,8720 keV,
a weak $\alpha$ - $\gamma$ - coincidence (E$_{\alpha}$\,=\,8451$\pm$11 keV, E$_{\gamma}$\,=\,342.1$\pm$1.1 keV)
was observed and interpreted as the decay $^{247m}$Md $^{\alpha}_{\rightarrow}$ $^{243}$Es$^{*}$ $^{\gamma}_{\rightarrow}$ 
$^{243}$Es(gs). On the basis of the decay energy balance for the ground state ($^{247g}$Md) and the isomeric state ($^{247m}$Md)
the excitation energy of the isomer was estimated as E$^{*}$ = 153 keV \cite{Hess22}.\\
The excitation energy of the isomer was modified in a recent study of S.Y. Xu et al. \cite{Xu25}. Using digital  electronics
on the basis of delayed coincidences between (E$_{\alpha}$=8416 keV, E$_{\gamma}$=209 keV) and CE the 7/2$^{+}$[633] level
in the daughter $^{243}$Es
populated by the E$_{\gamma}$=209 keV transition and located at E$^{*}$ = 10 keV in \cite{Hess22} was identified as an
isomeric state of T$_{1/2}$ = 5.0$\pm$0.5 $\mu$s and settled at E$^{*}$ = 85$\pm$16 keV. As the decay pattern of the isomeric
state remained unchanged, the excitation energy of $^{247m}$Md was reduced to E$^{*}$ = 92 keV.\\

{\bf $^{249}$Md}\\
In analogy to the case of $^{247}$Md also an 1/2$^{-}$[521] isomeric state could
be expected in $^{249}$Md. An indication of a low lying 1/2$^{-}$[521] state
in $^{249}$Md comes from the fact, that both states in $^{253}$Lr, 
tentatively assigned as 1/2$^{-}$[521] ($^{253}$Lr(1)) and 7/2$^{-}$[514]
($^{253}$Lr(2)) decay by unhindered $\alpha$ transitions, meaning that the
1/2$^{-}$[521] state should be populated by $\alpha$ decay of $^{253}$Lr(1)
\cite{Hess01}. It was also realized that $\alpha$ decays of $^{249}$Md
in the energy range E$_{\alpha}$ = (8050\,-\,8100) following $\alpha$ decay
of $^{253}$Lr(1) had a half-life T$_{1/2}$ = 1.5$^{+1.2}_{-0.5}$ s, 
while those, which followed decays of $^{253}$Lr(2) had a half-life
T$_{1/2}$ = 53$^{+27}_{-13}$ s. This was discussed as possibly $\alpha$
decay of an 1/2$^{-}$[521] isomeric state. (The $\alpha$ decays in that
energy range following decays of $^{253}$Lr(2) can be explained
due to energy summing of $\alpha$ particles and CE from the decay
process $^{249}$Md (7/2$^{-}$[514])  $^{\alpha}_{\rightarrow}$ $^{245}$Es (7/2$^{-}$[514]
$^{\gamma (200.4 keV)}_{\rightarrow}$ $^{245}$Es (9/2$^{+}$) 
$^{CE}_{\rightarrow}$ $^{245}$Es (7/2$^{+}$[633] \cite{Hess05}).
This finding, however, was not reproduced in a later study \cite{Hessb09}, so it had to 
be left open if indeed a 1/2$^{-}$[521]  isomeric state was existing in $^{249}$Md.
Strictly spoken, the $\alpha$ line indicated in earlier experiments was confirmed, but their
half-life was in agreement with the half-life of the $\alpha$ decays in the 'main line' and
in addition they were correlated to $\alpha$ decays from both longlived states in $^{253}$Lr. \\
New information on the existence of an isomeric state in $^{249}$Md came from a recent study
of P. Brionnet et al. \cite{BrioL25} who observed a weak $\alpha$ activity at E$_{\alpha}$ = 8350 keV
with a half-life of T$_{1/2}$ = 1.1$\pm$0.5 s, which they attributed to a 1/2$^{-}$[521] isomer.
It was interpreted as the decay $^{249}$Md(1/2$^{-}$[521] ) $\rightarrow$ $^{245}$Es(3/2$^{-}$[521] ),
the latter one assumed as the ground state. Based on the energy balance
for the decay of the ground state and isomeric
state and the assumption that the 7/2$^{+}$[633] level in $^{245}$Es populated by $\alpha - \gamma$ decay
of $^{249g}$Md is placed not higher than 30 keV above the ground state, the
excitation energy was settled at  E$^{*}$$\approx$(48-78) keV. \\
The interpretation of the isomeric decay to a transition 
1/2$^{-}$[521]$\downarrow$ $\rightarrow$ 3/2$^{-}$[521]$\uparrow$, however is ambiguous, as
it requires a spin flip. Such transitions are strongly hindered, which is in disagreement with the
relatively low hindrance factor of HF = 28 given by the authors.

\subsection{{\bf 6.2 Isomeric states in odd-mass lawrencium isotopes}}

{\bf $^{251}$Lr}\\
Information about decay of $^{251}$Lr is scarce. First identification of the isotope was reported 
by A. Lepp\"anen \cite{Lep05}, who observed a spontanous fission event following an $\alpha$ decay 
attributed to $^{255}$Db. The time difference between the $\alpha$ decay attributed to $^{255}$Db 
and the fission event was $\Delta$t($\alpha$-sf)\,=\,39 ms, resulting in a half-life of 
T$_{1/2}$ = 27$^{+113}_{-13}$ ms \cite{Lep05}.\\
Some detailed investigation of $^{251}$Lr was performed by T. Huang et al. \cite{HuS22}.
They observed two $\alpha$ - lines of E$_{\alpha 1}$ = 9246 $\pm$ 19 keV and 
E$_{\alpha 2}$ = 9210 $\pm$ 19 keV with slightly different half-lives of T$_{1/2}(\alpha 1)$
= 24.4$^{+7.0}_{-4.5}$ ms and T$_{1/2}(\alpha 2)$
= 42$^{+42}_{-14}$ ms. Theoretical half-lives according to \cite{PoI80} are T$_{calc}$ = 18 ms
for E$_{\alpha 1}$ and T$_{calc}$ = 23 ms
for E$_{\alpha 2}$. So both transitions have to be considered as unhindered. The energy of the
daughter $\alpha$ decays were  E$_{\alpha}$ = 8430 $\pm$ 13 keV, the half-lives were 
T$_{1/2}$ = 1.20$^{+0.52}_{-0.28}$ s for the daughter decays following  E$_{\alpha 1}$ 
and T$_{1/2}$ = 0.72$^{+0.98}_{-0.26}$ s for the daughter decays following  E$_{\alpha 2}$.
These decay properties are in-line with the known decay properties of $^{247g}$Md, but significantly
different from that of the isomeric state $^{247m}$Md (most intense $\alpha$ line: E$_{\alpha}$ =
8720 keV, T$_{1/2}$ = 0.23$\pm$0.03 s). Nevertheless T. Huang et al. attributed  E$_{\alpha 1}$ 
to the ground state to ground state decay of $^{251}$Lr ($^{251g}$Lr (7/2$^{-}$[514]) $\rightarrow$
$^{247g}$Md (7/2$^{-}$[514])) and E$_{\alpha 2}$ 
to the  decay of an isomer $^{251m}$Lr  located at E$^{*}$ = 117$\pm$27 keV into the isomeric state $^{247m}$Md ($^{251m}$Lr (1/2$^{-}$[521]) $\rightarrow$
$^{247m}$Md (1/2$^{-}$[521])). This interpretation, however, would require that $^{247m}$Md decays predominantly
by internal transition into the ground state $^{247g}$Md. Although this transition was not observed so far
in the direct investigation of $^{247}$Md \cite{Hess22}, it cannot be ruled out as the half-life for an
E3 transition (1/2$^{-}$ $\rightarrow$ 7/2$^{-}$) is expected to be $\approx$0.2 s according to 
the Weisskopf estimate \cite{Fire96}.\\

{\bf $^{253}$Lr}\\
The isotope $^{253}$Lr was first identified as the $\alpha$ - decay daughter of $^{257}$Db. Two $\alpha$ lines
(E$_{\alpha 1}$ = 8800 $\pm$ 20 keV (i$_{rel}$$\approx0.56$), E$_{\alpha 2}$ = 8722 $\pm$ 20 keV (i$_{rel}$ $\approx$0.44)) and a half-life of 
T$_{1/2}$ = 1.3$^{+0.6}_{-0.3}$ s were attributed to this isotope \cite{Hess85}. In a more detailed study 
\cite{Hess01} the results were essentially confirmed, it was, however, found, that the $\alpha$ decays stem from 
different levels in $^{253}$Lr, populated by $\alpha$ decay of different levels in $^{257}$Db. The assignment of the lines
to decay of the ground state and the isomeric state turned out to be rather difficult. The decay scenario of both isotopes,
$^{257}$Db and $^{253}$Lr is discussed in detail in \cite{Hess01}.\\
Further studies of the decay of $^{257}$Db performed at SHIP \cite{Streicher06,Hessb09} did not deliver 
new insight into the decay, level ordering and excitation energy of the isomeric state in $^{253}$Lr, which
is presently still uncertain. The results were recently confirmed by T. Huang et al. \cite{HuS22}. They identified another
$\alpha$ line (8660$\pm$20 keV), which was attributed to the decay of the ground state of $^{253}$Lr
into an 11/2$^{-}$ state in $^{249}$Md located at E$^{*}$ = (127$\pm$24) keV. The half-life of $^{253}$Lr(2)
was given as T$_{1/2}$ = 2.46$\pm$0.32 s, considerably longer than the value 
T$_{1/2}$ = 1.32$\pm$0.14 s, given in \cite{Hessb09}. A reanalysis of the data, however, showed that
the value given in \cite{Hessb09} underestimated the half-life and resulted in an improved value
of T$_{1/2}$ = 2.23$^{+0.22}_{-0.18}$ s. The suggested decay schemes from the SHIP experiments and
\cite{HuS22} are compared in fig.24.\\
In a recent study by P. Brionnet et al. \cite{BrioL25} the earlier results and conclusions were
confirmed. In addition the authors settle the isomeric state at E$^{*}$$\approx$(6-24) keV. \\

\begin{figure}
	\resizebox{0.8\textwidth}{!}{%
		\includegraphics{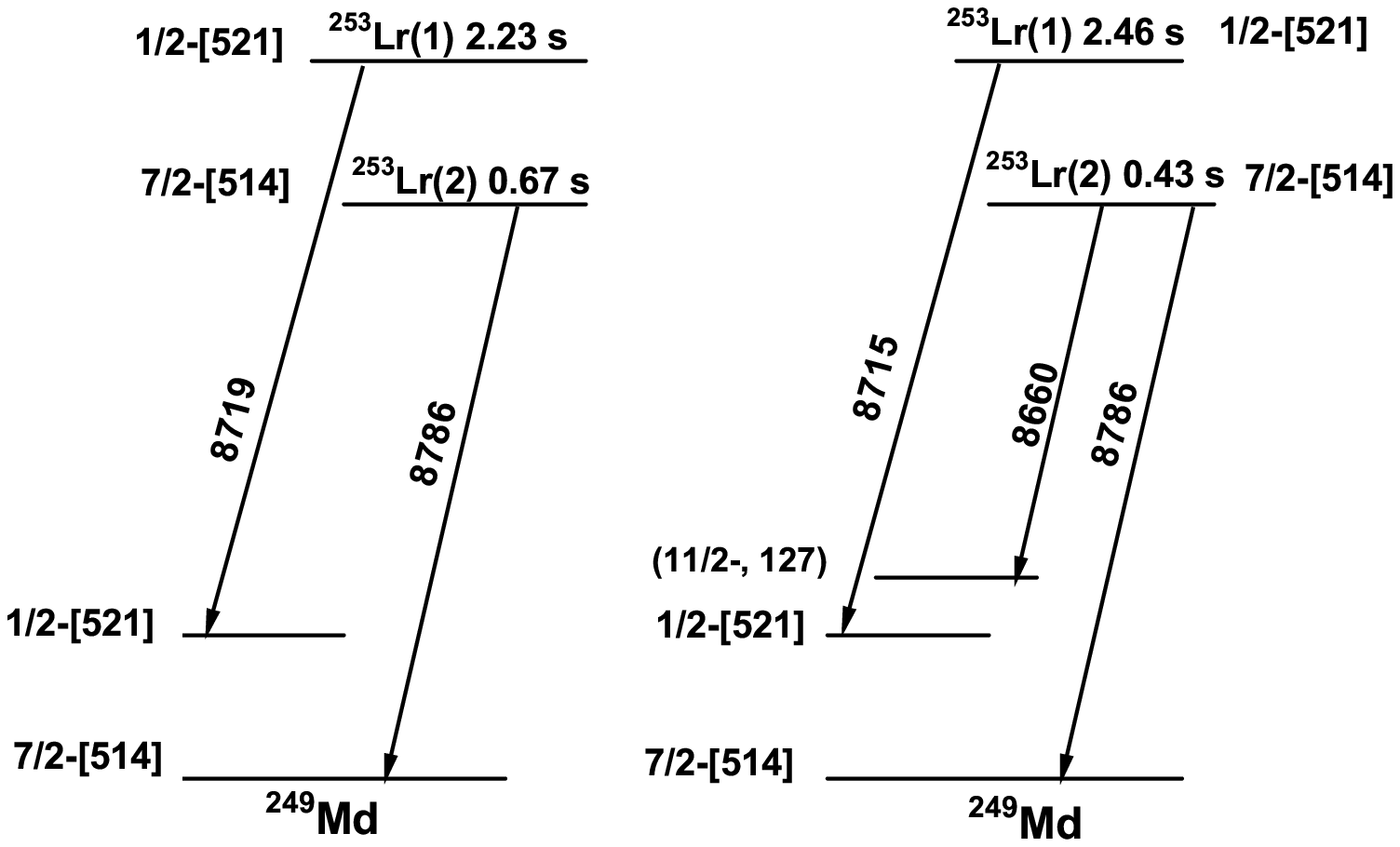}
	}
	\caption{Suggested decay scheme of $^{253g,253m}$Lr according to the results of 
		\cite{Hess01,Hessb09} (a) and \cite{HuS22} (b).}
	\label{fig:24}       
\end{figure}

\begin{figure}
	\resizebox{0.8\textwidth}{!}{%
		\includegraphics{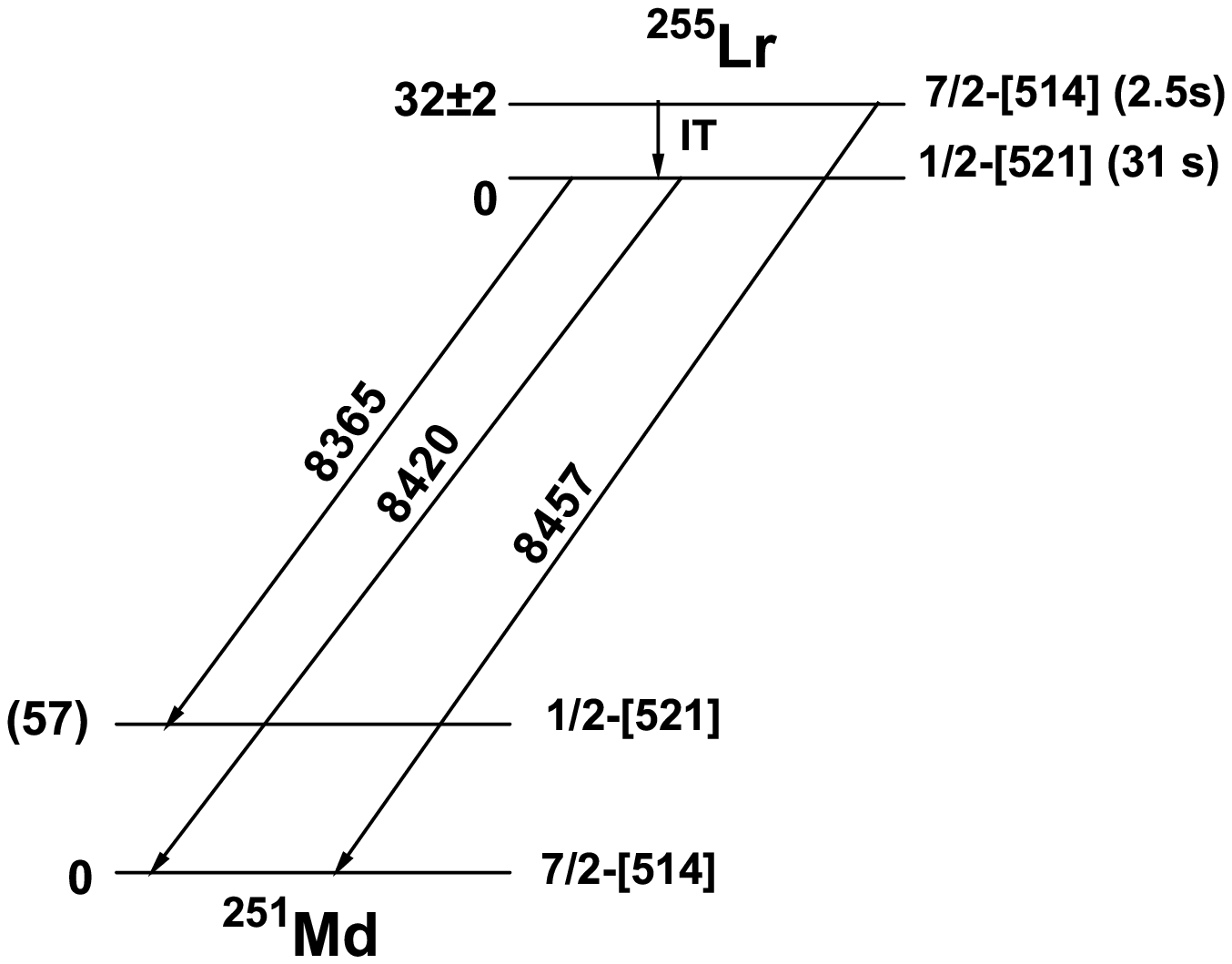}
	}
	\caption{Suggested decay scheme of $^{255g,255m1}$Lr according to the results of 
		\cite{Chat06,Kaleja20}}
	\label{fig:25}       
\end{figure}

{\bf $^{255}$Lr}\\
The isotope $^{255}$Lr was probably first observed by V.A. Druin \cite{Dru71}, who reported the observation of an $\alpha$ 
emitter (E$_{\alpha}$\,=\,8.38 MeV, T$_{1/2}$ $\approx$ 20 s) in bombardments of $^{243}$Am with $^{16}$O. The first definite assignment 
of an $\alpha$ emitter (E$_{\alpha}$\,=\,8370$\pm$18 keV, T$_{1/2}$ 22$\pm$5 s) to $^{255}$Lr was done by K. Eskola \cite{EsE71}. The data were
later confirmed by C.E. Bemis et al. \cite{BeD75}, who in addition observed an $\alpha$ line of E$_{\alpha}$\,=\,8410$\pm$18 keV,
with a relative intensity of i$_{rel}$ = 0.4. and gave a half-life of T$_{1/2}$ = 21.5$\pm$5.0 s. \\
Indication for the existence of a low lying isomeric state in $^{255}$Lr came from experiments performed
at the University of Jyv\"askyl\"a, Jyv\"askyl\"a, (Finland), later continued at GANIL, Caen (France) \cite{Chat06}
and at SHIP, GSI \cite{HesH04b} where an additional $\alpha$ line with a shorter half-life (E$_{\alpha}$ = 8457$\pm$2 keV,
T$_{1/2}$ = 2.53$\pm$0.13 s \cite{Chat06} or E$_{\alpha}$ = 8470$\pm$10 keV,
T$_{1/2}$ = 2.1$\pm$0.1 s \cite{HesH04b}) were observed. While the previously known transitions were
attributed to the decay of the 1/2$^{-}$[521] Nilsson level, the new activity was attributed to the decay of the
7/2$^{-}$[514] Nilsson level in both studies. However, while F.P. He\ss berger et al. attributed the 1/2$^{-}$[521]  level
to the isomeric state and the 7/2$^{-}$[514] level to the ground state, in analogy to $^{253}$Lr, A. Chatillon et al.
attributed on the basis of more detailed data the 1/2$^{-}$[521]  level
to the ground state and the 7/2$^{-}$[514] level to the isomeric state, which they located at E$^{*}$ = 37 keV. 
The latter interpretation was corroborated by further studies at SHIP \cite{AntH08}. While in the $\alpha$ decay
spectrum the E$_{\alpha}$ = 8373 keV - line (energy value from \cite{AntH08}) appeared to be much stronger than the 
E$_{\alpha}$ = 8467 keV - line (energy value from \cite{AntH08}), in correlation with decays of the 
high spin K isomer in $^{255}$Lr \cite{AntH08,HaL08} the E$_{\alpha}$ = 8467 keV - line appeared to be stronger.
This feature was interpreted in the following way: the decay of the K isomer preferably populates the state
of the higher spin (7/2$^{-}$[514]), which then decays either by $\alpha$ emission (8467 keV) or by
internal transition (IT) into the 1/2$^{-}$[521] ground state, then decaying by emission of E$_{\alpha}$ = 8373 keV $\alpha$ particles. \\
The excitation energy of the isomer was measured directly using SHIPTRAP. A value of  E$^{*}$ = 32$\pm$2 keV was obtained \cite{Kaleja20}. \\
A decay scheme is shown in fig.25\\

\vspace{5mm}

{\bf{$^{257}$Lr}}\\
First information on decay properties of $^{257}$Lr came from experiments
on direct production of this isotope by K. Eskola et al. \cite{EsE71} and from an experiment on synthesis
of $^{261,262}$Db (at this time named $^{261,262}$Ha) by A. Ghiorso et al. \cite{GhN71}, where 
$^{257}$Lr was produced by $\alpha$ decay of $^{261}$Db. While K. Eskola et al.
reported two $\alpha$ enegies (8.87$\pm$0.02 MeV, 8.81$\pm$0.02 MeV) and a half-life T$_{1/2}$ = 0.6$\pm$0.1 s,
A. Ghiorso et al. mentioned only the E$_{\alpha}$\,=\,8.87 MeV transition.\\
More detailed information on the decay properties was obtained in experiments performed
at SHIP, GSI where $^{257}$Lr was produced either by $\alpha$ decay of $^{261}$Db \cite{StH10}
or by EC decay of $^{257m,257g}$Rf \cite{StH10,Hess16a}. The results of the analysis are shown
in fig. 26a,b,c. Fig. 26a shows the energy distribution of the $^{257}$Lr $\alpha$ decays following
$\alpha$ decays of $^{261}$Db, figs. 26b and 26c show the energy distribution of 
$^{257}$Lr when produced by EC decay of $^{257}$Rf. While fig. 26b shows the energy
distribution of $^{257}$Lr  at 'indirect' production of $^{257}$Rf, i.e. by $\alpha$ decay of $^{261}$Sg,
where only the ground state $^{ 257g}$Rf is populated,
fig. 26c shows the energy
distribution of $^{257}$Lr  at 'direct' production of $^{257}$Rf via the reaction
$^{208}$Pb($^{50}$Ti,n)$^{257}$Rf, where $^{257m}$Rf and $^{257g}$Rf,
are produced at a ratio $\approx$2:1. From comparison of the spectra it is evident
that the occurrence of the 'low energy' line (E$_{\alpha}$ = 8811 keV) is essentilally
related to the production of $^{257g}$Rf, the occurrence of the 'high energy' line
with the production of $^{257m}$Rf and production by $\alpha$ decay of $^{261}$Db.
So it seems straightforward to assign the $\alpha$ lines to decays from different levels 
in $^{257}$Lr. This interpretion is supported by the half-life analysis,
For the  E$_{\alpha}$ = 8811 keV - line a value of T$_{1/2}$ = 0.20$^{+0.16}_{-0.06}$ s 
was obtained, for the E$_{\alpha}$ = 8878 keV - line a value of T$_{1/2}$ = 1.24$^{+0.85}_{-0.36}$ s.
As the groud-state of $^{257}$Rf is a low spin state (Nilsson configuration 1/2$^{+}$[620]), while
the isomer is a high spin state (Nilsson configuration 11/2$^{-}$[725], see sect. 5.6) and the ground state
of $^{261}$Db is predicted as 9/2$^{+}$[624], the level related to emission of the  E$_{\alpha}$ = 8811 keV
alpha particles is related to a low spin state, tentalively assigned as an 1/2$^{-}$[521] isomeric state, 
which is predicted as low lying level 
in lawrencium isotopes \cite{Park04}. \\
As the ground state of $^{261}$Db is predicted as 9/2$^{+}$[624] and the $\alpha$ decay of this isotope is only 
slightly hindered (HF $\approx$3, using a theoretical $\alpha$ half-life T$_{\alpha}$ = 0.37 s \cite{PoI80}) the 
$\alpha$ line connects transitions of the same Nilsson configuration. So the longer-lived level in $^{257}$Lr
may also assigned as 9/2$^{+}$[624]. It should be noticed, however, that the 7/2$^{-}$[514] Nilsson - state
is predicted as the ground state in $^{257}$Lr and as a low lying level in $^{261}$Db \cite{Park04} and,
considering the uncertainty of theoreical predictions, the 7/2$^{-}$[514] Nilsson - state could be 
ground state in both isotopes and the E$_{\alpha}$ = 8878 keV - line could represent a transition 
between these two states. As, however, this state is seemingly populated by EC decay of the
11/2$^{-}$[725] level in $^{257}$Rf and since the EC decay 11/2$^{-}$ $\rightarrow$ 9/2$^{+}$ is 
'single forbidden', while the  EC decay 11/2$^{-}$ $\rightarrow$ 7/2$^{-}$  is 'double forbidden'
\cite{MayKu70}, the 9/2$^{+}$[624] Nilsson level seems to be the better choice.\\
Decay patterns and level schemes are shown in fig.27.\\

\begin{figure}
	\resizebox{0.8\textwidth}{!}{%
		\includegraphics{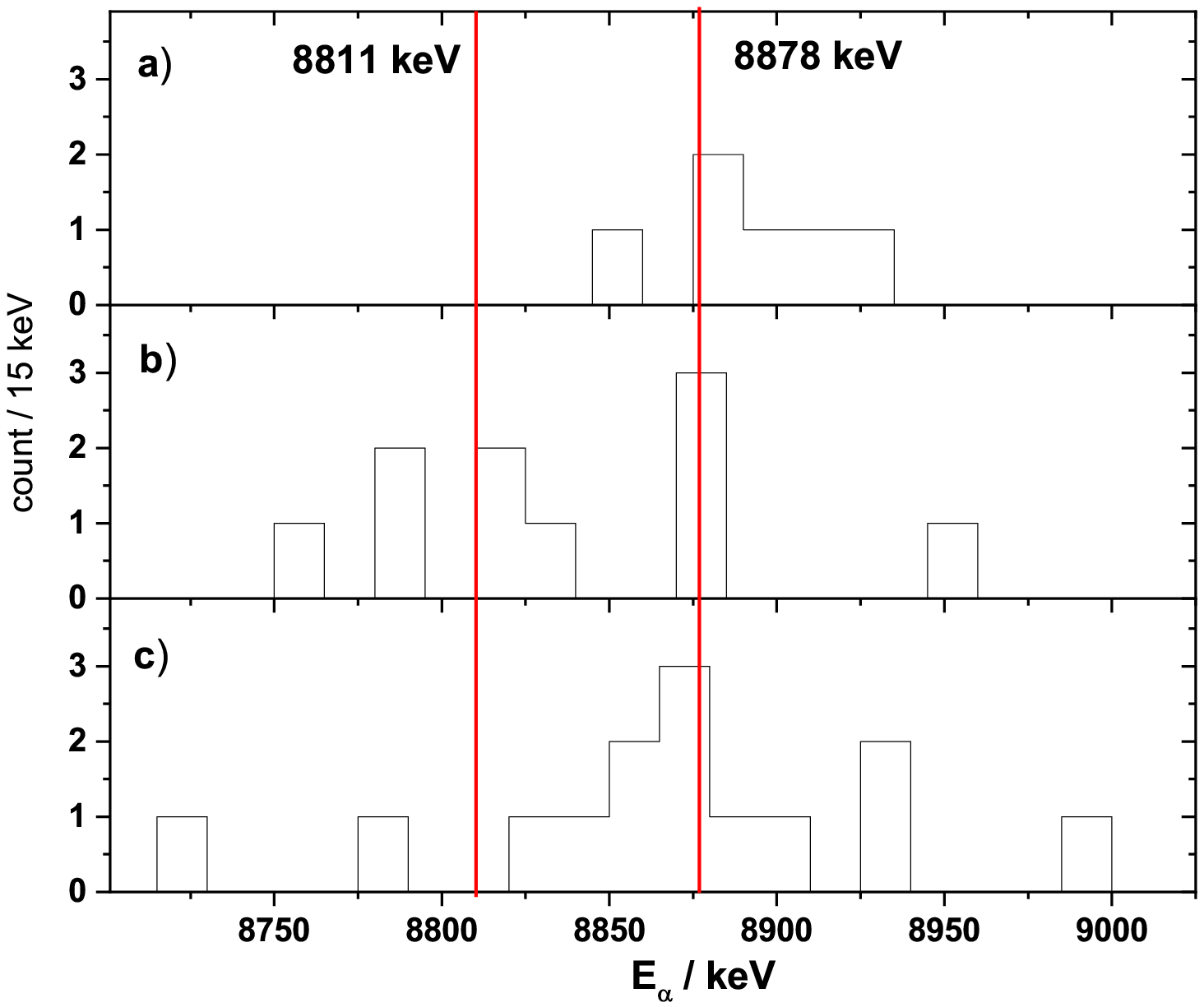}
	}
	\caption{$\alpha$ decays of $^{257g,257m}$Lr; a) $\alpha$ decays following $\alpha$ decays of $^{261}$Db; 
	b) $\alpha$ decays following EC decay of $^{257g}$Rf, produced by $\alpha$ decay of $^{261}$Sg;
    c) $\alpha$ decays following $^{257m,257g}$Rf, produced in the reaction $^{208}$Pb($^{50}$Ti,n)$^{257m,257g}$Rf.}
	\label{fig:26}       
\end{figure}

\begin{figure}
	\resizebox{0.8\textwidth}{!}{%
		\includegraphics{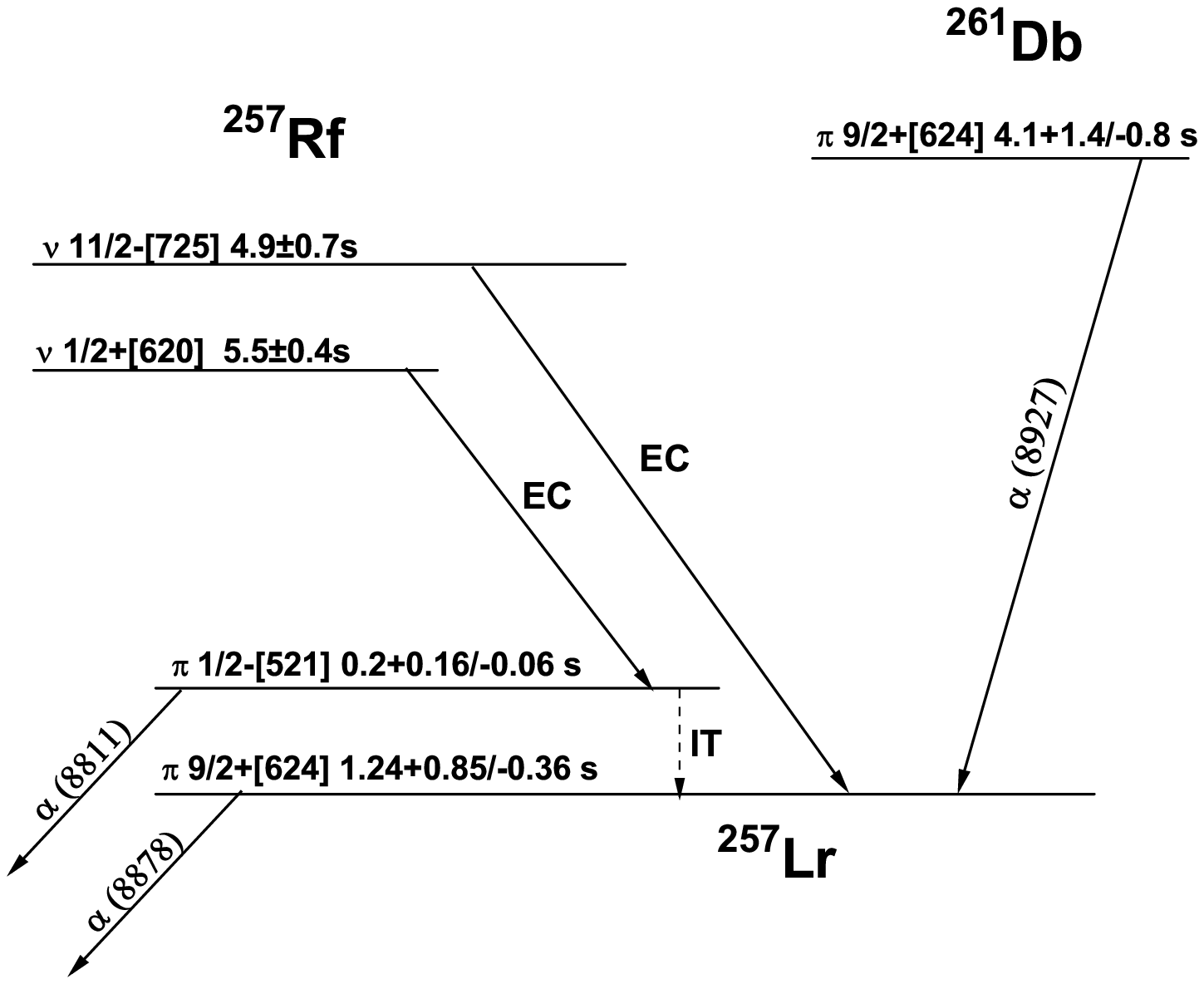}
	}
	\caption{Suggested population and decay of $^{257g,257m}$Lr}
	\label{fig:27}       
\end{figure}

\begin{figure}
	\resizebox{0.8\textwidth}{!}{%
		\includegraphics{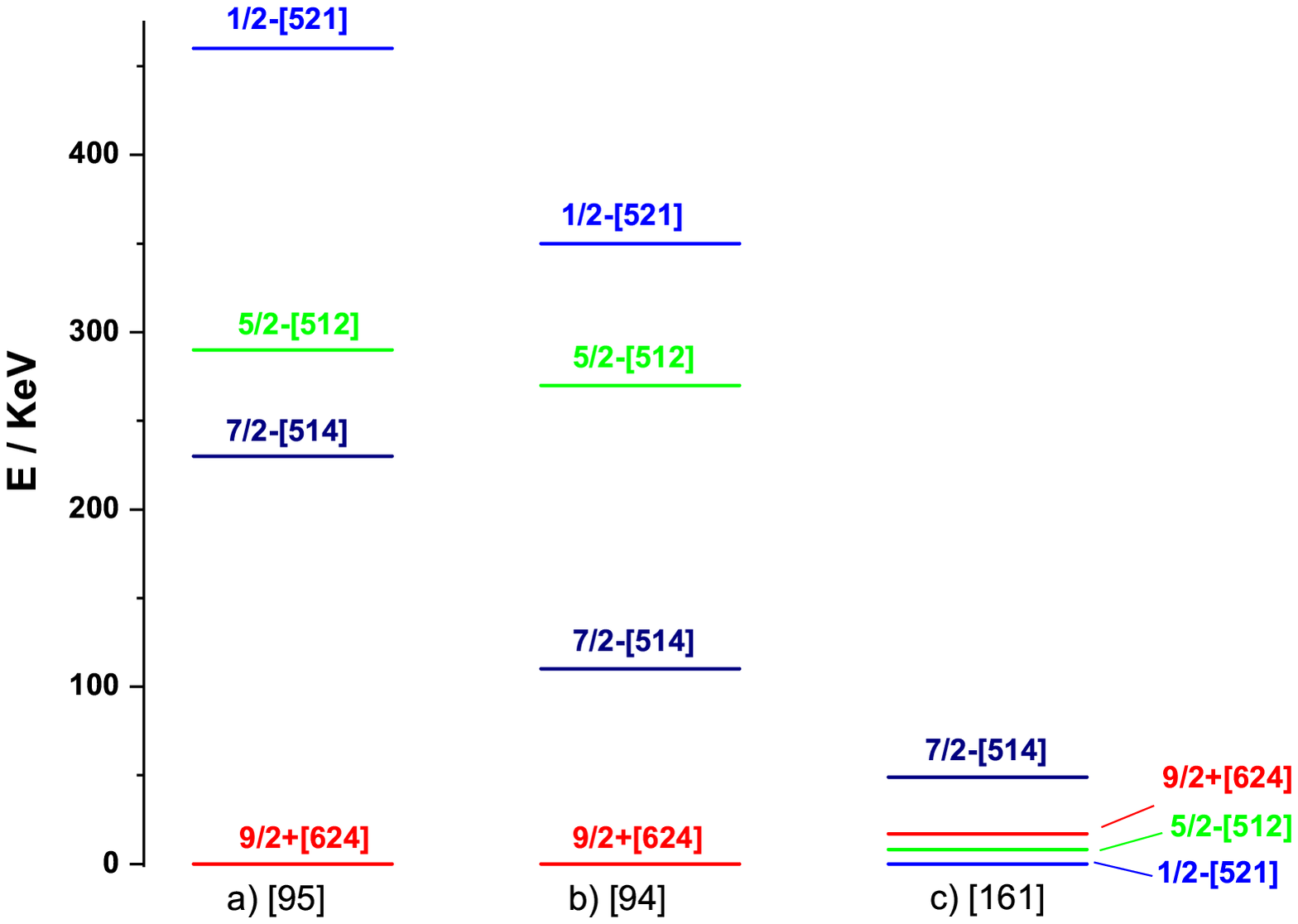}
	}
	\caption{Theoretical level schemes for $^{257}$Db, a) S. Cwiok et al.\cite{Cwiok94}, b) 
		A. Parkhomenko and A. Sobiczewski \cite{Park04},
		c) G.G. Adamian et al.\cite{AdaA10}.}
	\label{fig:28}       
\end{figure}

\begin{figure}
	\resizebox{0.8\textwidth}{!}{%
		\includegraphics{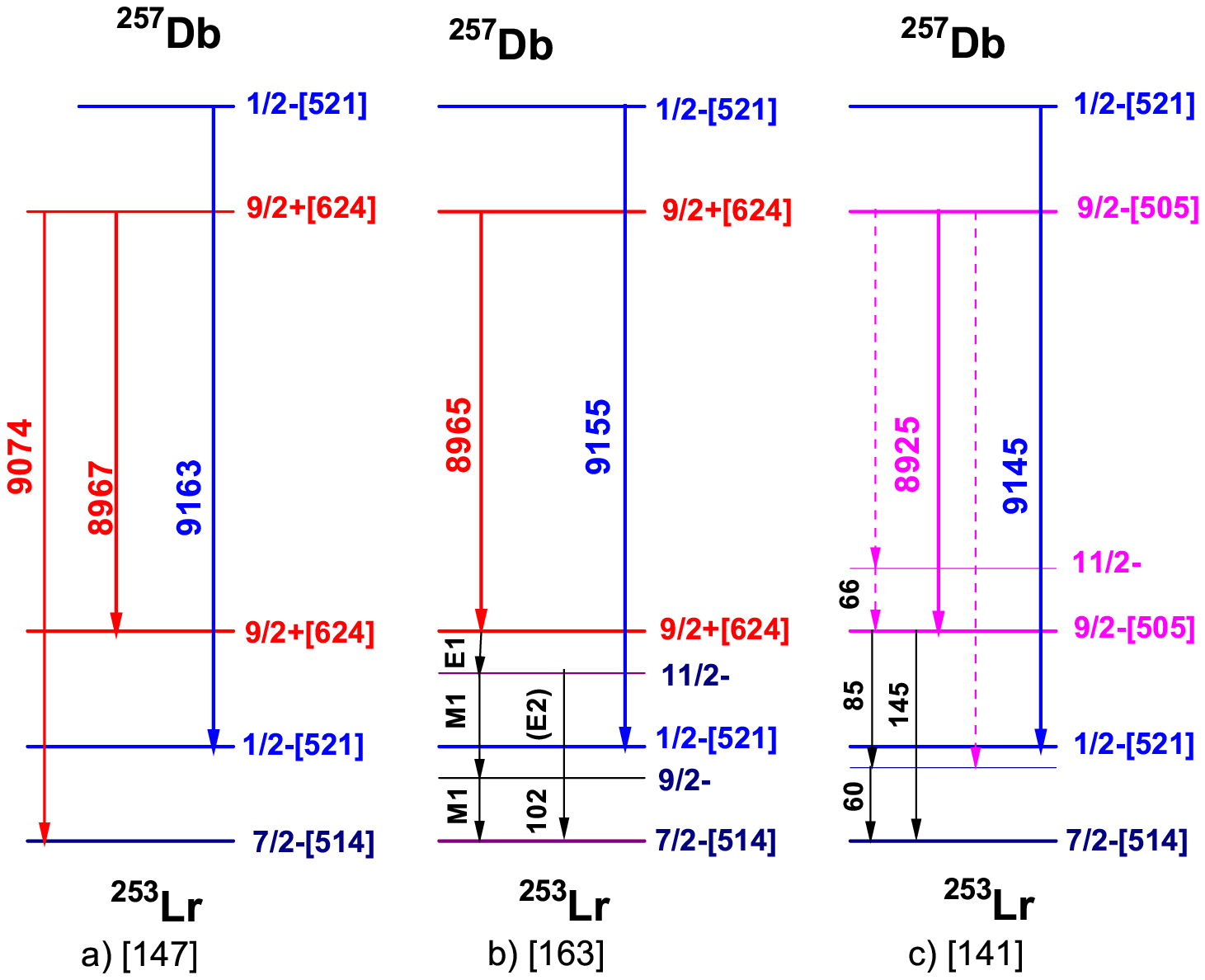}
	}
	\caption{Suggested decay schemes for $^{257m,257g}$Db, a) \cite{Hess01}, b) \cite{Hessb09},
		c) \cite{BrioL25}.}
	\label{fig:29}       
\end{figure}

\subsection{{\bf 6.3 Isomeric states in odd-mass dubnium isotopes}}

{\bf{$^{255}$Db}}\\
The first unambiguous identification of $^{255}$Db was reported by A. Lepp\"anen \cite{Lep05},
who observed three decays, one $\alpha$ decay (E$_{\alpha}$ = 9456$\pm$27 keV, $\Delta$t(ER-$\alpha$) = 56 ms,
and two SF events with time distances $\Delta$t(ER-SF(1))  = 99 ms and  $\Delta$t(ER-SF(2))  = 4 ms.
Specifically the value  $\Delta$t(ER-SF(2)), much shorter than  $\Delta$t(ER-SF(1))  and  $\Delta$t(ER-$\alpha$) 
suggests the existence of two 'longlived' (T$_{1/2}$ $>$ 1 ms) states in $^{255}$Db. New decay data were reported
by J.L. Pore et al. \cite{Pore24} who attributed 55 SF  events and
three $\alpha$ decays (9245$\pm$47, 9396$\pm$47, 9232$\pm$47 keV) with a half-life 
T$_{1/2}$ = 2.6$^{+0.4}_{-0.3}$ ms to . A longlived component was not observed. Thus the two longlived events from
\cite{Lep05} seem questionable and therefore presently there is no real evidence for a low lying isomeric \\
state in $^{255}$Db.\\

{\bf{$^{257}$Db}}\\
The existence of an isomeric state in$^{257}$Db was not straight
forward on the basis of the level scheme predicted at the time
of publication of \cite{Hess01},  suggessting
a level ordering 9/2$^{+}$[624] (gs), 7/2$^{-}$[514] (230 keV), 5/2$^{-}$[512] (290 keV),
1/2$^{-}$[521] (460 keV) \cite{Cwiok94} (see fig. 28a). The same level ordering, but
somewhat different energies were later predicted by A. Parkhomenko and A. Sobiczewski \cite{Park04}
(see fig. 28b).
In this scenario the low spin state 1/2$^{-}$ could decay via 
1/2$^{-}$[521] $^{E2}_{\rightarrow}$ 5/2$^{-}$[512] $^{M1}_{\rightarrow}$ 7/2$^{-}$[514] 
$^{E1}_{\rightarrow}$ 9/2$^{+}$[624] (ground state). Therefore tentatively
in \cite{Hess01} the 1/2$^{-}$[521] level was thus placed below
the 7/2$^{-}$[514] and  5/2$^{-}$[512] levels and the decay scheme shown in fig. 29a) was suggested.
In a further study at SHIP \cite{Streicher06,Hessb09} where a larger amount of decays was 
collected it was realized that the lines at E$_{\alpha}$ = 9066 keV (FWHM = 64 keV) and
E$_{\alpha}$ = 8965 keV (FWHM = 53 keV) were broader than the line at 
E$_{\alpha}$ = 9155 keV (FWHM = 28  keV) \cite{Hessberger25}. It was assumed that 
these lines were strongly influenced by energy summing with CE, specifically
the line at E$_{\alpha}$ = 9066 keV now rather interpreted as the energy sum
of the decay populating the 9/2$^{+}$[624] level and CE from the internal
transition into the 7/2$^{-}$[514] ground state of $^{253}$Lr than the decay
$^{257}$Db  (9/2$^{+}$[624]) $^{\alpha}_{\rightarrow}$$^{253}$Lr (7/2$^{-}$[514])
as such a transition requires a change of parity and in addition a spin flip, and is thus
strongly hindered (HF $>>$ 1000 \cite{SeaL90}. The assumption of a direct
decay  9/2$^{+}$[624]) $^{IC}_{\rightarrow}$ 7/2$^{-}$[514] was in contradiction
with the experimental data, as such a transition has E1 - multipolarity, is only little
converted, thus does not lead to broadening of the $\alpha$ lines, while on the
other hand only one $\gamma$ event (E$_{\gamma}$ = 102.2 keV  
was registered in coincidence with an $\alpha$ particle (E$_{\alpha}$ = 8941 keV),
whereas ten events were expected for an E1 transition. It thus was speculated,
that the 9/2$^{+}$[624] level might decay by a low energy  E1 transition into the
11/2$^{-}$ of the rotational band built up on the 7/2$^{-}$[514] ground state 
which then deexcites essentially by two M1 transitions into
the ground state as sketched in fig. 28b.\\
A different level ordering was predicted by G.G. Adamian et al. \cite{AdaA10} (see fig. 28c).
They predict the 1/2$^{-}$[521] as the ground state,  the 5/2$^{-}$[512] (E$^{*}$$\approx$8 keV)
and 9/2$^{+}$[624] (E$^{*}$$\approx$17 keV) levels as long-lived isomeric states
decaying at least partly by $\alpha$ emission.\\
Further results on the decay of $^{257m,257g}$Db were published by P. Brionnet et al. \cite{BrioL25}.
Essentially the results of \cite{Hess01,Hessb09} were confirmed, in addition three $\gamma$ events
at E$_{\gamma}$ = 145 keV in coincidence with E$_{\alpha}$\,=\,(8.90-8.95) MeV were observed. 
On the basis of Geant4 simulations they concluded that the speculation in \cite{Hessb09} of a 
decay pattern 9/2$^{+}$[624] $^{E1}_{\rightarrow}$ 11/2$^{-}$ $^{M1}_{\rightarrow}$ 9/2$^{-}$ $^{M1}_{\rightarrow}$ 7/2$^{-}$[514] 
is unlikely. To solve the 'problem' they suggested the ground state of $^{257}$Db rather being the Nilsson - level 
9/2$^{-}$[505] decaying essentially into the corresponding level in $^{253}$Lr and with smaller intensities into the
11/2$^{-}$ state of the rotational band built up on it and the 11/2$^{-}$ level built up on the ground state of $^{253}$Lr.
The weak point of such an assignment is the fact that none of the theoretical calculations predict the 
9/2$^{-}$[505] Nilsson level in odd-A odd-Z isotopes in elements Z$<$107 at excitation energies 
E$^{*}$$<$1 MeV.\\
To explain the low energy of the 9/2$^{-}$[505] level, the authors speculate on an influence of
higher order deformations, specifically hexadecapole deformation (denoted $\nu_{4}$ or $\beta_{4}$).
Calculations of R.R. Chasman et al. \cite{Chasman77} predict a strong increase of the energy
of the 9/2$^{+}$[624] and 1/2$^{-}$[521] levels when $\nu_{4}$ changes from
$\nu_{4}$ = 0.04 to $\nu_{4}$ = -0.04, while the energy of the 9/2$^{-}$[505] level slightly decreases. 
While at $\nu_{4}$ = -0.04 one obtains $\Delta$E(9/2$^{-}$,9/2$^{+}$) = 0.87 MeV, 
at $\nu_{4}$ = +0.04 one obtains $\Delta$E(9/2$^{-}$,9/2$^{+}$) = 2.08 MeV. 
The change of $\beta_{4}$ in dubnium isotopes is reproduced by the calculations.
Cwiok et al. \cite{Cwiok94} obtain a change from $\beta_{4}$ = 0.003 ($^{255}$Db) to
$\beta_{4}$ = -0.035 ($^{263}$Db),  A. Prakhomenko and A. Sobiczewski \cite{Park04} obtain a
change from $\beta_{4}$ = 0.008 ($^{253}$Db) to
$\beta_{4}$ = -0.036 ($^{263}$Db), but still in both calculations for none of the isotopes
the 9/2$^{-}$[505] level is located below 1 MeV. From this side the assignment of this
level to the ground state of $^{257}$Db is quite uncertain. More detailed investigations
are necessary to clarify the situation.\\
The decay scheme presented by P. Brionnet et al. \cite{BrioL25}  is shown in fig. 29c.
They settle the isomer at E$^{*}$$\approx$(59-89) keV and postulate besides 
the strong $\alpha$ transition $^{257}$Db(9/2$^{-}$[505]) $\rightarrow$  $^{253}$Lr(9/2$^{-}$[505])
two weaker ones  $^{257}$Db(9/2$^{-}$[505]) $\rightarrow$  $^{253}$Lr(11/2$^{-}$)
(member of the rotational band built up on the 9/2$^{-}$[505] level) and 
$^{257}$Db(9/2$^{-}$[505]) $\rightarrow$  $^{253}$Lr(9/2$^{-}$)
(member of the rotational band built up on the 7/2$^{-}$[514] ground state). 
At least the latter transition is unlikely, as it requires a spin-flip, thus should be
highly hindered and thus should not be oberserved on the basis of the
number of the total decays of $^{257}$Db. \\

\vspace{15mm}
\section{7. Isomers in odd-odd nuclei}

\subsection{{\bf 7.1 Isomeric states in odd-odd mendelevium isotopes}}

{\bf $^{246}$Md}\\
The isotope $^{246}$Md was first identified at SHIP, GSI in bombardments
of $^{209}$Bi targets with $^{40}$Ar projectiles by V. Ninov et al. \cite{NinH96}.
Two groups of $\alpha$ decays (E$_{\alpha}$ = (8500-8560) keV, 8740$\pm$20 keV)
and a half-life T$_{1/2}$ = 1.0$\pm$0.4 s were attributed to this isotope.\\
Later, more intense studies by S. Antalic et al. \cite{AntH10} showed that 
besides the known activity, now denoted as $^{246m1}$Md another
activity  ($^{246m2}$Md) had to be assigned. Reported \cite{AntH10} decay properties are:  \\
a)  $^{246m1}$Md:  (E$_{\alpha}$ = (8380-8640) keV, 8744$\pm$10 keV,
T$_{1/2}$ = 0.9$\pm$0.2 s. \\
b)  $^{246m2}$Md: E$_{\alpha}$ =  8178$\pm$10 keV,
T$_{1/2}$ = 4.4$\pm$0.8 s. \\
It was also found, that $^{246m1}$Md decays to a high extent by EC (b$_{EC}$$>$0.77)
accompanied by a high EC-delayed fission branch (b$_{ECDF}$$>$0.1).\\
No spin and parity assignments for $^{246m1}$Md and $^{246m2}$Md, also no (tentative)
assignment of both states to the ground state and the isomeric state were done.
Taking the $\alpha$ energy of E$_{\alpha}$ =  8740$\pm$10 keV (assuming ground state 
to ground state transition) for calculating the
decay Q-value $^{246}$Md and the sum of the
$\alpha$ decay energy (E$_{\alpha}$ =  8178$\pm$10 keV) and the energies of  the
coincident gammas (E$_{\gamma}$ = 252, 279 keV) (see fig.6 in \cite{AntH10}) for calculating
the decay Q-value of  $^{246m2}$Md one obtains values of 
Q($^{246m1}$Md) $\approx$ 8890 keV and 
Q($^{246m2}$Md) $\approx$ 8840 keV, which would suggest 
to attribute Q( $^{246m1}$Md)  to the ground state. Definitely it is
not known presently if these Q-values indeed represent the Q-value difference
between $^{246m1}$Md or $^{246m2}$Md and the ground state of $^{242}$Es.
So it is presently uncertain which of the activity is the ground state and which is 
the isomer.\\
In the neighbouring odd-even nucleus $^{245}$Fm low lying levels 7/2$^{-}$[743] and 1/2$^{+}$[631] \cite{Khuy20,ParS05}
are supposed or predicted for the the N=145th neutron,
while low lying proton levels in the neighbouring Z\,=\,101 nucleus $^{247}$Md are 7/2$^{-}$[514] and
1/2$^{+}$[521]. So the odd nucleon can couple to states I$^{\pi}$ = 1$^{-}$, 4$^{-}$, 4$^{+}$, and 7$^{+}$.
The spin differences  $\Delta$I(1$^{-}$, 4$^{-}$ or 4$^{+}$)  = 3,  $\Delta$I( 7$^{+}$, 4$^{-}$ or 4$^{+}$)  = 3, 
and  $\Delta$I( 1$^{-}$, 7$^{+}$) = 6, suggest indeed the existence of isomeric states, but presently 
it is not possible to assign safely spin and parity to $^{246m1}$Md or $^{246m2}$Md.
It should, however, be remarked that the large EC-delayed fission branch of $^{246m2}$Md indicates
that EC decay of this state populates notably high spin states in $^{246}$Fm, which can be
regarded as an indication that $^{246m2}$Md is a high spin state.\\

\begin{figure}
\resizebox{0.8\textwidth}{!}{%
\includegraphics{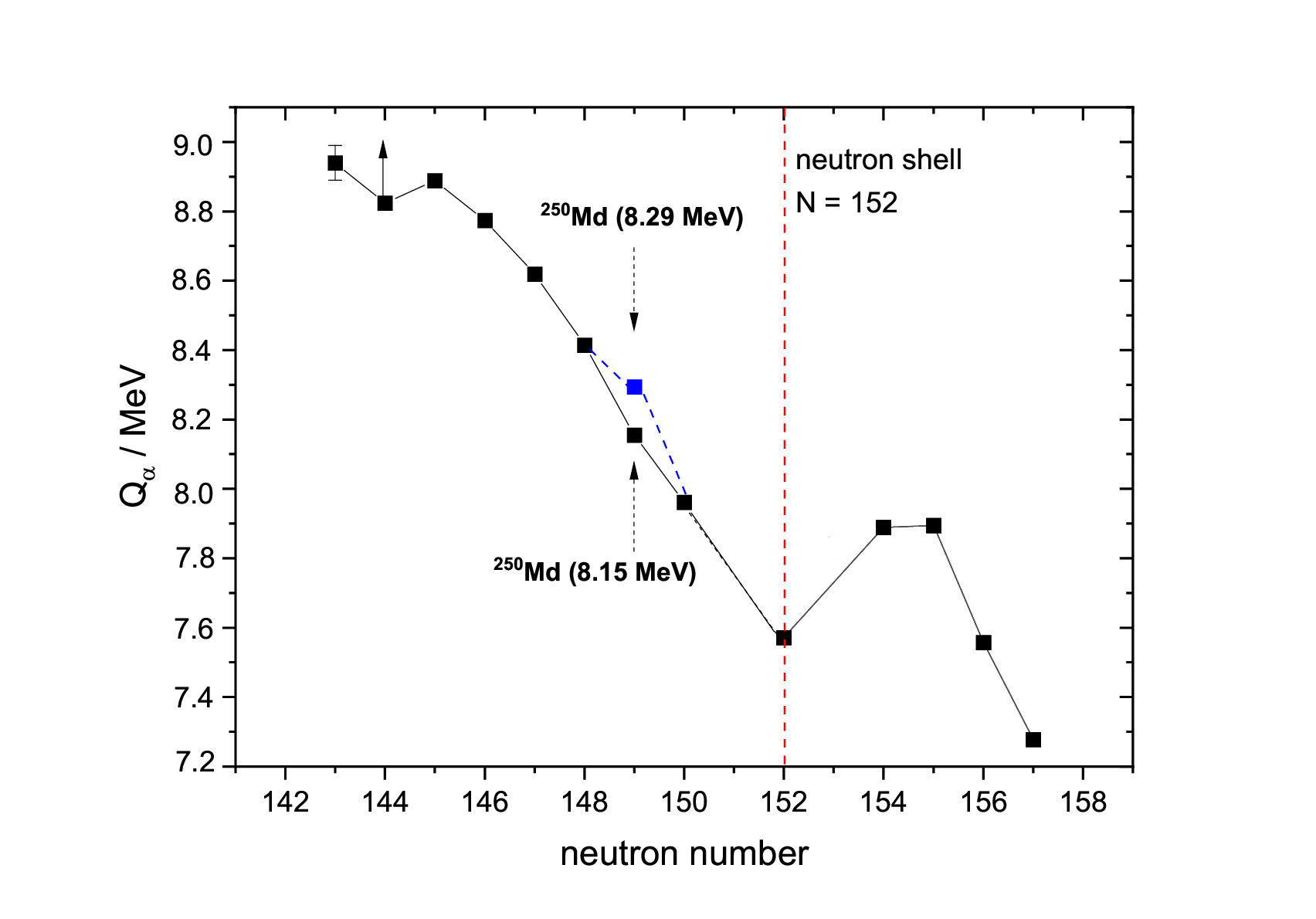}
}
\caption{Q$_{\alpha}$ systematics of mendelevium isotopes.}
\label{fig:30}       
\end{figure}

\begin{figure}
\resizebox{0.8\textwidth}{!}{%
\includegraphics{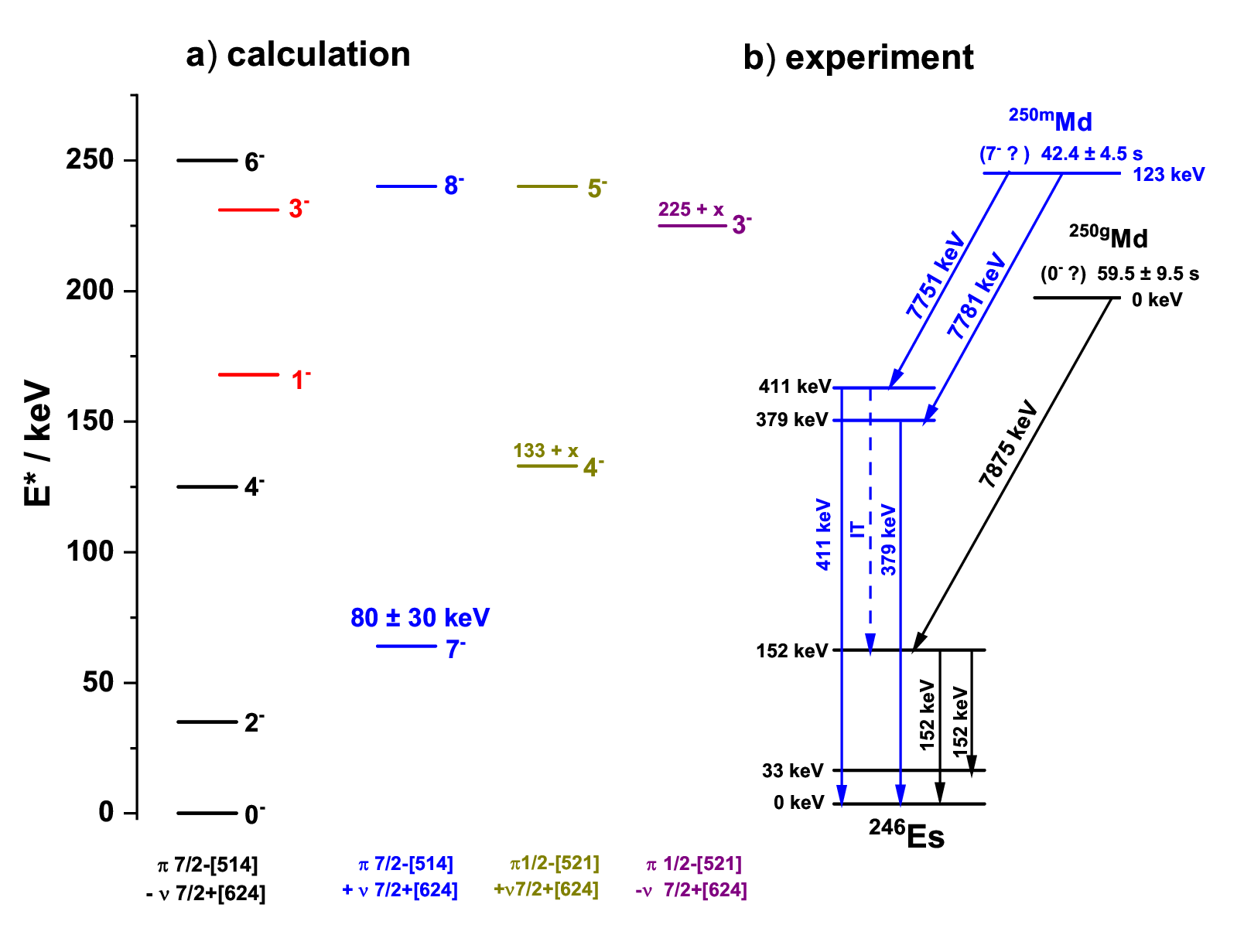}
}
\caption{a) Predicted low lying levels of $^{250}$Md(\cite{Sood2000}, b) experimentally (tentatively)  proposed decay schemes  of   $^{250m,250g}$Md \cite{Vosti19}. }
\label{fig:31}       
\end{figure}

{\bf $^{250}$Md}\\
The isotope $^{250}$Md was first reported by P. Eskola \cite{Eskola1973}
who produced it in irradiations $^{15}$N + $^{240}$Pu, $^{16}$O + $^{241}$Am
$^{12,13}$C + $^{243}$Am. She reported two $\alpha$ lines, E$_{\alpha}$ = 7.75$\pm$0.02,
7.82$\pm$0.02 MeV and a half-life T$_{1/2}$ = 52 $\pm$ 6 s.
The isotope was found to desintegrate predominantly by $\beta^{+}$, EC - decay.
The data were later confirmed by F.P. He\ss berger et al. \cite{Hess85} and by
S. Antalic et al. \cite{AntH08} by studying $\alpha$ decay of $^{254}$Lr or $^{258}$Db
( $^{209}$Bi($^{50}$Ti,n)$^{258}$Db) $\rightarrow$  $^{254}$Lr $\rightarrow$  $^{250}$Md or 
$^{209}$Bi($^{48}$Ca,3n)$^{254}$Lr $\rightarrow$  $^{250}$Md).
The discovery of isomeric states in $^{254}$Lr (see sect. 7.2) and $^{258}$Db (see sect. 7.3) 
suggested a possible existence of an isomeric state in  $^{250}$Md also decaying partly by $\alpha$ emission
and initiated a careful (re-)analysis of data accumulated in a detailed decay study of $^{258}$Db \cite{Vosti19}
 and measured at SHIP in the previous studies \cite{Hess85,AntH08,Hessb09}).\\
The analysis resulted in the identification of two $\alpha$ peaks (E$_{\alpha}$ = 7741 $\pm$ 24, 7781 $\pm$ 21 keV)
and a broad distribution of $\alpha$ decays at E$_{\alpha}$ $>$ 7.8 MeV. Both groups were found to have 
slightly different half-lives. For events E$_{\alpha}$ $<$ 7.8 MeV a value T$_{1/2}$ = 42.5 $\pm$ 4.5 s
was obtained, for events  E$_{\alpha}$ $>$ 7.8 MeV a value T$_{1/2}$ = 59.5 $\pm$ 9.5 s
was obtained.\\
A couple of $\gamma$ events was found in coincidence with $\alpha$ decays. The results are listed
in table 6.
The energy sums  E$_{\alpha}$ + E$_{\gamma}$ are\\
a) E = 8152 keV (7741+411) \\
b) E = 8160 keV (7781+379) \\
c) E = 8027 keV (7875 + 152) \\
The similar values in a) and b) suggest that the decays stem from the same level in $^{250}$Md, the higher
values in a) and b) than in c) suggest that these decays stem from a level having a higher excitation energy
than the one in c), which thus tentatively is assigned to the decay of the ground state. The assignment of a) and b) 
to an excited level in $^{250}$Md is corroborated by the systematics of $\alpha$ - decay Q-values of mendelevium
isotopes as shown in fig. 30. While the Q-value obtained from c) (Q = 8157 keV) fits quite well into the systematics
the one obtained from a) and b) (Q = 8282 keV) introduces a kink.
The decay scheme of $^{250}$Md derived from the experimental data is shown in fig. 31b.
As few events of E$_{\gamma}$ = 118.6, 151.9 keV were observed in coincidence with $\alpha$s E$<$7.8 MeV,
we assume also some internal transitions from the levels populated by the E$_ {\alpha}$ = 7741, 7781 keV to the
level(s) emitting the E = 118.6, 151.9 keV $\gamma$s as indicated in fig. 31b.  \\
Low lying states, shown in fig. 31a of $^{250}$Md have been calculated by P.C. Sood et al. \cite{Sood2000}.
As seen the I$^{\pi}$ = 7$^{-}$ state placed at E$^{*}$ = 80$\pm$30 keV was predicted as an isomeric state
decaying to large extent, by $\beta^{+}$/EC and/or $\alpha$ emission, which is in quite good agreement with 
the identification of the isomeric state at E$^{*}$ = 123 keV. \\

\begin{table}
\caption{$\alpha$ - $\gamma$ coincidences measured for $^{250}$Md}
\label{tab:6}       
\begin{tabular}{lll}
\hline\noalign{\smallskip}
\noalign{\smallskip}\hline\noalign{\smallskip}
E$_{\gamma}$/keV & E$_{\alpha}$/keV & number of events\\  
\hline\noalign{\smallskip}     
118.6 $\pm$ 0.4 & 7877 $\pm$ 2  & 2\\
&    7760 (1) & 1 \\
\hline\noalign{\smallskip}     
151.9 $\pm$ 0.6 & 7875 $\pm$ 8 & 4  \\
& 7766 $\pm$ 8 & 2  \\
& 7833* & 1 \\
& 7980** & 1 \\
\hline\noalign{\smallskip}     
379.5 $\pm$ 0.7 & 7781 $\pm$ 21 & 3  \\
\hline\noalign{\smallskip}     
411.4 $\pm$ 1.3 & 7741 $\pm$ 24 & 6  \\
\hline\noalign{\smallskip} 
  $\star$ observed during pulse-on period \\
  $\star$$\star$ observed as sum event in stop and box-detector\\
\end{tabular}
\\
\end{table}  

\vspace{10mm}

{\bf $^{254}$Md}\\
The isotope $^{254}$Md was first observed by P.R. Fields et al. \cite{Fields70}
in irradiations of an einsteinium target ($^{253}$Es content 99.4 per cent) with
$^{4}$He projectiles. Two EC activities with half-lives T$_{1/2}$\,=\,10$\pm$3 min
and T$_{1/2}$\,=\,28$\pm$8 min were attributed to this isotope. $\alpha$ decay was
not reported.\\
In a study at SHIP, GSI of K isomeric states in $^{254}$No \cite{Hes10}, the isotope $^{254}$Md was 
produced as a by-product via EC - days of $^{254}$No. It was shown by coincidences between 
fermium K X rays and $\gamma$ - events (E$_{\gamma}$ = 693.3, 648.3 keV)  that EC - decay populates 
the I$^{\pi}$ = 2$^{+}$ level at E$^{*}$ = 693.79 keV in $^{254}$Fm, by strongly converted
transitions from higher levels.
The I$^{\pi}$ = 2$^{+}$ level at E$^{*}$ = 693.79 keV in $^{254}$Fm is known from
$\beta^{-}$ decay studies of $^{254}$Es \cite{Fire96}. No new information about the isomeric state
$^{254m}$Md was obtained \cite{Hess24}.\\

{\bf $^{258}$Md}\\
First identification of an isomeric state was performed by
D.C. Hoffman et al. \cite{Hoff77,Hoff80}, who reported EC decay as the dominant decay mode
and a half-life T$_{1/2}$ = 43$\pm$3 min. \\
More detailed decay properties were given by  K.J. Moody et al. \cite{Moody93}.
They established the main decay channel as EC/$\beta^{+}$, with a branching
b$_ {EC/\beta^{+}}$ $\ge$ 0.7, and upper limits for other decay channels 
(b$_{\alpha}$ $\le$ 0.012, b$_{\beta^{-}}$ $\le$ 0.3,  b$_{SF}$ $\le$ 0.3).
Spin and parity were tentatively assigned as I$^{\pi}$ = 1$^{-}$, and a half-life 
T$_{1/2}$ = 57.0$\pm$0.9 min was reported. 
No value for the excitation energy was given so far.\\

\subsection{{\bf 7.2 Isomeric states in odd-odd lawrencium isotopes}}

{\bf $^{252}$Lr}\\
The isotope $^{252}$Lr was first identified as $\alpha$ - decay product of $^{256}$Db \cite{Hess01}; 
two $\alpha$ lines (E$_{\alpha 1}$ = 9018 $\pm$ 20 keV (i$_{rel}$ $\approx$ 0.75)) and 
(E$_{\alpha 2 }$ = 8974 $\pm$ 20 keV (i$_{rel}$ $\approx$ 0.25)) and a half-life 
T$_{1/2}$ = 0.36$^{+0.36}_{-0.11}$ s were attributed to this isotope. 
Some more events were reported from a decay study of $^{260}$Bh by S.L. Nelson et al. \cite{NeG08},
among them two events (E$_{\alpha}$ = 9.61 MeV,  E$_{\alpha}$ = 9.82 MeV) with significantly 
different energies. But both events were observed as 'sum' events measuring the energy loss
in the 'stop detector' and the residual energy in the 'box detector' surroundig the 'stop detector', which could 
lead to significantly different energies due to calibration purposes. The event of the higher energy was followed
by a $^{248}$Md decay of E$_{\alpha}$ = 8.46 MeV, significantly higher than the known $\alpha$ - decay 
energies of this isotope (E$_{\alpha}$ = 8.32, 8.36 MeV). So it was speculated if it could be due to
the decay of an isomeric state, but due to the low quality of the data, no definite conclusions were drawn.
A recent decay study of $^{256}$Db by J.L. Pore et al. \cite{Pore24a}, did not give a signature for
the existence of an isomeric state in $^{252}$Lr. \\

\begin{figure}
	\resizebox{0.8\textwidth}{!}{%
		\includegraphics{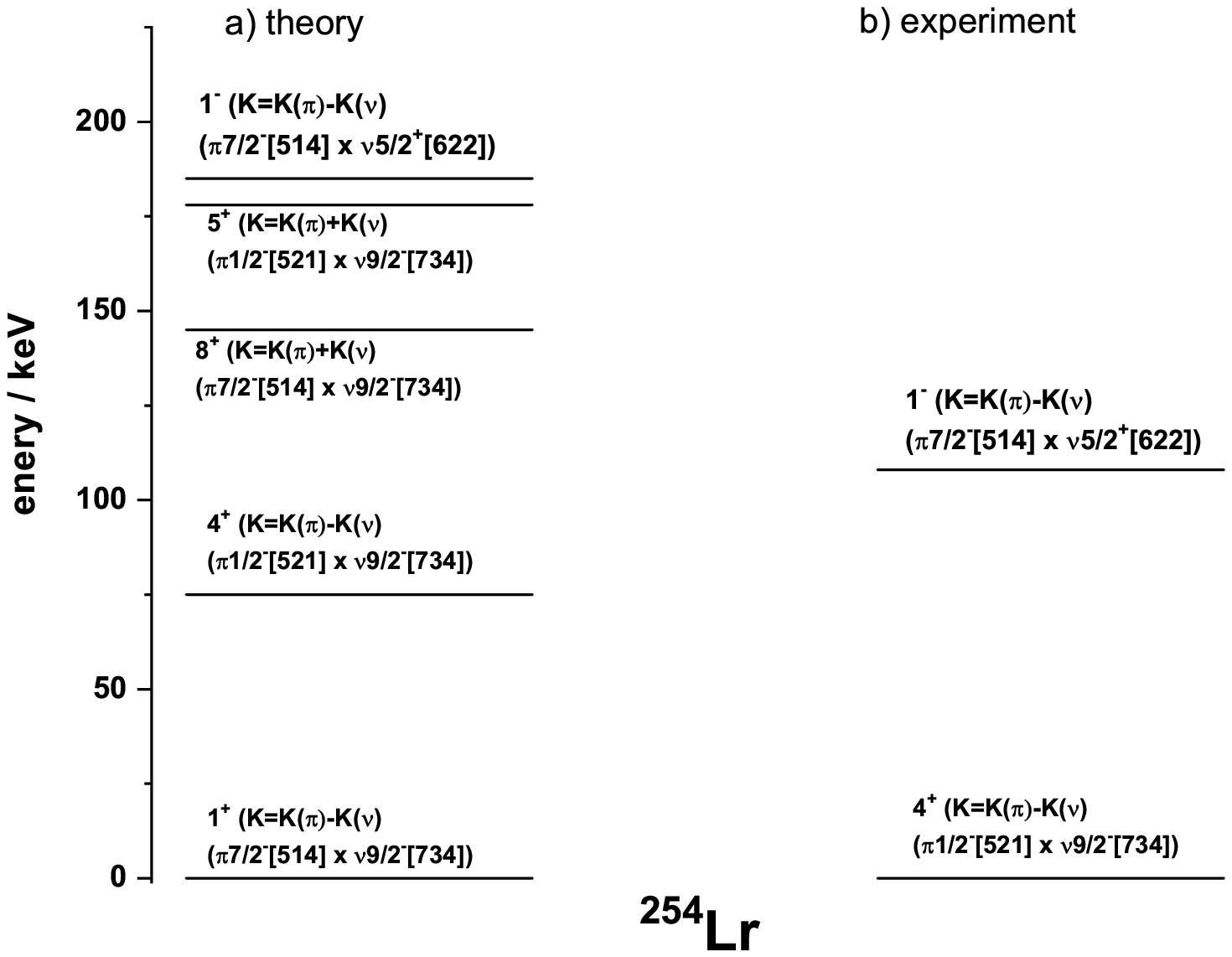}
	}
	\caption{Predicted (a) (\cite{Gow2019}) and (b) experimentally (tentatively) assigned \cite{Vosti19} low lying levels of $^{254}$Lr. }
	\label{fig:32}       
\end{figure}
{\bf $^{254}$Lr}\\
The isotope $^{254}$Lr was first synthesized by $\alpha$ decay of $^{258}$Db \cite{Hess85}. 
More detailed information on its decay properties was obtained from further decay studies of 
$^{258}$Db \cite{Hessb09,Vosti19} and direct production in the reaction $^{209}$Bi($^{48}$Ca,3n)$^{254}$Lr \cite{AntH08}.
The $\alpha$ spectrum turned out to be quite 'structureless', indicating a strong influence of energy summing
between $\alpha$ - particles and CE. Nevertheless some more detailed informations
on the decay of $^{254}$Lr were obtained from $\alpha$ - $\gamma$ measurements \cite{Vosti19}.
Information on the existence of a low lying isomeric state in $^{254}$Lr came from a detailed decay
study of $^{258}$Db \cite{Hessb09,Vosti19} which resulted in a notable population of a long
lived excited state in $^{254}$Lr by $\alpha$ decay of $^{258}$Db.
It was found on the basis of the energy balance, that the E = 9185 keV - transition  (in the former publication \cite{Hessb09} 
given as E = 9196 keV) of $^{258}$Db, 
did not populate the ground state of $^{254}$Lr. As this line was not found in coincidence with $\gamma$ rays or being influenced by energy summing with CE,
it was concluded that this transition must populate a level in $^{254}$Lr with a 'long' half-life ($>>$ 1 $\mu$s).
 Due to the above mentioned strong
influence of energy summing between $\alpha$ particles and CE $\alpha$ decays of $^{254m}$Lr and 
$^{254g}$Lr could not be disentangled, it just could be shown, that the $\alpha$ decays of $^{254m}$Lr 
are essentially concentrated in the energy range E$_{\alpha}$ $\approx$ (8430-8520) keV and in addition 
a small line at E$_{\alpha}$ = 8395 keV was indicated, while the energies of $\alpha$ particles attributed to the decay 
of $^{254g}$Lr were concentrated in  broader range E$_{\alpha}$ $\approx$ (8350-8520) keV. Most of the
$\gamma$ events could be attributed to the decay of levels in $^{250}$Md populated by $\alpha$ decay of
$^{254g}$Lr. It should be emphasized, that population of the isomeric state in $^{250}$Md at E$^{*}$ = 123 keV
by decay of $^{254g}$Lr could not be established so far. Also no detailed decay properties of $^{254m}$Lr 
were obtained so far.\\
The half-lives of the ground state and the isomeric state were measured as T$_{1/2}$ ($^{254g}$Lr) = 11.9 $\pm$ 0.9 s
and T$_{1/2}$ ($^{254m}$Lr) = 20.3 $\pm$ 4.2 s \cite{Vosti19}. On the basis of the decay properties of $^{258g,258m}$Db
the excitation energy of the isomer was determined as E$^{*}$ = 108 keV.  Existence and excitation energy of the 
isomeric state were confirmed by direct mass measurents of $^{254g,254m}$Lr, which resulted in E$^{*}$ = 107 $\pm$ 4 keV
for the isomer \cite{Kaleja20}.\\
Low energy levels (E$^{*}$ $<$ 250 keV) in $^{254}$Lr have been predicted by R. Gowrishankar et al.
on the basis of a 
'Two-Quasi-Particle-Rotor-Model' \cite{Gow2019}. The results 
are shown in fig. 32. The ground state is predicted as K$^{\pi}$\,=\,1$^{+}$ and an isomeric state  K$^{\pi}$\,=\,4$^{+}$
is predicted at E$^{*}$$\approx$75 keV, which is in quite fair agreement with the experimental
value of E$^{*}$ = 108 keV. Tentative spin and parity assigments are, however, different. Based on the experimental results 
the ground state was assigned as K$^{\pi}$\,=\,4$^{+}$, the isomeric state as K$^{\pi}$\,=\,1$^{-}$ (see fig. 35). This 
assignment was based on the assumed ground state configuration K$^{\pi}$\,=\,0$^{-}$ of $^{258}$Db and the low $\alpha$-decay hindrance factor
HF$\approx$30 for the transition $^{258g}$Db $\rightarrow$ $^{254m}$Lr 
which rather favors K$^{\pi}$\,=\,1$^{-}$ than K$^{\pi}$\,=\,4$^{+}$ as 
the latter configuration would require  an angular momentum change
$\Delta$K\,=\,3 and a change of the parity which requires a much larger hindrance factor (see sect. 7.1). \\
Here, however, two items should be considered: \\
a) the spin-parity assigmnent of $^{258}$Db is only tentative,\\
b) the calculations are based on the energies of low lying levels in the neighboring odd mass nuclei, 
in the respecting case $^{253}$No (N = 151) and $^{253}$Lr (Z = 103). The lowest Nilsson
levels in $^{253}$No are 9/2$^{-}$[734] for the ground state and  5/2$^{+}$[622] for a shortlived isomer at E$^{*}$\,=\,167 keV \cite{StH10}.
In $^{253}$Lr tentative assignments of the ground state (7/2$^{-}$[514]) and  1/2$^{-}$[521] 
for a low lying isomer are given in \cite{Hess01}. The energy of the isomer is experimentally not well established, for the calculations a value 
of 30 keV  was taken \cite{Gow2019},
while recently a value E$^{+}$\,$\approx$\,(6-24) keV was reported by P. Brionnet et al. \cite{BrioL25}.
It should be noted, however, that for the neighboring N = 152 isotope of lawrencium, $^{255}$Lr the ground state had been determined as the Nilsson level 1/2$^{-}$[521], while 7/2$^{-}$ was attributed to a low lying isomeric
state at E$^{*}$\,=\,37 keV \cite{Chat06}. Therefore, with respect to the uncertain starting conditions, 
the results of the calculations although not in 'perfect agreement' with the experimental results, are still
promising, and may be improved in future.\\

\subsection{{\bf 7.3 Isomeric states in odd-odd dubnium isotopes}}

\begin{figure}
	\resizebox{0.8\textwidth}{!}{%
		\includegraphics{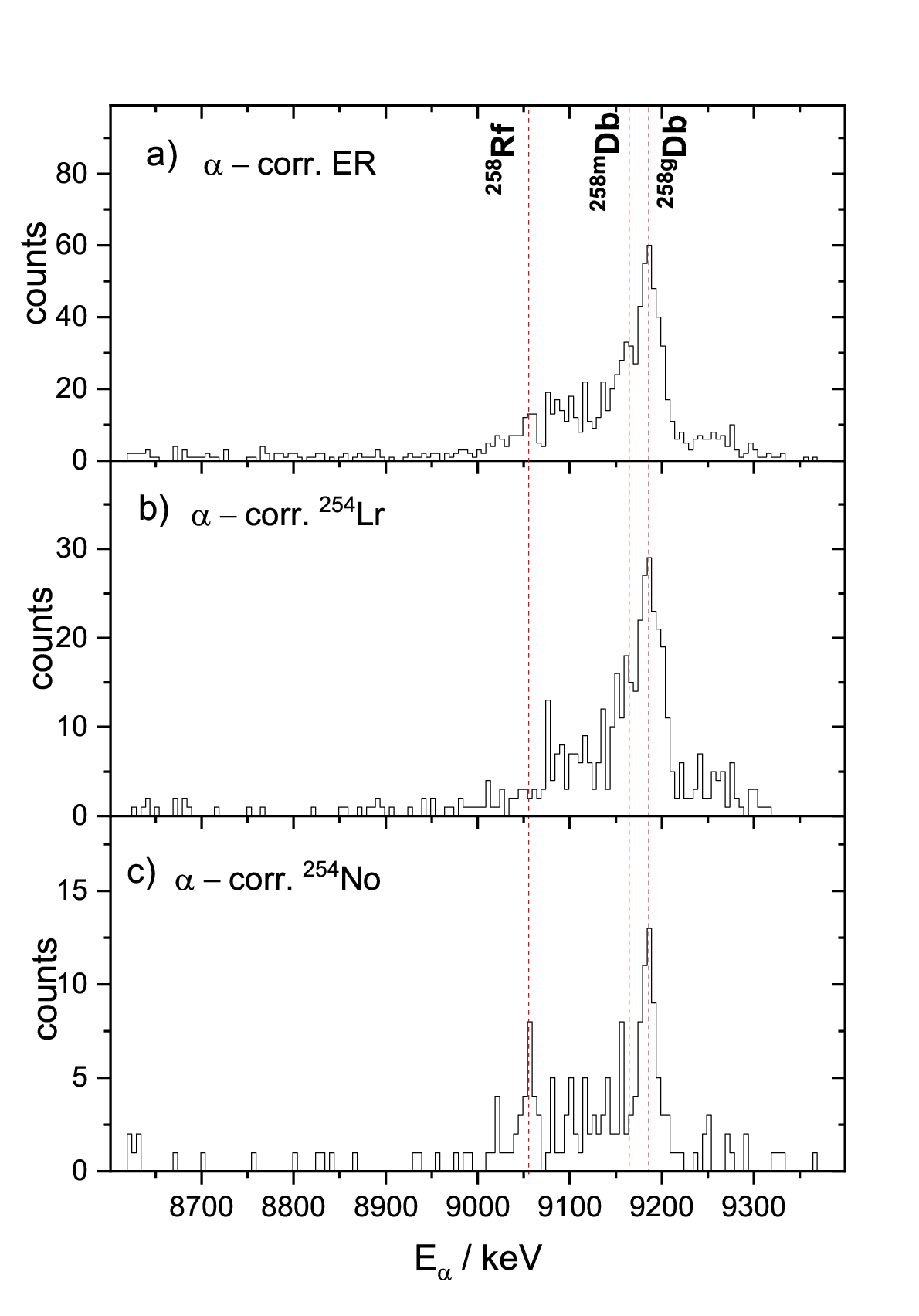}
	}
	\caption{$\alpha$ spectrum of $^{258}$Db; a) $\alpha$ decays correlated to ER; b) $\alpha$ decays 
	preceding $\alpha$ decays of $^{254}$Lr; c) $\alpha$ decays preceding $\alpha$ decays of $^{254}$No;
	in addition here also $\alpha$ decays of $^{258}$Rf, produced by EC decay of $^{258}$Db, are obseved.}
	\label{fig:33}       
\end{figure}

\begin{figure}
	\resizebox{0.9\textwidth}{!}{%
		\includegraphics{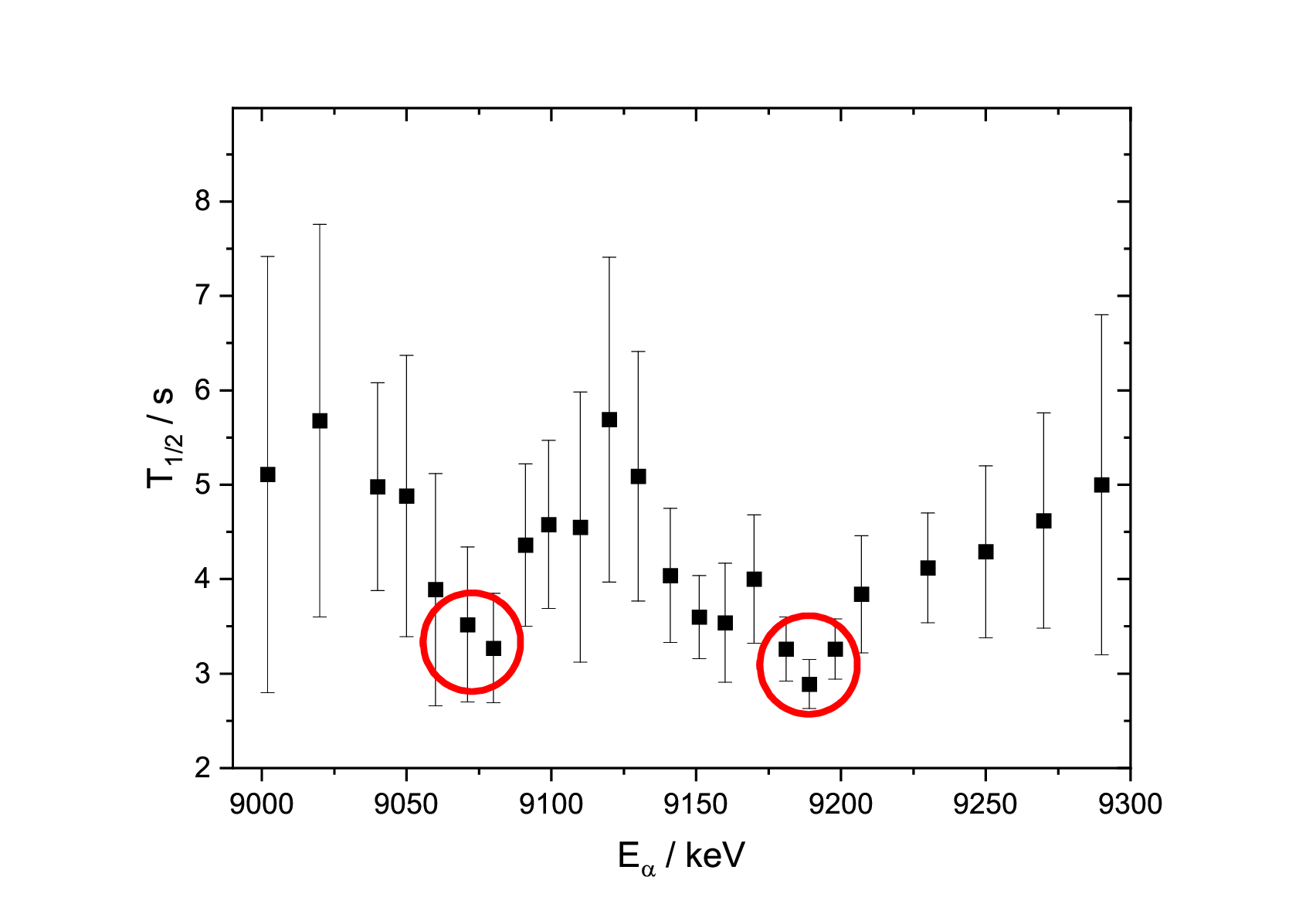}
	}
	\caption{Half-life distributions of $\alpha$ decays attributed to $^{258}$Db.}
	\label{fig:34}       
\end{figure}

\begin{figure}
	\resizebox{0.9\textwidth}{!}{%
		\includegraphics{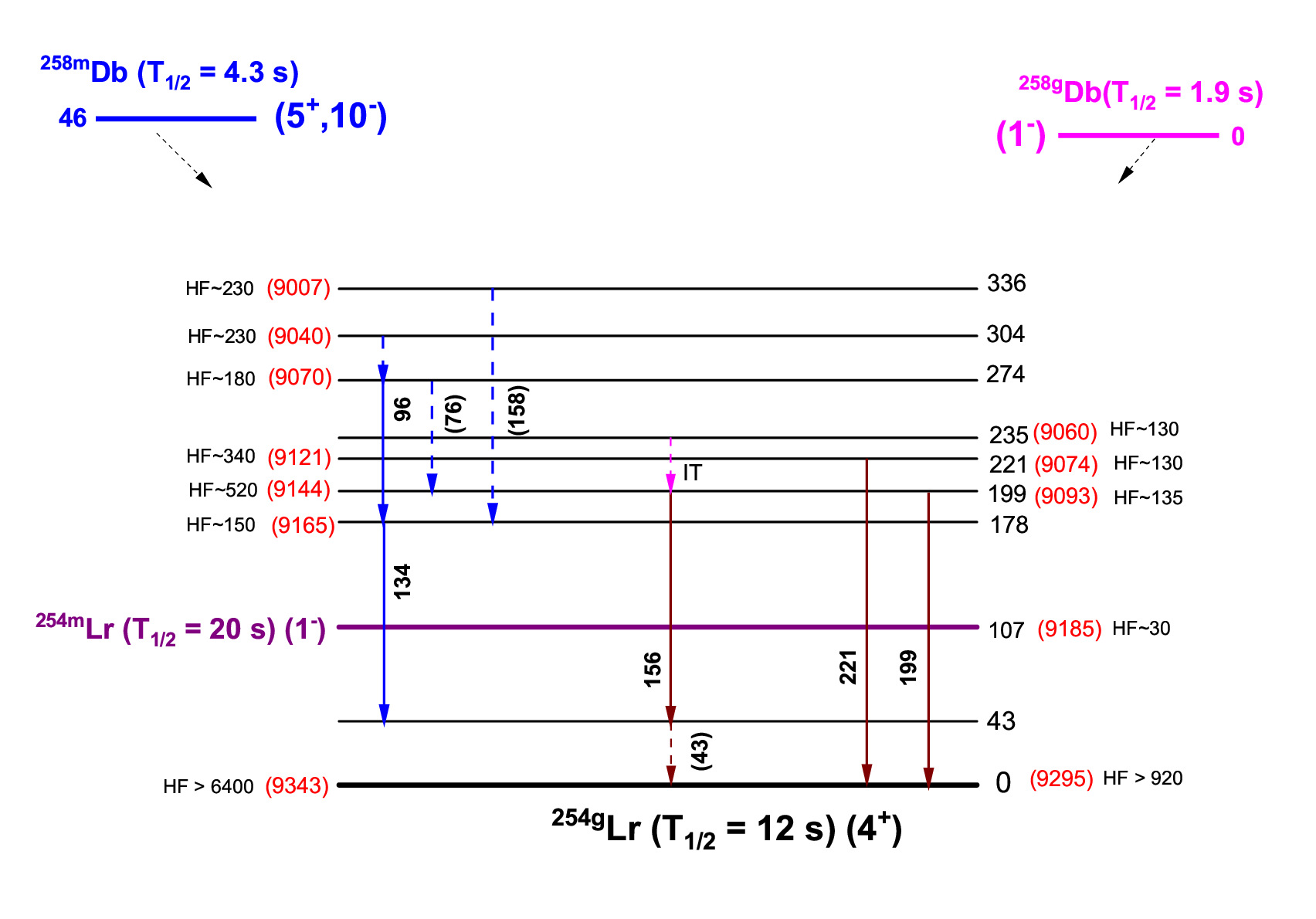}
	}
	\caption{Decay scheme of $^{258}$Db, partial level scheme of $^{254}$Lr as proposed in \cite{Vosti19}. }
	\label{fig:35}       
\end{figure}

{\bf $^{258}$Db}\\
$^{258}$Db is an illustrative example for the difficulties identifying an isomeric state
in case when the decay properties of the isomer and the ground state are similar.
Its identification thus will be discussed a bit more exhaustive.\\
The isotope $^{258}$Db was first identified at SHIP in the reaction 
$^{209}$Bi($^{50}$Ti,n)$^{258}$Db \cite{Hess85}. It was found to have a quite 
'complex' $\alpha$ - spectrum, a half-life T$_{1/2}$ = 4.4$^{+0.9}_{-0.6}$ s was
measured. Also SF with a comparable half-life was observed. As spontaneous fission
of odd-odd nuclei is strongly hindered the fission events were attributed to $^{258}$Rf,
a known spontaneously fissioning isotope, produced by EC decay of $^{258}$Db.
This interpretation was later proven by measuring K X-rays of rutherfordium (stemming 
from the EC process or converted internal transitions from excited levels in $^{258}$Rf
populated by the EC process) and SF events \cite{Hess16}.\\
First indication of an isomeric state in $^{258}$Db 
(and also in $^{254}$Lr) was obtained in an experiment 
performed at SHIP in 1997 from comparision of the spectra of $\alpha$  events of
$^{258}$Db followed by $\alpha$ decays of $^{254}$Lr or $^{254}$No \cite{Hessb09}. 
It was realized that the $\alpha$ line at E$_{\alpha}$ = 9166 keV was only present when the $\alpha$
decays of $^{258}$Db were followed by $\alpha$ decays of $^{254}$Lr, while 
the line at E$_{\alpha}$ = 9196 keV was seen when $\alpha$ decays of both
isotopes $^{254}$Lr and $^{254}$No followed. This feature was confirmed in a more 
detailed study \cite{Vosti19} (see fig. 33) and could be
interpreted in two ways:\\
a) the $\alpha$ line at E$_{\alpha}$ = 9196 keV followed the by decay of $^{254}$No
stems from $\alpha$ decay of $^{258}$Rf produced by EC decay of $^{258}$Db.
The $\alpha$ line itself had to be assumed to accidentally overlay with an
$\alpha$ transition energy of $^{258}$Db. Also a low EC branch of $^{254}$Lr 
and a so far not known high $\alpha$ branch of $^{258}$Rf had to be assumed.
The interpretation seemed tempting, but had to be rejected after $\alpha$ decay of 
$^{258}$Rf had been identified and the decay energy was measured 
as  E$_{\alpha}$ = 9.05 $\pm$ 0.03 MeV \cite{Gates2008}.\\
b) the second interpretation requires isomeric states in $^{258}$Db and also in $^{254}$Lr.
The $\alpha$ line at E$_{\alpha}$ = 9196 keV and at E$_{\alpha}$ = 9166 keV stem 
from different levels in $^{258}$Db and also populate different levels in $^{254}$Lr 
with the level populated by the line at E$_{\alpha}$ = 9196 keV having a significantly larger
EC - branch. The level populated by the E$_{\alpha}$ = 9196 keV transition may be the
excited one, decaying also notably by internal transition to the ground state.\\
The assumption of two different levels decaying by $\alpha$ emission was corroborated 
by the results from a decay study of $^{262}$Bh, for which two levels decaying by 
$\alpha$ emission were known \cite{Muen89}.
It was shown that the  E$_{\alpha}$ = 9196 keV - line was practically only preceded
by decays of $^{262}$Bh(2) \cite{Hessb09} (see also fig. 38 in sect.7.4).\\
As a result, a broad $\alpha$ energy distribution in the range E$_{\alpha}$ $\approx$ (9090-9353) keV
and a half-life T$_{1/2}$ = 4.3$\pm$0.5 s were attributed to $^{258}$Db(1), while  an $\alpha$ energy 
E$_{\alpha}$ = 9196 $\pm$ 10 keV
and a half-life T$_{1/2}$ = 1.9$\pm$0.5 s were attributed to $^{258}$Db(2).
It remained, however, unclear, which activity represented the ground state 
and which the isomer.\\
The situation could be clarified in a detailed study at SHIP, including also
$\alpha$ - $\gamma$ coincidence measurements \cite{Vosti19}.
An analysis of the half-lives of the broad energy distribution of the $\alpha$ decays,
subdivided in 20 keV- or 40 keV - bins is shown in fig. 34. Evidently two regions 
of lower half-lives are present, one at 
E$_{\alpha}$ $\approx$ 9070 keV and one at E$_{\alpha}$ $\approx$ 9180  keV.
Halflives of T$_{1/2}$ = 4.32 $\pm$ 0.42 s  for $^{258}$Db(1) 
(E$_{\alpha}$ $<$ 9040 keV, (9090-9170) keV, $>$9210 keV) and  
T$_{1/2}$ = 2.91 $\pm$ 0.24 s  for $^{258}$Db(2) 
(E$_{\alpha}$ (9065-9085) keV, (9175-9205) keV) are obtained.
We have here to note, that there might still be an overlay of $\alpha$s from
both states due to limited detector resolution and probable energy summing
between $\alpha$ - particles and CE that effect the half-life measurement.\\
The decay schemes for $^{258m}$Db and $^{258g}$Db and the partial level scheme of $^{254}$Lr 
derived from the measured data is shown in fig. 35.
Tentative spin and parity assignments of the ground state and the isomeric
 state were based on the EC decay properties of $^{258}$Db \cite{Hess16} and
the coupling rules of unpaired protons and neutrons in odd-odd nuclei as 
suggested by M.H. Brennan and A.M. Bernstein \cite{BrB60}. The latter
resulted in I$^{\pi}$ = 0$^{-}$, 5$^{+}$, 10$^{-}$  as low lying states 
\cite{Vosti19}. From time distributions of fission events of $^{258}$Rf produced by
EC decay of $^{258}$Db, somewhat different half-lifes resulted for
 events preceded by CE (T$_{1/2}$ = 4.4$\pm$1.0 s) or not preceded by CE 
(T$_{1/2}$ = 3.6$\pm$0.3 s). The first group was assigned to EC decays
 populating excited levels in $^{258}$Rf, the second group was interpreted
as a mixture of decays into the ground state of $^{258}$Rf (I$^{\pi}$ = 0$^{+}$)
 and decays into excited levels where the CE was not registered.
Consequently spin/parity  I$^{\pi}$ = 0$^{-}$ were attributed the state 
of the lower half-life, the ground state of $^{258}$Db and the spins 
I$^{\pi}$ = 5$^{+}$ or  10$^{-}$ to the isomeric state having the longer half-life.\\

\begin{figure}
	\resizebox{0.75\textwidth}{!}{%
		\includegraphics{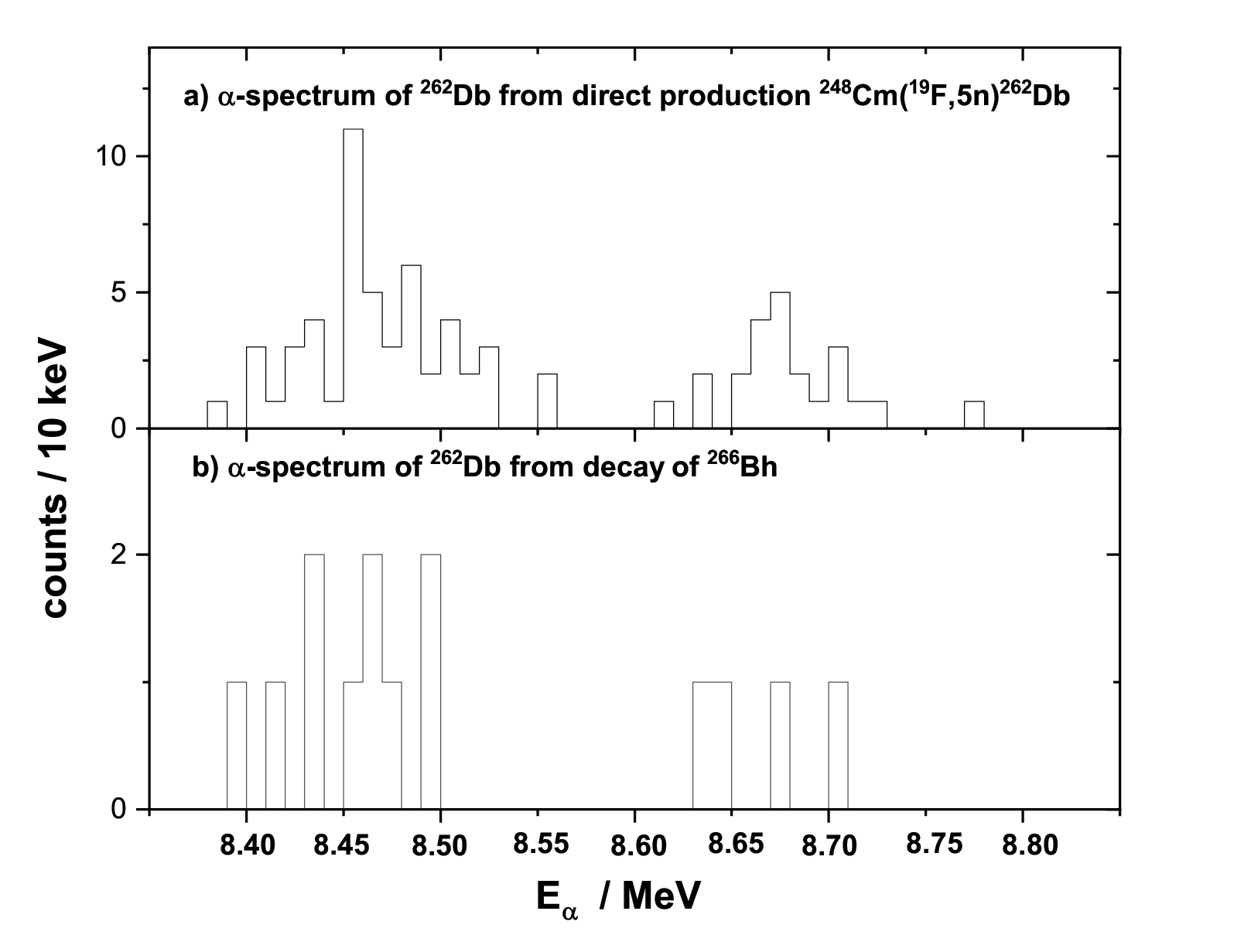}
	}
	\caption{$\alpha$ spectra of $^{262}$Db; a) spectrum taken in 'direct' production \cite{Haba14}; 
		b) spectrum taken from decay of $^{266}$Bh \cite{Haba20}.}
	\label{fig:36}       
\end{figure}

\vspace{10mm}

 {\bf $^{262}$Db}\\
 The observation of $^{262}$Db was first reported by A. Ghiorso et al. \cite{GhN71}, who reported a complex $\alpha$ spectrum
 with a 'main' line at E$_{\alpha}$ = 8.45 MeV and indication of another line at E$_{\alpha}$ = 8.66 MeV, which, however, was assumed
 to be overlayed by $\alpha$ decays of the daughter product $^{258}$Lr. Fission events with comparable half-lives were
 reported by J.R. Bemis et al. \cite{BeF77a} and V.A. Druin et al. \cite{DrB79}.\\
 In further experments $^{262}$Db was produced either directly by the reaction $^{248}$Cm($^{19}$F,5n)$^{262}$Db \cite{Dress99} or by decay of $^{266}$Bh
 \cite{Morita15,Wilk00,Morita09,Qin06}. All these experiments suffered from a low number of observed decays.\\
 Experiments with higher 'statistics' recently have been performed by H. Haba et al. who either produced it in the reaction
 $^{248}$Cm($^{19}$F,5n)$^{262}$Db \cite{Haba14} or by $\alpha$ decay of $^{266}$Bh \cite{Haba20}.
 The $\alpha$ spectra registered in both experiments are shown in figs. 36a) and 36b). As $\alpha$ decay energy
 regions of $^{262}$Db and $^{258}$Lr overlap in case of decay of $^{266}$Bh only data from triple correlations
 $^{266}$Bh $^{\alpha}_{\rightarrow}$ $^{262}$Db $^{\alpha}_{\rightarrow}$ $^{258}$Lr $^{\alpha}_{\rightarrow}$ are considered.
 Evidently there are two groups of $\alpha$ decays, E$_{\alpha 1}$\,=\,(8.35-8.55) MeV and E$_{\alpha 2}$\,=\,(8.60-8.80) MeV.
 Half-lives of both groups are T$_{1/2}$ = 35.5$^{+5.2}_{-4.2}$ s (E$_{\alpha 1}$) and T$_{1/2}$ = 21.8$^{+5.5}_{-3.6}$ s (E$_{\alpha 2}$).
 The difference is already outside the sum of the 1$\sigma$ error bars, and therefore both groups of $\alpha$ decays have to be
 considered to stem from the decay of different levels. This assumption is corroborated the analysis of 
 $\alpha$ - $\alpha$ correlations $^{266}$Bh $^{\alpha}_{\rightarrow}$ $^{262}$Db $^{\alpha}_{\rightarrow}$ 
 (see section 7.4 for a detailed discussion).

\subsection{{\bf 7.4 Isomeric states in odd-odd bohrium isotopes}}

\begin{figure*}
	\resizebox{0.75\textwidth}{!}{
		\includegraphics{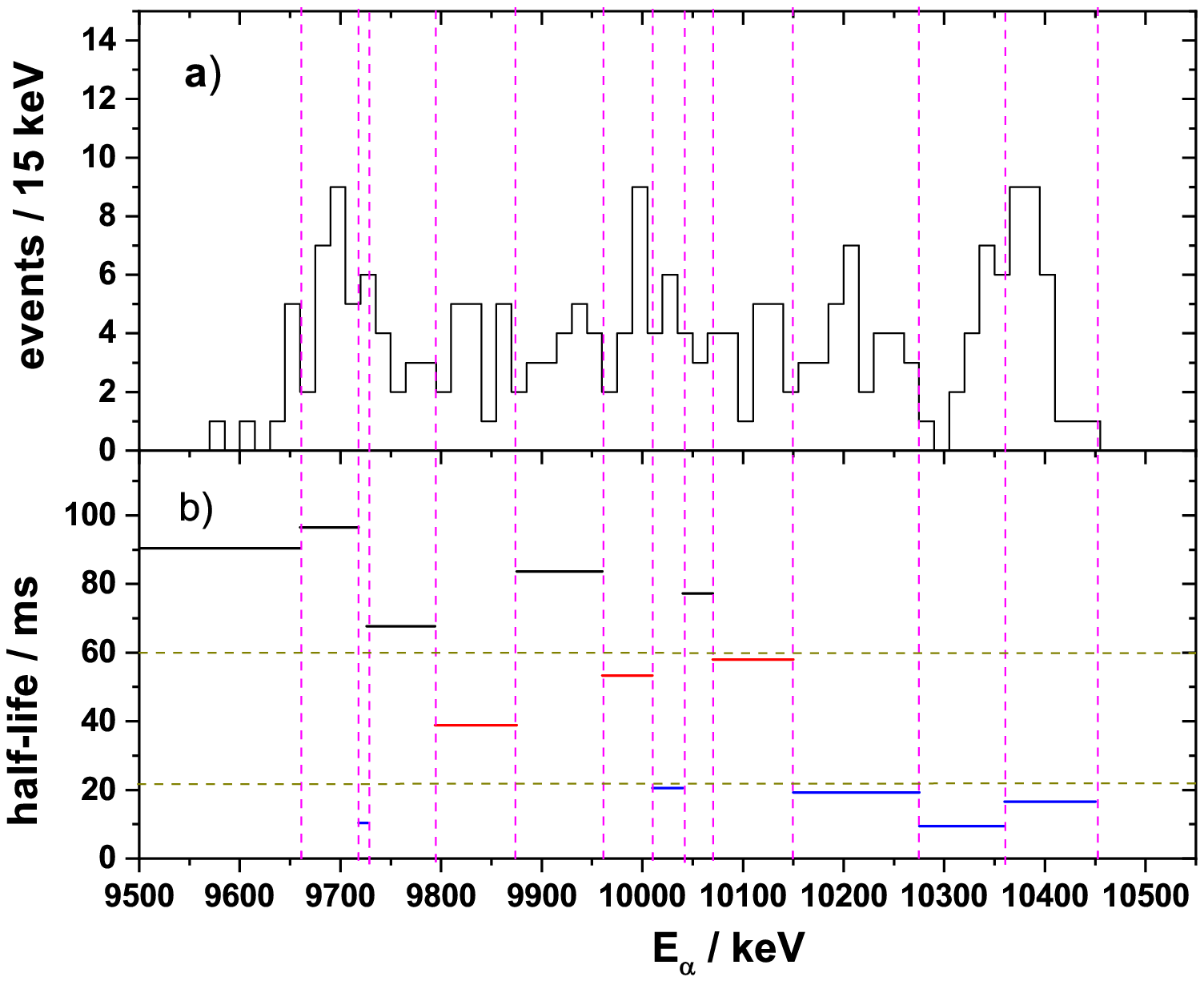}	
	}
	\caption{a) $\alpha$ spectrum of events attributed to the decay of $^{262}$Bh(1) or $^{262}$Bh(2);
		b) half-lives of $\alpha$ decays in the considered energy region; black: half-lives attributed to $^{262}$Bh(1), 
		blue: half-lives attributed to $^{262}$Bh(2),  red: no assignment. }
	
	\label{fig:37}       
\end{figure*}
\begin{figure*}
	\resizebox{0.75\textwidth}{!}{
		\includegraphics{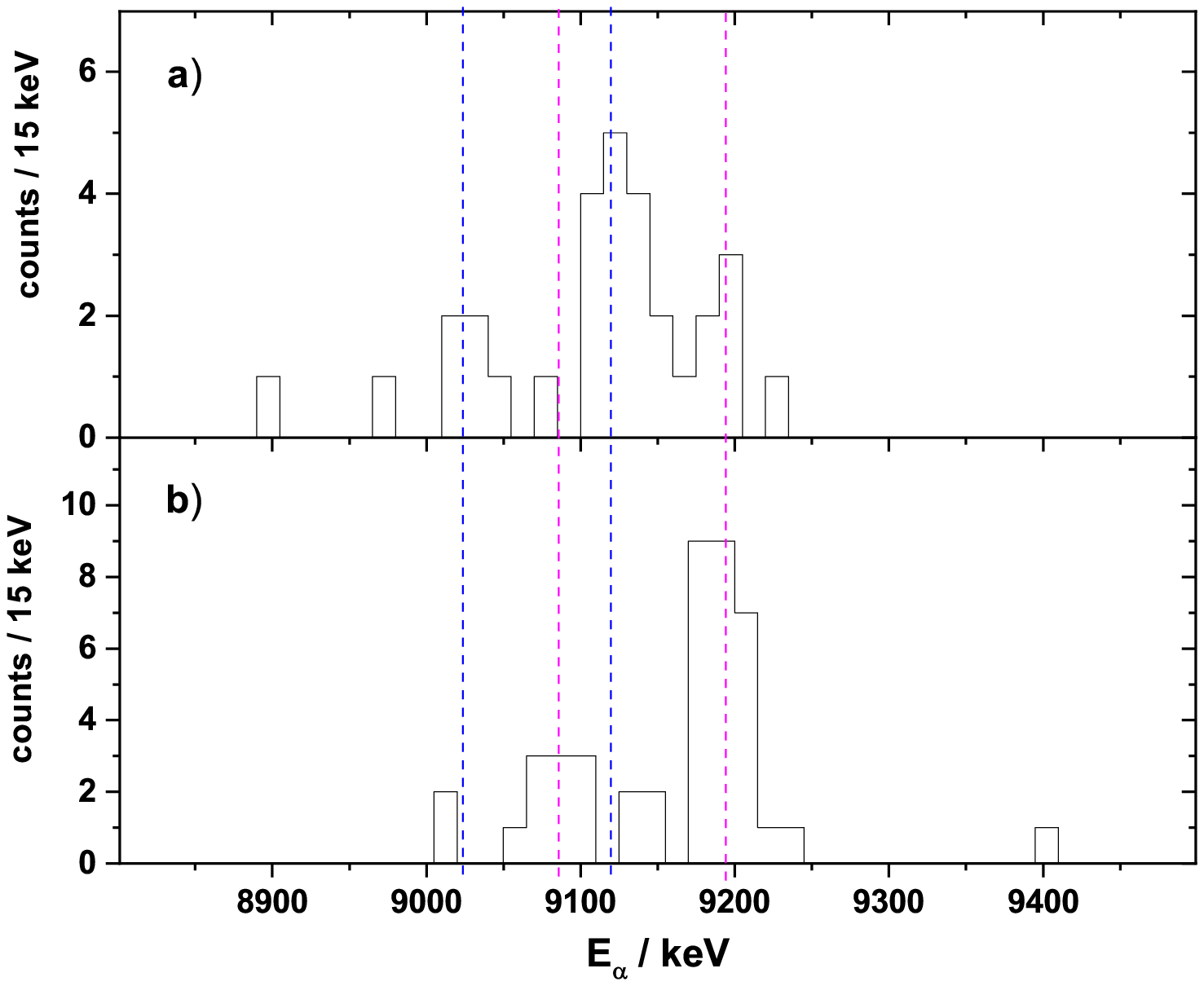}	
	}
	\caption{$\alpha$ spectra of $^{258}$Db events following $\alpha$ decays of $^{262}$Bh(1) (fig. 38a) or $^{262}$Bh(2) (fig. 38b). }
	
	\label{fig:38}       
\end{figure*}
\begin{figure*}
	\resizebox{0.75\textwidth}{!}{
		\includegraphics{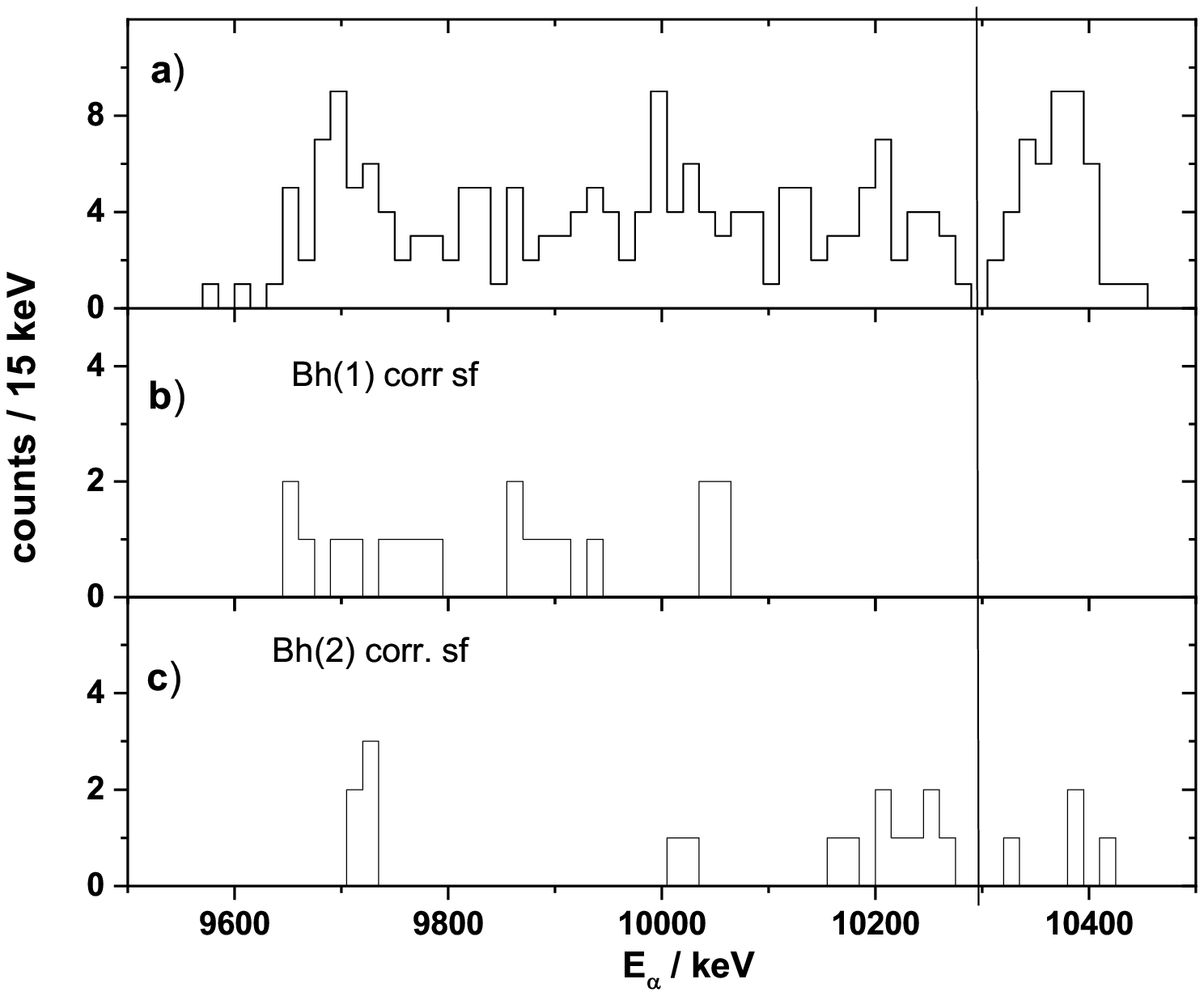}	
	}
	\caption{a) $\alpha$ spectrum of events attributed to the decay of $^{262}$Bh(1) or $^{262}$Bh(2);
		b) $\alpha$ spectrum of  $^{262}$Bh(1) events followd by sf;
		c) ) $\alpha$ spectrum of  $^{262}$Bh(2) events followd by sf.}
	
	\label{fig:39}       
\end{figure*}
\begin{figure*}
	\resizebox{0.75\textwidth}{!}{
		\includegraphics{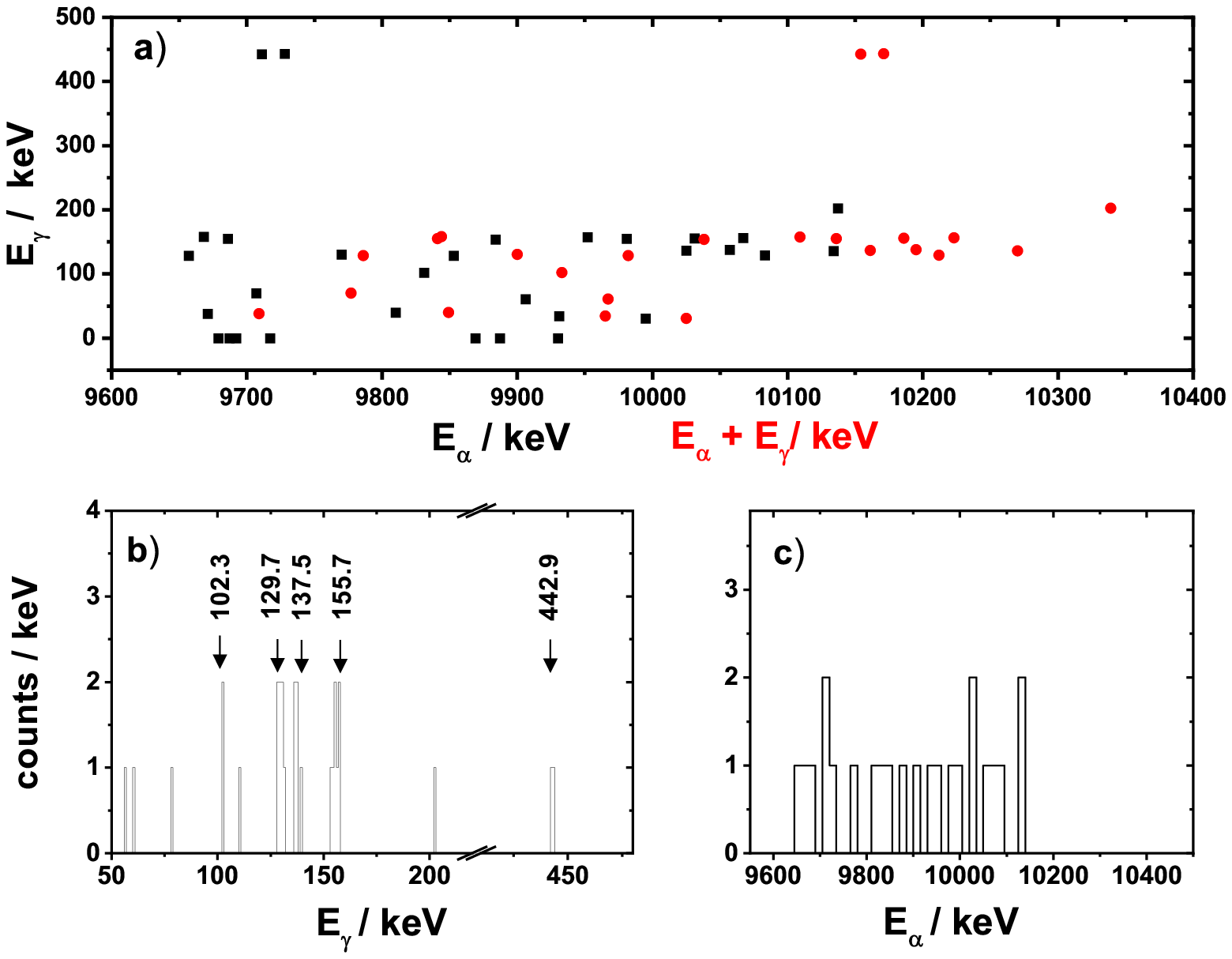}	
	}
	\caption{a) 2-dimensional plot of $\alpha$ - $\gamma$ - coincidences observed for the decay of $^{262}$Bh; black: coincidences E$_{\alpha}$ -  E$_{\gamma}$, 
		red: coincidences (E$_{\alpha}$+E$_{\gamma}$)   -  E$_{\alpha}$;
		b) $\gamma$ spectrum recorded in coincidence with $\alpha$ decays of  $^{262}$Bh;
		c) $\alpha$ spectrum recorded in coincidence with $\gamma$ events.}
	
	\label{fig:40}       
\end{figure*}
\begin{figure*}
	\resizebox{0.75\textwidth}{!}{
		\includegraphics{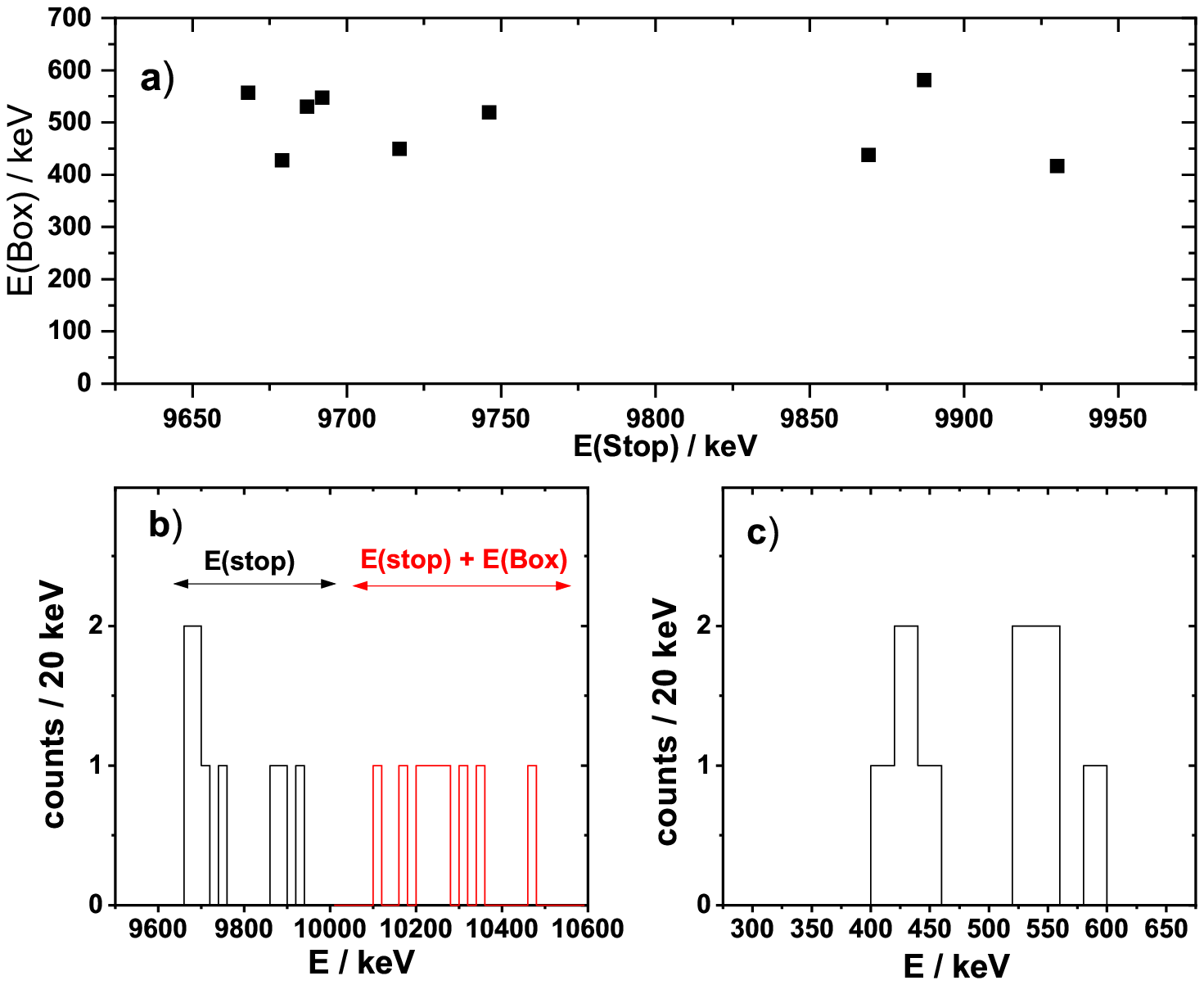}	
	}
	\caption{a) 2-dimensional plot of 'low energy' events registered in the 'box detector' and 
		coincident events registered in the 'stop detector';
		b) spectrum of $\alpha$ events recorded in coincidence with 'low' energy events in the 'box detector'; 
		black: energy recorded in the 'stop detector' (E(stop)), red: sum of energies recorded recorded in the 'stop' 
		and the 'box detector' (E(stop)+E(box));
		c) energy of the 'low' energy events recorded in the 'box' detector.}
	
	\label{fig:41}       
\end{figure*}

\textbf{$^{262}$Bh}\\
The isotope $^{262}$Bh was first synthesized at SHIP, GSI, by G. M\"unzenberg et al. \cite{MuH81}, which was regarded
as the discovery of element 107. At this time five decays with a mean energy E$_{\alpha}$ = 10376$\pm$35 keV and a half-life
T$_{1/2}$\,=\,4.7$^{+2.3}_{-1.6}$ ms and one decay with an energy  E$_{\alpha}$ = 9709$\pm$50 keV and a lifetime (time
difference between implantation of the evaporation residue and the $\alpha$ decay) of $\tau$ = 165 ms were observed. From the 
latter a half-life T$_{1/2}$ = 115$^{+231}_{-75}$ ms was extracted. Due to the significantly different half-lives it was 
speculated that the E$_{\alpha}$ = 9709$\pm$50 keV decay represents the ground state decay, the events of 
E$_{\alpha}$ = 10376$\pm$35 keV represent the decay of an isomeric state  \cite{MuH81}. Decay data of 
higher quality were obtained in succeeding experiments, either producing $^{262}$Bh directly in the
reaction $^{209}$Bi($^{54}$Cr,1n)$^{262}$Bh \cite{Hessb09,MuH81,MuA89} or by $\alpha$ decay of $^{266}$Mt \cite{MuA82,MuH88,HoH97}.
In these experiments the existence of an isomeric state was secured, but it turned out, that the $\alpha$ decay spectrum of $^{262m,262g}$Bh 
has a complicated structure ranging from E$_{\alpha}$ $\approx$ 9.6 MeV to  E$_{\alpha}$ $\approx$ 10.75 MeV, and decays of 
$^{262g}$Bh and $^{262m}$Bh overlay in energy. The SHIP results were confirmed by C.M. Folden III at the BGS, LBNL Berkeley
\cite{Folden06} and by T. Zhao et al. at the SHANS 2 separator at IMP, Lanzhou \cite{Zhao24}. 
The data are compared in table 7.\\

\begin{table}
	\caption{Half-lives and $\alpha$ energy ranges (in case of \cite{MuH81} $\alpha$ energies) and half-lives
	published for $^{262}$Bh(1) and  $^{262}$Bh(2. In \cite{Zhao24} the energy range (10.0\,-\,10.3) MeV
	is not assigned to $^{262}$Bh(1) or $^{262}$Bh(2).}
	\label{tab:3}       
	\begin{tabular}{lllll}
		\hline\noalign{\smallskip}
		\noalign{\smallskip}
		Refernce & $^{262}$Bh(1)       &            & $^{262}$Bh(2)        &   \\
                 & E$_{\alpha}$ / MeV  & T$_{1/2}$ /ms  &   E$_{\alpha}$ / MeV &  T$_{1/2}$ /ms \\
		\hline\noalign{\smallskip}  
	 \cite{MuH81}    &  9.704$\pm$0.050  & 115$^{+231}_{-75}$   & 10.376$\pm$0.035 &  4.7 $^{+231}_{-75}$ \\
	 \cite{MuA89}    & 9.74\,-\,10.06    & 102$\pm$26           & 10.24\,-\,10.37  & 12.4 $^{+8.1}_{-3.5}$ \\
	 \cite{HoH97}    & 9.75\,-\,10.15    & 64$^{+38}_{-18}$*    & 10.2\,-\,10.44   & 11.2 $^{+11.2}_{-3.7}$* \\
	 \cite{Hessb09}   & 9.69\,-\,10.01    & 83$\pm$14            & 9.83\,-\,10.37   & 22$\pm$4                \\
	 \cite{Folden06} & 9.65\,-\,10.08    & 84$^{+21}_{-1}$      & 10.15\,-\,10.35  & 9.6 $^{+3.6}_{-2.4}$  \\
	 \cite{Zhao24}   & 9.6\,-\,10.0     & 96$^{+18}_{-13}$    & 10.3\,-\,10.4    & 21$^{+8}_{-5}$ \\
	\end{tabular} \\
	
	* half-lives calculated  from the time differences given in table 2 in \cite{HoH97}. \\
\end{table}

In the following the decay of $^{262m,262g}$Bh will be discussed in some more detail. For this purpose
the data given in \cite{MuH81,MuA82,MuH88,MuA89,HoH97,Folden06,Hessb09,Zhao24} are
merged.\\
The energy distributions for all events from the different experiments registered with full energy release
in the 'stop detectors' and attributed to the decay of $^{262}$Bh are shown in fig. 37a.
It should be noted that the $\alpha$ energies might be influenced by energy summing with CE
and thus not represent transition energies between distinct levels. Nevertheless they can be 
divided into three groups on the basis of the lifetimes as shown in fig. 37b.\\
a) The events in group 1 have half-lives of T$_{1/2}$ $>$ 60 ms, and are found 
essentially in the energy range E$_{\alpha}$ $<$ 10 MeV. Evidently they represent
decays of the activity denoted as $^{262}$Bh(1) \cite{Hessb09}.\\
b) The events in group 2 have half-lives of T$_{1/2}$ $<$ 22 ms, and are found 
essentially in the energy range E$_{\alpha}$ $>$ 10.15 MeV. Evidently they represent
decays of the activity denoted as $^{262}$Bh(2) \cite{Hessb09}.\\
c) The events in group 3 have half-lives $T_{1/2}$ = (40-60) ms, probably a mixture of 
decays of  $^{262}$Bh(1) and  $^{262}$Bh(2). They are not considered in the further discussion.\\
The half-life for the events in group 1 attributed to $^{262}$Bh(1) is T$_{1/2}$ = 93.5$^{+12.3}_{-9.4}$ ms.
The $\alpha$ spectrum of the correlated $^{258}$Db decays is shown in fig. 38. Three lines are visible.\\
a) E$_{\alpha}$ = 9186 $\pm$ 3 keV, T$_{1/2}$ = 1.12$^{+0.73}_{-0.32}$ s.\\
b) E$_{\alpha}$ = 9125 $\pm$ 3 keV, T$_{1/2}$ = 2.23$^{+0.62}_{-0.40}$ s.\\
c) E$_{\alpha}$ = 9024 $\pm$ 13 keV, T$_{1/2}$ = 1.95$^{+1.19}_{-0.53}$ s.\\
Evidently the half-lives are inline with that of $^{258g}$Db, while the $\alpha$ line of 
E$_{\alpha}$ = 9186 $\pm$ 3 keV is compatible with the line at E$_{\alpha}$ = 9185 keV at direct production
populating
the isomeric state in $^{254}$Lr at E$^{*}$ = 108 keV. But no significant $\alpha$ lines 
at E$_{\alpha}$ = 9125 keV and E$_{\alpha}$ = 9024 keV have been observed at direct production. 
The half-life of the spontaneous fission events following the decay of $^{262}$Bh(1) is 
T$_{1/2}$ = 3.80$^{+1.32}_{-0.70}$ s. and thus inline with that of $^{258m}$Db.\\
The half-life of the events in group 2 ($^{262}$Bh(2)) is T$_{1/2}$ = 13.7$^{+1.6}_{-1.3}$ ms.
The correlated $\alpha$ events are shown in fig. 38b. It exhibits a prominent line of E$_{\alpha}$ = 9190$\pm$3 keV
having a half-life of T$_{1/2}$ = 2.11$^{+0.47}_{-0.33}$ s, which is in-line with the half-life of $^{258g}$Db
(T$_{1/2}$ = 2.17$\pm$0.36 s \cite{Vosti19}) and the energy of the $\alpha$ decay (E$_{\alpha}$ =9185 keV) populating
the isomeric state in $^{254}$Lr at E$^{*}$ = 108 keV. The energy of the $\alpha$ line at E$_{\alpha}$ = 9075 keV
is inline with the energy of the transition populating the level at E$^{*}$ = 221 keV in $^{254}$Lr. The half-life
obtained from all correlated events is slightly higher, T$_{1/2}$ = 2.76$^{+0.40}_{-0.31}$ s, that could indicate 
that the decays attributed to $^{262}$Bh(2) are either slightly contaminated with decays of $^{262}$Bh(1) or that 
$^{258m}$Db is populated by $\alpha$ decay of $^{262}$Bh(2) either directly or via internal transition after
$\alpha$ decay.\\
A completely different half-life is obtained for spontaneous fission events following decays of $^{262}$Bh(2). We
obtain a value T$_{1/2}$ = 4.92$^{+1.46}_{-0.92}$ s, rather inline with the half-life of $^{258m}$Db. 
The $\alpha$ spectra of $^{262}$Bh(1) and $^{262}$Bh(2) correlated to spontaneous fission events are shown 
in fig. 39b and fig. 39c, respectively. Obviously there is no 'real' connection between the 'general' $\alpha$ energy
distribution of $^{262}$Bh(1,2) (see fig. 39a) and those followed by spontaneous fission. 
Specifically one observes for $^{262}$Bh(1) a somewhat enhanced intensity of $\alpha$ events in the 
range E$_{\alpha}$ = (9840\,-\,9920) keV.  In total one obtains a ratio $\Sigma$($\alpha$ corr. SF) / $\Sigma$($\alpha$) = 0.40$\pm$0.15.\\
The situation is even more pronounced for $^{262}$Bh(2). In the quite narrow energy range E$_{\alpha}$ = (9718\,-\,9727) keV,
where all events have time distances $\Delta$t(ER-$\alpha$) $<$ 35 ms, only one out of six decay events of $^{262}$Bh(2) is 
correlated to an $\alpha$ decay of $^{258}$Db, the other events are followed by SF. In the region 
E$_{\alpha}$ $>$ 10150 keV one gets for the interval E$_{\alpha}$ = (10150\,-\,10300) keV a ratio
$\Sigma$($\alpha$ corr. SF) / $\Sigma$($\alpha$) = 0.28$\pm$0.14, while the interval E$_{\alpha}$ $>$ 10300 keV a
significantly lower ratio
$\Sigma$($\alpha$ corr. SF) / $\Sigma$($\alpha$) = 0.14$\pm$0.05 is observed.\\
The situation thus appears quite complicated. While energies and half-lives of alpha decays following the decays as well of
$^{262}$Bh(1) as of $^{262}$Bh(2) indicate that essentially the ground state of $^{258}$Db is populated, either directly
or via internal transitions, the half-lives of fission events following either $\alpha$ decays of $^{262}$Bh(1) or $^{262}$Bh(2) 
rather indicate that they stem from the decay of $^{258m}$Db (most probably from $^{258}$Rf after EC decay of $^{258m}$Db).
For $^{262}$Bh(2) decay one obtains significantly different rates of $\alpha$ decays to fission events for the different
$\alpha$ energy regions attributed to $^{262}$Bh(2) (E$_{\alpha}$ = (9718\,-\,9727) keV,  E$_{\alpha}$ = (10150\,-\,10300) keV,
E$_{\alpha}$ $>$10300 keV). Under these circumstances it seems impossible to establish reliable decay schemes for
$^{262}$Bh(1) and $^{262}$Bh(2) on the basis of the present data.\\
The $\alpha$ - $\gamma$ - coincidences are shown in fig. 40a-c. (It should be noteted, that in figs. 40a,c only 
$\alpha$ - $\gamma$ coincidences or  $\alpha$ - decays with full energy release of the $\alpha$ particles in the
'stop detector' are shown.) Fig. 40a) shows a two-dimensional plot of $\alpha$ - $\gamma$ coincidences. The black squares
represent coincidences E$_{\alpha}$ - E$_{\gamma}$, the red dots coincidences (E$_{\alpha}$+E$_{\gamma}$)  - E$_{\gamma}$.
Evidently the energy of the $\alpha$ particles reaches only up to E$_{\alpha}$ $\approx$ 10150 keV, as also seen somewhat 
clearer in fig. 40c. So it can be concluded, that practically all $\alpha$ - $\gamma$ coincidences stem from the decay
of $^{262}$Bh(1). (A few coincidences from decay of $^{262}$Bh(2) cannot be excluded.) This is corroborated by the
half-life of the coincidence events, which is T$_{1/2}$ = 96.0$^{+21.0}_{-14.6}$ ms. As also seen in fig. 40a (red dots), the sum 
E$_{\alpha}$ + E$_{\gamma}$ does not exceed (except for one event) the value E\,=\,10300 keV. This value is lower than the
value for the high-energy peak of $^{262}$Bh(2) of E\,$\approx$10380 keV. As both activities $^{262}$Bh(1) and
$^{262}$Bh(2) populate the ground state of $^{258}$Db, the higher decay energy of $^{262}$Bh(2) may indicate that it represents
the isomeric state in $^{262}$Bh, while $^{262}$Bh(1) represents the ground state. On the basis of the present data,
estimate of the excitation energy of the isomeric state is problematic. Tentatively, on the basis of the energy differences a value
of E$^{*}$ $\approx$ 150$\pm$50 keV is suggested.\\
As seen in fig. 40b, the $\gamma$ events exhibit five lines at E$_{gamma}$ = 102.25$\pm$0.15 keV (2 events), 
129.74$\pm$0.34 keV (7 events), 137.54$\pm$0.61 keV (5 events), 155.67$\pm$0.62 keV (7 events), 442.85$\pm$0.25 keV (2 events).
The energies of the lines at 129.74$\pm$0.34 keV and 137.54$\pm$0.61 keV are in-line with the 
K$_{\alpha 2}$ (129.88 keV) and K$_{\alpha 1}$ (137.35 keV)  x-ray lines of dubnium.\\
Finally some events with exceptional properties should be presented. Usually decays registered as sum events in the 'stop detector' and 'box detector',
exhibit a low energy (energy loss) signal  E$<$2 MeV in the 'stop detector' and a high energy signal in the 'box detector'. In the decay studies  of $^{262}$Bh 
altogether nine events with 'high' energies (9600\,-\,10000) in the 'stop detector' and 'low' energies (400\,-\,600) keV in the 'box detector' were
observed (see fig. 41a) Due to the low energy it is tempting to assign these events to conversion electrons. 
While the energies measured in the 'stop detector' are in the region of the $\alpha$ decay energies
of $^{262}$Bh(1), the sum energies in 'stop detector' + 'box detecor' are E$>$10100 keV and thus 
in the decay energy range of $^{262}$Bh(2)
(see fig. 41b).
An assignment to $^{262}$Bh(2) is supported by a half-life of T$_{1/2}$ = 28$^{+14}_{-7}$ ms.  
So it seems probable that $\alpha$ decay of $^{262}$Bh(2) also populates to notable extent an high energy level in $^{258}$Db (E$^{*}$$>$400 keV)
which essentially decays by internal conversion. However, conversion coefficients for transitions of low multipolarity 
like E1, E2, M1, M2 are low ($\alpha_{tot}$ $<1$) at $\Delta$E $>$ 400 keV 
(energy differece between initial and final level) \cite{Kib08}, while lifetimes for transitions with 
higher multipolarity are $>$10 $\mu$s \cite{Fire96} and thus CE are not observed in (prompt) coincidence with $\alpha$ particles.
So one my speculate about an E0 transition. In \cite{Vosti19} the ground state of $^{258}$Db 
tentatively was assigned as I$^{\pi}$ = 0$^{-}$ 
($\pi$1/2$^{-}$[521]$\downarrow$ $\otimes$ $\nu$1/2$^{+}$[620]$\uparrow$). Indeed in dubnium isotopes \cite{Park04}
and N\,=\,153 isotones \cite{ParS05} each one more single particle level at E$^{*}$ $<$ 400 keV is predicted
that can couple their spins to I$^{\pi}$ = 0$^{-}$ according to the coupling rules suggested by M.H.Brennan and A.M. Bernstein 
\cite{BrB60}, namely $\pi$7/2$^{-}$[514]$\downarrow$ $\otimes$ $\nu$7/2$^{+}$[633]$\uparrow$. Thus one can speculate 
that the CE stem fro an E0 transition 0$^{-}$ $\rightarrow$ 0$^{-}$.\\

\vspace{10mm}

\textbf{$^{264}$Bh}\\
The isotope $^{264}$Bh was observed by S. Hofmann et al. \cite{Hofm95,HoH02} as well as by K. Morita et al. \cite{Morita04a}
within the decay chain of $^{272}$Rg and was sythesized by Z.G. Gan et al. in bombardments of $^{243}$Am targets with
$^{26}$Mg projectiles \cite{Gan04}. The two-dimensional plot of decay times versus the $\alpha$ energies (from all experiments) is shown in fig. 42.
The time distribution is shown in the insert,
As seen, there are two events ((E$_{\alpha}$ = 9.30, $\Delta$t = 0.0036 ms), (E$_{\alpha}$ = 9.85, $\Delta$t = 0.0218 ms)) with extremely short decay times. Neglecting
these two events one obtains a half-life of T$_{1/2}$ = 1.06$^{+0.29}_{-0.17}$ s. As the probability to observe decays
with decay times $\Delta$t$<$0.025 s at this half-life is 0.016 while these two short-lived events represent a fraction of 0.08 of the
total number of the observed events it seems justified at first glance to assign them to a different activity, i.e. to the decay
of an isomeric state.  
However, one has to consider two items: first, the events were observed within the $\alpha$ decay chain of $^{272}$Rg.
Neither for the precurser $^{268}$Mt, nor for the $\alpha$ decay daughter $^{260}$Db, decay data ($\alpha$ energy, decay time)
were registered which were significantly different from those for the 'other' events. Second, the calculated \cite{PoI80} $\alpha$ 
half-life for the E$_{\alpha}$ = 9.85 MeV event is T$_{\alpha}$\,=\,8.35 ms, which is in-line with the time difference $\Delta$t = 21.8 ms,
the calculated $\alpha$  half-life for the E$_{\alpha}$ = 9.31 event is T$_{\alpha}$\,=\,274 ms, thus a factor of 75
longer than the measured  time difference $\Delta$t = 3.6 ms.\\
So, presently one can conclude that there is strong evidence for an isomeric state in $^{264}$Bh, but its existence has to be confirmed.\\

\begin{figure*}
	\resizebox{0.75\textwidth}{!}{
		\includegraphics{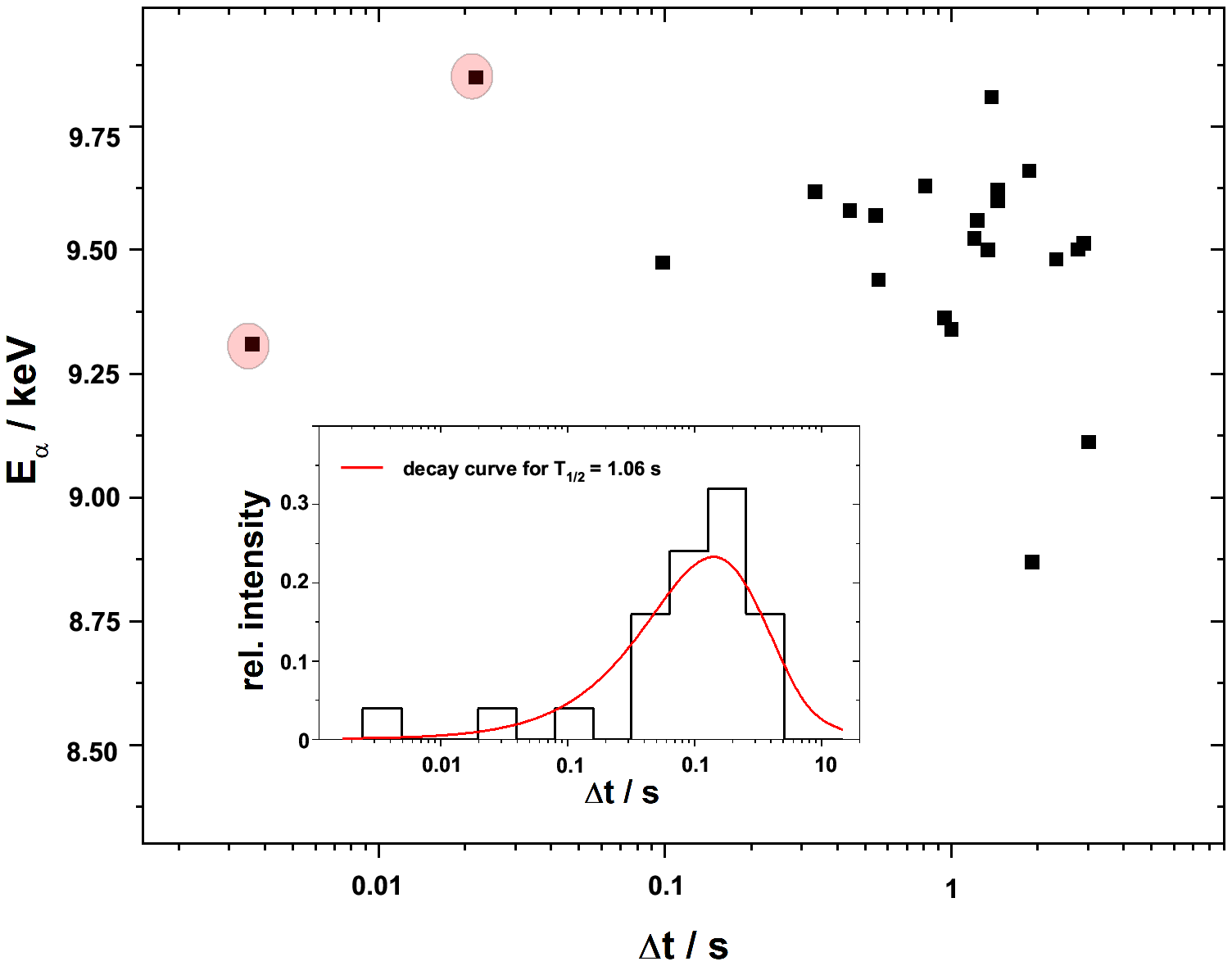}	
	}
	\caption{Two dimensional plot decay time ($\Delta$t) versus $\alpha$ for the decays of $^{264}$Bh
		(data from \cite{Hofm95,HoH02,Morita04a,Gan04}. The two events possibly stemminig from an isomeric state
		are marked by red circles; insert: time distribution of the events attributed to $^{264}$Bh; the red line
		represents the decay curve for a half-life T$_{1/2}$ = 1.06 s.
	}
	
	\label{fig:42}       
\end{figure*}

\begin{figure}
	\resizebox{0.85\textwidth}{!}{%
		\includegraphics{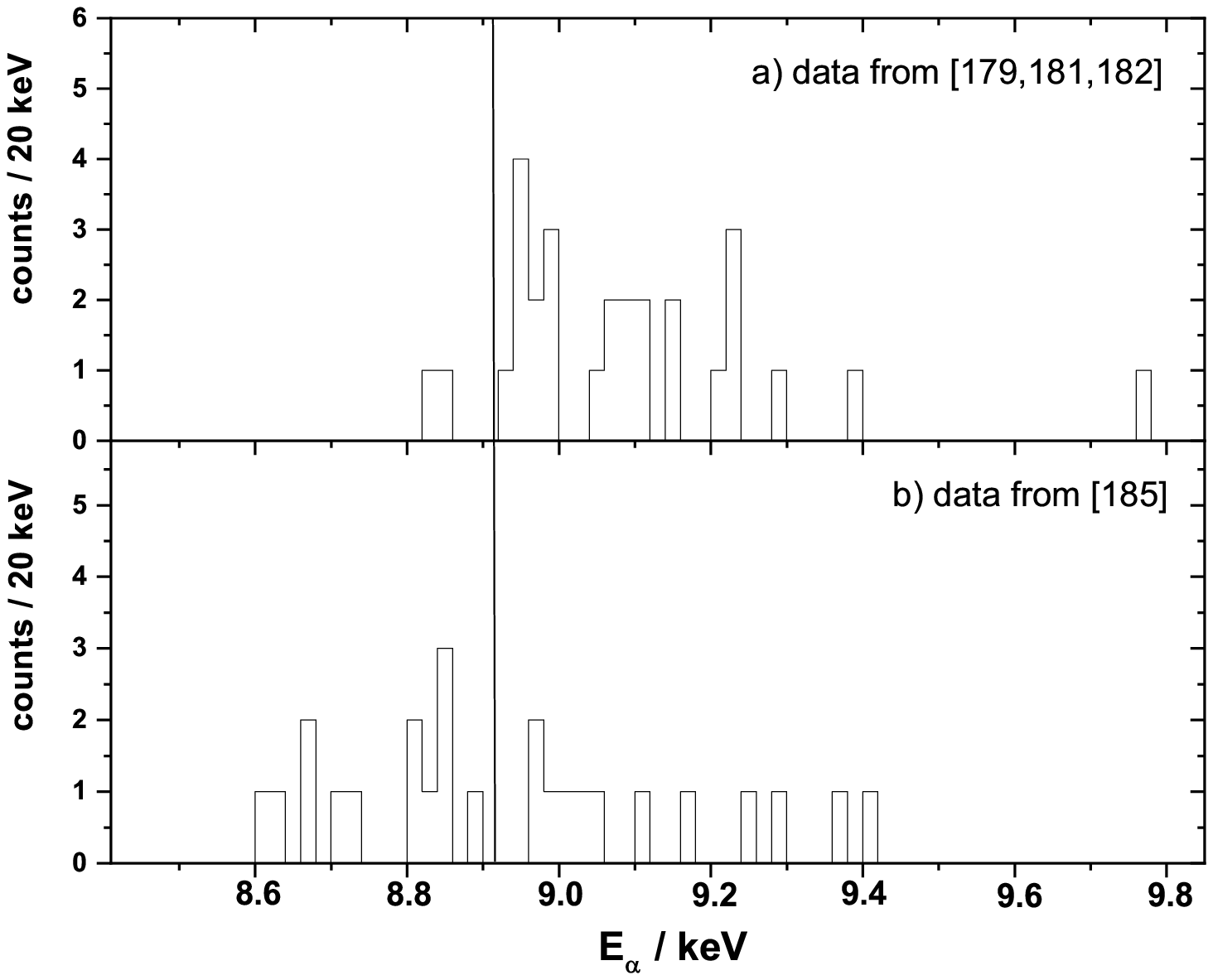}
	}
	\caption{$\alpha$ decay spectra of events attributed to $^{266}$Bh; a) data from \cite{Morita15,Wilk00,Morita09,Qin06},
	 b) data from \cite{Haba20}.} 
	\label{fig:43}       
\end{figure}

\begin{figure}
	\resizebox{0.85\textwidth}{!}{%
		\includegraphics{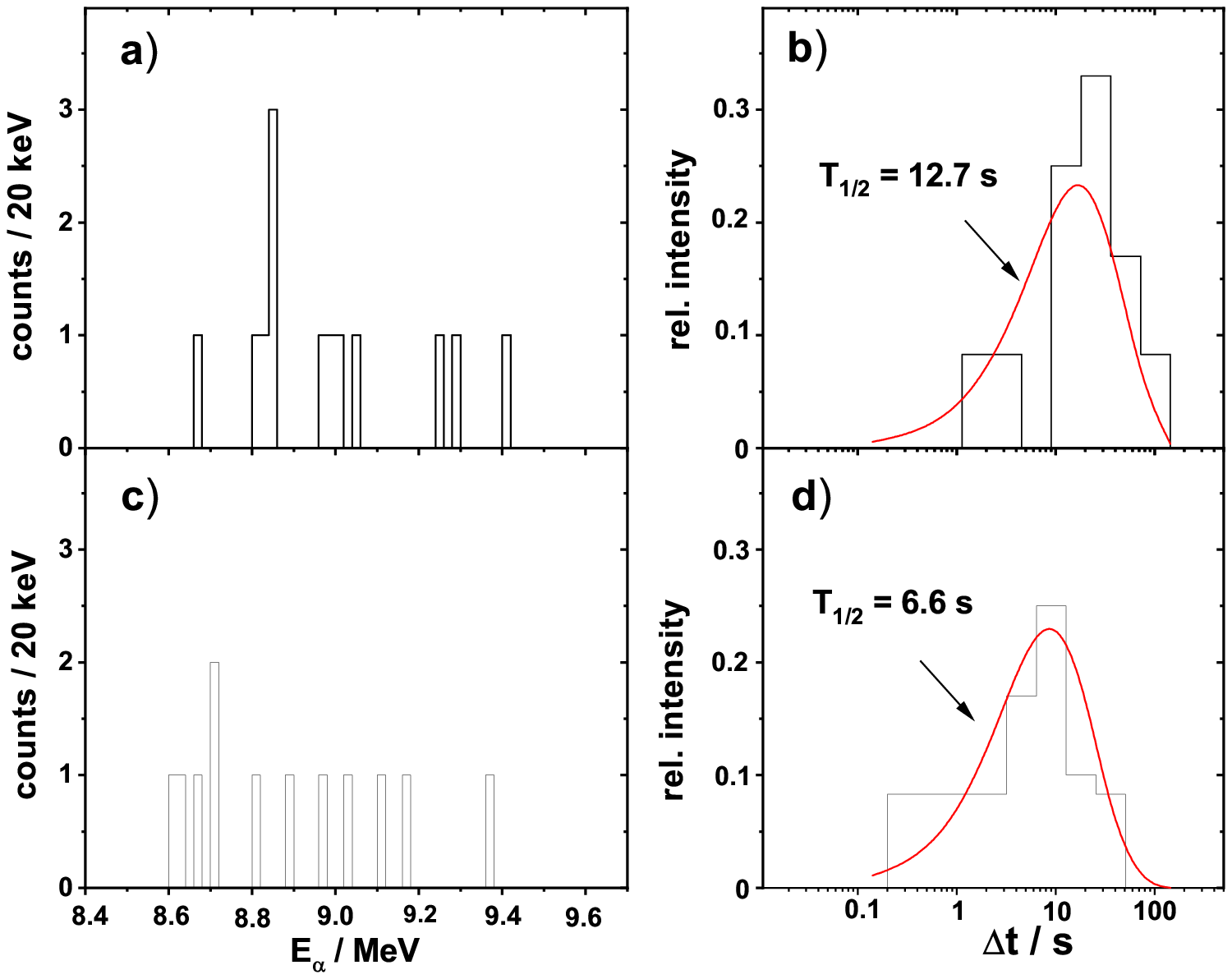}
	}
	\caption{$\alpha$ decay spectra and decay times of events attributed to $^{266}$Bh; a), b) energy and
		time distributions of events followed by spontaneous fission; c), d) energy and
		time distributions of events followed by $\alpha$ decays.} 
	\label{fig:44}       
\end{figure}

\begin{figure}
	\resizebox{0.80\textwidth}{!}{%
		\includegraphics{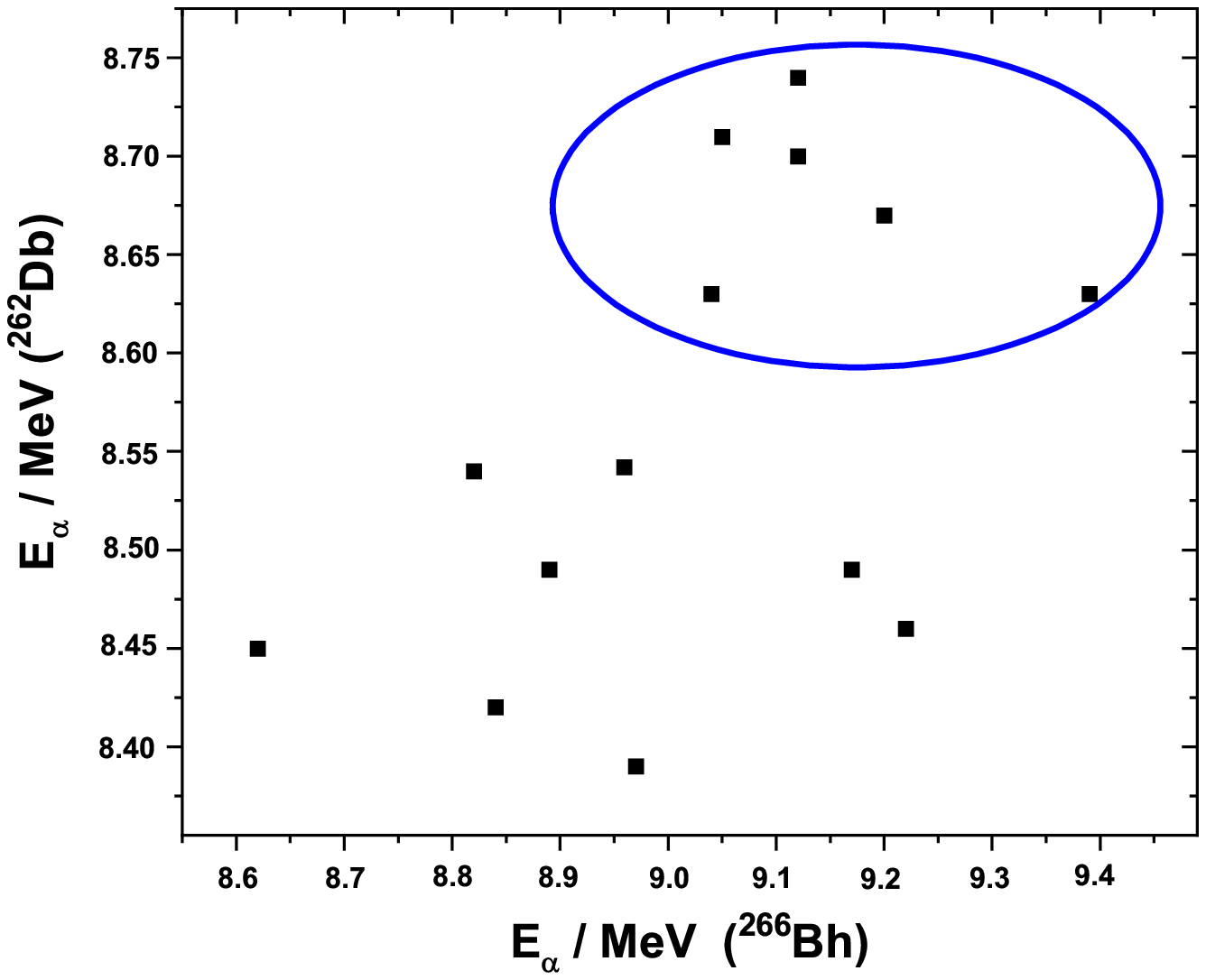}
	}
	\caption{$\alpha$ - $\alpha$ - correlations $^{266}$Bh - $^{262}$Db, requiring $\alpha$ decay or SF/EC of $^{258}$Lr.}
\label{fig:45}       
\end{figure}

\vspace{10mm}

\textbf{$^{266}$Bh}\\
Discovery of the isotope $^{266}$Bh was first reported by P.A. Wilk et al. \cite{Wilk00}, who produced it in the reaction $^{249}$Bk($^{22}$Ne,5n)$^{266}$Bh. They observed only
one decay chain. Further decay data were obtained by Z. Qin et al. \cite{Qin06}, who used the reaction $^{243}$Am($^{26}$Mg,3n)$^{266}$Bh and at RIKEN, where it was 
produced by K. Morita and coworkers either by the reaction $^{248}$Cm($^{23}$Na,5n)$^{266}$Bh \cite{Morita09} or within $\alpha$ decay chains interpreted to start from $^{278}$Nh.
The $\alpha$ spectrum summarizing the data from \cite{Wilk00,Qin06,Morita09,Morita15} is shown in fig. 43a. The obtained half-life is T$_{1/2}$ = 1.20$^{+0.65}_{-0.31}$ s.\\
A recent study of H. Haba et al. \cite{Haba20}, who produced the isotope in the reaction $^{248}$Cm($^{23}$Na,5n)$^{266}$Bh,
shows a completely different picture. They obeserved a low energy component at E$_{\alpha}$ $<$ 8.9 MeV, which was not present in the previous 
studies (see fig. 43a). Also they obtained a complete different half-life T$_{1/2}$ = 10.0$^{+2.6}_{-1.7}$ s. They mentioned also different half-lives of $\alpha$ decays
E$_{\alpha}$ $\ge$ 9.10 MeV (T$_{1/2}$ = 6.1$^{+3.6}_{-1.6}$ s) and those of E$_{\alpha}$ $<$ 9.10 MeV (T$_{1/2}$ = 11.8$^{+3.9}_{-2.3}$ s). They mentioned
that this could be an indication for an isomeric 
state, but also remarked, that both values were in agreement within the error bars.\\
In the present analysis somewhat different energy regions (E$>$9.0 MeV, E$<$9.0 MeV) were considered. They result in  
T$_{1/2}$ = 10.9$^{+4.0}_{-2.3}$ s for events E$_{\alpha}$ $<$ 9.0 MeV, 
T$_{1/2}$ = 4.4$^{+2.7}_{-1.2}$ s for events E$_{\alpha}$ $>$ 9.0 MeV (disregarding one event with an extreme long
correlation time $\Delta$t = 57.28 s).\\ 
Altogether Haba et al. attributed 25 $\alpha$ decays to $^{266}$Bh, 13 of them were followed by spontaneous fission, 12 by alpha decays. In seven cases triple correlations 
$^{266}$Bh $^{\alpha}_{\rightarrow}$ $^{262}$Db $^{\alpha}_{\rightarrow}$ $^{258}$Lr $^{\alpha}_{\rightarrow}$ 
were observed (see table 8).
Evidently there are differences as well for the energy as for the half-lives for $\alpha$ decays of $^{266}$Bh followed either by spontaneous fission or by 
$\alpha$ decay as seen in fig. 44. One obtains T$_{1/2}$ = 12.7$^{+5.2}_{-2.9}$ s for events followed by spontaneous fission, and T$_{1/2}$ = 6.6$^{+2.7}_{-1.5}$ s (fig. 44d) for events followed by $\alpha$ decay (fig.44b).\\
Additional information is obtained from $\alpha$ - $\alpha$ correlations (see table 8). As the $\alpha$ - energy distributions of $^{262}$Db and $^{258}$Lr
overlap, one has to restrict to triple correlations $^{266}$Bh $^{\alpha}_{\rightarrow}$ $^{262}$Db $^{\alpha}_{\rightarrow}$ $^{258}$Lr $^{\alpha}_{\rightarrow}$.
As seen in fig. 45, except of two cases, all $^{266}$Bh decays of E$_,{\alpha}$ $>$ 9.0 MeV 
are correlated to $^{262}$Db of E$>$8.60 MeV; the half-life extracted (from three events) is
T$_{1/2}$ = 1.5$^{+2.1}_{-0.6}$ s, similar to the value obtained from \cite{Morita15,Wilk00,Qin06}; the two events followed by $^{262}$Db of E$<$8.6 MeV
have considerably longer correlation times (8.81\,s, 21.17\,s), resulting in a half-life T$_{1/1}$ $\approx$ 11 s, while for all $^{266}$Bh followed by 
$^{262}$Db with E$_{\alpha}$ $<$ 8.6 MeV one obtains T$_{1/2}$ = 5.6$^{+3.8}_{-1.6}$ s.\\
In this context one also has to remark, that for both groups of $\alpha$ decays of $^{262}$Db different half-lives are obtained.\\
Under these considerations it seems quite evident, that two longlived states exist in $^{266}$Bh, that decay by
$\alpha$ emission into different long-lived states in $^{262}$Db, which may be characterized by the following
items.\\
$^{266}$Bh(1): E$_{\alpha}$ = (9.0-9.4 MeV), T$_{1/2}$ $\approx$ 1.5 s, decaying into $^{262}$Db(1),  E$_{\alpha}$ = (8.6-8.8 MeV),
T$_{1/2}$ = 21.8$^{+5.5}_{-3.6}$s having a small EC branch.\\
$^{266}$Bh(2): E$_{\alpha}$ $<$ 9.0 MeV, with a smaller fraction at E$>$9.0 MeV), T$_{1/2}$ $\approx$ 11 s, decaying into $^{262}$Db(2),  E$_{\alpha}$ = (8.35-8.55 MeV),
T$_{1/2}$ = 35.5$^{+5.2}_{-3.6}$s having a large EC branch.\\
To conclude: although it seems straight forward to postulate the existence of two longlived states in $^{266}$Bh decaying by
$\alpha$ emission, on the basis of present data detailed decay properties cannot be settled. This has to be left for further studies.\\

\begin{table}
	\caption{Decays of $^{266}$Bh reported by Z. Qin et al. \cite{Qin06}, K. Morita et al. from decay of $^{278}$Nh  \cite{Morita15}, 
		K. Morita et al. from 'direct' production \cite{Morita09}, H. Haba et al. \cite{Haba20} For \cite{Morita09,Haba20} only
		triple correlations $^{266}$Bh $^{\alpha}_{\rightarrow}$ $^{262}$Bh $^{\alpha}_{\rightarrow}$ $^{258}$Lr $^{\alpha}_{\rightarrow}$ 
		are given.}
	\label{tab:3}       
	\begin{tabular}{lllllll}
		\hline\noalign{\smallskip}
		\noalign{\smallskip}
		Reference & E$_\alpha$($^{266}$Bh) / MeV  & $\Delta$t/s & E$_\alpha$($^{262}$Db) / MeV  & $\Delta$t/s & E$_\alpha$($^{258}$Lr) / MeV  & $\Delta$t/s  \\
		\hline\noalign{\smallskip}
		\cite{Qin06}  &  8.981  & 1.13 & 8.459 & 33.62 &     &    \\
		\cite{Qin06}  &  9.071  & 0.79 & 8.604 & 34.14 &     &    \\
		\cite{Qin06}  &  8.981  & 1.13 & 8.459 &33.62 &     &    \\
		\cite{Qin06}  &  8.959  & 0.51 & 8.542 & 29.23 &  8.641   &  5.07    \\
		\cite{Qin06}  &  9.106  & 1.52 & 8.515 & 59.09 &     &    \\
		\hline\noalign{\smallskip}
		\cite{Morita15}  &  9.08  & 2.47 & (SF) & 40.9 &     &    \\
		\cite{Morita15}  &  9.77  & 1.31 & (SF) & 0.787 &     &    \\
		\cite{Morita15}  &  9.39  & 5.26 & 8.63 & 126. &  8.66   &  3.78   \\
		\hline\noalign{\smallskip}
		\cite{Morita09}  &  9.05  & - & 8.71$^{a}$ & 54.9 &  8.81   &  9.23   \\
		\cite{Morita09}  &  9.12$^{a}$  & - & 8.74$^{a}$ & 13.76 &  8.60   &  9.36   \\
		\cite{Morita09}  &  9.20  & - & 8.67 & 13.71 &  8.70$^{a}$   &  4.72   \\
		\cite{Morita09}  &  8.82  & - & 8.54 & 95.45 &  8.69   &  3.94   \\
		\cite{Morita09}  &  8.84  & - & 8.42 & 11.95 &  (SF)  &  27.22   \\
		\hline\noalign{\smallskip}
		\cite{Haba20} (9) &  8.89  & 5.17 & 8.49 & 58.93 &  8.64   &  8.83   \\
		\cite{Haba20} (13) &  9.12  & 0.92 & 8.70 & 10.29 &  8.59   &  7.05   \\
		\cite{Haba20} (20) &  8.62  & 2.24 & 8.45 & 89.48 &  8.63   &  1.24   \\
		\cite{Haba20} (22) &  8.97  & 10.23 & 8.39 & 2.72 &  8.62   &  5.89   \\
		\cite{Haba20} (27) &  9.17  & 8.81 & 8.49 & 19.75 &  8.61   &  5.58   \\
		\cite{Haba20} (28) &  9.04  & 0.33 & 8.63 & 9.07 &  8.59   &  4.64   \\
		\cite{Haba20} (32) &  9.22  & 21.17 & 8.46 & 19.27 &  8.34   &  4.00   \\
		\hline\noalign{\smallskip}		
		
	\end{tabular}
\end{table}

\subsection{{\bf 7.5 Isomeric states in odd-odd meitnerium isotopes}}
\vspace{5mm}
\textbf{$^{266}$Mt}\\
The isotope $^{266}$Mt was first observed in an irradiation of $^{209}$Bi with $^{58}$Fe performed at SHIP, GSI \cite{MuA82,MuW84}.
Only one decay chain has been observed in this experiment. More decay date were collected in follow-up experiments at SHIP \cite{MuH88,HoH97}. Since in the daughter nucleus $^{262}$Bh two states decaying by $\alpha$ emission had been identified \cite{MuA89}
a more detailed analysis of the $\alpha$($^{266}$Mt) - $\alpha$($^{262}$Bh) correlations was performed \cite{Hessb98}.  
The $\alpha$ spectra obtained for $^{266}$Mt and $^{262}$Bh are shown in fig. 46.
Evidently there is a small group of four decay events of $^{266}$Mt (E$_{\alpha}$$<$10600 keV) correlated in three cases to $^{262}$Bh decays 
of E$_{\alpha}$\,=\,(10370-10450) keV and in one case to E$_{\alpha}$\,=\,9902 keV. The half-life of these four events is 
T$_{1/2}$\,=\,1.24$^{+1.24}_{-0.41}$ ms, that of the correlated $^{262}$Bh events is T$_{1/2}$\,=\,10.5$^{+10.5}_{-3.5}$ ms,
while the half-life of the $^{266}$Mt decays E$_{\alpha}$$>$10600 keV is T$_{1/2}$\,=\,1.73$^{+0.87}_{-0.43}$ ms,
and the half-life of the correlated $^{262}$Bh events is T$_{1/2}$\,=\,49$^{+25}_{-12}$ ms.\\
These differences hint to the existence of an isomeric state in $^{266}$Mt. \\
\begin{figure*}
	\resizebox{0.90\textwidth}{!}{
		\includegraphics{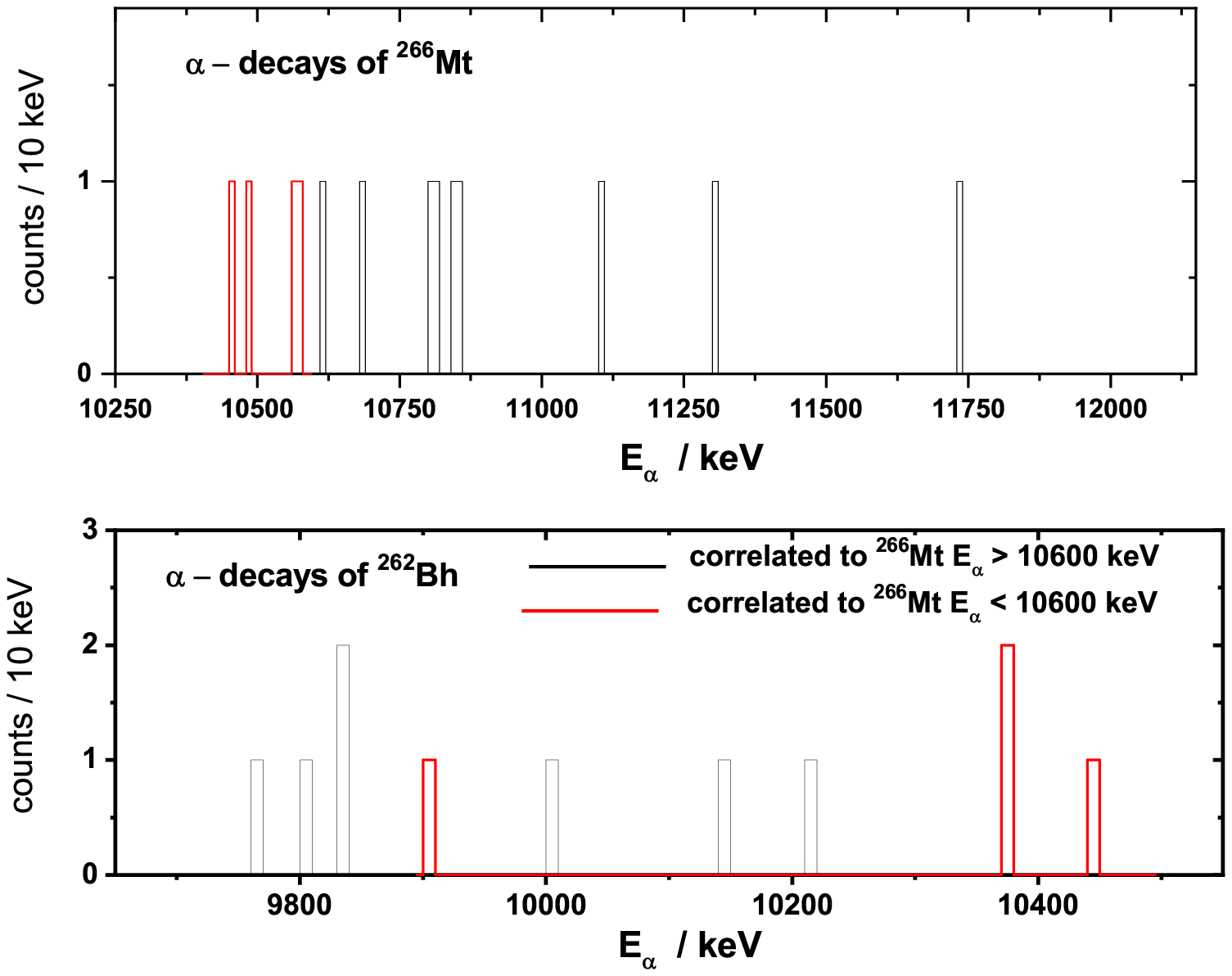}	
	}
	\caption{Spectra of $\alpha$ decays of $^{266}$Mt (upper figure) and $^{262}$Bh (lower figure) from the decay of $^{266}$Mt
	}
	
	\label{fig:46}       
\end{figure*}
\begin{figure*}
\resizebox{0.90\textwidth}{!}{
\includegraphics{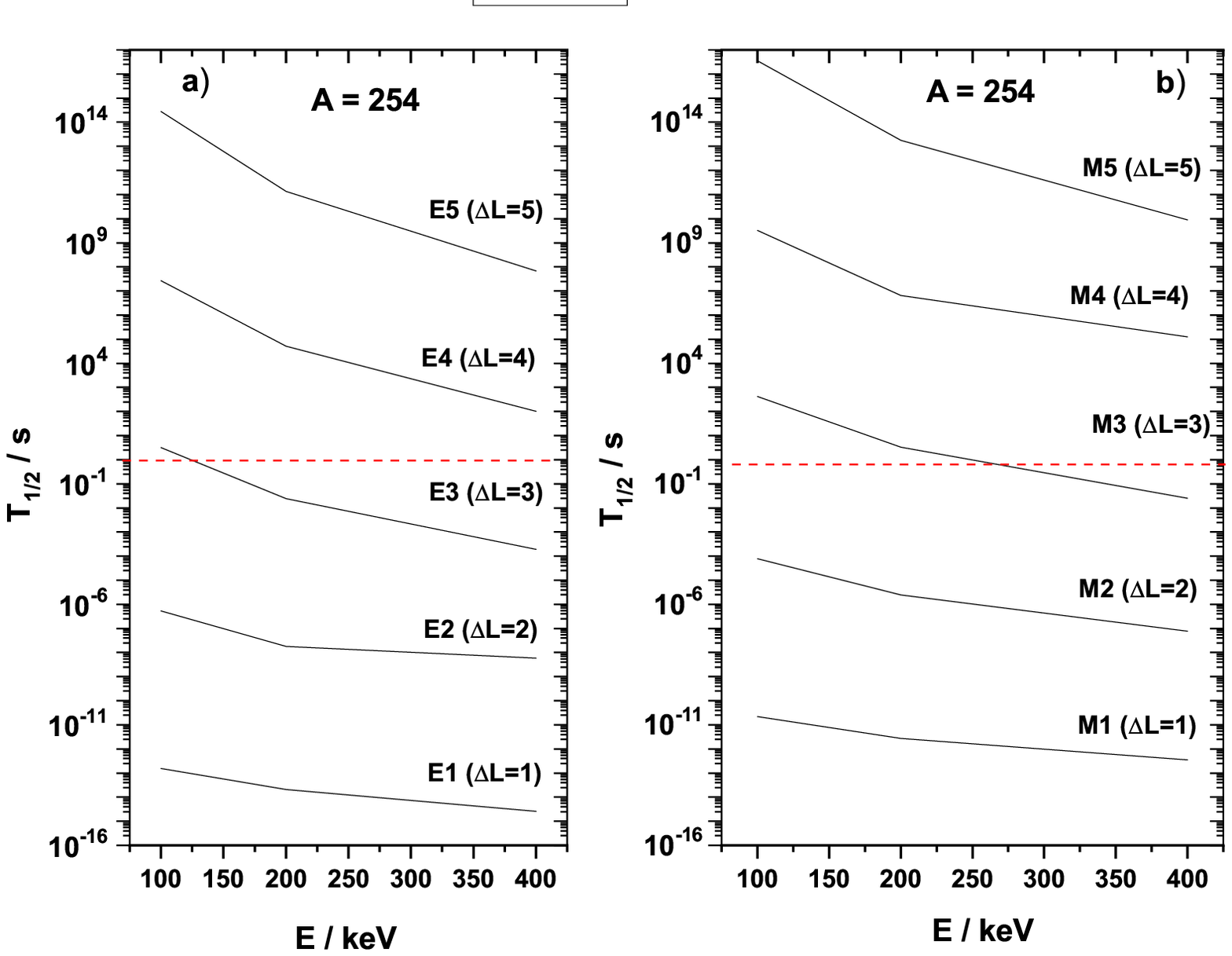}	
}
\caption{Weisskopf - lifetimes for different multipolarities of transitions.
}

\label{fig:47}       
\end{figure*}

\begin{figure*}
	\resizebox{0.90\textwidth}{!}{
		\includegraphics{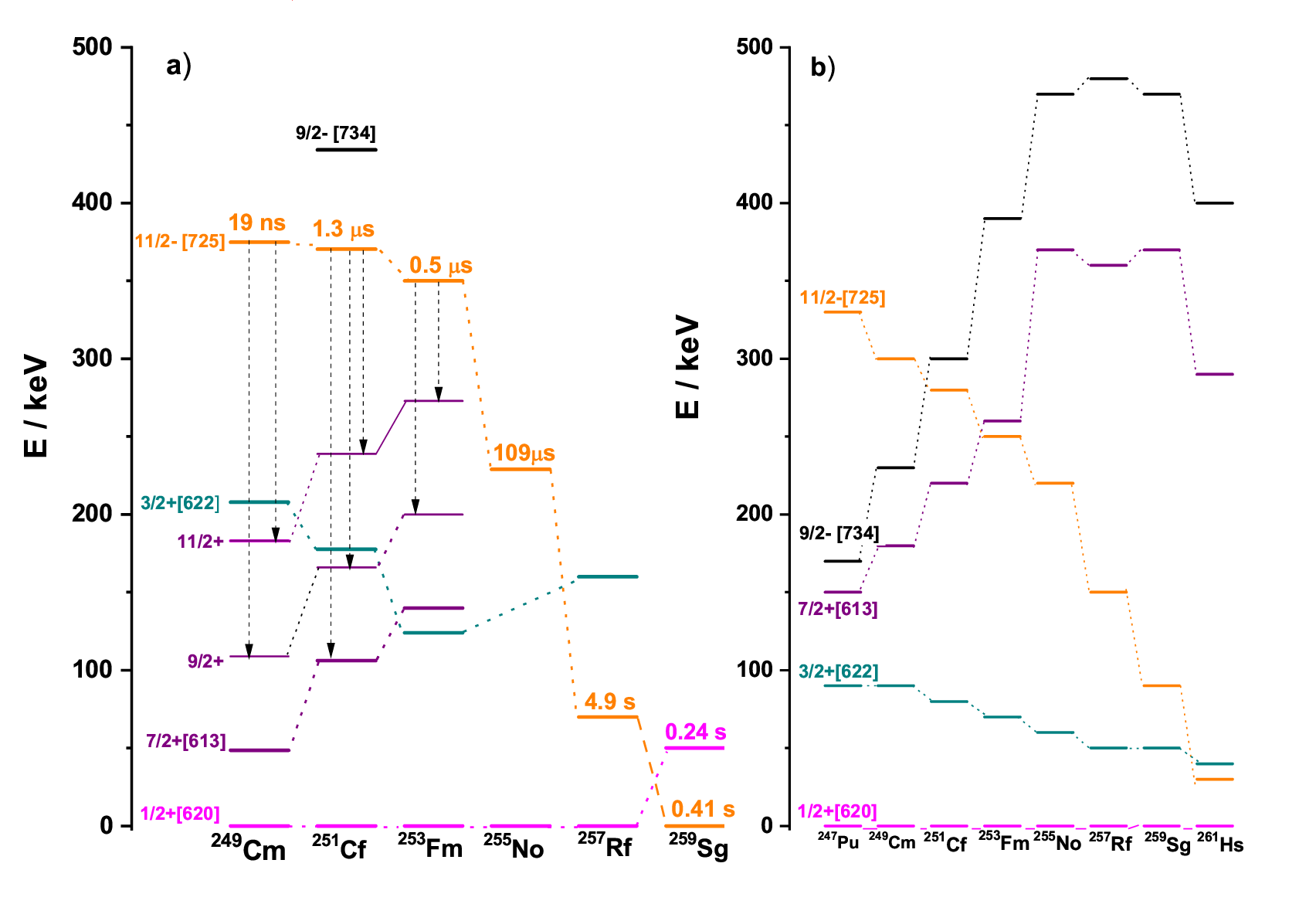}	
	}
	\caption{Systematics of 11/2$^{-}$[725] Nilsson levels in N\,=\,153 isotones and known $\gamma$ transitions.
	}
	
	\label{fig:48}       
\end{figure*}

\section{8. Systematics}

The existence of isomeric states is a wide spread phenomenon within the 'chart of nuclei' at Z$>$25, while for lighter
nuclei isomeric states are rather seldom. Except for seniority isomers and shape isomers their occurrence is strongly 
related to the angular momentum difference or the K difference between the isomeric state and lower lying states. Depending on the 
lifetime differences and the energy differences isomeric states decay by internal transitions, $\alpha$-, proton emission,
$\beta^{+}$, $\beta^{-}$ decay,  electron capture (EC) or
as in heaviest nuclei by spontaneous fission (SF).
Half-lives for nuclear states in dependence of energy differences (100\,-\,400 keV) and angular momentum differences to daughter states
using Weisskopf estimates \cite{Fire96} are shown in fig. 47.  As seen half-lives for transitions $\Delta$L $\le$ 2 are T$_{1/2}$ $<$ 10$^{-4}$ s.
Those isomers usually decay by internal transitions. At higher angular momentum differences other decay modes, 
$\beta^{+}$, $\beta^{-}$ decay,  electron capture (EC), $\alpha$ emission, and in few cases proton emission or
spontaneous fission (SF) have been observed. In general, besides internal transitions 
decay modes of isomers are in congruence with the ground state decay modes of
the nuclei. Thus besides internal transitions  $\beta^{+}$, $\beta^{-}$ decay,  electron capture (EC) dominate in the range
Z\,$<$\,100, although there are 'islands' of isomers decaying by $\alpha$ emission (e.g. nuclei 'slightly above' the N\,=\,82 
neutron shell at Z\,$\approx$(60-75), very neutron deficient isotops in the range Z\,$\approx$(75-90), nuclei around the
neutron shell N\,=\,126 at Z\,=\,(83-92).\\
In the transfermium region (Z\,$>$100) the existence of low lying long-lived isomers (T$_{1/2}$$>$1$\mu$s)
is a wide spread phenomenon caused 
by close vicinity of Nilsson levels of 'high' spins and 'low' spins. Here $\alpha$ decay seems to dominate on the basis of our 
present knowledge. One should, however, put emphasis on the latter half-sentence. $\alpha$ decay is 'quite easy' to detect
(as also spontaneous fission) and thus specifically in the case of low production rates in a couple of cases decay by internal transitions 
or  $\beta^{+}$, $\beta^{-}$ decay,  electron capture (EC) might not have been detected so far.
But also in case of $\alpha$ emission the decay properties (energies, half-lives) might be quite similar (e.g. $^{251}$No, $^{257}$Rf),
so from a 'direct' production ground state decay and isomeric decay hardly can be distinguised. So also study of the decay of the
($\alpha$ decay ) precurser is required for a clear identification.\\
Isomeric states can often be predicted on the basis of predicted level ordering in odd mass nuclei and are very helpful in (at least partly)
verifying or falsifying predicted level ordering. An example are the 5/2$^{+}$[622] isomers in the odd-mass N\,=\,151 isotones.
Theory predicts a level ordering  the 9/2$^{-}$[734] (ground state),  7/2$^{+}$[624],  5/2$^{+}$[622]. In that scenario the
5/2$^{+}$[622] level could decay into the ground state via  5/2$^{+}$[622]  $^{M1}_{\rightarrow}$  7/2$^{+}$[624]  
$^{E1}_{\rightarrow}$  9/2$^{-}$[734], i.e. no isomer would exist. From experimental side, however, the  5/2$^{+}$[622]
level is placed below the  7/2$^{+}$[624], so the  5/2$^{+}$[622] level will decay into the groud-state via an M2 - transition 
(with some E3 admixture) into the  9/2$^{-}$[734] ground state leading to isomers with half-lives of some tens of seconds.\\
Principally one can
distinguish between 'normal' spin isomers and K isomers, where decay into rotational band members with low spin differences
is hindered by a large K difference to the bandhead. Such cases occur, if the K number of the emitting (mother) level
is larger than that of the populated (daughter) level. Examples are found for the 11/2$^{-}$[725] isomeric states 
in the N\,=\,153 nuclei. The situation is shown in fig. 48. In $^{249}$Cm, $^{251}$Cf, $^{253}$Fm, and probably also
in $^{255}$No \cite{Kessaci24}, although the decay mode is not well established in that case, the 11/2$^{-}$[725] isomeric
state decays predominantly by K-hindered E1 - transitions into the 9/$^{+}$ and 11/2$^{+}$ members of the rotational band
built up on the 7/2$^{+}$[622] Nilsson level ($\Delta$K\,=\,2). Due to the steep decrease of the 11/2$^{-}$[725] level
towards increasing atomic number in $^{257}$Rf it becomes the first Nilsson above the 1/2$^{+}$[620] ground state, leading 
to an increase to $\Delta$K\,=\,5 between mother and daughter level, resulting in an increase of the half-life in the order of 50000 
and a change of the decay mode from internal transitions to $\alpha$ and EC decay \cite{Hess16a}. In $^{259}$Sg 
the 11/2$^{-}$[725] level finally becomes the ground state, while the 1/2$^{+}$[620] Nilsson level becomes the isomeric state,
thus the isomer changes its character from a K isomer to a 'normal' spin isomer. \\

While in odd-mass nuclei the existence of isomeric states can be predicted or justified on the basis of predicted single-particle levels
(Nilsson levels) the situation is not so easy in odd-odd nuclei. Some ideas about existence of isomeric states can be obtained from
low lying levels in odd-mass isotopic nuclei ('proton levels') and low lying levels in odd-mass isotonic nuclei ('neutron levels') and applying
the coupling rules suggested by M.H. Brennan and A.M. Bernstein \cite{BrB60}. To obtain more reliable information on existence and excitation 
energies of isomeric states, however, rigorous calculations are required. Those are existing so far only in very few cases. But those 
calculations are highly desired, as 
also in odd-odd nuclei the existence of long-lived isomeric states in the region of transfermium isotopes is a wide spread phenomenon. 
Presently such states are safely identified or likely to exist in $^{246,250,254,258}$Md, $^{254}$Lr, $^{258,262}$Db, 
$^{262,264,268}$Bh, $^{266}$Mt. But certainly in much more nuclei low lying isomeric states exist and still wait to be
identified. Identification of isomers is often complicated in case of $\alpha$ decay due to broad $\alpha$ energy distributions,
overlap of $\alpha$ 
energy regions and similar half-lives. Helpful are $\alpha$ - $\gamma$ - coincidence measurements, which, however, often suffer from
low count rates.\\
Spin and parity assignments are only vague or not possible. While in odd-mass nuclei such assigments at least tentatively 
can be done on the basis of decay properties, systematics and/or calculated level schemes, calculations are hardly available for odd-odd nuclei.
Two examples, $^{250}$Md and $^{254}$Lr have been discussed in this review showing encouraging results.\\
As understandig the structure of isomeric states also in odd-odd nuclei is certainly of high importance for understanding
the structure of heaviest nuclei and improving predictions of the next spherical proton and neutron shells above
$^{208}$Pb (Z\,=\,82, N\,=\,126), it seems of scientific relevance, also to focus as well from experimental as
from theoretical side, on investigation of low lying isomeric states in odd-odd nuclei in the transfermium region.\\

\section{9. Conclusions}
The review of low lying isomeric states in the heaviest element region has to be undersood as 
a status report on the basis of available data, which means that some statements and conclusions
presented here may have to be modified, when more detailed studies are performed and data of higher
quality are available. That is seemingly unavoidable. In this sence this review should be regared as
what it is, as a snapshot of present knowledge, showing, what is known, what is not known, what is still
speculation. It thus should act rather as a guide for further investigations.\\

\section{10. Appendix - Tables}

In the following tables compilations of claims of isomers in heaviest are given. Some of the cases 
as discussed above are uncertain or even
dubious, but this does not rule out that these isomers nevertheless may exist. For this reason they are
included in the following tables and it should be left for further investigations to prove or to disprove
their existence. IT as decay mode is only given if it is identified or if there is strong evidence. 
In nuclei where there is no evidence found so far, it might be a strong decay branch nevertheless.
For 'space reasons' only the strongest $\alpha$ transitions are given.  \\
'this work' means that conclusions were drawn in this work on the basis of previously published data
as discussed in the text.
\begin{table}
	\caption{Isomeric states in odd-mass even- Z nuclei; $\alpha$ energies given in MeV}
	\label{tab:3}       
	\begin{tabular}{lllllll}
		\hline\noalign{\smallskip}
		\noalign{\smallskip}
		Isotope & Neutron  & E$^{*}$ / keV & T$_{1/2}$ & decay mode & configuration & reference \\ 
		&  number &  &  &   &   &  \\  
		\hline\noalign{\smallskip}  
		$^{235}$U   & 143 & 0.0768 & $\approx$25 min & IT & 1/2$^{+}$[631] & \cite{Fire96} \\
		\hline\noalign{\smallskip}  
		$^{237}$Pu   & 143 & 145.544 & 0.18$\pm$0.2 s & IT & 1/2$^{+}$[631] & \cite{Fire96} \\
		\hline\noalign{\smallskip}  
		$^{239}$U   & 147 & 133.799 & 0.78$\pm$0.04 $\mu$s & IT & 1/2$^{+}$[631] & \cite{Fire96} \\
		\hline\noalign{\smallskip}  
		$^{241}$Pu   & 147 & 161.6 & 0.88$\pm$0.05 $\mu$s & IT & 1/2$^{+}$[631] & \cite{Fire96} \\
		\hline\noalign{\smallskip}  
		$^{243}$Cm   & 147 & 87 & 1.08$\pm$0.05 $\mu$s & IT & 1/2$^{+}$[631] & \cite{Fire96} \\
		\hline\noalign{\smallskip}  
		$^{247}$Fm   & 147 & 45$\pm$7 & 5.1$\pm$0.2 s & $\alpha$(8.18) & 1/2$^{+}$[631] & \cite{Hess06} \\
		\hline\noalign{\smallskip}  
		$^{243}$Pu   & 149 & 383.64 & 0.33$\pm$0.03 $\mu$s & IT & 1/2$^{+}$[631] & \cite{Fire96} \\
		\hline\noalign{\smallskip}  
		$^{245}$Cm   & 149 & 355.95 & 0.29$\pm$0.02 $\mu$s & IT & 1/2$^{+}$[631] & \cite{Fire96} \\
		\hline\noalign{\smallskip}   
		$^{251}$No   & 149 & 105$\pm$3 & 1.02$\pm$0.03 s & $\alpha$(8.668) & 1/2$^{+}$[631] & \cite{Hess06,Kaleja20} \\
		\hline\noalign{\smallskip}  
		$^{245}$Pu   & 151 & 311 & 0.33$\pm$0.03 $\mu$s & IT & 1/2$^{+}$[620] & \cite{Asai11} \\
		\hline\noalign{\smallskip}  
		$^{247}$Cm  & 151 & 227.38 & 26.3$\pm$0.3 $\mu$s & IT & 5/2$^{+}$[622] & \cite{Ahmad03} \\
		\hline\noalign{\smallskip}  
		$^{247}$Cm  & 151 & 404.9 & 100.6$\pm$0.6 ns & IT & 1/2$^{+}$[620] & \cite{Ahmad03} \\
		\hline\noalign{\smallskip}  
		$^{249}$Cf  & 151 & 144.96 & 45$\pm$5 $\mu$s & IT & 5/2$^{+}$[622] & \cite{Fire96} \\
		\hline\noalign{\smallskip}  
		$^{249}$Cf  & 151 & 416.8 & $<$150 ns  & IT & 1/2$^{+}$[620] & this work\\
		\hline\noalign{\smallskip}  
		$^{251}$Fm  & 151 & 199.9$\pm$0.3 & 21$\pm$3 $\mu$s  & IT & 5/2$^{+}$[622] & \cite{Hess06a}\\
		\hline\noalign{\smallskip}  
		$^{251}$Fm  & 151 & 392 & 22 ns  & IT & 1/2$^{+}$[631] & \cite{Asai11} \\
		\hline\noalign{\smallskip}  
		$^{253}$No  & 151 & 167 & 24$\pm$3 $\mu$s & IT & 5/2$^{+}$[622] & \cite{Hess07,LoH07} \\
		\hline\noalign{\smallskip} 
		$^{255}$Rf  & 151 & $\approx$135 & 50$\pm$17 $\mu$s  & IT & 5/2$^{+}$[622] & \cite{AntH15} \\
		\hline\noalign{\smallskip}
		$^{249}$Cm  & 153 & 48.74 & $\approx$23 $\mu$s  & IT & 7/2$^{+}$[613] & \cite{Fire96} \\
		\hline\noalign{\smallskip}  
		$^{249}$Cm  & 153 & 375 & 19$\pm$2 ns  & IT & 11/2$^{-}$[725] & \cite{Ishii08} \\
		\hline\noalign{\smallskip}  
		$^{251}$Cf  & 153 & 106.3 & 38$\pm$1 ns  & IT & 7/2$^{+}$[613] & \cite{Ahmad71} \\
		\hline\noalign{\smallskip}  
		$^{251}$Cf  & 153 & 370.39 & 1.3$\pm$0.1 $\mu$s  & IT & 11/2$^{-}$[725] & \cite{Ahmad71} \\
		\hline\noalign{\smallskip}  
		$^{253}$Fm  & 153 & (341-361) & 0.56$\pm$0.06 $\mu$s  & IT & 11/2$^{-}$[725] & \cite{AntH11} \\
		\hline\noalign{\smallskip}  
		$^{255}$No  & 153 & (240-300) & 109$\pm$9 $\mu$s  & IT & 11/2$^{-}$[725] & \cite{Bronis22} \\
	     		                &         & $\approx$200 & 86$\pm$8 $\mu$s  &   &  & \cite{Kessaci22} \\
	    \hline\noalign{\smallskip}   		                
		$^{257}$Rf   & 153 & 74 & 4.37$\pm$0.05 s & $\alpha$(9.023($\approx$0.8)), IT($\approx$0.14 & 11/2$^{-}$[725] &  \cite{Hess97,HaL22} \\
		&  &  &  &  EC($\approx$0.046),SF($\approx$0.004) & &   \\
		\hline\noalign{\smallskip}   
		$^{259}$Sg   & 153 & $\approx$90 & 226$\pm$27 ms & $\alpha$(9.55, 0.97), SF & 1/2$^{+}$[620] & \cite{AntH15} \\
		\hline\noalign{\smallskip}   
		$^{261}$Sg   & 155 & $\approx$200 & 9.0$^{+2.0}_{-1.5}$ $\mu$s & IT & 11/2$^{-}$[725] & \cite{Berry10} \\

	\end{tabular}
\end{table}  

\begin{table}
	\caption{Isomeric states in odd-mass odd- Z nuclei}
	\label{tab:3}       
	\begin{tabular}{lllllll}
		\hline\noalign{\smallskip}
		\noalign{\smallskip}
		\hline\noalign{\smallskip}
		Isotope & N  & E$^{*}$ / keV & T$_{1/2}$ & decay mode & configuration & reference \\  
		
		\hline\noalign{\smallskip}   
		$^{245}$Md   &  144 & ? & 0.9$\pm$0.25 ms &  SF &  1/2$^{-}$[521] &  \cite{NinH96} \\
		\hline\noalign{\smallskip}  
		$^{247}$Md   & 146 & 92 & 0.23$\pm$0.03 s & $\alpha$(8.72(0.8$\pm$0.02)), SF(0.2$\pm$0.02 IT & 1/2$^{-}$[521]  & 
		\cite{Hess22,HuS22,Xu25} \\
		 \hline\noalign{\smallskip}  
		  $^{249}$Md   & 148 & 48-78 & 1.1$\pm$0.5 s & $\alpha$(8.35) & 1/2$^{-}$[521]  & \cite{BrioL25} \\
		\hline\noalign{\smallskip}   
		$^{251}$Lr   & 148 & 117$\pm$27 & 42$^{+42}_{-14}$ ms & $\alpha$(9.21) & (1/2$^{-}$[521])  & \cite{HuS22} \\
		\hline\noalign{\smallskip}   
		$^{253}$Lr   & 150 & 6-24 & 2.23$^{+0.22}_{-0.10}$ s & $\alpha$(8.72), SF(0.08$\pm$0.05) & (1/2$^{-}$[521])  & \cite{Hess01,HuS22,BrioL25}\\
		\hline\noalign{\smallskip}  
		$^{255}$Lr   & 152 & 32$\pm$2 & 2.53$\pm$0.13  s & $\alpha$(8.46), IT & 7/2$^{-}$[514] & \cite{Chat06,AntH08,Kaleja20} \\
		\hline\noalign{\smallskip}  
		$^{257}$Lr   & 154 & ? &  0.2$^{+0.16}_{-0.06}$ s   & $\alpha$(8.81), IT  & (1/2$^{-}$[521]) & \cite{Hess16a} \\
		\hline\noalign{\smallskip}
		$^{257}$Db   & 152 & 59-89 &  1.50$^{+0.19}_{-0.15}$ s   & $\alpha$ & (1/2$^{-}$[521]) & \cite{Hess01,BrioL25} \\
		\hline\noalign{\smallskip}   
		
	\end{tabular}
\end{table}

\begin{table}
	\caption{Isomeric states in odd-odd nuclei; $\alpha$ energies are given in MeV.}
	\label{tab:3}       
	\begin{tabular}{lllllll}
		\hline\noalign{\smallskip}
		\noalign{\smallskip}\hline\noalign{\smallskip}
		Isotope &  & E$^{*}$ / keV & T$_{1/2}$ & essential decay Mode  & spin/parity & reference \\   
		\hline\noalign{\smallskip}     
		$^{250}$Md   &  149 & 123 & 42.5$\pm$4.5 s & $\alpha$(7.7-7.8),EC & (7$^{-}$) &  \cite{Vosti19} \\
		\hline\noalign{\smallskip} 
		$^{254}$Md   & 153   & ? & 28 min & EC &   &  \cite{Fire96} \\
		\hline\noalign{\smallskip}     
		$^{258}$Md   &  157 & ? &  57$\pm$0.9 min & EC ($\ge$0.7) & (1$^{-}$) &  \cite{Fire96} \\
		\hline\noalign{\smallskip}     
		$^{254}$Lr   & 149 & 107$\pm$4 & 20.3$\pm$4.2 & $\alpha$(8.43-8.52), EC, IT? & (1$^{-}$) & \cite{Vosti19,Kaleja20} \\
		\hline\noalign{\smallskip}   
		$^{258}$Db   & 153 & 51 & 4.3$\pm$0.5 s & $\alpha$ (9.07-9.17), EC, IT? & (5$^{+}$, 10$^{-}$)  & \cite{Hessb09,Vosti19} \\
		\hline\noalign{\smallskip}   
		
	\end{tabular}
\end{table}  

\begin{table}
	\caption{Even-Z odd-N nuclei, where it is not established which state is ground state and which state the isomer;
		this work: data are obtained from reanalysis of the published data; $\alpha$ enegies are given in MeV}
	
	\label{tab:3}       
	\begin{tabular}{llllll}
		\hline\noalign{\smallskip}
		\noalign{\smallskip}\hline\noalign{\smallskip}
		Isotope & N &   T$_{1/2}$ & essential decay  & spin / parity & reference \\  
		\hline\noalign{\smallskip}   
		$^{253}$Rf(1)     & 149 &  52.8$\pm$4.4 $\mu$s &  SF&  7/2$^{+}$[624] &  \cite{LopH22a,Khuy21} \\
		$^{253}$Rf(2)     & 149 &  9.9$\pm$1.2 ms &  SF, $\alpha$(9.21,9.31)&  1/2$^{+}$[631] &  \cite{LopH22a,Khuy21} \\
		\hline\noalign{\smallskip}     
		$^{259}$Rf(1)   & 155 & 2.25$^{+0.75}_{-0.45}$  s &  $\alpha$(8.89 MeV) &  &  this work \\
		$^{259}$Rf(2)   & 155 &  $\approx$3 s &  $\alpha$(8.77) &  & this work \\  
		\hline\noalign{\smallskip}
		$^{261}$Rf(1)   & 157 & 55$\pm$9 s &  $\alpha$(8.27 MeV) & (9/2$^{-}$[615], 11/2$^{-}$[725]) &  this work, \cite{Asai05}\\
		$^{261}$Rf(2)   & 157 & 3.8$^{+0.72}_{-0.52}$ s &  $\alpha$(8.51 MeV), SF  &  (3/2$^{+}$[622], 1/2$^{+}$[620])  &  this work, \cite{Asai05} \\
		\hline\noalign{\smallskip}
		$^{263}$Sg(1)   &  157 & 0.33$^{+0.10}_{-0.06}$ s & $\alpha$(9.25) &   &   this work \\
	    $^{263}$Sg(2)   &  157 &  0.79$^{+0.37}_{-0.19}$ s & $\alpha$(9.05) &   &  this work \\
		\hline\noalign{\smallskip}
    	$^{265}$Sg(1)   &  159 &  8.9$^{+2.7}_{-1.9}$ s & $\alpha$(8.85) &   &  \cite{DuT08}\\
    	$^{265}$Sg(2)   &  159 & 16.2$^{+4.7}_{-3.5}$ s & $\alpha$(9.25) &   &  \cite{DuT08} \\
     	\hline\noalign{\smallskip}
     	$^{267}$Sg(1)   &  161 &  465$^{+465}_{-155}$ s & $\alpha$($\approx$8.27) &   &  \cite{DvB08,OgU24},this work\\
     	$^{267}$Sg(2)   &  161 & 84$^{+51}_{-23}$ s & SF &   &  \cite{DvB08,OgU24},this work \\
		\hline\noalign{\smallskip}
		$^{265}$Hs(1)   &  157 &  1.9$\pm$0.2 ms & $\alpha$(10.28-10.57) &   & \cite{Hessb09} \\
		$^{265}$Hs(2)   &  157 &  0.3$^{+0.2}_{-23}$ ms & $\alpha$(10.538) &   &  \cite{Hessb09} \\
		
		\hline\noalign{\smallskip}
		$^{271}$Hs(1)   &  163 &  7.1$^{+8.4}_{-2.5}$ ms & $\alpha$(9.05) &   &   \cite{OgU24} \\
		$^{271}$Hs(2)   &  163 &  46$^{+56}_{-16}$ ms & $\alpha$(9.34) &   &   \cite{OgU24}   \\
		\hline\noalign{\smallskip}
		$^{271}$Ds(1)   &  161 &  1.93$^{+0.49}_{-0.30}$ ms & $\alpha$(10.68-10.74 ) & (9/2$^{+}$[615]) & this work  \\
		$^{271}$Ds(2)$\star$   &  161 &  69$^{+52}_{-21}$ ms & $\alpha$(10.71) &  & this work \\
		\hline\noalign{\smallskip} 
			\end{tabular}
		$\star$ $^{271}$Ds(2) could also be a K isomer

\end{table}

\begin{table}
	\caption{odd-odd nuclei, where it is not established which state is ground state and which state the isomer;
		this work: data are obtained from reanalysis of the published data.}
	
	\label{tab:3}       
	\begin{tabular}{llllll}
		\hline\noalign{\smallskip}
		\noalign{\smallskip}\hline\noalign{\smallskip}
		Isotope & N &   T$_{1/2}$ & essential decay  & spin / parity & reference \\  
		\hline\noalign{\smallskip}     
		$^{246}$Md(1)   & 145 & 4.4$\pm$0.4 s &  $\alpha$(1.28 (0.23), EC(0.77) &  &  \cite{AntH10} \\
		$^{246}$Md(2)   & 145 &  0.9$\pm$0.2 s &  $\alpha$(8.38-8.64,8.74), EC(?) &  &  \cite{AntH10} \\
		\hline\noalign{\smallskip}   
		$^{262}$Db(1)   & 157 & 35.5$^{+5.2}_{-3.6}$ s &  $\alpha$(8.35-8.55) &  &  this work \\
		$^{262}$Db(2)   & 157 & 21.8$^{+5.5}_{-3.6}$  &  $\alpha$(8.6-8.8)  &  &  this work \\
		\hline\noalign{\smallskip}
		$^{262}$Bh(1)   &  155 &  93.5$^{+12.3}_{9.4}$ ms & $\alpha$(9.6-10.2) &   &  this work \\
		$^{262}$Bh(2)   &  155 &  13.7$^{+1.6}_{-1.3}$ ms & $\alpha$ (10.2-10.5) &  &   this work \\
		\hline\noalign{\smallskip}
		$^{266}$Bh(1)   &  159 &  $\approx$ 1.5 s & $\alpha$(9.0-9.4) &   &  this work \\
		$^{266}$Bh(2)   &  159 &  $\approx$ 11 s & $\alpha$($<$9.0) &   &   this work \\
		\hline\noalign{\smallskip}
		$^{266}$Mt(1)   &  157 & 1.24$^{+1.24}_{-0.41}$ ms & $\alpha$($<$10.6) &   &   this work\\
		$^{266}$Mt(2)   &  157 &  1.73$^{+0.87}_{-0.43}$ ms & $\alpha$($>$10.6) &   &  this work\\

		\hline\noalign{\smallskip}
		
	\end{tabular}
\end{table}

\section{Achnowledgement}
The author thanks Z. Zhang for providing him a list of the 'single decays' from the decay study of $^{262}$Bh by 
T. Zhao et al. \cite{Zhao24}.

\pagebreak
%

\end{document}